\documentclass[12pt]{book}
\pdfoutput=1
\usepackage{./auxi/jheppub}
\usepackage[utf8]{inputenc}

\usepackage[ngerman, english]{babel}
\usepackage[T1]{fontenc}
\usepackage[titletoc]{appendix}
\usepackage{titletoc}
\usepackage{graphicx}
\usepackage{subcaption} % to use subfigure environment (pics side by side)
\usepackage[dvipsnames]{xcolor}
\usepackage{amsmath,amssymb,amsthm}
\usepackage{bm}
\usepackage{float}
\usepackage[format=plain,labelfont={bf,it},textfont=it]{caption}
\usepackage{multirow}
\usepackage{slashed} % for Feynman slashed notation
\usepackage{mathtools} % for subequation numbering
\usepackage{tabularx}
\usepackage{braket}
\usepackage{bbold}
\usepackage{enumitem}
\usepackage{tikz}
\usetikzlibrary{positioning,arrows,calc}
\usetikzlibrary{arrows.meta}
\usetikzlibrary{decorations.pathmorphing}
\usetikzlibrary{decorations.markings}
\usetikzlibrary{decorations.pathreplacing,calligraphy}
\usetikzlibrary{decorations.pathreplacing}
\usetikzlibrary{decorations.pathmorphing, patterns,shapes}
\usetikzlibrary{shapes.geometric}
\usetikzlibrary{patterns}
\tikzset{
  snake it/.style={
    decorate, 
    decoration=snake,
    segment length=3
  }
}
\usepackage{xparse}
\usepackage{pdfpages} %for inserting pdfs in the file
\usepackage[makeroom]{cancel}
\usepackage{pgfplots}
\usetikzlibrary{datavisualization}
\usetikzlibrary{datavisualization.formats.functions}
\usetikzlibrary{decorations.pathreplacing,calligraphy}
\usetikzlibrary{decorations.pathreplacing}
\usetikzlibrary{decorations.pathmorphing, patterns,shapes}
\usepackage{titlesec} % for changing chapter style
\usepackage{fancyhdr} % for changing header style
\usepackage{emptypage} % removes header and page number in empty pages
\usepackage{chngcntr} % for changing numbering of equations to chapter
\usepackage{etoolbox} % for making chapter title to uppercase in ToC
\usepackage{amsmath}

\usepackage{dsfont} % for displaying the blackboard 1 properly
\usepackage{tensor}
\usepackage{algpseudocode}
\usepackage{diagbox}
\usepackage{pifont}% http://ctan.org/pkg/pifont
\newcommand{\cmark}{\ding{51}}%
\newcommand{\xmark}{\ding{55}}%
\usetikzlibrary{shapes.geometric, arrows.meta, positioning}
\definecolor{DarkBlueGrey}{RGB}{76,94,107}
\definecolor{MediumBlueGrey}{RGB}{110,135,153}
\definecolor{LightBlueGrey}{RGB}{134,163,184}
\definecolor{WCOrange}{RGB}{242,146,29}
\definecolor{SCRed}{RGB}{179,48,48}
\definecolor{VertexColor}{RGB}{242,146,29}
\definecolor{GluonColor}{RGB}{255,172,172}
\definecolor{SEColor}{RGB}{134,163,184}

\definecolor{BGBox}{RGB}{255,254,230}
\definecolor{PlaneColor}{RGB}{230,230,230}
\definecolor{BlobColor}{RGB}{190,180,230}

\DeclareMathOperator{\Disc}{Disc}

\newcommand{\Li}{{\normalfont\text{Li}}}
\newcommand{\vev}[1]{\langle\, #1 \, \rangle}

\def\veps{\varepsilon}

\newcommand{\Op}{\mathcal{O}}
\newcommand{\Oh}{\hat{\mathcal{O}}}
\newcommand{\Dh}{\hat{\Delta}}
\newcommand{\bh}{\hat{b}}

\def\Em{{\mathcal{E}}}

\def\Mm{{\mathcal{M}}}

\def\Om{{\mathcal{O}}}
\def\Pm{{\mathcal{P}}}

\def\Wm{{\mathcal{W}}}

\newcommand{\op}[1]{\boldsymbol{#1}}
\usepackage{bbm}

\def\veps{\varepsilon}

\newcommand\zb{{\bar{z}}}

\newcommand{\co}{\mathcal{O}}

\newcommand{\cO}[1]{\mathcal{O}_{#1}(x_{#1})}
\newcommand{\phu}{\varphi}

\definecolor{VertexColor}{RGB}{242,146,29}
\newif\ifstartcompletesineup
\newif\ifendcompletesineup
\pgfkeys{
    /pgf/decoration/.cd,
    start up/.is if=startcompletesineup,
    start up=true,
    start up/.default=true,
    start down/.style={/pgf/decoration/start up=false},
    end up/.is if=endcompletesineup,
    end up=true,
    end up/.default=true,
    end down/.style={/pgf/decoration/end up=false}
}
\pgfdeclaredecoration{complete sines}{initial}
{
    \state{initial}[
        width=+0pt,
        next state=upsine,
        persistent precomputation={
            \ifstartcompletesineup
                \pgfkeys{/pgf/decoration automaton/next state=upsine}
                \ifendcompletesineup
                    \pgfmathsetmacro\matchinglength{
                        0.5*\pgfdecoratedinputsegmentlength / (ceil(0.5* \pgfdecoratedinputsegmentlength / \pgfdecorationsegmentlength) )
                    }
                \else
                    \pgfmathsetmacro\matchinglength{
                        0.5 * \pgfdecoratedinputsegmentlength / (ceil(0.5 * \pgfdecoratedinputsegmentlength / \pgfdecorationsegmentlength ) - 0.499)
                    }
                \fi
            \else
                \pgfkeys{/pgf/decoration automaton/next state=downsine}
                \ifendcompletesineup
                    \pgfmathsetmacro\matchinglength{
                        0.5* \pgfdecoratedinputsegmentlength / (ceil(0.5 * \pgfdecoratedinputsegmentlength / \pgfdecorationsegmentlength ) - 0.4999)
                    }
                \else
                    \pgfmathsetmacro\matchinglength{
                        0.5 * \pgfdecoratedinputsegmentlength / (ceil(0.5 * \pgfdecoratedinputsegmentlength / \pgfdecorationsegmentlength ) )
                    }
                \fi
            \fi
            \setlength{\pgfdecorationsegmentlength}{\matchinglength pt}
        }] {}
    \state{downsine}[width=\pgfdecorationsegmentlength,next state=upsine]{
        \pgfpathsine{\pgfpoint{0.5\pgfdecorationsegmentlength}{0.5\pgfdecorationsegmentamplitude}}
        \pgfpathcosine{\pgfpoint{0.5\pgfdecorationsegmentlength}{-0.5\pgfdecorationsegmentamplitude}}
    }
    \state{upsine}[width=\pgfdecorationsegmentlength,next state=downsine]{
        \pgfpathsine{\pgfpoint{0.5\pgfdecorationsegmentlength}{-0.5\pgfdecorationsegmentamplitude}}
        \pgfpathcosine{\pgfpoint{0.5\pgfdecorationsegmentlength}{0.5\pgfdecorationsegmentamplitude}}
}
    \state{final}{}
}

\tikzset{
corner/.style={line width=1pt,dashed,draw=black,dash pattern=on 6pt off 4pt},
scalar/.style={line width=1pt,draw=black},
gluon/.style={line width=1pt,decorate, draw=GluonColor,
    decoration={complete sines,aspect=0,amplitude=1.25mm,segment length=1.5mm,start up,end up}},
gluontwo/.style={line width=1pt,decorate, draw=GluonColor,
    decoration={complete sines,aspect=0,amplitude=.7mm,segment length=1mm,start up,end up}},
ghost/.style={line width=1pt,loosely dotted,draw=black},
wilson/.style={line width=2pt,draw=black},
 }
\NewDocumentCommand\semiloop{O{black}mmmO{}O{above}}
{%
\draw[#1] let \p1 = ($(#3)-(#2)$) in (#3) arc (#4:({#4+180}):({0.5*veclen(\x1,\y1)})node[midway, #6] {#5};)
}

\newcommand\x{.75}

\newcommand\vertexsize{.07}

\newcommand{\DiagramB}{\begin{tikzpicture}[baseline={([yshift=-.5ex]current bounding box.center)}]
\draw[scalar] (0,0) -- (1.5,0);
\draw[scalar] (.75,.25) circle (.25);
\draw[fill=white] (0,0) circle (.075);
\draw[fill=white] (1.5,0) circle (.075);
\draw[fill=VertexColor,VertexColor] (.75,0) circle (\vertexsize);
\end{tikzpicture}}

\newcommand{\AnalyticStructureToyModel}{\begin{tikzpicture}[baseline={([yshift=-.5ex]current bounding box.center)}]
\draw[->, decorate, decoration={zigzag, segment length=6, amplitude=3},SCRed] (0,0) -- (1,0);
\draw[fill=SCRed,SCRed] (1,0) circle (.075);
\node[SCRed] at (1,-.5) {$-1$};
\draw[scalar] (1,0) -- (4,0);
\draw[scalar] (4,0) -- (5,0);
\draw[scalar,-Latex] (6,0) -- (6.2,0);
\draw[scalar,-Latex] (3,-2) -- (3,2);
\node[] at (5.7,1.7) {$E$};
\draw[scalar] (5.4,1.9) -- (5.4,1.45);
\draw[scalar] (5.4,1.45) -- (6,1.45);
\draw[fill=SCRed,SCRed] (3,0) circle (.075);
\draw[->, decorate, decoration={zigzag, segment length=6, amplitude=3},SCRed] (5,0) -- (6,0);
\draw[fill=SCRed,SCRed] (5,0) circle (.075);
\node[SCRed] at (5,-.5) {$1$};
\end{tikzpicture}}

\newcommand{\rdiscontinuity}{\begin{tikzpicture}[baseline={([yshift=-.5ex]current bounding box.center)}]
\draw[->, decorate, decoration={zigzag, segment length=6, amplitude=3},SCRed] (3,-2) -- (3,-1);
\draw[fill=SCRed,SCRed] (3,1) circle (.075);
\node[SCRed] at (4,1) {$i (\tau+n)$};
\draw[scalar] (0,0) -- (4,0);
\draw[scalar] (4,0) -- (5,0);
\draw[scalar,-Latex] (5,0) -- (6.2,0);
\draw[scalar,-Latex] (3,2) -- (3,2.2);
\draw[scalar] (3,-1) -- (3,1);
\node[] at (5.7,1.7) {$r$};
\draw[scalar] (5.4,1.9) -- (5.4,1.45);
\draw[scalar] (5.4,1.45) -- (6,1.45);
\draw[->, decorate, decoration={zigzag, segment length=6, amplitude=3},SCRed] (3,1) -- (3,2);
\draw[fill=SCRed,SCRed] (3,-1) circle (.075);
\node[SCRed] at (1.7,-1) {$i (-\tau+n)$};
\end{tikzpicture}}
\newcommand{\PolyakovLoop}{\begin{tikzpicture}[baseline={([yshift=-.5ex]current bounding box.center)}]
\draw[scalar] (0,0) ellipse (.5cm and 1cm);
\draw[wilson,Red] (1.5,0) ellipse (.5cm and 1cm);
\draw[ghost] (3,0) ellipse (.5cm and 1cm);
\draw[scalar] (0,1) -- (3,1);
\draw[scalar] (0,-1) -- (3,-1);
\draw[scalar,-Latex] (2,1.2) -- (3.2,1.2);
\node[] at (2.6,1.6) {\footnotesize $\vec{x}$};
\draw[scalar,-Latex] (4.95-1.25,-.5) .. controls (5.05-1.25,-.16) and (5.05-1.25,.16) .. (4.95-1.25,.5);
\node[] at (5.25-1.25,0) {\footnotesize $\tau$};
\node[Red] at (1.5,1.35) {$\Pm$};
\node[Red] at (1.5,-1.35) {\footnotesize $\vec{x}=0$};
\path[clip] (3,-1) rectangle (5,2);
\draw[scalar] (3,0) ellipse (.5cm and 1cm);
\end{tikzpicture}}
\newcommand{\WilsonLine}{\begin{tikzpicture}[baseline={([yshift=-.5ex]current bounding box.center)}]
\draw[scalar] (0,0) ellipse (.5cm and 1cm);
\draw[ghost] (3,0) ellipse (.5cm and 1cm);
\draw[scalar] (0,1) -- (3,1);
\draw[wilson,Red] (0,-1) -- (3,-1);
\draw[scalar,-Latex] (.9,1.2) -- (2.1,1.2);
\node[] at (1.5,1.6) {\footnotesize $\vec{x}$};
\draw[scalar,-Latex] (4.95-1.25,-.5) .. controls (5.05-1.25,-.16) and (5.05-1.25,.16) .. (4.95-1.25,.5);
\node[] at (5.25-1.25,0) {\footnotesize $\tau$};
\node[Red] at (1.5,-1.35) {\footnotesize $\phantom{\vec{x}=0}$};
\node[Red] at (-.75,-1) {$\Wm$};
\node[Red] at (3.75,-1) {\footnotesize $\tau=0$};
\path[clip] (3,-1) rectangle (5,2);
\draw[scalar] (3,0) ellipse (.5cm and 1cm);
\end{tikzpicture}}

\newcommand{\KMSCollinear}{\begin{tikzpicture}[baseline={([yshift=-.5ex]current bounding box.center)}]
\draw[scalar] (0,0) ellipse (.5cm and 1cm);
\draw[wilson,Red] (1.5,0) ellipse (.5cm and 1cm);
\draw[fill=MediumBlueGrey,MediumBlueGrey] (3.425,.5) circle (.1);
\draw[fill=MediumBlueGrey,MediumBlueGrey] (3.425,-.5) circle (.1);
\draw[ghost] (3,0) ellipse (.5cm and 1cm);
\draw[scalar] (0,1) -- (4.25,1);
\draw[scalar] (0,-1) -- (4.25,-1);
\draw[scalar,-Latex] (1.5,-1.35) -- (3,-1.35);
\node[] at (2.25,-1.75) {$|\vec{x}|$};
\draw[scalar,-Latex] (4.95,-.5) .. controls (5.05,-.16) and (5.05,.16) .. (4.95,.5);
\node[] at (5.25,0) {$\tau$};
\node[Red] at (1.5,1.35) {$\Pm$};
\node[MediumBlueGrey] at (3.75,-.475) {$\phi$};
\node[MediumBlueGrey] at (3.75,.525) {$\phi$};
\path[clip] (3,-1) rectangle (5,2);
\draw[scalar] (3,0) ellipse (.5cm and 1cm);
\draw[fill=MediumBlueGrey,MediumBlueGrey] (3.425,.5) circle (.1);
\draw[fill=MediumBlueGrey,MediumBlueGrey] (3.425,-.5) circle (.1);
\path[clip] (4.25,-1) rectangle (5,2);
\draw[scalar] (4.25,0) ellipse (.5cm and 1cm);
\end{tikzpicture}}

\newcommand{\OrderOfOPE}{\begin{tikzpicture}[baseline={([yshift=-.5ex]current bounding box.center)}]
\draw[scalar,double,double distance between line centers=.25em] (-2,0) -- (0,0);
\node[] at (-1.45,-.5) {$\mathbf{2}$};
\draw[scalar,double,double distance between line centers=.25em] (0,1) -- (0,-1);
\node[] at (.5,0) {$\mathbf{1}$};
\draw[ghost] (0,0) circle (1);
\draw[fill=MediumBlueGrey,MediumBlueGrey] (0,1) circle (.2);
\draw[fill=MediumBlueGrey,MediumBlueGrey] (0,-1) circle (.2);
\draw[wilson,Red] (-2.5,0) ellipse (.5cm and 1.25cm);
\end{tikzpicture}}

\newcommand{\ScalarPropagator}{\begin{tikzpicture}[baseline={([yshift=-.5ex]current bounding box.center)}]
\draw[scalar] (0,0) -- (1.25,0);
\draw[fill=white] (0,0) circle (.075);
\draw[fill=white] (1.25,0) circle (.075);
\end{tikzpicture}}
\newcommand{\DefectFeynmanRuleTwo}{\begin{tikzpicture}[baseline={([yshift=-.5ex]current bounding box.center)}]
\draw[scalar,Red] (-.15,0) ellipse (.15 and .5);
\draw[scalar] (0,0) -- (.75,0);
\draw[LightBlueGrey,fill=LightBlueGrey] (0,0) circle (.05);
\end{tikzpicture}}
\newcommand{\FourVertex}{\begin{tikzpicture}[baseline={([yshift=-.35ex]current bounding box.center)}]
\draw[scalar] (0,0) -- (1,1);
\draw[scalar] (0,1) -- (1,0);
\draw[fill=LightBlueGrey,LightBlueGrey] (.5,.5) circle (.05);
\end{tikzpicture}}

\newcommand{\phiGFF}{\begin{tikzpicture}[baseline={([yshift=-.5ex]current bounding box.center)}]
\draw[scalar,Red] (0,0) ellipse (.15 and .5);
\draw[scalar] (.15,0) -- (1,0);
\draw[fill=white] (1,0) circle (.1);
\draw[LightBlueGrey,fill=LightBlueGrey] (.15,0) circle (.05);
\end{tikzpicture}}

\newcommand{\phisquaredGFFOne}{\begin{tikzpicture}[baseline={([yshift=-.5ex]current bounding box.center)}]
\draw[scalar,Red] (0,0) ellipse (.15 and .5);
\draw[scalar] (1-.25,0) circle (.25);
\draw[fill=white] (1,0) circle (.1);
\end{tikzpicture}}
\newcommand{\phisquaredGFFTwo}{\begin{tikzpicture}[baseline={([yshift=-.5ex]current bounding box.center)}]
\draw[scalar,Red] (0,0) ellipse (.15 and .5);
\draw[scalar] (.15,.25) -- (1,.05);
\draw[scalar] (.15,-.25) -- (1,-.05);
\draw[fill=white] (1,0) circle (.1);
\draw[LightBlueGrey,fill=LightBlueGrey] (.13,.25) circle (.05);
\draw[LightBlueGrey,fill=LightBlueGrey] (.13,-.25) circle (.05);
\end{tikzpicture}}

\newcommand{\phiEpsExpTwo}{\begin{tikzpicture}[baseline={([yshift=-.5ex]current bounding box.center)}]
\draw[scalar,Red] (0,0) ellipse (.15 and .5);
\draw[scalar] (.15,0) -- (1.5,0);
\draw[fill=white] (1.5,0) circle (.1);
\draw[scalar] (.75,.25) circle (.25);
\draw[LightBlueGrey,fill=LightBlueGrey] (.15,0) circle (.05);
\draw[LightBlueGrey,fill=LightBlueGrey] (.75,0) circle (.05);
\end{tikzpicture}}
\newcommand{\phiEpsExpThree}{\begin{tikzpicture}[baseline={([yshift=-.5ex]current bounding box.center)}]
\draw[scalar,Red] (0,0) ellipse (.15 and .5);
\draw[scalar] (.15,0) -- (1.5,0);
\draw[scalar] (.15,.25) -- (.75,0);
\draw[scalar] (.15,-.25) -- (.75,0);
\draw[fill=white] (1.5,0) circle (.1);
\draw[LightBlueGrey,fill=LightBlueGrey] (.15,0) circle (.05);
\draw[LightBlueGrey,fill=LightBlueGrey] (.13,.25) circle (.05);
\draw[LightBlueGrey,fill=LightBlueGrey] (.13,-.25) circle (.05);
\draw[LightBlueGrey,fill=LightBlueGrey] (.75,0) circle (.05);
\end{tikzpicture}}

\newcommand{\AnalyticStructureThermal}{\begin{tikzpicture}[baseline={([yshift=-.5ex]current bounding box.center)}]
\draw[scalar] (1,0) -- (2,0);
\draw[scalar] (4,0) -- (5,0);
\draw[scalar,-Latex] (6,0) -- (6.2,0);
\draw[scalar,-Latex] (3,-2) -- (3,2);
\node[] at (5.7,1.7) {$w$};
\draw[scalar] (5.4,1.9) -- (5.4,1.45);
\draw[scalar] (5.4,1.45) -- (6,1.45);
\draw[->, decorate, decoration={zigzag, segment length=6, amplitude=3},SCRed] (0,0) -- (1,0);
\draw[fill=SCRed,SCRed] (1,0) circle (.075);
\node[SCRed] at (1,-.5) {$-1/r$};
\draw[fill=SCRed,SCRed] (2,0) circle (.075);
\node[SCRed] at (2,-.5) {$-r$};
\draw[->, decorate, decoration={zigzag, segment length=6, amplitude=3},SCRed] (2,0) -- (4,0);
\draw[fill=SCRed,SCRed] (4,0) circle (.075);
\node[SCRed] at (4,-.5) {$r$};
\draw[->, decorate, decoration={zigzag, segment length=6, amplitude=3},SCRed] (5,0) -- (6,0);
\draw[fill=SCRed,SCRed] (5,0) circle (.075);
\node[SCRed] at (5,-.5) {$1/r$};
\draw[fill=SCRed,SCRed] (3,0) circle (.075);
\end{tikzpicture}}

\newcommand{\Timeline}{
\begin{tikzpicture}[
  timeline/.style={
    draw, thick, -{Latex[length=2mm]}, line width=0.6pt
  },
  event/.style={
    rectangle, draw=black, fill=white, rounded corners, text width=6.5cm, align=left, font=\small
  },
  year/.style={
    font=\bfseries
  }
]

\draw[scalar] (-1,0) -- (4,0);
\draw[scalar,dashed] (4,0) -- (5,0);
\draw[timeline] (5,0) -- (13,0);

\node[year] at (-1,0.5) {1950};
\node[year] at (6.5,0.5) {1998};
\node[year] at (13.2,0.6) {2018};

\filldraw[black] (0,0) circle (2.5pt) node[anchor=north]{\cite{Matsubara:1955ws}};

\filldraw[black] (0.75,0) circle (2.5pt) node[anchor=north]{\cite{Kubo:1957mj}};

\filldraw[black] (0.9,0) circle (2.5pt) node[anchor=south]{\cite{Martin:1959jp,Schwinger:1960qe}};

\filldraw[black] (1.7,0) circle (2.5pt) node[anchor=north]{\cite{Chew:1961zza,Keldysh:1964ud}};

\filldraw[black] (2.9,0) circle (2.5pt) node[anchor=south]{\cite{Ferrara:1973yt,Polyakov:1974gs}};

\filldraw[black] (5.5,0) circle (2.5pt) node[anchor=north]{\cite{Boyd:1996bx}};

\filldraw[black] (6.5,0) circle (2.5pt) node[anchor=north]{\cite{Maldacena:1997re,Witten:1998zw}};

\filldraw[black] (8,0) circle (2.5pt) node[anchor=south]{\cite{Policastro:2002se,Son:2002sd}};

\filldraw[black] (10,0) circle (2.5pt) node[anchor=south]{\cite{Rattazzi:2008pe}};

\filldraw[black] (11,0) circle (2.5pt) node[anchor=north]{\cite{El-Showk:2011yvt,El-Showk:2012cjh}};

\filldraw[black] (11.3,0) circle (2.5pt) node[anchor=south]{\cite{Katz:2014rla}};

\filldraw[black] (12.2,0) circle (2.5pt) node[anchor=south]{\cite{Caron-Huot:2017vep}};

\filldraw[black] (12.5,0) circle (2.5pt) node[anchor=north]{\cite{Iliesiu:2018fao}};

\end{tikzpicture}}

\newcommand{\Phasediagramintro}{\begin{tikzpicture}[x=0.6pt,y=0.6pt,yscale=-1.30,xscale=1.30]
\draw[very thick, ->] (50,10)--(50,-200);
\draw[very thick, ->] (0,0)--(400,0);

\draw [very thick,  Magenta] (200, 0) to[out=-110,in=30] (180,-20);
\draw [very thick, dotted, Magenta] (180, -20) to[out=30,in=20] (170,-25);
\draw (20,-200) node [anchor=north west][inner sep=0.75pt]  [font=\Large]  {$T$};
\draw (400,0) node [anchor=north west][inner sep=0.75pt]  [font=\Large]  {$g$};
\draw (210,10) node [anchor=north east][inner sep=0.75pt]  [font=\large]  {$g_c$};
\draw (300,-55) node [anchor=north west][inner sep=0.75pt]  [font=\normalsize]  {$\text{\textcolor{blue}{Quantum criticality}}$};
\draw[thick, -stealth] (295, -40)--(205,-3);
\draw (60,-100) node [anchor=north west][inner sep=0.75pt]  [font=\small]  {$\text{\textcolor{Magenta}{T>0 phase transition}}$};
\draw[thick, -stealth] (125, -75)--(165,-30);
\draw (300,-155) node [anchor=north west][inner sep=0.75pt]
[font=\normalsize] {$\text{\textcolor{red}{CFT at finite temperature}}$};
\draw[thick, -stealth] (295, -140)--(215,-102.5);
\draw[very thick, dashed,red] (200,0)--(200,-200);
\filldraw[blue] (200,0) circle (2.5pt) node[anchor=north]{};
\end{tikzpicture}}

\newcommand{\Phasediagram}{\begin{tikzpicture}[x=0.6pt,y=0.6pt,yscale=-1.30,xscale=1.30]
\draw[very thick, ->] (50,10)--(50,-200);
\draw[very thick, ->] (0,0)--(400,0);
\draw (20,-200) node [anchor=north west][inner sep=0.75pt]  [font=\Large]  {$T$};
\draw (400,0) node [anchor=north west][inner sep=0.75pt]  [font=\Large]  {$g$};
\draw (226,5) node [anchor=north east][inner sep=0.75pt]  [font=\large]  {$\left(g^\star, 0\right)$};
\draw (300,-55) node [anchor=north west][inner sep=0.75pt]  [font=\normalsize]  {\boxed{$\text{\textcolor{blue}{CFT at zero T}}$}};
\draw[thick, -stealth] (295, -40)--(205,-3);
\draw (60,-100) node [anchor=north west][inner sep=0.75pt]  [font=\small]  {\boxed{$\text{\textcolor{blue}{(Possible) critical line}}$}};
\draw[thick, -stealth] (125, -75)--(165,-30);
\draw (300,-155) node [anchor=north west][inner sep=0.75pt]
[font=\normalsize] {\boxed{$\text{\textcolor{red}{Thermal CFT}}$}};
\draw [thick, decorate, decoration = {calligraphic brace,mirror}] (210,-10) --  (210,-195);
\draw[thick, -stealth] (295, -140)--(215,-102.5);
\draw[very thick, dashed,red] (200,0)--(200,-200);
\filldraw[blue] (200,0) circle (2.5pt) node[anchor=north]{};
\draw [very thick,  blue] (200, 0) to[out=-110,in=30] (180,-20);
\draw [very thick, dotted, blue] (180, -20) to[out=30,in=20] (170,-25);
\end{tikzpicture}}
\newcommand{\ContourThermaltau}{\begin{tikzpicture}[baseline={([yshift=-.5ex]current bounding box.center)}]
\draw[scalar] (1,0) -- (1.925,0);
\draw[scalar] (4,0) -- (5,0);
\draw[scalar,-Latex] (6,0) -- (6.2,0);
\draw[scalar,-Latex] (3,0.075) -- (3,2);
\node[] at (5.7,1.7) {$\xi'$};
\draw[scalar] (5.4,1.9) -- (5.4,1.45);
\draw[scalar] (5.4,1.45) -- (6,1.45);
\draw[scalar] (0,0) -- (0.925,0);
\draw[scalar] (2,0) -- (2.925,0);
\draw[scalar] (3.075,0) -- (4,0);
\draw[scalar] (5.075,0) -- (6,0);
\draw[fill=MediumBlueGrey,MediumBlueGrey] (4,1.25) circle (.075);
\node[MediumBlueGrey] at (4.7,1.25) {$\xi$};
\draw[scalar] (4,1.25) circle (.4);
\draw[scalar] (0.7,0.1) -- (0.7,-2);
\draw[scalar] (1.3,0.1) -- (1.3,-2);
\draw[scalar] (0.7,.1) arc (180:0:.3);
\draw[scalar] (1.7,0.1) -- (1.7,-2);
\draw[scalar] (2.3,0.1) -- (2.3,-2);
\draw[scalar] (1.7,.1) arc (180:0:.3);
\draw[scalar] (2.7,0.1) -- (2.7,-2);
\draw[scalar] (3.3,0.1) -- (3.3,-2);
\draw[scalar] (2.7,.1) arc (180:0:.3);
\draw[scalar] (3.7,0.1) -- (3.7,-2);
\draw[scalar] (4.3,0.1) -- (4.3,-2);
\draw[scalar] (3.7,.1) arc (180:0:.3);
\draw[scalar] (4.7,0.1) -- (4.7,-2);
\draw[scalar] (5.3,0.1) -- (5.3,-2);
\draw[scalar] (4.7,.1) arc (180:0:.3);
\draw[->, decorate, decoration={zigzag, segment length=6, amplitude=3},SCRed] (1,-2) -- (1,0);
\draw[fill=SCRed,SCRed] (1,0) circle (.075);
\draw[->, decorate, decoration={zigzag, segment length=6, amplitude=3},SCRed] (3,-2) -- (3,0);
\draw[fill=SCRed,SCRed] (3,0) circle (.075);
\draw[->, decorate, decoration={zigzag, segment length=6, amplitude=3},SCRed] (2,-2) -- (2,0);
\draw[fill=SCRed,SCRed] (2,0) circle (.075);
\draw[->, decorate, decoration={zigzag, segment length=6, amplitude=3},SCRed] (4,-2) -- (4,0);
\draw[fill=SCRed,SCRed] (4,0) circle (.075);
\draw[->, decorate, decoration={zigzag, segment length=6, amplitude=3},SCRed] (5,-2) -- (5,0);
\draw[fill=SCRed,SCRed] (5,0) circle (.075);
\end{tikzpicture}}

\newcommand{\ContourThermaltauandw}{\begin{tikzpicture}[baseline={([yshift=-.5ex]current bounding box.center)}]
\draw[scalar] (1,0) -- (1.925,0);
\draw[scalar] (4,0) -- (5,0);
\draw[scalar,dashed,red] (2.5,-2) -- (2.5,2);
\draw[scalar,dashed,red] (3.5,-2) -- (3.5,2);
\draw[scalar,-Latex] (6,0) -- (6.2,0);
\draw[scalar,-Latex] (3,0.075) -- (3,2);
\node[] at (5.7,1.7) {$\xi$};
\fill [domain=0:4, red!50]   (3,0.075) -- (3,2);
\draw[scalar] (5.4,1.9) -- (5.4,1.45);
\draw[scalar] (5.4,1.45) -- (6,1.45);
\draw[scalar] (0,0) -- (0.925,0);
\draw[scalar] (2,0) -- (2.925,0);
\draw[scalar] (3.075,0) -- (4,0);
\draw[scalar] (5.075,0) -- (6,0);
\draw[->, decorate, decoration={zigzag, segment length=6, amplitude=3},SCRed] (1,-2) -- (1,0);
\draw[fill=SCRed,SCRed] (1,0) circle (.075);
\draw[->, decorate, decoration={zigzag, segment length=6, amplitude=3},SCRed] (3,-2) -- (3,0);
\draw[fill=SCRed,SCRed] (3,0) circle (.075);
\draw[->, decorate, decoration={zigzag, segment length=6, amplitude=3},SCRed] (2,-2) -- (2,0);
\draw[fill=SCRed,SCRed] (2,0) circle (.075);
\draw[->, decorate, decoration={zigzag, segment length=6, amplitude=3},SCRed] (4,-2) -- (4,0);
\draw[fill=SCRed,SCRed] (4,0) circle (.075);
\draw[->, decorate, decoration={zigzag, segment length=6, amplitude=3},SCRed] (5,-2) -- (5,0);
\draw[fill=SCRed,SCRed] (5,0) circle (.075);
\end{tikzpicture}}

\usepackage{dsfont}
\usepackage{float}
\usepackage{pgfplots}
\pgfplotsset{compat=1.14}

\let\oldbfseries=\bfseries
\let\oldmdseries=\mdseries
\let\oldnormalfont=\normalfont
\renewcommand{\bfseries}{\oldbfseries\boldmath}
\renewcommand{\mdseries}{\oldmdseries\unboldmath}
\renewcommand{\normalfont}{\oldnormalfont\unboldmath}

\makeatletter
\newlength{\apb@width}
\newcommand{\autoparbox}[2][c]{\settowidth{\apb@width}{#2}\parbox[#1]{\apb@width}{#2}}

\makeatother

\def\Em{{\mathcal{E}}}

\def\Mm{{\mathcal{M}}}

\def\Om{{\mathcal{O}}}
\def\Pm{{\mathcal{P}}}

\def\zb{{\bar{z}}}

\def\veps{\varepsilon}

\newcommand{\beq}{\begin{equation}}
\newcommand{\eeq}{\end{equation}}

\definecolor{nicegreen}{rgb}{0.1,0.6,0.1}

\newmuskip\pFqskip
\mathchardef\pFcomma=\mathcode`,

\makeatletter
\renewcommand*\env@matrix[1][\arraystretch]{%
  \edef\arraystretch{#1}%
  \hskip -\arraycolsep
  \let\@ifnextchar\new@ifnextchar
  \array{*\c@MaxMatrixCols c}}
\makeatother

\begin{document}
\begin{titlepage}
	\begin{center}
	\textbf{
		\vspace{0.5cm}
		\begin{Huge} 
			Thermal effects in conformal field theories
		\end{Huge}
		\\[6.5cm]    %
			Dissertation\\
			zur Erlangung des Doktorgrades\\
			an der Fakultät für Mathematik, Informatik und Naturwissenschaften\\
			Fachbereich Physik\\
			der Universität Hamburg 
			\\[4cm]
			Vorgelegt von\\
			Alessio Miscioscia
			\\[3cm]
			Hamburg\\
			2025
		}
	\vfill
	\end{center}
%	\newpage
	\null\thispagestyle{empty}\newpage
%	\emptypage
	
	\vspace*{\fill}\thispagestyle{empty}
	\begin{minipage}{0.6\textwidth}
		\begin{flushleft} \large
			\begin{normalsize}
				Gutachter/innen der Dissertation: 
				\\[2\baselineskip]
				Zusammensetzung der Pr\"ufungskommission:
				\\[5\baselineskip]
				Vorsitzende/r der Pr\"ufungskommission:
				\\[\baselineskip]
				Datum der Disputation:
				\\[\baselineskip]
				Vorsitzender Fach-Promotionsausschusses PHYSIK:
				\\[\baselineskip]
				Leiter des Fachbereichs PHYSIK:
				\\[\baselineskip]
				Dekan der Fakult\"at MIN:
			\end{normalsize}
		\end{flushleft}
	\end{minipage}
	\hfill
	\begin{minipage}{0.4\textwidth}
		\begin{flushright} \large
			\begin{normalsize}
				Prof. Dr. Elli Pomoni \\
				Prof. Dr. Volker Schomerus
				\\[\baselineskip]
				Prof. Dr. Gregor Kasieczka \\
				Prof. Dr. Elli Pomoni \\
				Prof. Dr. Michael Potthoff \\
				Prof. Dr. Volker Schomerus \\
                Prof. Dr. Timo Weigand
				\\[\baselineskip]
				Prof. Dr. Michael Potthoff
				\\[\baselineskip]
				31.07.2025
				\\[\baselineskip]
			    Prof. Dr. Markus Drescher
				\\[\baselineskip]
				Prof. Dr. Wolfgang J. Parak
				\\[\baselineskip]
			    Prof. Dr.-Ing. Norbert Ritter
			\end{normalsize}
		\end{flushright}
	\end{minipage}
\end{titlepage}	
\newpage\null\thispagestyle{empty}%\newpage
%\emptypage

\section*{Zusammenfassung}
Konforme Feldtheorien (CFTs) sind spezielle Klassen von Quantenfeldtheorien, die Anwendungen finden, die von kritischen Phänomenen bis hin zu Theorien der Quantengravitation über die Holographie reichen. Das Verständnis thermischer Effekte in CFTs ist von zentraler Bedeutung: Kritikalität wird im Labor bei endlicher Temperatur untersucht, und aus holographischer Perspektive entspricht die Untersuchung thermischer CFTs der Untersuchung von Schwarzen Löchern im Anti-de-Sitter-Raum. In dieser Arbeit untersuchen wir die Kinematik und Dynamik von CFTs bei endlicher Temperatur, indem wir gebrochene und ungebrochene Symmetrien analysieren und verschiedene analytische und numerische Bootstrap-Ansätze auf Korrelationsfunktionen bei endlicher Temperatur anpassen. Diese Methoden sind nicht-perturbativ gültig und können an exakt lösbaren Modellen, wie freien Theorien und zweidimensionalen Systemen, sowie im Vergleich mit perturbativen Rechnungen überprüft werden. Die Hauptanwendungen dieser Arbeit betreffen Ein- und Zweipunktfunktionen sowie die freie Energiedichte in den $\mathrm{O}(N)$-Modellen in drei Dimensionen, mit besonderem Fokus auf die dreidimensionalen Ising-, XY- und Heisenberg-Modelle ($N=1,2,3$).

\section*{Abstract}
Conformal Field Theories (CFTs) are special classes of quantum field theories that find applications ranging from critical phenomena to theories of quantum gravity via holography. Understanding thermal effects in CFTs is crucial: criticality is experimentally probed at finite temperature, and, from the holographic perspective, the study of thermal CFTs is dual to the study of black holes in Anti-de Sitter space. In this thesis, we explore the kinematics and dynamics of finite-temperature CFTs by analyzing broken and unbroken symmetries and adapting various analytical and numerical bootstrap approaches to finite-temperature correlation functions. These methods are non-perturbatively valid and can be tested against exactly solvable models, such as free theories and two-dimensional systems, as well as compared with perturbative calculations. The main applications discussed in this thesis concern one- and two-point functions and the free energy density in the $\mathrm{O}(N)$ models in three dimensions, with particular focus on the 3d Ising, XY, and Heisenberg models ($N=1,2,3$).

\newpage\null\thispagestyle{empty}%\newpage
\section*{Acknowledgment}
I would first like to thank my supervisor, Elli Pomoni, for her help, guidance, discussions, and support during these three years of my Ph.D. experience; in particular, for guiding me through the vast area of theoretical physics research while allowing me the freedom to express my own ideas, personality, and style.

 Special thanks also go to Giuseppe Mussardo for many discussions, suggestions, collaborations, and his constant support.

 During my Ph.D., I greatly benefited from the wonderful environment of Hamburg, and DESY in particular. I thank the entire group for their support and for many precious discussions. In particular, I would like to thank Volker Schomerus for several insightful conversations.

 I am also grateful to Till, Craig, J\"org, Carlos, Davide Bononi, Davide Polvara, Gabriele, Ziwen, Paul, Fiona, Torben, Antonio, Apratim, Albert, Sara, Fabio, Tabea, Francesco, Guido, Rigers, and Margherita for contributing to create the beautiful working environment at DESY.

 Special thanks go to my wonderful office mates Lorenzo, Deniz, and Federico, with whom I happily shared my daily life over the past three years.

 Finishing my Ph.D. would not have been possible without my fantastic collaborators. I would like to thank, in particular, Enrico for years of discussions on finite-temperature systems; Julien for the days spent in the library bootstrapping finite-temperature correlation functions; and Deniz, Florent, Carlos, M\'at\'e, and G\'abor for the many collaborations and fruitful discussions.

Finally, I would like to thank my parents, Angela and Vincenzo, for their endless support throughout my years of study, including my Ph.D. journey.

\newpage\null\thispagestyle{empty}%\newpage
%%%%%typos25/08/2024%%%%%%%%%%%
% -5.36: there is a wrong square 
%%%%%%%%%%%%%%%%%%%%%%%%%%%%%%%
\topskip0pt
\vspace*{\fill}
\section*{Author's contributions}
This thesis is based on the following works by the author:
\begin{itemize}
    \item[$\star$] Broken (super) conformal Ward identities at finite temperature \cite{Marchetto:2023fcw};
    \item[$\star$] Sum rules $\&$ Tauberian theorems at finite temperature \cite{Marchetto:2023xap};
    \item[$\star$] Conformal line defects at finite temperature \cite{Barrat:2024aoa};
    \item[$\star$] The thermal bootstrap for the critical O(N) model \cite{Barrat:2024fwq};
    \item[$\star$] The analytic bootstrap for thermal correlators \cite{NewAnalytic};
\end{itemize}
The author has also contributed to the following papers:
\begin{itemize}
    \item[$\star$] Multicriticality in Yang-Lee edge singularity \cite{Lencses:2022ira};
    \item[$\star$] $\mathcal P \mathcal T$ breaking and RG flows between multicritical Yang-Lee fixed points \cite{Lencses:2023evr};
    \item[$\star$] Ginzburg-Landau description for multicritical Yang-Lee models \cite{Lencses:2024wib};
    \item[$\star$] Constraints on RG Flows from Protected Operators \cite{Baume:2024poj};
    \item[$\star$] Around (quantum) criticality at finite temperature \cite{future1};
\end{itemize}\vspace*{\fill}

\newpage

%Comment out next line if we want indentation
\setlength{\parindent}{0pt}

{
%Comment out next line if we want subsubsections in toc
\setcounter{tocdepth}{2}
\tableofcontents
}
\newpage

\setlength{\parskip}{0.1in}

\chapter{Introduction}
\label{sec:Introduction}
In the last century, the two most significant revolutions in our understanding of the universe were undoubtedly the discoveries of quantum mechanics and the theory of relativity. Intriguingly, the former emerged as a consequence of experiments on the thermal radiation of hot bodies~\cite{planck1901}, while the latter arose from pure theoretical insight~\cite{einstein1905,einstein1916}. Quantum mechanics describes nature at very small scales, where quantum effects are essential to understanding the microscopic structure of the universe. In contrast, special and general relativity provide extremely accurate predictions for the motion of massive bodies, black hole physics and gravitational waves.

Neglecting gravitational interactions—which are believed to be the weakest of the fundamental forces~\cite{Arkani-Hamed:2006emk}—theoretical physicists have succeeded in unifying special relativity with the principles of quantum mechanics. The result of this unification is described in the language  of \textit{Quantum Field Theory} (QFT), the most advanced and accurate framework we have for describing natural phenomena. Remarkably, QFT provides a unified language to account for a wide range of phenomena, from condensed matter and statistical physics to high-energy physics and the microscopic laws of nature.

In both low- and high-energy contexts, it is crucial to understand what constitutes a consistent QFT. In this sense, one can introduce the formal concept of \textit{theory space}. Quantum field theories are not isolated points in this space; rather, thanks to the framework of the renormalization group (RG), it is possible to describe how a physical system evolves with the energy scale.

Imagine we have a microscopic description of a physical system, but we are interested in its behavior at larger distances—equivalently, at lower energies. To achieve this, one applies the rules of renormalization to change the effective description, modifying the relevant degrees of freedom and coupling constants, while still describing the same physical system at different energy scales \cite{Kadanoff:1966wm}. This process defines trajectories in the theory space called RG flows \cite{Wilson:1983xri,Wilson:1974mb}.

Fixed points of the renormalization group play a particularly important role: at these points, the Poincaré symmetry of a relativistic quantum theory is enhanced to include scale invariance, and often full conformal invariance. In Chapter~\ref{chap:basics}, we will review the main features of conformal invariance and its connection with the fixed points of the renormalization group.
Because of the deep connection between conformal invariance and RG fixed points, conformal field theories (CFTs) are extremely important in the modern understanding of theoretical physics, as they describe universality classes of critical phenomena.

CFTs also play a crucial role in our attempts to quantize gravitational interactions. In fact, in the context of holography, a very important and well-tested duality was proposed by J.~Maldacena between a conformal field theory in $d$ dimensions and a quantum gravity theory in $(d+1)$-dimensional Anti-de Sitter space~\cite{Maldacena:1997re}. In the following sections, we will review the basic concepts and explain why this duality is particularly powerful in the strong coupling limit of the CFT, where gravity becomes (semi-)classical. This fundamental correspondence can therefore be used in two complementary ways: we can learn about strongly coupled conformal field theories (and, as we will explain, even more general systems) by studying (semi-)classical gravity in AdS, or we can approach the problem of quantizing gravity from the CFT side. Both directions are crucial in modern research, as they bridge two of the most important challenges in theoretical physics: understanding non-perturbative aspects of quantum field theories and the quantization of gravity.

In this thesis, we will consider thermal effects in conformal field theories. The study of finite-temperature effects is extremely relevant in both high- and low-energy physics. In particular, criticality always occurs at non-zero temperature in laboratory experiments, and the analysis of thermal quantities enables direct contact with experimental results. Furthermore, in the holographic context, introducing finite temperature in a CFT corresponds to the presence of a black hole in Anti-de Sitter space~\cite{Witten:1998zw}, making the study of thermal CFTs potentially insightful for black hole physics.

We will now further motivate the importance of finite-temperature effects in conformal field theories for both low-energy physics and holography. In this Chapter we will then then briefly review the existing literature on the subject and explain the novelty of this thesis before presenting a summary of the thesis.

\section{Motivation}
In this section, we explain the two main contexts in which finite-temperature conformal field theories play a crucial role. The first is the theory of quantum criticality, while the second is holography and black hole physics.

\subsection{Quantum criticality and finite-temperature effects}
We describe the notion of quantum criticality and its connection with finite-temperature conformal field theories.

\begin{figure}[h]
\centering
\includegraphics[width=75mm]{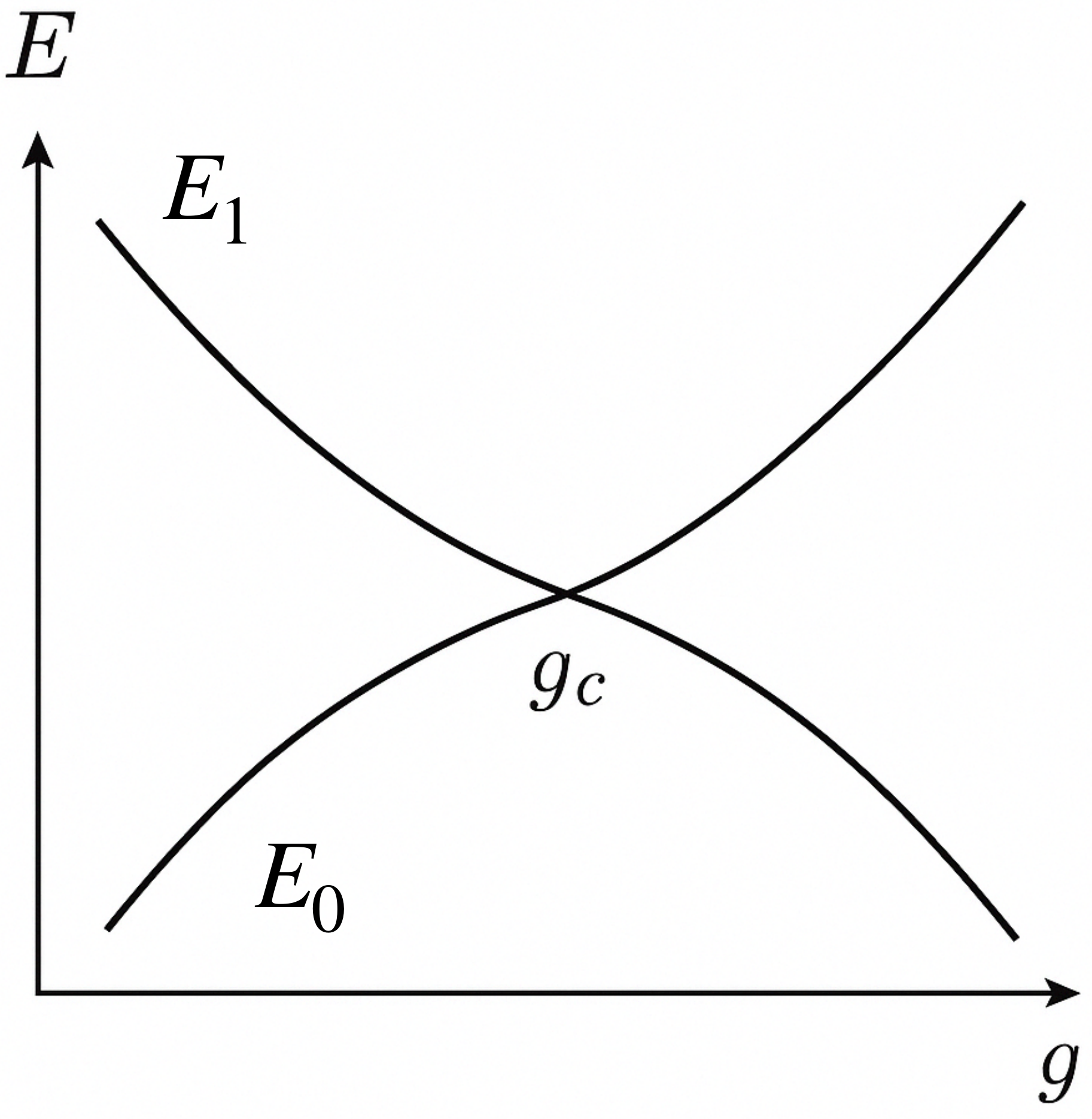}
\caption{Schematic representation of level crossing between the ground state and the first excited state.}
\label{fig:levelcrossing}
\end{figure}

Let us consider a Hamiltonian $\op H \equiv \op H(g)$, where $g$ is a coupling constant of the theory.\footnote{A more standard way of writing this, which may be more familiar to the reader, is $\op H = \op H_0 + g \op H_1$, where $\op H_0$ is the Hamiltonian at a fixed point of the renormalization group or, in the case where $g = 0$ and the theory is non-interacting, corresponds to a free Hamiltonian. The implicit assumption is that $\op H_0$ and $\op H_1$ commute, so that they can be simultaneously diagonalized.} We now consider the ground state energy of the system: we expect this physical quantity to be a smooth function of $g$. However, there may be values of $g$, say $g_c$, at which the first excited state becomes the ground state and vice versa, due to a crossing of energy levels, as shown in Fig.~\ref{fig:levelcrossing}. This typically results in non-analyticities in the ground state\footnote{The discussion here is very general, and the Hamiltonian may describe either a quantum field theory or a lattice system. However, these non-analyticities usually emerge in the continuum limit: if the Hamiltonian under consideration describes a lattice system, one may need to take the continuum limit to observe the non-analytic structure of the ground state.}. Such level crossings are usually accompanied by qualitative changes in the system and its correlation functions, such as spontaneous symmetry breaking \cite{Cardy:1996xt,Mussardo:2020rxh}.

In this thesis are not interested in generic phase transitions, but in those where the characteristic energy scale of the system vanishes as the coupling $g$ approaches $g_c$. Let us denote this energy scale by $\Delta$: it can be defined as the energy of the first excited state above the ground state. In general, we expect\footnote{There are examples with different behavior. For instance, in the \textit{Kosterlitz--Thouless} transition, the characteristic energy scale behaves exponentially close to the critical point. For the purposes of this thesis, we will not discuss such cases in detail, and refer instead to~\cite{Itzykson:1989sx,Itzykson:1989sy} for a comprehensive treatment.}
\begin{equation}
    \Delta \propto |g - g_c|^{z \nu} \ ,
\end{equation}
where $z\nu$ is called the \textit{critical exponent}. This crucial quantity characterizes the nature of the phase transition and is not expected to depend on the microscopic details of the original Hamiltonian $\op H(g)$.

In second-order phase transitions, the typical length scale of the system, $\xi$, which characterizes the exponential decay of correlation functions, diverges. 
By definition, the ratio between the characteristic time scale $\Delta^{-1}$ and the correlation length $\xi$ is governed by the \textit{dynamical critical exponent} $z$, so that
\begin{equation}
    \xi \sim \Delta^{-1/z} \propto |g - g_c|^{-\nu} \ .
\end{equation}

Since the correlation length diverges, the theory at $T = 0$ is scale invariant. In most cases, scale invariance is enhanced to conformal invariance, as we will discuss in Chapter~\ref{chap:basics}. We will focus on this latter scenario, as it is the one relevant for this thesis.

Since quantum criticality is probed in the ground state, strictly speaking, it takes place at zero temperature. Nonetheless, all experiments are necessarily performed at non-zero temperature. Some observables, such as the critical exponents, remain closely related to the low-temperature behavior in the regime $|g - g_c| \ll 1$. For an in-depth discussion, including experimental references, we refer to the excellent book by S.~Sachdev~\cite{Sachdev:2011fcc}.

\begin{figure}[htb]
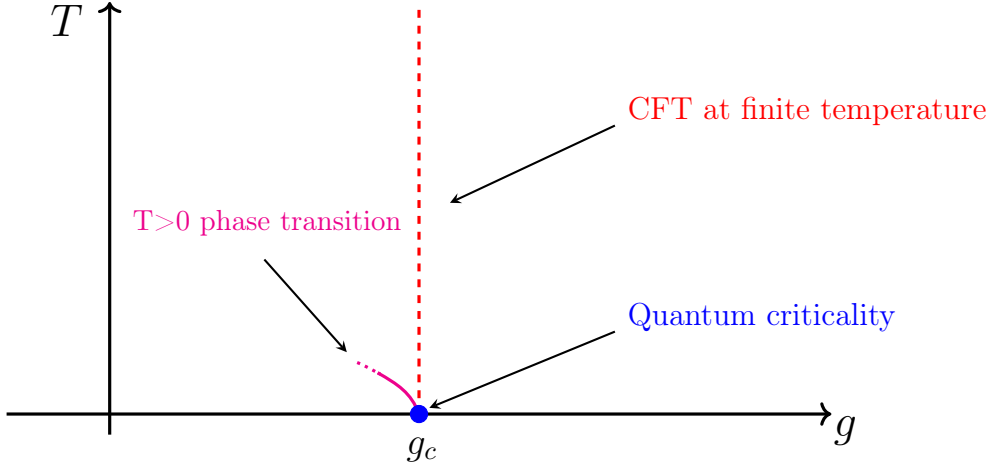

\centering
\Phasediagramintro
\caption{A schematic representation of the $(g,T)$ plane, where $T$ is the temperature and $g$ is a coupling constant of the theory. The quantum phase transition sits at $g = g_c$ and $T = 0$. For $T>0$, there may be additional phase transitions described by conformal field theories. The points in the phase diagram corresponding to thermal effects on the $T = 0$ CFT are indicated with the red dashed line.}
\label{fig:SchemeIntro}
\end{figure}

It is natural, therefore, to ask what the possible scenarios are for the phase diagram when $T > 0$ and $g \sim g_c$. There are two possibilities:
\begin{itemize}
    \item[$\star$] The thermodynamic singularity is present only at $T = 0$, and all physical quantities are analytic at $T > 0$ near $g \sim g_c$. In this case, the critical point $g = g_c$ (and $T = 0$ in the phase diagram shown in Fig.~\ref{fig:SchemeIntro}) is isolated.
    \item[$\star$] There exists a critical line at $T > 0$, terminating at the quantum critical point at $T = 0$.
\end{itemize}

Since the correlation length diverges at criticality, the theory is scale invariant. In most cases, this symmetry is further enhanced to full conformal invariance, as discussed in Chapter~\ref{chap:basics}. We will consider this scenario, which is the one relevant for the rest of the thesis.

There is a distinguished line in the phase diagram shown in Fig.~\ref{fig:SchemeIntro}, given by $g = g_c$ and arbitrary $T \neq 0$, represented by the red dashed line. Since the theory at $T = 0$ is scale invariant, and temperature is the only energy scale on this line, the theory is effectively described by thermal perturbations of the quantum critical point. In this regime, all physical quantities are expected to be smooth functions of $T$. As a result, no phase transition occurs along this line. The aim of this thesis is to study and compute physical observables in this region, which is also relevant for experimental applications.

Even from this qualitative presentation, one can already use physical intuition to anticipate several non-trivial features of thermal effects in conformal field theories.
As already observed, phase transitions typically distinguish between two distinct phases of a material. One of these two phases usually exhibits spontaneous symmetry breaking; this is called the \textit{ordered phase}, while the phase in which the symmetry is preserved is called the \textit{disordered phase}.

From the thermodynamic relation
\begin{equation}
    F = E - T S \ ,
\end{equation}
we expect that, at high temperature, the entropy term dominates. Physically, this implies that systems at high temperature tend to be in a disordered phase. Indeed, this behavior is observed in the majority of physical models, including the case of the $\mathrm{O}(N)$ models we will describe in Chapter~\ref{chap:ONmodel}.

This expectation remains valid even at low temperatures along the line $g = g_c$, where every temperature is equivalent due to scale invariance. The general expectation is therefore that, in a CFT with thermal effects, symmetries remain unbroken.

However, recent investigations have shown that this is not always the case. In fact, explicit examples have been constructed in which a system remains in an ordered phase even at arbitrarily high temperature~\cite{Chai:2020onq,Komargodski:2024zmt,Han:2025eiw,Hawashin:2024dpp}.

\subsection{Holography, black holes and black branes}
We discuss here another application to finite temperature CFTs and its connection to black hole physics via holography.

One of the most important open questions in theoretical physics is understanding the nature of quantum gravity and identifying the correct tools to describe it. Our current understanding of physics lacks a definitive answer to this question. However, significant progress—especially regarding the formal aspects of quantum gravity—has been made thanks to a remarkably precise mathematical relation between gravity in $(d+1)$-dimensional Anti-de Sitter space (AdS$_{d+1}$) and conformal field theories (CFTs) in $d$ dimensions. At the level of symmetries, one can readily observe that the isometry group of AdS$_{d+1}$ matches exactly the conformal group in $d$ dimensions.

This relation is formulated in the context of holography, first introduced in the seminal paper~\cite{Susskind:1994vu}, where the author observed that the number of degrees of freedom of a gravitational system in a volume $V$ scales with the area of its boundary $\partial V$. This led to the conjecture that gravitational systems can be viewed as a \textit{hologram} of a theory defined on the boundary of space. Although this idea might initially seem close to science fiction, it is remarkable that it finds a rigorous realization in the precise correspondence between gravity in AdS and CFTs on its boundary.

This duality was precisely formulated by J.~Maldacena in his famous work~\cite{Maldacena:1997re}. In its strongest form, the AdS/CFT correspondence posits a duality between superstring theory on a background of the form $\text{AdS}_{d+1} \times \mathcal{M}_{9-d}$ and a superconformal field theory living on the $d$-dimensional boundary of AdS. The correspondence can be formulated at the level of the partition function:
\begin{equation}
    Z_{\text{CFT}_d} = Z_{\text{string}/\text{AdS}} \ .
\end{equation}

Since its proposal in 1997, many non-trivial checks and explicit computations have provided strong evidence in support of this duality. For a more detailed account of the duality, its precise formulation, and the dictionary mapping correlators between gravity and CFT, we refer to the standard reviews and textbooks~\cite{Nastase:2015wjb,Erdmenger:2018xqz,Ammon:2015wua,DHoker:2002nbb}.

\begin{figure}[htb]
\centering
\includegraphics[width=105mm]{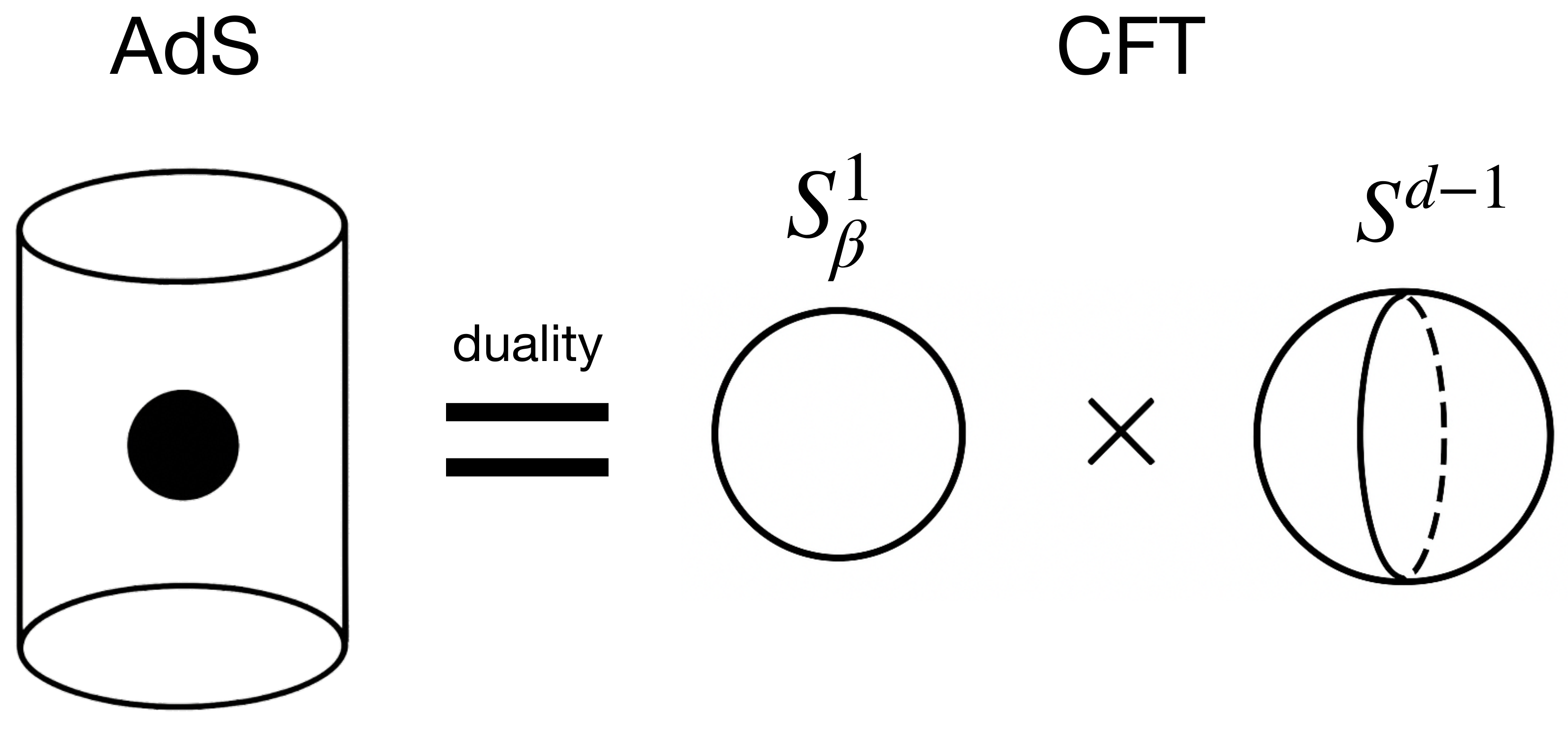}
\caption{Schematic representation of the holographic duality for a black hole in AdS and the CFT at finite volume and finite temperature on the boundary.}
\label{fig:AdSCFT}
\end{figure}

The duality is also believed to be robust under deformations. The study of relevant deformations of the boundary CFT has become a topic of great interest~\cite{Bianchi:2001kw,Skenderis:2002wp}. In this thesis, we are interested in thermal effects in CFTs, which, in the context of AdS/CFT, correspond to introducing black holes in AdS~\cite{Witten:1998qj} (a schematic representation is depicted in Fig. \ref{fig:AdSCFT}). Let us explain this point more precisely.

We consider a CFT defined on $S^1_\beta \times S^{d-1}$, where $\beta = 1/(k_B T)$. As we will explain in Chapter~\ref{chap:basics}, the thermal effects are encoded in the $S^1_\beta$ factor, while the spatial sphere $S^{d-1}$ effectively puts the theory at finite volume. What is the gravitational dual of this setup? There are two possible metric solutions:
\begin{itemize}
    \item[$\star$] \textit{Thermal AdS}, with metric
    \begin{equation}
        ds^2 = \left(1 + \frac{r^2}{L^2} \right) d\tau^2 + \left(1 + \frac{r^2}{L^2} \right)^{-1} dr^2 + r^2 d\Omega_{d-1}^2 \ ,
    \end{equation}
    where $\tau = i t$ is the Euclidean time compactified as $\tau \sim \tau + \beta$, and $L$ is the AdS radius;
    
    \item[$\star$] \textit{AdS-Schwarzschild black hole}, with Euclidean metric
    \begin{equation}
        ds^2 = f(r) d\tau^2 + f(r)^{-1} dr^2 + r^2 d\Omega_{d-1}^2 \ ,
    \end{equation}
    where
    \begin{equation}
        f(r) = 1 - \frac{\mu}{r^{d-2}} + \frac{r^2}{L^2} \ ,
    \end{equation}
    and $\mu$ is related to the mass of the black hole. The temperature of the boundary theory is obtained by requiring regularity of the Euclidean solution at the horizon $r = r_h$, where $f(r_h) = 0$.
\end{itemize}

The thermal AdS geometry simply corresponds to thermal effects on both sides of the duality. The AdS-Schwarzschild black hole case, on the other hand, encodes richer physics. One may ask why the CFT at the boundary is at finite temperature in this case. The answer comes from black hole thermodynamics: due to Hawking radiation~\cite{Hawking:1975vcx}, black holes are inherently thermal objects. The CFT at the boundary inherits this thermal behavior.

However, black holes are not stable at all temperatures. One can show that there is a minimum temperature below which the AdS-Schwarzschild black hole is unstable. For sufficiently low temperatures, the dominant gravitational saddle is thermal AdS. For higher temperatures, both saddle points contribute, and the dominant one is selected by comparing their free energies. Typically, at high temperatures, the AdS-Schwarzschild solution dominates. The transition between these two phases is known as the \textit{Hawking--Page phase transition}. This transition corresponds to the confinement/deconfinement phase transition of the CFT on the boundary, and the order parameter is an extended object known as the \textit{Polyakov loop}. The latter can be described as a line defect wrapping the thermal circle; we will discuss similar setups in Chapter~\ref{sec:defectsBO}.

In this thesis, we will not study CFTs at finite volume, but always work in the infinite-volume limit. This corresponds to sending the radius $R$ of $S^{d-1}$ to infinity, or more precisely, taking the limit $R/\beta \to \infty$, which can also be interpreted as a high-temperature limit. On the gravity side, this corresponds to sending $\mu \to \infty$.

Thus, the study of thermal effects in conformal field theories can yield insight into black hole physics (in AdS) and their quantum properties. Conversely, the gravity dual can be used to extract information about thermal correlators at strong coupling. A particularly striking recent example is the proposed analytic structure of thermal two-point functions in momentum space~\cite{Dodelson:2023vrw}, which reveals the precise analytic behavior of these correlators at large coupling.

\section{State of the art: from Matsubara to the thermal bootstrap}
Finite-temperature effects in quantum field theories have been studied since the early days of the subject, due to their crucial importance in describing systems relevant to low-energy physics and condensed matter phenomena. A major breakthrough in both conceptual understanding and computational technique was made by Matsubara~\cite{Matsubara:1955ws}, who showed that thermal effects in quantum field theory can be incorporated by considering the theory on the Euclidean manifold $S^1_\beta \times \mathbb{R}^{d-1}$. Here, the compact circle $S^1_\beta$ corresponds to imaginary time with periodicity $\beta = 1/(k_B T)$, where $T$ is the temperature. This identification leads to a discrete spectrum of frequencies—now known as \textit{Matsubara frequencies}—which play a central role in finite-temperature quantum field theory.

Even today, this so-called Matsubara formalism remains the standard framework for studying quantum systems in thermal equilibrium and it is also the formalism adopted for this thesis. Its power lies in the fact that it allows one to adapt many techniques from Euclidean quantum field theory to the thermal setting. Consequently, perturbative computations—often involving high-loop expansions—have been successfully performed in a wide variety of physical systems. These include, for example, effective field theories for superconductivity, as well as studies of the quark-gluon plasma relevant for heavy-ion collisions and early universe cosmology (for details we refer to the classical reviews \cite{Laine:2016hma,Fetter:1971,Shuryak:2004cy} and references thereof).

However, as in the zero-temperature case, non-perturbative effects are typically difficult to access. The estimation and control of such effects remain a major challenge, especially in strongly coupled theories. Moreover, as we will stress in Chapter~\ref{chap:basics}, perturbation theory at finite temperature may be significantly more subtle and technically involved than its zero-temperature counterpart. Just after the idea of Matsubara, Keldysh and Schwinger \cite{Schwinger:1960qe,Keldysh:1964ud} extended his work by introducing a complex time direction in which real time allows also to study (real) time evolution of correlation functions at finite temperature.

For a long time, the only non-perturbative tool available for studying quantum field theories at finite temperature was lattice field theory. Remarkable progress has been made in this direction, especially in the context of quantum chromodynamics (QCD). For instance, high-precision lattice simulations have enabled a detailed characterization of the thermodynamics of QCD, including the deconfinement transition and the equation of state at finite temperature. We refer the reader to~\cite{Boyd:1996bx,Borsanyi:2010cj} for important results in this context.

With the advent of the AdS/CFT correspondence, it became possible to compute thermal correlation functions of strongly coupled quantum field theories via their dual gravitational description~\cite{Son:2002sd,Policastro:2002se}. In this framework, thermal effects in the boundary theory are encoded in the presence of black holes or black branes in the bulk geometry. In this context, precisely in \cite{El-Showk:2011yvt}, emerged the idea that the KMS condition could be used to constrain thermal correlation functions.

A first attempt to combine Monte Carlo simulations with conformal field theory techniques to study thermal effects in CFTs was presented in~\cite{Katz:2014rla}.
A significant breakthrough came only in 2017, with the introduction of a modern approach to thermal correlation functions in conformal field theories~\cite{Iliesiu:2018fao}. This work extended the conformal bootstrap program—originally developed for zero-temperature correlators—to the finite-temperature setting. The bootstrap philosophy aims to use general principles such as self-consistency, unitarity, and symmetry to constrain, and in some cases compute, non-perturbative observables in quantum field theories. The bootstrap idea was first proposed in the 1960s to constrain the S-matrix of quantum field theories~\cite{Chew:1961zza}, and was later extended to conformal field theories~\cite{Ferrara:1973yt,Polyakov:1974gs,Rattazzi:2008pe,El-Showk:2014dwa,El-Showk:2012cjh}. We will review these ideas in Chapter~\ref{chap:basics}.
In~\cite{Iliesiu:2018zlz}, the authors successfully applied the bootstrap approach to the computation of thermal correlation functions in the three-dimensional Ising model.

\hspace{1 cm}

\begin{figure}[h!]
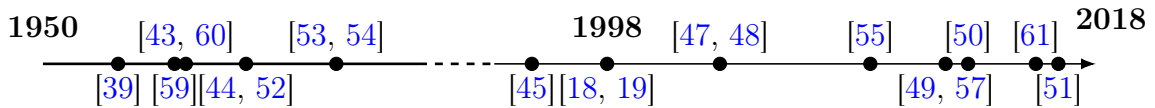

\centering
\Timeline
\caption{Milestones in the development of thermal quantum field theory and benchmarks in the bootstrap.}
\label{fig:thermalQFTtimeline}
\end{figure}

\section{Summary and structure of the thesis}
This thesis presents novel bootstrap--based methods—both analytical and numerical--which allow us to efficiently study strongly coupled systems and obtain new predictions. We now conclude this introduction by summarizing the main results and outlining the structure of the thesis. The structure and main results are outlined as follows:

\paragraph{Basics: CFT and finite-temperature effects}  
We begin with an overview of CFTs and finite-temperature quantum field theories in Chapter~\ref{chap:basics}. The discussion of CFT is self-contained but not exhaustive, as the goal is to introduce the main tools we will use later, such as the Operator Product Expansion and the state-operator correspondence. We also highlight the main differences between conformal field theories at $T = 0$ and $T > 0$, particularly focusing on the symmetries, as encoded in the Ward identities.

In addition, we present an overview of the available methods for computing the dynamical data of a CFT, namely the conformal dimensions and three-point function coefficients. We place particular emphasis on the conformal bootstrap program, since this approach will later be adapted to study thermal correlators. This part of the review is meant to be conceptual rather than technical, aiming to convey the main ideas behind the method.

In Chapter~\ref{chap:basics}, we also review the fundamentals of quantum field theory at finite temperature, with special attention to the Matsubara formalism. We introduce the Kubo–Martin–Schwinger (KMS) condition, which plays a central role in the thermal bootstrap framework developed in this thesis. Finally, we comment on perturbation theory and the infrared problems that can arise at finite temperature, illustrating these issues with the example of a scalar field theory with quartic potential—an important case for applications to the $\mathrm{O}(N)$ model discussed in Chapter~\ref{chap:ONmodel}.

\paragraph{Kinematics at finite temperature}  
The first step in characterizing a physical system is to study its kinematical constraints, which are generally determined by the symmetries of the theory. In Chapter~\ref{chap:ThermalKIno}, we present a rigorous discussion of thermal kinematics, with particular emphasis on one- and two-point functions. Our strategy is to analyze the Ward identities associated with the conformal group and to identify which symmetries are broken at finite temperature. While translations and spatial rotations remain unbroken, dilatations, Lorentz boosts, and special conformal transformations are broken. This analysis leads to a consistent set of broken and unbroken Ward identities, which can be used to constrain the kinematics of the theory.

The main result of this chapter is the form of the Operator Product Expansion (OPE) for the two-point function of identical scalar operators at finite temperature, which is presented in equation~\ref{eq:thermalOPEreview}.

The same strategy can be extended to superconformal field theories. In this context, we derive the broken Ward identities associated with supersymmetry and superconformal symmetry, and show that $R$-symmetry remains preserved at the global level.

The chapter also includes a discussion of the analytic structure of thermal correlators and its applications. In particular, we analyze the inversion formula, dispersion relations, and certain results concerning momentum-space correlation functions. Both the inversion formula and dispersion relations provide analytic tools for computing thermal OPE coefficients and thermal two-point functions, assuming the analytic properties of the correlators are known. Similarly, in momentum space, there exists a notion of dispersion relation which allows one to recover the thermal two-point function from its position-space structure.

We also comment on the notion of OPE in momentum space and show that such an expansion exists as an asymptotic series. These discussions build upon the work in~\cite{Iliesiu:2018fao,Alday:2020eua,Manenti:2019wxs}, but also include original results and insights developed in this thesis.

\paragraph{Dynamics and bootstrap for thermal correlators}  
In Chapter~\ref{ref:dynamics}, we show how the OPE decomposition derived in Chapter~\ref{chap:ThermalKIno} can be used to formulate a bootstrap problem for thermal OPE coefficients. The key observation is that equation~\ref{eq:thermalOPEreview} is not trivially invariant under the KMS condition. Therefore, imposing KMS invariance on the OPE expansion provides nontrivial constraints on the thermal OPE data.

To make this idea concrete, we expand around the KMS fixed point, obtaining an infinite set of linear constraints on the thermal OPE coefficients. These equations are presented in equation~\eqref{eq:sumrule}, and further simplified in the special case where the two operators are not spatially separated (see equation~\eqref{eq:BootstrapFor1D}).

We comment on why these equations alone are not sufficient to fully solve the thermal bootstrap problem. Nevertheless, since they are derived from an expansion around the KMS fixed point, it is natural to consider complementary limits. In particular, we derive an asymptotic estimate for the thermal OPE coefficients associated with heavy operators. We first present a heuristic—but physically transparent—derivation, and then discuss how it can be made rigorous under suitable assumptions.

We then combine the sum rules with the heavy-operator asymptotics to develop a novel hybrid numerical and analytical method to compute thermal OPE coefficients non-perturbatively. The idea is to solve the sum rules for the light operators’ OPE coefficients, while approximating the contributions from heavy operators using their asymptotic behavior. This method is tested in both four-dimensional free scalar theory and the two-dimensional Ising model.

We also make use of the dispersion relations introduced in Chapter~\ref{chap:ThermalKIno} to set up an analytic bootstrap approach for thermal correlation functions. This method is particularly effective in perturbative regimes. The key observation is that the result of the dispersion relation is not KMS invariant: one can therefore impose KMS symmetry to reconstruct the missing pieces of information not captured by the dispersion formula. This can be done in two ways: the first is a hybrid analytic/numerical method based on specific assumptions about the functional form of the missing part of the correlator. The second is a purely analytic approach in which one adds \emph{images} to the dispersion result in order to restore KMS invariance. Both strategies yield correct and exact results in free field theory and in two-dimensional test cases. We comment on the relation between these approches and the Operator Product Expansion in momentum space.

Finally, we discuss the special case of the two-point function with zero spatial separation between the operators. We show that its analytic form can be reconstructed, up to a constant, from its analytic structure.

\paragraph{Conformal line defects at finite temperature}  
The setup developed in Chapters~\ref{chap:ThermalKIno} and~\ref{ref:dynamics} can be enriched by introducing line defects. In Chapter~\ref{sec:defectsBO}, we focus on the case of a conformal line defect wrapping the thermal circle. A conformal line defect at zero temperature is defined as a defect that does not store energy and can be described as a one-dimensional CFT localized along the defect.

We study the kinematics of such configurations and show that bulk operators can acquire nontrivial one-point functions in the presence of the defect. In particular, we analyze the structure of these one-point functions and their Operator Product Expansion (OPE). Special attention is devoted to the one-point function of the stress-energy tensor, from which we extract thermodynamic quantities such as the free energy, energy density, and entropy density of the system.

We also formulate a bootstrap problem for this setup by constructing sum rules and deriving asymptotics for heavy operators, following the strategy outlined in Chapter~\ref{ref:dynamics}.

As an explicit example, we present the case of a \textit{magnetic line defect} in a generalized free theory, where all computations can be carried out analytically.

\paragraph{The $\mathrm O(N)$ models at finite temperature} In Chapter~\ref{chap:ONmodel}, we apply the methods developed in the previous chapters to study the $\mathrm{O}(N)$ models in $2 \le d \le 4$. In particular, by making use of the hybrid numerical/analytical method introduced in Chapter~\ref{ref:dynamics}, we study the 3D Ising model, the XY model, and the Heisenberg model in a non-perturbative framework. All of these models are experimentally realized, and extensive Monte Carlo simulations are available at zero temperature. 

In the case of the 3D Ising model, our results can be compared with previous numerical estimates~\cite{Iliesiu:2018zlz,PhysRevE.79.041142}, showing good agreement. For the XY and Heisenberg models, our results represent new predictions that can be tested in the future through numerical simulations and, hopefully, experiments. The main results are summarized in Figs.~\ref{Fig:IsingResults},~\ref{fig:OPE_Coefficients} and Tables~\ref{tab:3dIsingResults},~\ref{tab:3do2o3Results}.

We also present the evolution of the free energy across dimensions ($3 \le d < 4$), using zero-temperature data obtained from the $\varepsilon$-expansion.

Furthermore, we apply the analytic bootstrap methods introduced in Chapter~\ref{ref:dynamics} to compute the finite-temperature two-point function of fundamental scalars to first order in the $\varepsilon$-expansion. The result perfectly matches the prediction of the inversion formula and provides corrections beyond it. We also find excellent agreement with a direct perturbative computation. Additionally, we comment on the connection between this approach and the Operator Product Expansion in momentum space.

We also initiated a mixed analytic-numerical approach, in which results obtained from numerical methods serve as input for the analytic bootstrap. This combination of techniques has already yielded new predictions in the 3D Ising model, such as thermal two-point functions that we aim to compare with Monte Carlo results in the near future.

Finally, we present new results from the $\varepsilon$-expansion and the large-$N$ expansion for the magnetic line defect in the $\mathrm{O}(N)$ model, and discuss their physical implications.

\paragraph{Conclusions and Outlook}  
In Chapter~\ref{sec:ConclusionsAndOutlook}, we conclude by summarizing the main results of this thesis and highlighting several directions for future research.

 Our analysis opens the door to further applications, especially in strongly coupled systems where conventional methods are limited.

Several interesting extensions of this work can be envisioned. First, one could further develop the thermal bootstrap framework by exploring higher-point functions and spinning operators. Second, an important direction involves the study of deformations of thermal CFTs: by perturbing the theory with relevant operators, one can investigate the interplay between thermal effects and RG flow.
Another crucial step would be the comparison of our predictions with Monte Carlo simulations and experimental results, particularly in systems where finite-temperature behavior plays a central role.
Finally, it would be natural to apply the bootstrap approach to holographic CFTs. Given the strong coupling nature of these theories, the thermal bootstrap may provide new insights into quantum gravity and the thermodynamics of black holes in AdS. In particular, studying thermal correlators in CFTs with large central charges could further test the correspondence between thermal field theory and black hole physics.

\chapter{Basics on CFT and finite temperature effects}\label{chap:basics}
The main goal of this thesis is to set up a bootstrap problem for finite-temperature effects in Conformal Field Theories (CFTs). Before addressing this topic, it is crucial to review the basic aspects of conformal field theories in order to establish notation and highlight the main differences between CFTs at zero and finite temperature.

The CFT review that follows will focus on the essential and fundamental aspects relevant to this thesis. More complete and well-written reviews are already available in the literature \cite{DiFrancesco:1997nk,Rychkov:2016iqz,Simmons-Duffin:2016gjk,Poland:2016chs,Qualls:2015qjb,Cardy:1996xt,Mussardo:2020rxh,Poland:2018epd}. This brief summary is based on those books and reviews.
\section{A CFT primer}\label{eq:CFTprimer}
The modern perspective on Quantum Field Theories (QFTs) is formulated in terms of RG flows \cite{Wilson:1971bg,Wilson:1971dh}. In particular, QFTs are understood as interpolating, in coupling space, between two RG fixed points.

\begin{figure}[h]
\centering
\includegraphics[width=55mm]{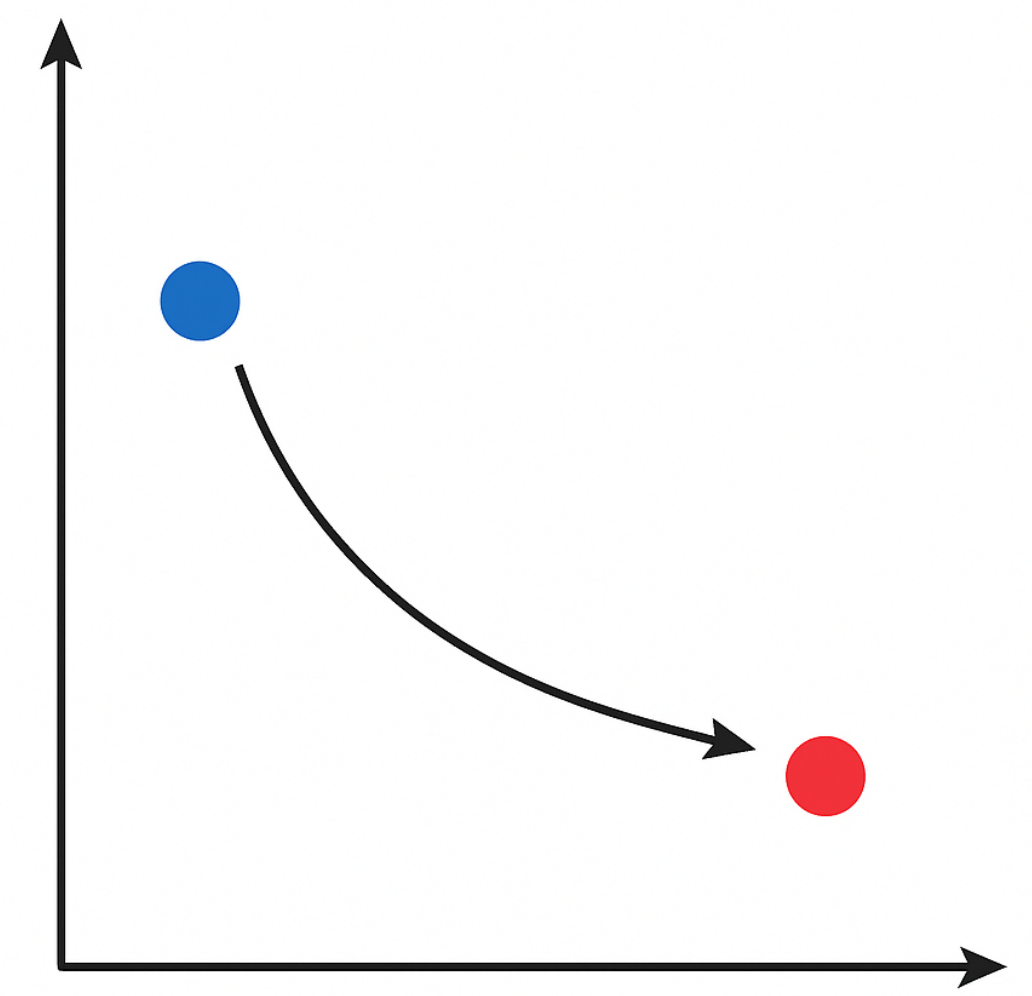}
\caption{Schematic RG flows in parameter space. The dots represent RG fixed points: we focus on the flow between the blue (UV) point and the red (IR) fixed point.}
\label{fig:RGscheme}
\end{figure}

At RG fixed points, the correlation length of the theory—or, in other words, the inverse mass scale—is either zero or infinite. If we focus only on local operators, this restricts the fixed points to two possible types:
\begin{itemize}
    \item[$\star$] \textit{Trivial theories}: these are massive theories considered below the mass scale.\footnote{If this concept is extended to include non-local operators, the theory may still possess such operators in its Hilbert space. In that case, it could correspond to a topological theory.}
    \item[$\star$] \textit{Scale-invariant theories}: these are theories containing local operators and exhibiting an additional symmetry, as they cannot possess any intrinsic scale.
\end{itemize}

In this thesis, we focus on the second possibility. Scale invariance implies that the theory is invariant under the transformations
\begin{equation}
    x \to \lambda x \ , \hspace{1 cm} g_{\mu \nu} \to \lambda^2 g_{\mu \nu} \ .
\end{equation}
For \(\lambda = 1+\epsilon\) with \(\epsilon \ll 1\), the metric variation is given by \(\delta g_{\mu \nu} = 2 \epsilon \delta_{\mu \nu}\), assuming we are working with \(g_{\mu \nu} = \delta_{\mu \nu}\). Therefore, under infinitesimal scale transformations, the variation of the formal action is
\begin{equation}
    \delta \mathcal{A}  = \int d^{d}x \, T_{\mu \nu} \delta g^{\mu \nu} \propto \int d^{d}x \, T_{\, \mu}^{\mu} \ .
\end{equation}
This implies that the trace of the stress tensor must be expressible as a total derivative:
\begin{equation}
    T_{\, \mu}^{\mu} = \partial_\mu V^{\mu} \ ,
\end{equation}
where \(V^\mu\) is known as the \textit{virial current}.
 Generically, we cannot prove more than this, and in a scale-invariant theory one typically has \(T_{\, \mu}^{\mu} \neq 0\). However, the majority of fixed points of the renormalization group are believed to satisfy \(V^{\mu} = \partial_\nu L^{\nu \mu}\), which enhances scale invariance to conformal invariance—a concept we will explain in the following. Note that if the virial current can be written as a total divergence, it is possible to \textit{improve} the stress-energy tensor to obtain a traceless one.

It is worth emphasizing that at a strongly coupled fixed point, the existence of a virial current is highly non-trivial, since the stress tensor has a protected scaling dimension \(\Delta_T = d\), and by dimensional analysis, \(\Delta_V = d - 1\). Therefore, in order to have a fixed point that is scale-invariant but not conformally invariant, the spectrum must contain a spin-one operator with exact scaling dimension \((d - 1)\). This is extremely non-trivial, as the anomalous dimension of the virial current must vanish.

In \(d = 2\), it is possible to prove that scale invariance is always enhanced to conformal invariance in unitary models, while examples of scale-invariant but non-conformal theories exist in non-unitary settings \cite{Riva:2005gd}. The situation in higher dimensions is still an open question, enriched by several beautiful physical arguments and examples \cite{Komargodski:2011xv,Dymarsky:2013pqa,Dymarsky:2014zja,Nakayama:2010zz,Nakayama:2011tk,Nakayama:2013is,Nakayama:2016cyh,Nakayama:2020bsx,Gimenez-Grau:2023lpz,Nakayama:2024jwq,Luty:2012ww}. 

In this thesis, we will not further comment on this problem and will assume that the theory is conformally invariant.

\subsection{Conformal generators and Ward identities}
Note that if the trace of the stress-energy tensor vanishes operatorially, then the symmetry algebra is enlarged, since any infinitesimal transformation of the type \(\delta g_{\mu \nu} = f(x)\delta_{\mu \nu}\) leaves the formal action\footnote{Note that the notion of \textit{formal action} does not assume the existence of a Lagrangian or being in a perturbative regime; it is a general, non-perturbative concept.} invariant.
 Indeed,
\begin{equation}
    \delta \mathcal{A} \propto \int d^{d}x \, f(x) T_{\, \mu}^{\mu} = 0 \ .
\end{equation}
Transformations that act in this way are called \textit{Weyl transformations}. 

Nevertheless, we are not interested in the full class of Weyl transformations, since in general a Weyl transformation maps flat space into a curved one. As our physical theory is defined in flat space, we restrict attention to those Weyl transformations that preserve flatness. These correspond to coordinate transformations of the form
\begin{equation}
    x^\mu \to x^\mu + \epsilon^\mu(x) \ ,
\end{equation}
such that \(\delta g_{\mu \nu} = \partial_\mu \epsilon_\nu + \partial_\nu \epsilon_\mu \equiv f(x) \delta_{\mu \nu}\). The solution is that, in \(d > 3\), the solutions of this differential equation are:
\begin{itemize}
    \item[$\star$] \textbf{Translations}: \(\epsilon^\mu = \text{const.}\), \(f(x) = 0\);
    \item[$\star$] \textbf{Rotations}: \(\epsilon^\mu = x^\nu \omega_{[\nu \mu]}\), \(f(x) = 0\);
    \item[$\star$] \textbf{Dilatation}: \(\epsilon^\mu = \lambda x^\mu\), \(f(x) = 2 \lambda\);
    \item[$\star$] \textbf{Special conformal transformation}: \(\epsilon^\mu = 2(a \cdot x)x^\mu - x^2 a^\mu\), \(f(x) = a \cdot x\), for a generic vector \(a^\mu\);
\end{itemize}

In \(d = 2\), the situation is slightly different, and we will discuss this in the following sections of the thesis.

It is also possible to formulate conformal symmetry in terms of an algebra generated by:
\begin{itemize}
    \item[$\star$] \textbf{Translations}: \(\op P_\mu = i \partial_\mu\);
    \item[$\star$] \textbf{Rotations}: \(\op M_{\mu \nu} = i (x_\mu \partial_\nu - x_\nu \partial_\mu)\);
    \item[$\star$] \textbf{Dilatation}: \(\op D = i x^\mu \partial_\mu\);
    \item[$\star$] \textbf{Special conformal transformations}: \(\op K_\mu = i (2 x_\mu x^\nu \partial_\nu - x^2 \partial_\mu)\);
\end{itemize}

These generators satisfy the usual Poincaré algebra:
\begin{equation}
    [\op M_{\mu \nu}, \op M_{\rho \sigma}] = -i (\delta_{\mu \rho} \op M_{\nu \sigma} \pm \text{permutations}); \hspace{1cm} [\op M_{\mu \nu}, \op P_\rho] = i (\delta_{\nu \rho} \op P_\mu - \delta_{\mu \rho} \op P_\nu) \ ,
\end{equation}
along with the new conformal relations:
\begin{equation}
    [\op D, \op P_\mu] = -i \op P_\mu \ , \hspace{1cm} [\op D, \op K_\mu] = i \op K_\mu \ ,
\end{equation}
\begin{equation}
    [\op P_\mu, \op K_\nu] = 2i (\delta_{\mu \nu} \op D - \op M_{\mu \nu}) \ .
\end{equation}
It turns out that this algebra, also called the \textit{conformal algebra}, is isomorphic to \(\mathrm{SO}(d+1,1)\), which is also the algebra generating Lorentz transformations in \(\text{Mink}_{d+1,1}\). The N\"oether procedure associates to each of the transformations listed above a corresponding conserved current. 

Therefore, for a generic generator \(\op G_a\) associated with the current \(J_a^\mu\), it is possible to write the Ward identity:
\begin{equation}
    i \sum_{i} \delta(x_i - y) \langle \mathcal{O}_1(x_1) \ldots (\op G_a \mathcal{O}_i(x_i)) \ldots \mathcal{O}_n(x_n) \rangle = \frac{\partial}{\partial y^\mu} \langle J_a^\mu(y) \mathcal{O}_1(x_1) \ldots \mathcal{O}_n(x_n) \rangle \ .
\end{equation}

This equation holds for any symmetry generator and can be used to constrain correlation functions. In the case of the conformal group, we apply this identity to show that it strongly constrains the structure of correlators.

\paragraph{One-point functions}

Let us start with the simplest possible correlation function: the vacuum expectation value of an operator in the theory. Translation invariance implies
\begin{equation}
    \partial_\mu \langle \mathcal{O}(x) \rangle = 0 \ ,
\end{equation}
which means that \(\langle \mathcal{O} \rangle\) is constant. The absence of a scale implies that this constant must be zero by dimensional analysis, and this can be formally shown using dilatation invariance.

Under scale transformations, the operator transforms as
\begin{equation}
    \mathcal{O}(\lambda x) = \lambda^{\Delta_{\mathcal{O}}} \mathcal{O}(x) \ , \hspace{1 cm} \Rightarrow \hspace{1 cm} \op D \mathcal{O}(0) = \Delta_{\mathcal{O}} \mathcal{O}(0) \ ,
\end{equation}
where \(\Delta_{\mathcal{O}}\) is the conformal dimension of the operator \(\mathcal{O}\).

We thus obtain
\begin{equation}
   \op D \langle \mathcal{O} \rangle = \left(\cancel{x^\mu \partial_\mu} + \Delta_{\mathcal{O}}\right) \langle \mathcal{O} \rangle = 0 \ ,
\end{equation}
which implies that either \(\Delta_{\mathcal{O}} = 0\) or \(\langle \mathcal{O} \rangle = 0\).

In unitary theories, this means that the only operator that can have a non-zero one-point function is the identity operator \(\op 1\), whose conformal dimension is \(\Delta_{\op 1} = 0\), and whose vacuum expectation value is typically normalized to \(\langle \op 1 \rangle = 1\). Any other operator with non-zero conformal dimension must have a vanishing one-point function.

\paragraph{Two-point functions}

We can also use Ward identities to fix the form of the two-point function between scalar fields. In fact, Poincaré invariance constrains it to the form
\begin{equation}
    \langle \mathcal{O}_1(x_1) \mathcal{O}_2(x_2) \rangle = f(|x_1 - x_2|) \ ,
\end{equation}
where we consider scalar operators for simplicity.

The Ward identity associated with dilatations further constrains the correlation function to satisfy
\begin{equation}
    \left(x_1^\mu \partial_\mu^1 + x_2^\mu \partial_\mu^2 + \Delta_1 + \Delta_2 \right) f(|x_1 - x_2|) = 0 \ ,
\end{equation}
which is solved by
\begin{equation}
    \langle \mathcal{O}_1(x_1) \mathcal{O}_2(x_2) \rangle = \frac{C}{|x_1 - x_2|^{\Delta_1 + \Delta_2}} \ .
\end{equation}

This is not yet the end of the story. In fact, conformal invariance under special conformal transformations imposes the additional constraint
\begin{equation}
    \langle (\op K_\mu \mathcal{O}_1(x_1)) \mathcal{O}_2(x_2) \rangle + \langle \mathcal{O}_1(x_1) (\op K_\mu \mathcal{O}_2(x_2)) \rangle \propto (\Delta_1 - \Delta_2) C = 0 \ ,
\end{equation}
which implies that either \(C = 0\) or \(\Delta_1 = \Delta_2\). Therefore, a non-vanishing two-point function must satisfy \(\Delta_1 = \Delta_2\).

Let us comment on the constant \(C\), which remains unfixed. This constant cannot be determined by symmetry alone, as it depends on the normalization of the operators. Indeed, it is always possible to rescale an operator as \(\tilde{\mathcal{O}} = \lambda \mathcal{O}\), which rescales the two-point function by \(\lambda^2\). For this reason, \(C\) can be interpreted as the normalization of the operator \(\mathcal{O}\), and it is customary to set \(C = 1\).

This result can be straightforwardly extended to spinning operators by considering a scalar quantity obtained by contraction, e.g., \(\mathcal{O}^{\mu_1 \ldots \mu_n} \xi_{\mu_1} \ldots \xi_{\mu_n}\). The argument then proceeds exactly as in the scalar case, and the two-point function is again completely fixed.

However, some care must be taken with the interpretation of the normalization constant \(C\): not all operators allow it to be interpreted merely as a normalization. The main and most important example is the stress tensor. In that case, Ward identities impose that the two-point function coefficient is directly related to the three-point function of the stress tensor (to be discussed later), and its value defines the central charge \(C_T\) of the theory.

\paragraph{Three-point functions}

It is also possible to constrain three-point functions of local operators using only conformal symmetry. We will not discuss the detailed procedure here and refer the reader to \cite{Rychkov:2016iqz,Simmons-Duffin:2016gjk} for a complete treatment. Crucially, the three-point function of three scalar fields can be written as
\begin{equation}
    \langle \mathcal{O}_1(x_1) \mathcal{O}_2(x_2) \mathcal{O}_3(x_3) \rangle = \frac{f_{\mathcal{O}_1 \mathcal{O}_2 \mathcal{O}_3}}{|x_1 - x_2|^{\alpha_{123}} |x_1 - x_3|^{\alpha_{132}} |x_2 - x_3|^{\alpha_{231}}} \ ,
\end{equation}
where
\begin{equation}
    \alpha_{ijk} = \Delta_{\mathcal{O}_i} + \Delta_{\mathcal{O}_j} - \Delta_{\mathcal{O}_k} \ .
\end{equation}

The coefficient \(f_{\mathcal{O}_1 \mathcal{O}_2 \mathcal{O}_3}\) is again unfixed by symmetry. However, it is crucially different from the normalization constant in the two-point function. While it is also sensitive to the definitions and rescalings of the operators, once the normalizations of the operators appearing in the two-point functions are fixed, the three-point coefficient \(f_{\mathcal{O}_1 \mathcal{O}_2 \mathcal{O}_3}\) becomes a well-defined, physical quantity. Therefore, it is more than just a normalization constant: it is a dynamical parameter that characterizes the conformal field theory.

Higher-point functions can be decomposed into (infinite) sums of two- and three-point functions. However, in order to show this explicitly, we will need additional tools.

\subsection{Radial quantization and OPE}

We now show that there is a one-to-one correspondence between operators and states in conformal field theories. There are several approaches to demonstrate this; for the purposes of this thesis, we adopt the cylinder interpretation. The idea is to consider the metric on \(\mathbb{R} \times S^{d-1}\), which is related, via a Weyl transformation, to the flat space metric. In fact,
\begin{equation}
    ds_{\mathbb{R}^d}^2 = r^2 \left( \frac{dr^2}{r^2} + ds_{S^{d-1}}^2 \right) = e^{2\tau} ds_{\mathbb{R} \times S^{d-1}}^2 \ ,
\end{equation}
where \(r = e^\tau\). In these coordinates, dilatations become translations in \(\tau\), since \(r \to \lambda r\) corresponds to \(\tau \to \tau + \log \lambda\).

This implies that the \textit{time evolution} on the cylinder is generated by \(e^{- \op D \tau}\), or equivalently, the Hamiltonian on the cylinder is
\begin{equation}
    \op H_{\text{cyl}} = \frac{\op D}{R} \ ,
\end{equation}
where \(R\) is the radius of the sphere, introduced by dimensional analysis.

This means that we can associate to any energy eigenstate on the cylinder a local operator in the CFT via the identity
\begin{equation}
    |\Delta_{\mathcal{O}} \rangle = \mathcal{O}(0) |0\rangle \ .
\end{equation}

Unitarity then imposes the condition
\begin{equation}
    \langle \Delta | \Delta \rangle \ge 0 \ ,
\end{equation}
which leads to the unitarity bounds:
\begin{equation}
    \Delta \ge \frac{d-2}{2} \quad \text{(for scalars)} \ , \qquad
    \Delta \ge d - 2 + J \quad \text{(for spinning operators)} \ .
\end{equation}

These bounds are satisfied in all unitary conformal field theories.

The operator spectrum of a CFT is organized in specific patterns. Consider an operator \(\mathcal{O}(0)\); one can construct infinitely many states by acting with derivatives:
\begin{equation}
    |\Delta, n\rangle = \left. \partial^n \mathcal{O}(x) \right|_{x = 0} |0\rangle \ .
\end{equation}

The conformal dimensions of these states follow from the commutator \([ \op D, \op P_\mu ] = i \op P_\mu\), and satisfy \(\Delta_n = \Delta_{\mathcal{O}} + n\). In fact,
\begin{equation}
    |\Delta\rangle \overset{\op P_\mu}{\to} |\Delta + 1\rangle \overset{\op P_\mu}{\to} |\Delta + 2\rangle \overset{\op P_\mu}{\to} \cdots \ .
\end{equation}

Similarly, from \([ \op D, \op K_\mu ] = -i \op K_\mu\), we find that \(\op K_\mu\) lowers the conformal dimension by one unit:
\begin{equation}
    0 \overset{\op K_\mu}{\leftarrow} |\Delta\rangle \overset{\op K_\mu}{\leftarrow} |\Delta + 1\rangle \overset{\op K_\mu}{\leftarrow} \cdots \ .
\end{equation}

Note that while we can always apply \(\op P_\mu\) repeatedly to generate higher-dimension states, unitarity bounds prevent lowering the dimension indefinitely. Therefore, acting many times with \(\op K_\mu\) must eventually give a null vector. This allows us to distinguish: \textit{primary operators}, for which \(\op K_\mu |\mathcal{O}\rangle = 0\), \textit{descendant operators}, for which \(\op K_\mu |\mathcal{O}\rangle \neq 0\).

\begin{figure}[h]
\centering
\includegraphics[width=45mm]{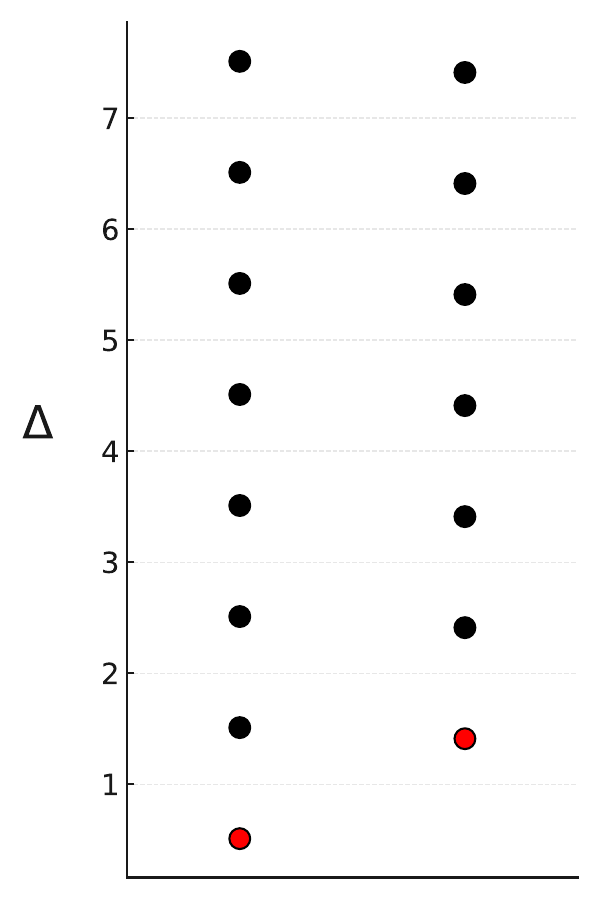}
\caption{Example of conformal multiplets in a conformal field theory (here shown for \(\sigma\) and \(\epsilon\) in the 3D Ising model). In red, the primary fields; in black, their conformal descendants.}
\label{fig:multiplets}
\end{figure}

Given a primary operator, all its conformal descendants can be generated by acting with \(\op P_\mu\). These sets of states form what are called \textit{conformal multiplets}. An example is shown in Fig.~\ref{fig:multiplets}.

Let us now consider a generic state \(|\Psi\rangle\), and expand it in energy eigenstates:
\begin{equation}
    |\Psi\rangle = \sum_n c_n |\mathcal{E}_n\rangle \ .
\end{equation}

An example is the state generated by acting with two scalar fields:
\begin{equation}
    \phi(x) \phi(0) |0\rangle = \sum_{\mathcal{O}} c_{\mathcal{O}}(x) \, \mathcal{O}(0) |0\rangle \ .
\end{equation}

This defines an expansion of the product of two operators in terms of other local operators in the theory: this is the \textit{Operator Product Expansion} (OPE).

Since two-point functions are diagonal, it is easy to show that
\begin{equation}
    c_{\mathcal{O}} \propto f_{\mathcal{O} \phi \phi} \ .
\end{equation}

This expansion also holds inside any correlation function, meaning that the OPE is valid operatorially.

In particular, the OPE between two scalar operators takes the form
\begin{equation}
    \phi(x) \phi(0) = \sum_{\mathcal{O}} f_{\phi \phi \mathcal{O}} \, |x|^{\Delta - 2\Delta_\phi - J} \, x_{\mu_1} \cdots x_{\mu_J} \, \mathcal{O}^{\mu_1 \ldots \mu_J}(0) \ .
\end{equation}

\subsubsection{Higher-point functions and the conformal data}

Observe that the OPE contains only information about two- and three-point functions—specifically, the conformal dimensions of the operators in the theory and the three-point function coefficients, also known as \textit{structure constants}. However, this is sufficient to reconstruct any \(n\)-point function, since one can use the OPE recursively to reduce an \(n\)-point function to an \((n{-}1)\)-point function. To see this, observe that
\begin{equation}
    \langle \mathcal{O}_1(x) \mathcal{O}_2(y) \mathcal{O}_3(z) \ldots \mathcal{O}_n(w) \rangle = \sum_{\mathcal{O} \in \mathcal{O}_1 \times \mathcal{O}_2} c_{\mathcal{O}}(x) \langle \mathcal{O}(y) \mathcal{O}_3(z) \ldots \mathcal{O}_n(w) \rangle \ ,
\end{equation}
where \(c_{\mathcal{O}}\) contains the three-point coefficient \(f_{\mathcal{O}_1 \mathcal{O}_2 \mathcal{O}}\) and a kinematic factor which is not relevant for the purpose of this argument.

By iterating this procedure, one can write an expansion for any \(n\)-point function entirely in terms of the conformal spectrum and structure constants. This implies that the full content of all correlation functions in the theory is encoded in the set of conformal dimensions, spins, and structure constants—collectively referred to as the \textit{conformal data}. Schematically,
\begin{equation}
    \{\langle \mathcal{O}_1(x_1) \ldots \mathcal{O}_n(x_n) \rangle\} \quad \Leftrightarrow \quad \left\{ \left\{ \Delta_{\mathcal{O}}, J_{\mathcal{O}} \right\}, f_{\mathcal{O}_1 \mathcal{O}_2 \mathcal{O}_3} \right\} \equiv \textit{conformal data} \ .
\end{equation}

In the presence of additional global symmetries, the conformal data must be extended to include the corresponding charges.

From the point of view of the cylinder interpretation, the conformal data correspond to knowledge of the Hamiltonian’s eigenstates and of how any physical state decomposes in terms of them.

Note that constraints on the conformal data may not fully characterize the theory. For instance, there exist examples of CFTs with the same spectrum but different structure constants \cite{Henriksson:2022dnu}.

\subsection{The special case of \( d = 2 \)}

Two dimensions constitute a very special case, and we will not cover all the features of two-dimensional conformal field theories in this thesis. For more complete treatments, see for instance \cite{DiFrancesco:1997nk,Mussardo:2020rxh}.

The crucial point is that we can formulate the theory using complex coordinates:
\begin{equation}
    z = x + i y \ , \hspace{1cm} \overline{z} = x - i y \ ,
\end{equation}
so that the flat metric reads \( ds^2 = dz \, d\overline{z} \). In these coordinates, the conformal transformations studied above generalize to
\begin{equation}
    z \to f(z) \ , \hspace{1cm} \overline{z} \to \overline{f}(\overline{z}) \ .
\end{equation}

It is easy to observe that the infinitesimal transformations \( \delta z \propto z^k \) correspond to conformal transformations. In particular, the cases \(k = 0, 1, 2\), i.e. the constant, linear, and quadratic terms, reproduce the conformal transformations already encountered in the \(d > 2\) case. However, in \(d = 2\), also all terms with \(k > 2\) correspond to valid conformal transformations.

This is not a paradox: the transformations that go beyond the M\"obius group are not globally well-defined on the entire complex plane. We will not delve into this technical point here, but it is important to emphasize that the structure described above implies the existence of an \emph{infinite-dimensional} symmetry algebra in \(d = 2\) \cite{Belavin:1984vu}.

All the previous discussion remains valid in \(d = 2\), but the theory is subject to much stronger constraints due to this enhanced symmetry.

\subsection{Solving the dynamics}

The discussion on CFT presented so far in this thesis has been entirely non-perturbative and mathematically well defined. However, it is natural to ask how one can compute the conformal data necessary to access any correlation function of a CFT. Before turning to the conformal bootstrap problem, which is the focus of this thesis, we first enumerate some alternative approaches:

\begin{itemize}
    \item[$\star$] \textit{Monte Carlo simulations}: This is a numerical method in which one typically constructs a lattice regularization of the theory and then, in a completely non-perturbative manner and without further assumptions, is able to numerically compute simple correlation functions. From these, conformal dimensions and structure constants can be extracted. The literature on this method is extensive; for example, for conformal dimensions and structure constant of various $\mathrm O(N)$ models (in particular $N = 1,2,3$) we refer to \cite{Hasenbusch:2010,Campostrini:2002cf,Hasenbusch:1999cc}; Its downsides include high computational cost for high-precision results, and the need for careful extrapolation to the continuum limit, which may not be straightforward.

    \item[$\star$] \textit{High-temperature lattice expansions}: From the point of view of lattice models, it is possible to develop a high-temperature expansion organized in powers of \(1/T\). This expansion typically diverges as the temperature approaches a \textit{critical temperature}, but certain critical exponents (i.e., conformal dimensions) can still be extracted from it. See for example \cite{Kramers:1941kn,Kramers:1941PartII,Butera:1997ak} for applications to the Ising model.

    \item[$\star$] \textit{Exact renormalization group approach}: This approach starts from the lattice and uses RG transformations directly, requiring that at the fixed point the Hamiltonian remains invariant under RG flow. In general, the resulting Hamiltonian is very complicated in terms of the original lattice variables and involves, in principle, an infinite number of terms. A truncation is typically necessary to obtain concrete results. This idea was already implemented in Wilson’s seminal paper in 1975 for the 2D Ising model \cite{Wilson:1974mb}.

    \item[$\star$] \textit{Hamiltonian truncation}: Assuming we know the UV fixed point of an RG flow—that is, we know its conformal data—it is possible to deform the \textit{formal action} of the theory by
    \begin{equation}
        \mathcal{A} = \mathcal{A}_{UV} + \lambda \int d^d x \, \Phi(x) \ ,
    \end{equation}
    where \(\Phi\) is a relevant operator. This action defines an RG flow. On the sphere, the Hamiltonian becomes
    \begin{equation}
        \op H = \frac{\op D}{R} - \lambda \int_{S_R^{d-1} \times \mathbb{R}} d^d x \, \Phi(x) \ ,
    \end{equation}
    and its matrix elements can be expressed solely in terms of the conformal dimensions and structure constants via expressions like \(\langle \Psi | \Phi | \Psi \rangle\). This method has been successfully applied to many massive theories, starting from the seminal paper by Yurov and Zamolodchikov \cite{Yurov:1989yu}, and more recently to the 2D Ising model via the flow from the free massless scalar to the Ising CFT \cite{Rychkov:2014eea,Rychkov:2015vap,Lencses:2024wib}. The generalization to higher dimensions is not straightforward, though progress has been recently made \cite{Elias-Miro:2020qwz}.

    \item[$\star$] \textit{Fuzzy sphere regularization}: Direct numerical exploration of conformal invariance has traditionally been difficult due to mathematical and conceptual challenges. Recently, a new approach based on non-commutative geometry—called \textit{fuzzy sphere regularization}—was introduced, allowing the study of the 3D Ising model in a new setting \cite{Zhu:2022gjc,Hu:2023xak}. This method has shown great promise due to its lower computational cost compared to methods like Monte Carlo, and it has produced first-time predictions for some observables in 3D Ising.

    \item[$\star$] \textit{\(\varepsilon\)-expansion}: In the early days of RG and phase transitions, Wilson and Fisher introduced an analytical method for computing physical quantities in CFTs based on a Ginzburg-Landau approach and dimensional expansion \cite{Wilson:1971dc}. The idea is to start from the upper critical dimension, where the theory is weakly coupled, and treat the spacetime dimension as a perturbative parameter. As we will discuss in Chapter~\ref{chap:ONmodel}, consider \(N\) scalar fields \(\varphi_i\) deformed by a quartic potential \(V \sim \lambda (\varphi_i \varphi_i)^2\) in \(d=4\), where the interaction is marginally irrelevant and does not generate an RG flow. Indeed, the beta function has only one (trivial) zero at \(\lambda = 0\). However, the same theory is known to describe the critical \(\mathrm{O}(N)\) model. In \(d = 4 - \varepsilon\), the beta function develops two zeros: one at \(\lambda = 0\), and a non-trivial one at \(\lambda \sim \varepsilon\), which defines a weakly coupled fixed point. Perturbation theory in \(\varepsilon\) can then be used to compute corrections and extract physical quantities. This method provides analytical control, often to high orders in \(\varepsilon\), in regimes inaccessible to Hamiltonian truncation or Monte Carlo. However, setting \(\varepsilon = 1\) or \(2\) to access physical dimensions like \(d = 3\) or \(2\) typically yields less accurate results and requires sophisticated resummation techniques, such as Padé or Borel analysis \cite{Giombi:2019upv,Giombi:2024zrt}.
\end{itemize}

\vspace{0.3cm}

In this thesis, the approach we adopt is the \textit{bootstrap}. The conformal bootstrap program was initiated independently by Ferrara, Grillo, and Gatto \cite{Ferrara:1973yt}, and by Polyakov \cite{Polyakov:1974gs}. These seminal works proposed a non-perturbative (or, as Polyakov emphasized, \textit{non-Hamiltonian}) approach, noting that in conformal field theory, a non-trivial constraint arises from the associativity of the OPE. Concretely,
\begin{equation}
    \langle \mathcal{O}_1 \mathcal{O}_2 \mathcal{O}_3 \mathcal{O}_4 \rangle = \sum_{\mathcal{O}} f_{\mathcal{O}_1 \mathcal{O}_2 \mathcal{O}} f_{\mathcal{O}_3 \mathcal{O}_4 \mathcal{O}} \, \text{block}(\mathcal{O}) = \sum_{\mathcal{O}'} f_{\mathcal{O}_1 \mathcal{O}_4 \mathcal{O}'} f_{\mathcal{O}_2 \mathcal{O}_3 \mathcal{O}'} \, \text{block}(\mathcal{O}') \ .
\end{equation}
This is the so-called \textit{crossing equation}, reminiscent of \(s\)- and \(t\)-channel equivalence in S-matrix theory. The precise forms of the \textit{blocks} is not necessary for this discussion.

Ferrara, Grillo, and Gatto focused on computing the conformal blocks, organizing them around primary operators and developing techniques such as the \textit{shadow formalism}. Polyakov instead used the crossing equation to derive dynamical results, including non-trivial predictions for anomalous dimensions at the Wilson-Fisher fixed point in \(d = 4 - \varepsilon\). Mack later made key contributions, including a rigorous proof of OPE convergence \cite{Mack:1975jr,Mack:1976pa}.

The first full solutions of the crossing equations appeared in 2D CFTs in the celebrated work of Belavin, Polyakov, and Zamolodchikov \cite{Belavin:1984vu}. They analytically solved theories with a finite number of Virasoro primaries—the so-called \textit{minimal models}—and introduced the term \textit{conformal bootstrap}.

It was only much later that significant results were obtained in higher dimensions. In a landmark paper, Rattazzi et al. \cite{Rattazzi:2008pe} proposed combining crossing symmetry with unitarity (\(f_{\mathcal{O}_1 \mathcal{O}_2 \mathcal{O}_3}^2 \geq 0\)) to obtain rigorous bounds on conformal dimensions. Remarkably, these bounds often exhibit \textit{kinks}, which are believed to correspond to actual physical theories—most notably the 3D Ising model \cite{El-Showk:2012cjh,El-Showk:2014dwa}.

Later, Simmons-Duffin and collaborators showed that considering multiple correlators simultaneously could constrain the space of allowed conformal data into small \textit{islands}, thus yielding not only bounds but also predictions \cite{Kos:2014bka,Simmons-Duffin:2015qma}. This method now provides extremely precise and rigorous determinations of conformal data for theories such as the 3D Ising model (see \cite{Chang:2024whx} for the latest results). While the outcome is formally a bound, the precision is often such that these are effectively treated as predictions.

Alternative approaches were also developed, notably by Gliozzi and collaborators \cite{Gliozzi:2013ysa,Gliozzi:2014jsa,Gliozzi:2016cmg}. Their method involves a severe truncation of the OPE and enforcement of crossing symmetry. While this approach lacks the rigor of the numerical bootstrap and does not require unitarity, it has been applied to non-unitary theories like the Lee-Yang model in \(d = 2,3,4,5\), where the approach explained before fails. Even in unitary theories, its simplicity makes it appealing despite lower precision, and research in this direction remains active.

Finally, the conformal bootstrap has also led to important analytical insights. Notably, one can extract asymptotic information about operators with large spin or large dimension directly from crossing symmetry \cite{Fitzpatrick:2012yx,Komargodski:2012ek}. In this context, we emphasise the role of analytic assumptions on the correlation functions, which made it possible to derive the \textit{inversion formula} \cite{Caron-Huot:2017vep,Simmons-Duffin:2017nub,Mazac:2018qmi} as well as certain \textit{dispersion relations} \cite{Caron-Huot:2020adz}. These formulae have led to analytic predictions, particularly for operators of large spin. Both the inversion formula and the dispersion relation make use of the analytical assumptions on the CFT correlators and CFT data to compute approximate analytical results from the crossing relation. The most famous result that can be obtained using these tools is that asymptotic large-spin operators appearing in the OPE of two scalars have vanishing anomalous dimensions. More precisely, in the OPE of two scalars $\phi \times \phi$, there are operators whose classical forms are $\phi \partial^J \Box^n \phi$, denoted as $[\phi \phi]_{n,J}$, and whose dimensions are given by
\begin{equation}
    \Delta_{[\phi \phi]_{n,J}} \overset{J \to \infty}{=} 2\Delta_\phi + 2n + J + \mathcal{O}\left(\frac{1}{J}\right)\ .
\end{equation}
The physics behind this result is the following: there is a limit of the crossing equation in which the contribution of the identity operator dominates. One can show that, in the crossed channel, an infinite tower of operators is required to compensate for the UV divergence generated by the identity. These operators can be identified precisely with the $[\phi \phi]_{n,J}$ operators, and their conformal dimensions, as given above, can be computed using either the inversion formula or dispersion relations, with the identity contribution as input. Corrections at large but finite spin arise due to the presence of the lightest operator in the OPE after the identity.

In this thesis, both the numerical approach and analytic bootstrap methods will be developed at finite temperature. In Section~\ref{eq:bootstrapporoblem}, we will provide all the details about the finite-temperature bootstrap problem. In Section~\ref{sec:numericalProblem}, we will explain why the numerical approach of~\cite{El-Showk:2012cjh,El-Showk:2014dwa} cannot be straightforwardly generalized to the finite-temperature case, and we will instead adapt a version of the Gliozzi method, improved by some analytical techniques. In~\cite{Iliesiu:2018fao}, an inversion formula for thermal two-point functions was already proposed, and in~\cite{Alday:2020eua}, a dispersion relation was also derived. All of this is reviewed and extended in Section~\ref{sec:analyticprop}, and used concretely to make predictions about thermal dynamics in Section~\ref{sec:analyticbootstrap}.

\section{Finite temperature effects in quantum field theory}
\subsection{Finite-temperature field theory: basic methods}

 The goal of this thesis is to formulate a non-perturbative bootstrap problem for thermal effects in conformal field theories. Before addressing this topic, let us briefly recall, in this section, some basics of thermal quantum field theory.

\vspace{0.2cm}

A finite-temperature quantum system is defined through the thermal density matrix:
\begin{equation}
    \rho = e^{-\beta \op H} \ ,
\end{equation}
where \(\beta = 1/T\) is the inverse temperature \footnote{In this thesis we will work in units where the Boltzmann constant is set to $k_B = 1$.}. The partition function is then given by
\begin{equation}
    Z(\beta) = \operatorname{Tr}e^{-\beta \op H} = \sum_{\mathcal E} e^{-\beta \mathcal E} \langle \mathcal E|\mathcal E \rangle = \sum_{\mathcal E} e^{-\beta \mathcal E} \ ,
\end{equation}
where in the last step we have chosen an orthonormal basis of energy eigenstates, \(\langle \mathcal E | \mathcal E \rangle = 1\).

Thermal correlation functions are defined as traces over the Hilbert space:
\begin{multline}
    \langle \mathcal O_1(x_1)\ldots \mathcal O_n(x_n) \rangle_\beta = Z^{-1}(\beta) \operatorname{Tr}\left(e^{-\beta \op H} \mathcal O_1(x_1)\ldots \mathcal O_n(x_n)\right) = \\
    = Z^{-1}(\beta) \sum_{\mathcal E} e^{-\beta \mathcal E} \langle \mathcal E | \mathcal O_1(x_1)\ldots \mathcal O_n(x_n)|\mathcal E\rangle \ ,
\end{multline}
where we use the notation \(\langle \cdot \rangle_\beta\) to denote thermal expectation values. From this expression, one sees that finite-temperature correlators can be written as infinite sums of zero-temperature matrix elements.

There are several standard approaches to describe and compute observables at finite temperature. In this thesis, we will work in the \textit{Matsubara}, or \textit{imaginary time}, formalism. However, let us briefly review all the main methods:

\begin{itemize}
    \item[$\star$] \textbf{Thermofield double (TFD)}: The idea is to consider a doubled Hilbert space, \(\mathcal H_{TFD} = \mathcal H \otimes \mathcal H\), and define the TFD state as
    \begin{equation}
        \ket{TFD}= Z^{-1/2}(\beta)\sum_{\mathcal E}e^{-\beta \mathcal E/2} \ket{\mathcal E}\otimes \ket{\mathcal E} \ .
    \end{equation}
    This state plays the role of a thermal vacuum, in the sense that thermal correlators can be expressed as
    \begin{equation}
        \langle \mathcal O_1(x_1) \ldots \mathcal O_n(x_n) \rangle_\beta = \langle TFD|  \mathcal O_1(x_1) \ldots \mathcal O_n(x_n) |TFD\rangle \ ,
    \end{equation}
    where the operators act only on the first copy of the Hilbert space, i.e. \(\mathcal O \equiv \mathcal O \otimes \op 1\).

    \item[$\star$] \textbf{Matsubara formalism}: This approach relies on the observation that \(e^{-\beta \op H}\) corresponds to time evolution by imaginary time \(t = i \beta\). One can prove that $\tau \equiv \tau +\beta$ and therefore, introducing imaginary time \(\tau\), one considers the system in Euclidean signature, compactified on a thermal circle. In this setup, the QFT is placed on \(\mathbb S^1_\beta \times \mathbb R^{d-1}\), and temperature effects are encoded in the geometry.

    \item[$\star$] \textbf{Keldysh-Schwinger formalism}: This method incorporates both imaginary and real-time evolution. The imaginary time component captures the thermal effects (similarly to Matsubara), but by considering complex contours in the time plane, one can study real-time dynamics. Unlike Matsubara, which assumes equilibrium, the Keldysh-Schwinger formalism allows the treatment of non-equilibrium systems.
\end{itemize}

\subsection{Matsubara formalism and KMS}
We consider in this thesis the system at finite temperature using the Matsubara formalism, as proposed in \cite{Matsubara:1955ws}. Practically, this amounts to considering imaginary time evolution \( t = i \tau \), so that
\begin{equation}\label{eq:densitymat}
    \mathcal O(\tau) = e^{\tau \op H} \mathcal O e^{-\tau \op H} \ .
\end{equation}

We now restrict to \( 0 < \tau < \beta \), for reasons that will become clear shortly. Let us consider the two-point function of bosonic operators:
\begin{multline}
    g(\tau_1,\tau_2) = -Z^{-1}(\beta)\operatorname{Tr}\left(e^{-\beta \op H}e^{\tau_2 \op H} \mathcal O_2 e^{-\tau_2 \op H} e^{\tau_1\op H} \mathcal O_1 e^{-\tau_1 \op H}\right) =\\
    =  -Z^{-1}(\beta)\operatorname{Tr}\left(e^{-\beta \op H} \mathcal O_2 e^{-(\tau_2-\tau_1) \op H}\mathcal O_1 e^{(\tau_2-\tau_1) \op H}\right) \ ,
\end{multline}
where we have assumed \( \tau_2 - \tau_1 > 0 \). Defining \( \tau = \tau_1 - \tau_2 \), we obtain
\begin{multline}
    g(\tau) = Z^{-1}(\beta) \operatorname{Tr}\left(e^{-\beta \op H} \mathcal O_2 e^{\tau \op H} \mathcal O_1 e^{-\tau \op H}\right) 
    = Z^{-1}(\beta) \operatorname{Tr}\left(\mathcal O_2 e^{\tau \op H} \mathcal O_1 e^{-(\beta + \tau) \op H}\right) = \\
    = Z^{-1}(\beta) \operatorname{Tr}\left(e^{-\beta \op H} \mathcal O_1 e^{-(\tau + \beta) \op H} \mathcal O_2 e^{(\tau + \beta) \op H} \right) = g(\tau + \beta) \ .
\end{multline}

This shows that the function \( g(\tau) \) is periodic in imaginary time with period \( \beta \), i.e.,
\begin{equation}
    g(\tau) = g(\tau + \beta) \ .
\end{equation}
This condition is known as the \textit{Kubo-Martin-Schwinger} (KMS) condition \cite{Kubo:1957mj,Martin:1959jp}. Because of this periodicity, we are effectively studying correlation functions on the geometry \( S_\beta^1 \times \mathbb R^{d-1} \), with periodic boundary conditions for bosonic fields.

An analogous computation can be performed for fermions. However, due to Fermi-Dirac statistics, the thermal boundary conditions for fermions are anti-periodic, introducing a minus sign in the corresponding relation.

\smallskip

Note that the derivation of the KMS condition here relies on the definition of the thermal density matrix in equation~\eqref{eq:densitymat}. From an axiomatic perspective, however, this is not necessary. In fact, it is more common to define a thermal quantum field theory as a theory satisfying the \textit{Wightman axioms} (or the \textit{Osterwalder-Schrader axioms} in Euclidean signature), together with the KMS condition as an additional axiom. 

In this algebraic approach, the KMS condition plays the fundamental role of characterizing thermal equilibrium. We will see later that this condition is the key equation from which we will derive predictions and results for thermal effects in conformal field theories. 

For a classic reference on the axiomatic framework, see \cite{streater2000pct}, and for recent developments concerning conformal symmetry within this context, see \cite{Kravchuk:2020scc,Kravchuk:2021kwe}.

\subsection{Comments about perturbation theory}
All the discussion so far has been completely non-perturbative and does not even rely on the existence of a Lagrangian formulation of the theory. Of course, if a Lagrangian formulation is available and the theory is weakly coupled, one can attempt to study the system using standard perturbation theory by replacing the manifold \( \mathbb R^d \) with the thermal manifold \( S_\beta^1 \times \mathbb R^{d-1} \).

In this context, the massive propagator in momentum space is exactly the same as at zero temperature:
\begin{equation}
    G(p) = \frac{1}{\omega_n^2 + \vec p^2 + m^2} \ ,
\end{equation}
where \( \omega_n = 2\pi n/\beta \), and the Fourier transform reflects the fact that we are compactifying time on a circle:
\begin{equation}
    G(\tau, \vec x) = \sum_{n = -\infty}^\infty e^{i \omega_n \tau} \int \frac{d^{d-1}p}{(2\pi)^{d-1}} e^{i \vec p \cdot \vec x} G(p) \ .
\end{equation}

Using the Poisson resummation formula, one can show that
\begin{equation}
    G(\tau, \vec x) = \sum_{n = -\infty}^\infty \tilde G(\tau - n\beta, \vec x) \ ,
\end{equation}
where \( \tilde G(\tau, \vec x) \) is the Fourier transform of \( G(p) \) in \( \mathbb R^d \). Therefore,
\begin{equation}
    G(\tau, \vec x) = \sum_{n = -\infty}^\infty G_0(\tau - n\beta, \vec x) \ ,
\end{equation}
where \( G_0 \) denotes the zero-temperature propagator. The formula above is commonly referred to as the \textit{method of images}. Using this method, one can compute any correlation function perturbatively using standard Feynman diagrams and loop computations on the thermal manifold \( \mathbb S_\beta^1 \times \mathbb R^{d-1} \). Nevertheless, there are some crucial differences that must be considered.

\paragraph{The infrared problem.}
It may happen that a theory which is weakly coupled at zero temperature becomes strongly coupled at finite temperature. A main focus of this thesis will be the \( \mathrm{O}(N) \) models, whose Ginzburg-Landau description involves a quartic potential, as we shall explain later. For now, let us illustrate the potential issues of perturbation theory at finite temperature in the case of the \( \phi^4 \) theory in \( d = 4 \). Consider a Lagrangian with potential
\begin{equation}
    V = \frac{1}{2} m^2 \phi^2 + \frac{1}{4!} \lambda \phi^4 \ .
\end{equation}

Since thermal effects are equivalent to studying the QFT on the thermal geometry \( S_\beta^1 \times \mathbb R^{d-1} \), one can apply Kaluza-Klein reduction, yielding the effective Lagrangian:
\begin{equation}\label{eq:lagphi4}
    \mathcal L = \frac{1}{2}(\partial \phi_0)^2 + \frac{\lambda}{\beta 4!} \phi_0^4 + \sum_{n = 1}^\infty \left[ \partial \phi_n \partial \overline{\phi}_n + \frac{4\pi^2 n^2}{\beta^2} |\phi_n|^2 + \frac{\lambda}{2\beta} |\phi_n|^2 \phi_0^2 \right] \ .
\end{equation}
This Lagrangian is now defined in \( d = 3 \). We omitted further self-interactions among the Kaluza-Klein modes. The fields \( \phi_n \) are massive, with mass \( m_n = 2\pi n/\beta \), while \( \phi_0 \) is massless and features a \( \phi_0^4 \) interaction. At zero temperature, the \( \phi^4 \) operator is marginally irrelevant and the theory is infrared free. However, in \( d = 3 \), the same operator becomes relevant, and the infrared dynamics below the scale \( \lambda/\beta \) is governed by strong coupling effects.

Hence, the dynamics of \( \phi_0 \) may drive the theory to strong coupling in the IR, introducing an infrared problem absent at zero temperature. In this specific case, however, we are saved from this problem thanks to the last term in the Lagrangian in~\eqref{eq:lagphi4}. As we will also discuss for the \( \mathrm O(N) \) model in Chapter~\ref{chap:ONmodel}, a one-loop computation shows that \( \phi_0 \) acquires a thermal mass:
\begin{equation}
    m_\text{th}^2 \propto \frac{\lambda}{\beta^2} \ ,
\end{equation}
which is sufficiently large to ensure that the theory remains weakly coupled, as long as \( \lambda \ll 1 \). In this regime, the thermal mass lies above the strong coupling scale, and perturbation theory remains self-consistent. This ensures that the perturbative computations we will present in Chapter \ref{chap:ONmodel} will be consistent and can be used and can be compared with other non-perturbative methods. 

For further details on this example, see \cite{Kapusta:1989tk,Bellac:2011kqa,Chai:2020zgq}.

\section{Summary of the chapter}

In this chapter, we reviewed the basics of conformal field theories and thermal quantum field theories, focusing on the key concepts and results that will be used in the following. In particular, in Section~\ref{eq:CFTprimer}, we presented an introduction to conformal invariance, explaining its physical origin in terms of fixed points of the renormalization group flow, and briefly discussed the distinction between scale-invariant and conformally invariant theories. 

We then focused on conformal field theories, introducing the conformal group, its generators, and the conformal algebra. The consequences of conformal invariance are encoded in the conformal Ward identities, from which we derived the form of two- and three-point functions. Crucially, we also explained why one-point functions vanish for any operator in a conformal field theory except the identity. This will play a fundamental role when studying thermal effects, where one-point functions are generically non-zero at finite temperature.

Radial quantization can be viewed as the study of a conformal field theory on the geometry $\mathbb R \times S^{d-1}$. In this picture, the Hamiltonian corresponds to the dilatation operator, and one can construct a one-to-one map between states and (local) operators of the CFT. This framework justifies the operatorial expansion of a product of two operators into an infinite sum of local operators, known as the operator product expansion (OPE):
\begin{equation}
    \mathcal O_1(x) \mathcal O_2(y) = \sum_{\mathcal O} c(x) \mathcal O(y) \ .
\end{equation}
Using the OPE, any correlation function can be decomposed in terms of conformal data: the scaling dimensions and the three-point function coefficients.

This conformal data can be rigorously bounded and, in some cases, computed via the conformal bootstrap program. This approach constrains the data by imposing the associativity of the OPE in a four-point function. We also commented on other numerical and analytical techniques for computing the conformal data.

We then introduced finite-temperature effects in quantum field theory. Thermal effects were defined via a thermal density matrix $\rho = e^{-\beta \op H}$, where $\beta$ is the inverse temperature and $\op H$ the Hamiltonian. We outlined the main approaches to thermal field theory, focusing on the Matsubara (imaginary-time) formalism, which amounts to placing the CFT on the geometry $S^1_\beta \times \mathbb R^{d-1}$, with periodic (anti-periodic) boundary conditions for bosons (fermions). The resulting periodicity condition along the thermal circle is known as the KMS condition, and it will play a fundamental role in the remainder of this thesis, acting as the thermal analogue of OPE associativity in the bootstrap framework.

We also briefly discussed perturbation theory at finite temperature, providing explicit rules for the propagators and noting the possible emergence of infrared problems. Importantly, these issues are negligible in $\phi^4$ theory for $2 \le d \le 4$ (we explicitly considered $d = 4$). This is relevant for applications to the $\mathrm O(N)$ model, where the $\phi^4$ theory provides the Ginzburg–Landau description of the Ising fixed point.

\chapter{Finite temperature effects in CFT: kinematics}\label{chap:ThermalKIno}
\section{Thermal data and OPE}
As discussed previously in Chapter~\ref{chap:basics}, conformal field theories provide a well-established framework in which a model is fully characterized by its conformal data. This data consists of the spectrum of scaling operators—equivalently, the energy eigenstates on the sphere—and the operator product expansion (OPE) structure constants.  
Schematically, disregarding charges under global symmetries, the conformal data is given by:
\begin{equation}
   \text{conformal data} \sim \left\{\{\Delta_{\mathcal O}, J_{\mathcal O}\},f_{\mathcal O \mathcal O' \mathcal O''}\right\} \ .
\end{equation} 

When considering finite-temperature effects, we expect conformal symmetry to be broken. This is because temperature introduces a scale, implying that the theory is no longer scale-invariant. Consequently, one might anticipate that the minimal set of data required to characterize a thermal CFT is larger than in the zero-temperature case.

However, this expectation is conceptually incorrect. In fact, by employing the real-time formalism, any thermal correlation function can, in principle, be expressed in terms of an infinite number of zero-temperature correlation functions. This follows from the fact that, using the thermal density matrix $\rho = e^{-\beta \op H}$, we can always write
\begin{multline}
    \langle \mathcal O_1(x_1)\ldots \mathcal O_n(x_n) \rangle_\beta = \frac{\operatorname{Tr}\left[e^{-\beta \op H}\mathcal O_1(x_1)\ldots \mathcal O_n(x_n)\right]}{\operatorname{Tr}(e^{-\beta \op H})} = \\
    = \frac{\sum_{\alpha} e^{-\beta \mathcal E_\alpha} \langle \mathcal E_\alpha |  \mathcal O_1(x_1)\ldots \mathcal O_n(x_n) |\mathcal E_\alpha \rangle}{\sum_{\alpha}e^{-\beta \mathcal E_\alpha}} \ .
\end{multline}

Nonetheless, it is very challenging to make practical use of this fact when studying thermal effects in CFTs. To understand why, it suffices to consider the CFT formulated on $\mathbb R \times S_R^{d-1}$. In this setup, radial quantization maps the dilatation operator in flat space to the Hamiltonian on the sphere, $\op H = \op D/R$, where $R$ is the radius of the $(d-1)$-sphere. As a consequence, it becomes natural to study thermal correlation functions in this setting via
\begin{equation}\label{eq:Sd1}
    \langle \mathcal O_1(x_1)\ldots \mathcal O_n(x_n) \rangle_{\beta,R} = \frac{\sum_{\mathcal O} e^{-\beta \frac{\Delta_{\mathcal O}}{R}} \langle \mathcal O | \mathcal O_1(x_1)\ldots \mathcal O_n(x_n) |\mathcal O \rangle_R}{\sum_{\mathcal O} e^{-\beta \frac{\Delta_{\mathcal O}}{R}} \langle \mathcal O| \mathcal O \rangle_R} \ ,
\end{equation}
where we defined $\langle \cdot \rangle_{\beta,R}$ as thermal correlation functions of the CFT on $\mathbb R \times S^{d-1}$, and $\langle \cdot \rangle_R$ as zero-temperature correlation functions on the same geometry.

To be very concrete, let us consider a scalar operator $\phi$ and study its thermal one-point function. The equation above simplifies to
\begin{equation}
    \langle \phi \rangle_\beta = \lim_{R \to \infty} \frac{\sum_{\mathcal O} e^{-\beta \frac{\Delta_{\mathcal O}}{R}} \langle \mathcal O | \phi |\mathcal O \rangle_R}{\sum_{\mathcal O} e^{-\beta \frac{\Delta_{\mathcal O}}{R}} \langle \mathcal O |\mathcal O \rangle_R} \ .
\end{equation}
Notice that, in the limit where we consider thermal effects in a CFT on flat space, the numerator of thermal correlation functions involves an infinite sum of three-point function coefficients. This sum is regulated by potential divergences arising from the thermal partition function in the denominator.

Therefore, performing this computation requires exact knowledge of an infinite set of conformal data, which is beyond reach for any generic CFT, particularly in dimensions $d > 2$.

\medskip

\begin{figure}[t]
\centering
 \includegraphics[width=99mm]{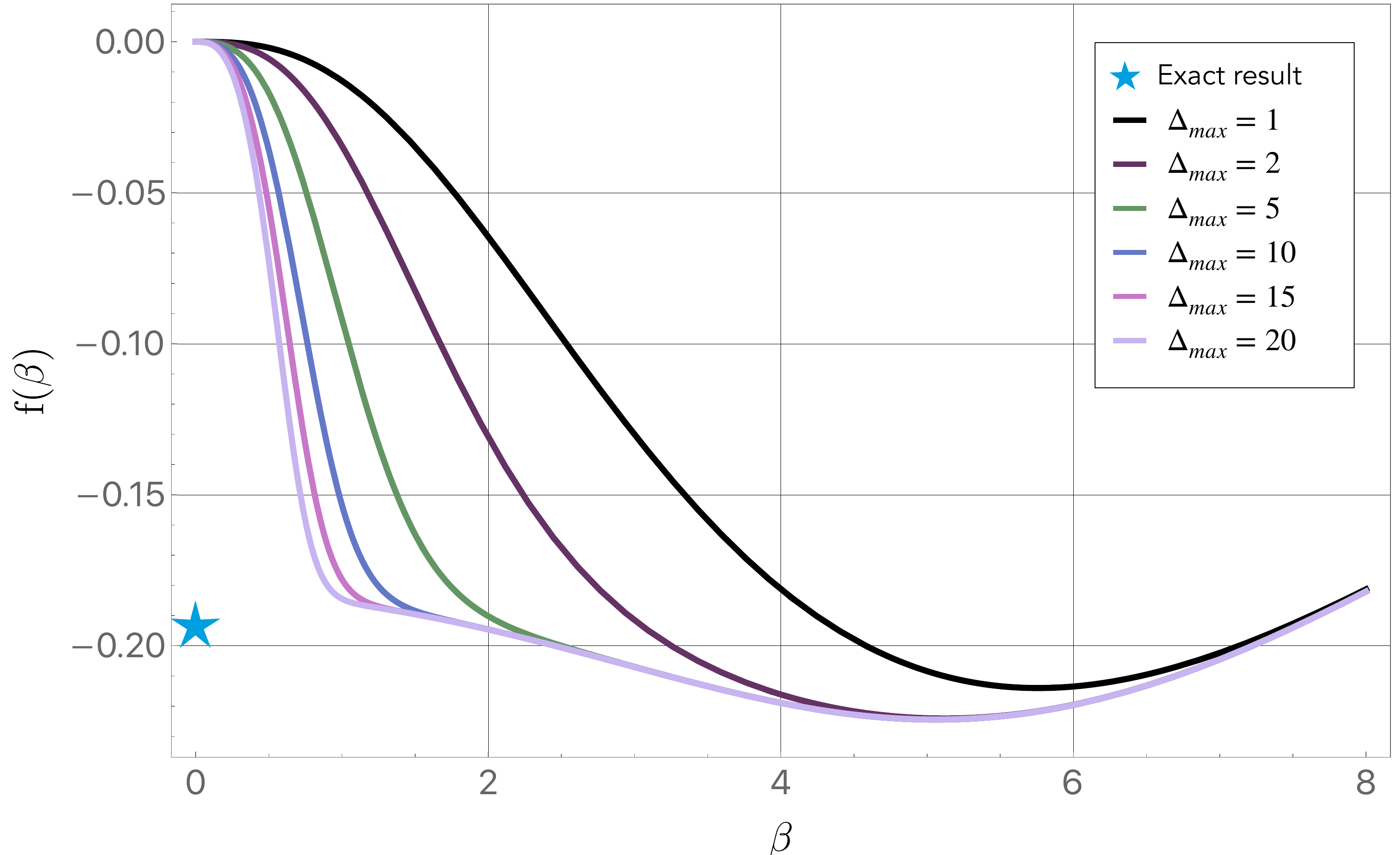}
    \caption{The function $f(\beta)$ for various cutoffs $\Delta_\text{max}$, compared with the exact expectation as $\Delta_\text{max}\to \infty$.}
    \label{fig:FiniteVf}
\end{figure}

To appreciate how difficult this limit is to take, one can, for instance, compute the free energy from the thermal partition function of the theory on $\mathbb R \times S_R^{d-1}$, as done in Appendix A of~\cite{Iliesiu:2018fao}. Consider, for example, the partition function of a free theory in $d=3$, given by
\begin{equation}
    Z(q = e^{-\beta}) = \prod_{i = 0}^\infty \left(1 - q^{i+1/2}\right)^{-2i-1} \ ,
\end{equation}
where we normalize the radius of $S^{d-1}_R$ to $R = 1$. The formula above can be truncated at some power of $q$, corresponding to computing the partition function up to a conformal dimension $\Delta_\text{max}$. Using this expansion, we define a function which approximates the free energy density in the infinite temperature limit:
\begin{equation}
    f(\beta) = \frac{1}{8 \pi^2} \beta^3 \frac{\partial}{\partial \beta} \log Z(e^{-\beta}) \ .
\end{equation}
The result for different values of $\Delta_\text{max}$ is shown in Fig.~\ref{fig:FiniteVf}.

In particular, observe that the only way to recover the correct free energy density on $\mathbb R^d$ is to take the limit $\beta \to 0$, which gives a nonzero result only if $\Delta_\text{max} \to \infty$.

\medskip

For the reasons outlined above, it is convenient to distinguish between two cases:  
(i) thermal effects on the CFT defined on $\mathbb R \times S^{d-1}_R$, where the expansion in Eq.~\eqref{eq:Sd1} is a useful tool, at least at low temperature, and has been recently explored in the literature \cite{Gobeil:2018fzy,Alkalaev:2024jxh,Buric:2024kxo}; and  
(ii) thermal effects in flat space, i.e., on $\mathbb R^d$.  

In the latter case, it is more efficient—as we partially show in this thesis—to work with an enlarged set of \textit{thermal CFT data}. As explained in the previous chapter, one way to introduce thermal effects in a QFT is to consider the theory on $S_\beta^1 \times \mathbb R^{d-1}$, where $\beta = 1/T$ is the inverse temperature. Thus, the relevant \textit{thermal conformal data} corresponds to the minimal data needed to characterize a generic CFT on this \textit{thermal manifold}. This different approach is also used in $\mathbb R \times S^{d-1}_R$ at high temperature for example for the $\mathrm O(N)$ model at large $N$ \cite{David:2024pir}.

\medskip

To identify this new data, recall that in a zero-temperature CFT, two- and three-point functions fully characterize the theory. This is because the Operator Product Expansion (OPE) can be iteratively applied to construct any $n$-point function (for $n > 3$), reducing it to a sum of two- and three-point data. This procedure can also be generalized to the finite temperature case, since the manifold $S^1 \times \mathbb R^{d-1}$ is conformally flat\footnote{Conformally flat here means that there exists a conformal map locally identifying the thermal geometry with $\mathbb R^d$. This map is not globally defined (unless $d = 2$), so one cannot use it to map correlators from flat space to the thermal manifold.}.

In fact, in conformally flat manifolds, the OPE is still convergent in a finite region, and operatorially it remains identical to that at zero temperature. Concretely, let us examine the two-point function of two identical scalar operators. The OPE reads:
\begin{equation}
    \phi(x) \times \phi(0) = \sum_{\mathcal O^{\mu_1\ldots \mu_J}} f_{\phi \phi \mathcal O} |x|^{\Delta - 2\Delta_\phi - J} x_{\mu_1}\ldots x_{\mu_J} \mathcal O^{\mu_1\ldots \mu_J} \ ,
\end{equation}
and hence in correlation functions (within the convergence radius) we obtain:
\begin{equation}\label{eq:OPEincorr}
    \langle \phi(x) \phi(0) \rangle =  \sum_{\mathcal O^{\mu_1\ldots \mu_J}} f_{\phi \phi \mathcal O} |x|^{\Delta - 2\Delta_\phi - J} x_{\mu_1}\ldots x_{\mu_J} \langle \mathcal O^{\mu_1\ldots \mu_J} \rangle \ .
\end{equation}
At zero temperature, the expression above collapses to a single term, since only the identity operator has a nonzero one-point function. This follows from dilatation invariance, as already anticipated in Chapter \ref{chap:basics}.

At finite temperature, when correlators are computed on $S_\beta^1 \times \mathbb R^{d-1}$, equation~\eqref{eq:OPEincorr} still holds. However, since dilatation symmetry is broken, one-point functions of scaling operators may be non-zero. This leads to an effective expansion of the two-point function. Higher-point functions can similarly be expanded using the OPE, meaning that—alongside the standard conformal data—thermal one-point functions must be included.  

Thus, the complete thermal conformal data is:
\begin{equation}
   \text{thermal conformal data} \sim \left\{\{\Delta_{\mathcal O}, J_{\mathcal O}\},f_{\mathcal O \mathcal O' \mathcal O''},\langle \mathcal O^{\mu_1\ldots \mu_J}\rangle_\beta\right\} \ .
\end{equation} 

In the next section, we show that thermal one-point functions reduce effectively to a single coefficient per operator—the new thermal data, which is central to the developments in this thesis.

\medskip

Before proceeding with the discussion of broken and unbroken symmetries, let us revisit the issue of OPE convergence at finite temperature. Until now, we have simply assumed the OPE converges locally when the operators are sufficiently close. But how close is “sufficient”?

This can be quantified by considering periodicity along the thermal circle. In unitary theories, when two operators are very close, the two-point function behaves as:
\begin{equation}
    \langle \phi(\tau, \vec x) \phi(0) \rangle_\beta \sim \frac{1}{\left(\tau^2 + \vec x^2\right)^{2\Delta_\phi}} \ .
\end{equation}
This behavior is dictated by the leading contribution from the identity operator in the OPE. Due to periodicity in the thermal circle, we also have:
\begin{equation}
    \langle \phi(\beta - \tau, \vec x) \phi(0) \rangle_\beta \sim \frac{1}{\left(\tau^2 + \vec x^2\right)^{2\Delta_\phi}} \ .
\end{equation}
However, in this regime, no operator in the OPE accounts for the divergence, which must then signal the breakdown of OPE convergence. Therefore, we conclude that the OPE converges only when the Euclidean distance between the two operators satisfies:
\begin{equation}
    \tau^2 + \vec x^2 < \beta^2 \ .
\end{equation}
On the other hand, within this region, the correlator is expected to be finite and free of other singularities. The rigorous proof of this is non-trivial~\cite{Iliesiu:2018fao,Katz:2014rla} and will be discussed in the final sections of this chapter. The conclusion is that the convergence radius of the OPE at finite temperature is set by $\beta$.

\section{Broken and unbroken symmetries}
    In this section, we extract a Ward identity (or a set thereof) for each symmetry transformation on the thermal manifold $\mathcal M_\beta = S_\beta^1\times \mathbb R^{d-1}$. In general, we do not expect these symmetries to be preserved. For instance, dilatations are expected to be broken due to the insertion of a scale in the theory, represented by the temperature $T$ (hereafter appearing through its inverse $\beta = 1/T$). 

During the derivation, the non-triviality of the geometry leads to modifications of the standard flat-space Ward identities. We interpret these modifications as \emph{symmetry breakings}. Although the focus of this thesis is on the thermal manifold, the logic of this section can be extended to a theory on a generic manifold $\mathcal M$. In fact, the same framework was used in~\cite{Marchetto:2023fcw} to compute broken Ward identities on the geometry $T^2\times \mathbb R^{d-2}$.

\medskip

Let us consider a generic CFT on $\mathbb R^{d}$. Observables in this theory are constructed from $n$-point correlation functions:
\begin{equation}
    \Braket{\cO{1}\dots \cO{n}} \ .
\end{equation}
Due to the invariance under the global conformal group, these observables satisfy specific symmetry properties. We are interested in how such properties are modified when the correlation functions are computed on a non-trivial manifold $\mathcal M$, where conformal symmetry is generically expected to be broken.\footnote{This can be understood as the breaking of some covariantly constant conformal Killing vectors on a manifold that is not conformally equivalent to $\mathbb R^d$. In this work we consider specific manifolds that retain some unbroken symmetries, such as $S^1\times \mathbb R^{d-1}$ or $T^2\times \mathbb R^{d-2}$.}

\begin{equation}
    \Braket{\cO{1}\dots \cO{n}}_{\mathcal M} \ , \label{eq: ncorrfunct}
\end{equation}
where $\braket{\dots}_{\mathcal M}$ denotes correlation functions evaluated on the manifold $\mathcal M$, with the implicit understanding that the operators satisfy appropriate boundary conditions tied to the geometry of $\mathcal M$.\footnote{For example, on the thermal manifold $\mathcal{M}_\beta = S^1_\beta \times \mathbb R^{d-1}$, bosonic operators are periodic and fermionic operators are anti-periodic along the thermal circle.}

\medskip

The derivation of Ward identities in flat space is standard in the literature (see e.g.~\cite{DiFrancesco:1997nk,Papadodimas:2009eu}). Let us start from a theory defined by a formal action $\mathcal A$, and consider an infinitesimal symmetry transformation of the fields $\varphi' (x) = \varphi(x) -i \omega_a \op G_a \varphi(x)$, where
\begin{equation}
   i \omega_{a}\op G_a \varphi(x) = \omega_{a}\frac{\delta x^\mu}{\delta \omega_a}\partial_\mu \varphi - \omega_{a}\frac{\delta \mathcal F}{\delta \omega_a} \ . \label{eq: symmtrans} 
\end{equation}
Here, $\mathcal F$ is a function that parameterizes the field variation under the generator $\op G_a$.

In Section~\ref{sec: Conformal group}, we will focus on the conformal group, whose action on generic correlators is known and summarized in Table~\ref{tab: currents and generators}. These results can be easily extended to the superconformal case.

\medskip

The correlation function \eqref{eq: ncorrfunct} can be written as a path integral over the fundamental fields:
\begin{equation}
    \Braket{\cO{1}\dots \cO{n}}_{\mathcal M} = \frac{1}{\mathcal Z}\int [d \phu] \  \mathcal O_1(x_1) \ldots \mathcal O_n(x_n) e^{-\mathcal A[\varphi]} \ ;
\end{equation}
and since the fields $\phu$ are just integration variables, we can perform the transformation \eqref{eq: symmtrans} as a change of variables:
\begin{multline}\label{eq: Correlator identity}
     \Braket{\cO{1}\dots \cO{n}}_{\mathcal M} = \\ = \frac{1}{\mathcal Z}\int [d \phu']\ \left (\mathcal O_1(x_1)+\delta \mathcal O_1(x_1)\right) \ldots \left (\mathcal O_n(x_n)+\delta \mathcal O_n(x_n) \right) e^{-\mathcal A'[\varphi']} \ ,
\end{multline}
where $\delta \mathcal O_{i}(x_i)$ denotes the variation of the $i$-th local operator under the transformation.

Assuming invariance of the theory under rigid symmetry transformations, the transformed action on a generic manifold $\mathcal M$ can be written as:
\begin{equation} \label{eq: S transf}
    \mathcal A'[\varphi'] =  \mathcal A[\varphi]+ \int_{\mathcal M}d^d x  \sqrt{g} \ 
    \nabla_\mu \omega_a(x) J_a^{\mu}(x) \ ,
\end{equation}
where $g$ is the determinant of the metric on $\mathcal M$, and $J_a^\mu$ is the current associated with the symmetry generated by $\op G_a$.

In a Lagrangian theory, these currents can be written explicitly using the canonical stress-energy tensor:
\begin{equation}
    J_a^\mu = \frac{\delta x^\mu}{\delta \omega_a}\mathcal L - \frac{\partial \mathcal L}{\partial (\partial _\mu \varphi)} \partial_\nu \varphi \frac{\delta x^\nu }{\delta \omega_a} + \sum_{\phu}\frac{\partial \mathcal L}{\partial (\partial_\mu \varphi)}\frac{\delta \mathcal F}{\delta \omega_a} = T_{\text{can.}}^{\mu \nu}\frac{\delta x_\nu}{\delta \omega_a} + \sum_{\phu}\frac{\partial \mathcal L}{\partial (\partial_\mu \varphi)}\frac{\delta \mathcal F}{\delta \omega_a} \ .
\end{equation}

When restricted to conformal transformations, these expressions can be rewritten in terms of the symmetric, traceless stress-energy tensor $T^{\mu \nu}$.\footnote{As in the flat case~\cite{DiFrancesco:1997nk}, the broken Ward identities are unaffected by replacing the canonical stress tensor with the traceless one; any differences due to improvement terms correspond to divergences that are removed upon normal ordering.} Explicit expressions for these currents are collected in Table~\ref{tab:ExplicitThermalContributions}.

\medskip

By inserting the transformed action into~\eqref{eq: Correlator identity} and expanding to first order in the infinitesimal parameters $\omega_a$, we obtain:
\begin{multline}\label{eq: Noether1}
    \sum_i  \left \langle \mathcal O_1(x_i) \ldots \delta \mathcal O_i(x_i) \ldots \mathcal O_n(x_n) \right \rangle_{\mathcal M} =  \left \langle \delta \mathcal A \  \mathcal O_1(x_1) \ldots \mathcal O_n(x_n) \right \rangle_{\mathcal M} \\
    = \int_{\mathcal M} d^d x \sqrt{g} \ \nabla_\mu \omega_a (x) \left \langle J_a^\mu (x) \mathcal O_1(x_1) \ldots \mathcal O_n(x_n) \right \rangle_{\mathcal M} \ .
\end{multline}

Recalling that $\delta \mathcal O(x) = -i \omega_a(x) \op G_a \mathcal O(x)$ and taking functional derivatives with respect to $\omega_b(y)$, the identity becomes:
\begin{multline}
    -i \sum_{i} \delta(x_i - y) \left \langle \mathcal O_1(x_1) \ldots \op G_a \mathcal O_i(x_i) \ldots \mathcal O_n(x_n) \right \rangle_{\mathcal M} = \\
    = \int_{\mathcal M} d^d x \sqrt{g} \ \partial_\mu \delta(x - y) \left \langle J_a^\mu(x) \mathcal O_1(x_1) \ldots \mathcal O_n(x_n) \right \rangle_{\mathcal M} \ .
\end{multline}

Switching to derivatives in $y$ and simplifying gives the \textit{unintegrated broken Ward identity}:
\begin{multline}\label{eq:UnintegratedBWI}
    i \sum_i \delta(x_i - y) \left \langle \mathcal O_1(x_1) \ldots \op G_a \mathcal O_i(x_i) \ldots \mathcal O_n(x_n) \right \rangle_{\mathcal M} = \\
    = \nabla_\mu^y \left \langle J_a^\mu(y) \mathcal O_1(x_1) \ldots \mathcal O_n(x_n) \right \rangle_{\mathcal M} \ .
\end{multline}

    It is important to note that the current on the right-hand side of equation~\eqref{eq:UnintegratedBWI} acts iteratively on all operators in the correlator. For example, specializing to the thermal manifold $\mathcal M_\beta = S^1 \times \mathbb R^{d-1}$, the broken dilatation Ward identity for a one-point function becomes:
\begin{equation}\label{eq:Noether1}
    i \delta(x - y) \left \langle [\op D, \mathcal O](x) \right \rangle_\beta = \frac{\partial}{\partial y^\mu} \left[ (y - x)_\nu \left \langle T^{\mu\nu}(y) \mathcal O(x) \right \rangle_\beta \right] \ ,
\end{equation}
where from now on we denote thermal correlators as $\langle \cdot \rangle_\beta = \langle \cdot \rangle_{\mathcal M_\beta}$. 

This relation reflects the interpretation of $\op D$ as an integrated charge over a $(d-1)$-dimensional surface centered on the operator $\mathcal O$. This is consistent with the definition of the generator as the integral of the corresponding current:
\begin{equation} \label{eq: dil example}
    \op D \, \mathcal O(x) = \int d^{d-1}y \, (x - y)_\nu T^{0\nu}(y) \mathcal O(x) \ .
\end{equation}

While the unintegrated Ward identities are quite explicit, they can be simplified upon specifying the geometry. As previously mentioned, we now specialize to the thermal manifold $\mathcal M_\beta = S_1^\beta \times \mathbb R^{d-1}$. Integrating both sides of equation~\eqref{eq:Noether1} yields:
\begin{multline}\label{eq: general manifold Ward identity}
    i \sum_i \left \langle \mathcal O_1(x_1) \ldots \op G_a \mathcal O_i(x_i) \ldots \mathcal O_n(x_n) \right \rangle_{\beta} = \\
    = \int_0^\beta d y_0 \int_{\mathbb R^{d-1}} d^{d-1} y \, \frac{\partial}{\partial y^\mu} \left \langle J_a^\mu(y) \mathcal O_1(x_1) \ldots \mathcal O_n(x_n) \right \rangle_{\beta} \ .
\end{multline}

In flat space $\mathbb R^d$, the boundary term resulting from the divergence vanishes (under appropriate fall-off conditions), and one recovers the standard, unbroken Ward identities:
\begin{equation}
    i \sum_i \left \langle \mathcal O_1(x_1) \ldots \op G_a \mathcal O_i(x_i) \ldots \mathcal O_n(x_n) \right \rangle_{\mathbb R^d} = 0 \ . \label{eq: flat WI}
\end{equation}

On a generic manifold $\mathcal M$, however, the boundary term is generally nonzero and must be interpreted as a \emph{breaking term} correcting the Ward identity.

\medskip

To make this analysis explicit, we parametrize the thermal circle $S^1_\beta$ using the coordinate $\tau \in [0, \beta)$, with the identification:
\begin{equation}
    \tau \equiv \tau + \beta \ .
\end{equation}

We impose periodic or antiperiodic boundary conditions on the fields, depending on their spin: bosons are periodic, fermions are antiperiodic. (This is an important point for supersymmetry breaking.)

The integral on the right-hand side of~\eqref{eq: general manifold Ward identity} splits into two contributions. The first, along the thermal circle, is:
\begin{equation}\label{eq: boundary term 1}
    \int_{\mathbb R^{d-1}} d^{d-1} y \ \left\langle \left[J_a^0(\beta, \vec y) - J_a^0(0, \vec y) \right] \mathcal O_1(x_1) \ldots \mathcal O_n(x_n) \right\rangle_\beta \ .
\end{equation}

The second is the contribution from the spatial boundary at large radius:
\begin{equation}\label{eq: boundary term 2}
    \lim_{R \to \infty} \int_0^\beta d\tau \int_{S^{d-2}} d\Omega \ R^{d-2} n_i \left\langle J_a^i(\tau, R, \Omega) \mathcal O_1(x_1) \ldots \mathcal O_n(x_n) \right\rangle_\beta \ ,
\end{equation}
where $n_i$ is the unit normal vector on the spatial boundary, and $d\Omega$ the measure on the $(d-2)$-sphere.

\medskip

This term can be studied using the clustering property:
\begin{equation}\label{eq: clastering 1}
    \lim_{R \to \infty} \left\langle J_a^i(\tau, R, \Omega) \mathcal O_1(x_1) \ldots \mathcal O_n(x_n) \right\rangle_\beta = \left\langle J_a^i(\tau, R, \Omega) \right\rangle_\beta \left\langle \mathcal O_1 \ldots \mathcal O_n \right\rangle_\beta \ .
\end{equation}

Because translation invariance is unbroken, one-point functions on the thermal manifold must be constant in space and $\tau$, so the integral in~\eqref{eq: boundary term 2} is divergent unless $\langle J_a^i \rangle_\beta = 0$. This gives a possible infrared divergence proportional to the volume:
\begin{equation}
    \langle \mathcal O_1 \ldots \mathcal O_n \rangle_\beta \times \text{Vol}[\mathcal M_\beta] \ . \label{eq: bdy IR div}
\end{equation}

Hence, this term must be renormalized or discarded, and in what follows we simply set it to zero.

\medskip

Returning to the first boundary term, we define the operator
\begin{equation}\label{eq: breaking term definition}
    \Gamma_a^\beta(\vec y) = J_a^0(\beta, \vec y) - J_a^0(0, \vec y) \ ,
\end{equation}
so that the breaking term reads:
\begin{equation}\label{eq: breaking term of the Ward identity}
    \int_{\mathbb R^{d-1}} d^{d-1} y \ \left\langle \Gamma_a^\beta(\vec y) \mathcal O_1(x_1) \ldots \mathcal O_n(x_n) \right\rangle_\beta \ .
\end{equation}

From this, we immediately see that translations and spatial rotations are unbroken, since the corresponding currents are periodic along the thermal circle. On the other hand, dilatations, time-like rotations, and special conformal transformations are generically broken due to the non-periodicity of their currents. Supersymmetry is also broken, due to the anti-periodicity of the supercurrent.

\medskip

Thus, the final form of the broken Ward identities reads:
\begin{equation}\label{eq:GeneralBWI}
    i \sum_i \left\langle \mathcal O_1(x_1) \ldots \op G_a \mathcal O_i(x_i) \ldots \mathcal O_n(x_n) \right\rangle_\beta = \int_{\mathbb R^{d-1}} d^{d-1} y \ \left\langle \Gamma_a^\beta(\vec y) \mathcal O_1 \ldots \mathcal O_n \right\rangle_\beta \ .
\end{equation}

Among all symmetries, the broken Ward identity for dilatations carries special significance: it captures the breaking of scale invariance and can be interpreted as a non-perturbative version of the Callan–Symanzik equation. In fact, the presence of a new scale—namely the temperature—modifies the dilatation identity to:
\begin{equation}\label{eq:KallaS}
    \left( \op D + \beta \frac{\partial}{\partial \beta} \right) \left\langle \mathcal O_1(x_1) \ldots \mathcal O_n(x_n) \right\rangle_\beta = 0 \ .
\end{equation}
A derivation of this equation can be found in Appendix A of~\cite{Marchetto:2023fcw}, but it can also be understood heuristically as a dimensional analysis statement: a scale transformation can be compensated by rescaling the temperature.

\medskip

\begin{table}[h!]
    \centering
    \renewcommand{\arraystretch}{1.5}
    \begin{tabular}{|c|c|}
        \hline
        Symmetry & Infinitesimal transformation: $i\omega_{a} \op G_{a}\cO{i}$ \\
        \hline
        \hline
        Translations & $a^{\mu}\partial_\mu \cO{i}$ \\
        \hline
        Dilatations & $b\left( x^\mu \partial_\mu + \Delta_i \right)\cO{i}$ \\
        \hline
        Rotations & $c^{\mu\nu}\left( -x_\mu \partial_\nu + x_\nu \partial_\mu + i \op S_{\mu\nu} \right) \cO{i}$ \\
        \hline
        Special conf. tr. & $d^\mu \left( -x^2 \partial_\mu + 2 x_\mu x^\nu \partial_\nu + 2 x_\mu \Delta_i - 2i x^\nu \op S_{\mu\nu} \right) \cO{i}$ \\
        \hline
    \end{tabular}
    \caption{Symmetries of the global conformal group and their action on a local operator $\cO{i}$ in an irreducible Lorentz representation~\cite{DiFrancesco:1997nk}. $a^\mu$, $d^\mu$ are vectors; $b$ is a scalar; $c^{\mu\nu}$ is an antisymmetric tensor; $\op S_{\mu\nu}$ is the spin operator.}
    \label{tab: currents and generators}
\end{table}

    Explicit checks of equation \eqref{eq:GeneralBWI} with the breaking terms of Table \ref{tab:ExplicitThermalContributions} were performed in two-dimensional CFTs and free theory in four space-time dimensions.
 \subsection{Conformal Group at finite temperature}\label{sec: Conformal group}
    By plugging the general broken Ward identity~\eqref{eq:GeneralBWI} into the explicit breaking terms associated with the symmetries of the conformal group (whose corresponding operators are listed in Table~\ref{tab:ExplicitThermalContributions}), it is possible to derive concrete constraints on thermal correlation functions in a generic CFT at finite temperature.

In Table~\ref{tab:ExplicitThermalContributions}, we report—for each symmetry in the conformal group—the explicit expression for the corresponding current and the associated breaking term $\Gamma^\beta_a(x)$, computed via equation~\eqref{eq: breaking term definition}. For notational clarity, we introduce the shorthand:
\begin{equation}
    \mathcal{O}_{1 \cdots n}(\bm{x}) = \mathcal{O}_1(x_1) \cdots \mathcal{O}_n(x_n) \ .
\end{equation}

\begin{table}[h!]
    \centering
    \renewcommand{\arraystretch}{1.5}
    \begin{tabular}{|c|c|c|}
        \hline
        Symmetry & Currents: $\omega_a J_a^\mu$ & Breaking terms: $\Gamma^\beta_a(0, \vec x)$ \\
        \hline \hline
        Translations & $a_\nu T^{\mu\nu}$ & $0$ \\
        \hline
        Dilatations & $b x_\nu T^{\mu\nu}$ & $\beta T^{00}(0, \vec x)$ \\
        \hline
        Spatial rotations & $-c_i^{\ j} x_j T^{\mu i} + c_i^{\ j} x_j T^{\mu i}$ & $0$ \\
        \hline
        Boosts & $-c_i^{\ 0} \tau T^{\mu i} + c_i^{\ 0} \tau T^{\mu i}$ & $\beta T^{0i}(0, \vec x)$ \\
        \hline
        S.c.t. on $S^1_\beta$ & $d(-x^2 \delta_{0\nu} + 2\tau x_\nu) T^{\mu\nu}$ & $\beta\left[(\beta - 2\tau) T^{00}(0, \vec x) + 2(x_i - y_i) T^{0i}(0, \vec x)\right]$ \\
        \hline
        S.c.t. on $\mathbb R^{d-1}$ & $d^i (-x^2 \delta_{i\nu} + 2 x_i x_\nu) T^{\mu\nu}$ & $\beta\left[(\beta + 2\tau) T^{0j}(0, \vec x) + 2(x_i - y_i) T^{ij}(0, \vec x)\right]$ \\
        \hline
    \end{tabular}
    \caption{Explicit expressions for $\Gamma^\beta_a(x)$ associated with the symmetries of the global conformal group. The coordinates $(\tau, \vec y)$ specify the insertion point of the breaking term. Note that thermal effects explicitly break the manifest $SO(d)$ symmetry, but preserve $SO(d-1)$ invariance.}
    \label{tab:ExplicitThermalContributions}
\end{table}

\paragraph{Translations.}
The Ward identities associated with translations (in both time and space) remain unbroken at finite temperature. This results in the constraints:
\begin{equation}\label{eq: unbroken translations}
    \sum_{r=1}^{n} \frac{\partial}{\partial \tau_r} \Braket{\mathcal{O}_{1 \cdots n}(\bm{x})}_\beta = 0 \ , \qquad
    \sum_{r=1}^{n} \frac{\partial}{\partial x_r^i} \Braket{\mathcal{O}_{1 \cdots n}(\bm{x})}_\beta = 0 \ .
\end{equation}

\paragraph{Dilatations.}
The dilatation Ward identity is broken by the presence of a finite temperature, which introduces a scale into the theory. This is captured by the broken identity:
\begin{equation}\label{eq: dilatation BWI}
    \sum_{r=1}^{n} \left(\tau_r \frac{\partial}{\partial \tau_r} + x_r^i \frac{\partial}{\partial x_r^i} + \Delta_r \right)
    \Braket{\mathcal{O}_{1 \cdots n}(\bm{x})}_\beta =
    \beta \int d^{d-1} y \left\langle T^{00}(0, \vec y) \mathcal{O}_{1 \cdots n}(\bm{x}) \right\rangle_\beta \ ,
\end{equation}
where $\Delta_r$ denotes the scaling dimension of the $r$-th operator.

\paragraph{Rotations.}
From Table~\ref{tab:ExplicitThermalContributions}, we see that only the Lorentz boosts are associated with nontrivial breaking terms. Spatial rotations yield unbroken Ward identities:
\begin{equation}\label{eq: unbr rot wi}
    \sum_{r=1}^{n} \left(x_{r,i} \frac{\partial}{\partial x_r^j} - x_{r,j} \frac{\partial}{\partial x_r^i} - i \op S_{ij}^r \right)
    \Braket{\mathcal{O}_{1 \cdots n}(\bm{x})}_\beta = 0 \ ,
\end{equation}
where $\op S_{\mu\nu}^r$ acts on the spin indices of the $r$-th operator. Conversely, the boosts lead to the broken Ward identity:
\begin{equation}\label{eq: rotation BWI}
    \sum_{r=1}^{n} \left(\tau_r \frac{\partial}{\partial x_r^i} - x_{r,i} \frac{\partial}{\partial \tau_r} + i \op S_{i0}^r \right)
    \Braket{\mathcal{O}_{1 \cdots n}(\bm{x})}_\beta =
    \beta \int d^{d-1} y \ \Braket{T^{0i}(0, \vec y) \mathcal{O}_{1 \cdots n}(\bm{x})}_\beta \ .
\end{equation}

\paragraph{Special Conformal Transformations.}
We now turn to the special conformal transformations (SCTs), whose Ward identities are broken by temperature. For SCTs along the thermal circle $S^1_\beta$, the broken identity reads:
\begin{multline}\label{eq: sct1 gen}
    \sum_{r=1}^{n} \left[ (\tau_r^2 - x_r^i x_r^i) \frac{\partial}{\partial \tau_r}
    + 2 \tau_r x_r^i \frac{\partial}{\partial x_r^i}
    + 2 \tau_r \Delta_r
    + 2i x_r^i \op S^r_{i0} \right] \Braket{\mathcal{O}_{1 \cdots n}(\bm{x})}_\beta = \\
    = \sum_{r=1}^{n} \beta (\beta - 2\tau_r) \int d^{d-1} y \left[ 
    \Braket{T^{00}(0, \vec y) \mathcal{O}_{1 \cdots n}(\bm{x})}_\beta +
    (x_i - y_i) \Braket{T^{0i}(0, \vec y) \mathcal{O}_{1 \cdots n}(\bm{x})}_\beta
    \right] \ .
\end{multline}

Finally, for SCTs along the spatial directions of $\mathbb R^{d-1}$, we obtain:
\begin{multline}\label{eq: sct2 gen}
    \sum_{r=1}^{n} \left[
    (\tau_r^2 + x_r^j x_r^j) \frac{\partial}{\partial x_r^i}
    - 2 x_{r,i} x_r^\mu \frac{\partial}{\partial x_r^\mu}
    - 2 x_{r,i} \Delta_r
    + 2i \tau_r \op S_{i0}^r
    + 2i x_r^j \op S_{ij}^r \right]
    \Braket{\mathcal{O}_{1 \cdots n}(\bm{x})}_\beta = \\
    = \sum_{r=1}^{n} \beta (\beta + 2\tau_r) \int d^{d-1} y \left[
    \Braket{T^{00}(0, \vec y) \mathcal{O}_{1 \cdots n}(\bm{x})}_\beta +
    (x_i - y_i) \Braket{T^{0i}(0, \vec y) \mathcal{O}_{1 \cdots n}(\bm{x})}_\beta
    \right] \ .
\end{multline}

   \subsection{Hamiltonian and momentum} \label{ssec: ham and mom}

\begin{figure}[htb]
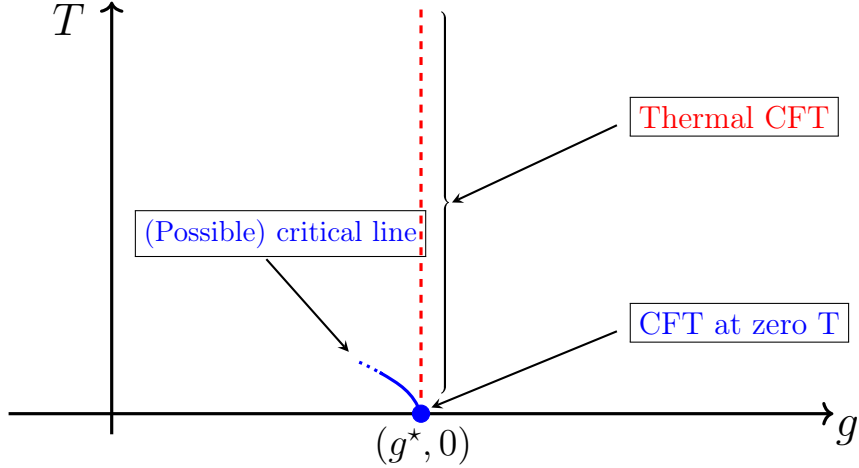

\centering
\Phasediagram
\caption{A schematic representation of the $(g,T)$ plane, where $T$ is the temperature and $g$ is a coupling constant of the theory. The zero-temperature CFT sits at the critical point $(g^\star,0)$, represented by a blue dot. In this paper, we study the thermal CFT, which corresponds to the QFT lying along the vertical dashed red line above the critical point. For clarity, a pictorial representation of a hypothetical line of critical points near the zero-temperature fixed point is also included, to distinguish these from the thermal CFT.}
\label{fig:Scheme}
\end{figure}

In this section, we derive the operatorial expressions for the Hamiltonian and the momentum at finite temperature. The energy spectrum obtained by diagonalizing such a Hamiltonian corresponds to the spectrum of the theories lying on the vertical axis above the critical point at zero temperature, as depicted in Fig.~\ref{fig:Scheme}.

\vspace{0.3em}

It is instructive to observe that the broken Ward identities \eqref{eq: dilatation BWI} and \eqref{eq: rotation BWI} acquire a compelling physical interpretation upon introducing the Hamiltonian and the spatial momentum operators\footnote{The minus sign arises from our use of Euclidean signature, where $T_\text{Lorentzian}^{00} = i^2 T_\text{Euclidean}^{00} = -T_\text{Euclidean}^{00}$. Similarly, the factor of $i$ in front of the momentum stems from $T_\text{Lorentzian}^{0i} = i T_\text{Euclidean}^{0i}$.}:
\begin{equation}\label{eq: Hamiltonian and momentum}
    H = -\int_{\mathbb R^{d-1}} d^{d-1}x \ \left(T^{00}(\tau,\vec x)+\frac{d-1}{d} \frac{b_T}{\beta^{d}} \right)\ , \qquad  
    P^i = i \int_{\mathbb R^{d-1}} d^{d-1}x \ T^{0i}(\tau,\vec x) \ ,
\end{equation}
where~\cite{Iliesiu:2018fao}
\begin{equation}
    \left \langle T^{00} \right\rangle_\beta = - \frac{d-1}{d} \frac{b_T}{\beta^{d}} \ .
\end{equation}
This relation, describing the one-point function of the stress-energy tensor, follows directly from the dilatation Ward identity \eqref{eq: dilatation BWI} evaluated on one-point functions.

Using translational invariance for one-point functions, equation~\eqref{eq: dilatation BWI} becomes:
\begin{equation}\label{eq: D= H}
    \op H = -\frac{1}{\beta} \op D - E_0 \ ,
\end{equation}
where $\op H$ is the Hamiltonian of the finite temperature theory, which should be interpreted as a $(d{-}1)$-dimensional interacting QFT at finite temperature~\cite{Iliesiu:2018fao,Benjamin:2023qsc}. The energy levels of this theory can be expressed in terms of the conformal spectrum of the $d$-dimensional CFT. Using the operator-state correspondence, and a basis $\ket{i} = \mathcal O_i \ket{0}$ (with $\mathcal O_i$ a scaling operator), the matrix elements of the Hamiltonian are:
\begin{equation}\label{eq: D= H expli}
    \langle i | \op H | j \rangle_\beta = - \frac{\Delta_i}{\beta} \delta_{ij} - E_0 \ ,
\end{equation}
where $\{ \Delta_i \}$ denotes the set of conformal dimensions of the operators at zero temperature.

Equation~\eqref{eq: D= H expli} reveals that the zero-temperature conformal data encodes the finite-temperature energy spectrum—that is, the spectrum of the theory on the vertical red axis in Fig.~\ref{fig:Scheme}. A clarification for attentive readers: a CFT at finite temperature is neither a critical point nor a fixed point of the RG flow, as dilatation invariance is explicitly broken. In general, all couplings run with temperature. However, one may consider fixing the theory's couplings to their critical values at zero temperature and then study the resulting theory as a function of temperature. In this context, temperature becomes a scale, and observables (such as the thermal mass) vary accordingly.

The operatorial expressions for the Hamiltonian and momentum in this nontrivial quantum system, parametrized by $\beta$, can thus be understood as consequences of the broken Ward identities. From the above, the energy levels at finite temperature take the form\footnote{\label{footnote Lorentzian} The minus sign is due to the Euclidean signature. In Lorentzian signature, the first term becomes positive.}:
\begin{equation}\label{eq: FT energy}
    E_n = -\frac{\Delta_n}{\beta} - E_0 \ ,
\end{equation}
where $\Delta_n$ are ordered from lowest to highest. Unitarity implies that the minimal $\Delta_n$ corresponds to the identity operator ($\Delta=0$). $E_0$ is the energy of the thermal vacuum and incorporates the Casimir energy of the system\footnote{The Casimir energy at finite temperature can be interpreted as the contribution of the thermal gas: while at $T=0$ we have $\langle T^{00} \rangle = 0$, at finite temperature $\langle T^{00} \rangle_\beta = - \left(1 - \frac{1}{d} \right) \frac{b_T}{\beta^d}$, giving an energy density $\mathcal E_0^{\text{gas}} = - \frac{d-1}{d} \frac{b_T}{\beta^{d-1}}$~\cite{Iliesiu:2018fao}.}.

Equation~\eqref{eq: FT energy} is consistent with dimensional analysis: since $\beta$ is the only scale, $E \propto 1/\beta$\footnote{Except for $E_0$, which is proportional to $\text{Vol}(\mathcal M_\beta)$ times a dimensionful constant.}. The nontriviality lies in the precise coefficient. 

As previously discussed in~\cite{El-Showk:2011yvt}, the broken dilatation Ward identity can be modified by a temperature derivative term:
\begin{equation} \label{eq: elshowk}
    \left( \op D + \beta \frac{\partial}{\partial \beta} \right)
    \langle \mathcal O_1(x_1) \ldots \mathcal O_n(x_n) \rangle_\beta = 0 \ ,
\end{equation}
where $\beta \partial_\beta$ acts effectively as a Hamiltonian and is directly identified with $- \op D / \beta$, validating the previous identification.

\paragraph{Momentum.}
Analogously, the boost operator $\op L_i$ is related to the momentum operator via the broken Ward identity~\eqref{eq: rotation BWI}:
\begin{equation}\label{eq: S= P}
    \op P_i = \frac{i}{\beta} \op L_i \ ,
\end{equation}
indicating that the dilatation and boost operators generate time and space translations, respectively, on the thermal circle $S^1_\beta$.

\vspace{0.5em}

As noted in~\cite{Iliesiu:2018fao}, the definitions of $H$ and $P$ in equation~\eqref{eq: Hamiltonian and momentum} are natural when interpreting the system thermally. However, in a Kaluza-Klein perspective—where a spatial, not temporal, direction is compactified—it is more natural to define:
\begin{align}
    H &= -\int_0^\beta d\tau \int_{\mathbb R^{d-2}} d^{d-2}x \left( T^{11}(\tau, \vec x) + \frac{1}{d} \frac{b_T}{\beta^d} \right) \ , \\
    P^i &= i \int_0^\beta d\tau \int_{\mathbb R^{d-2}} d^{d-2}x \ T^{1i}(\tau, \vec x) \ , \qquad i = 0,2,3,\ldots
\end{align}

\vspace{0.5em}

There is also a third natural possibility we do not explore here, namely a finite-volume CFT. In the language of this thesis, such a theory lives on the manifold $\mathbb R \times S_R^{d-1}$, with time along $\mathbb R$ and space compactified on a $(d{-}1)$-sphere. In $d=2$, the Hamiltonian and momentum are well understood~\cite{Cardy:1989da,Cardy:1991kr,Mussardo:2020rxh}, and follow from conformal maps between $\mathbb C$ and the cylinder $\mathbb R \times S^1$. In higher dimensions, similar constructions exist~\cite{Iliesiu:2018fao,Hogervorst:2013sma,Hogervorst:2014rta}, and equations~\eqref{eq: D= H}, \eqref{eq: S= P} closely resemble their finite-volume counterparts (modulo Euclidean conventions), suggesting a deeper relation between thermal and finite-volume CFTs.

This relation is well understood in $d=2$ via modular invariance~\cite{Cardy:1986ie,DiFrancesco:1997nk,Mussardo:2020rxh}, but remains unclear in higher dimensions. As proposed in~\cite{Berg:2023pca,Downing:2023uuc}, one might attempt to generalize modular invariance to higher dimensions by considering the CFT on $S^1_\beta \times S_R^{d-1}$. This geometry interpolates between finite-volume and thermal limits, depending on whether $R \to \infty$ or $\beta \to \infty$, respectively.

Furthermore, the consistency of the theory on the thermal manifold may itself be seen as a form of generalized modular invariance~\cite{El-Showk:2011yvt}, as the same Hamiltonian structure arises in both finite-temperature and finite-volume settings.

\vspace{0.5em}

In conclusion, the Hamiltonian of a finite-temperature CFT can be written in terms of the dilatation operator of the zero-temperature theory, and the momentum in terms of the boost operator. This suggests that some thermal observables can, in principle, be accessed via zero-temperature CFT data, through the broken Ward identities.

\paragraph{Thermal mass.}
A central observable of interest is the \textit{thermal mass}. Although not a physical mass, it controls the exponential decay of two-point functions at large spatial separation:
\begin{equation}
    \langle \phi(\tau, \vec x) \phi(0) \rangle_\beta \simeq \langle \phi(0) \rangle_\beta^2 + \mathcal O\left(e^{-m_\text{th} |\vec x|}\right) \ .
\end{equation}
For $\vec x = (x, 0, 0, \ldots)$, this can be expressed as:
\begin{equation}
    \langle \phi(0) \phi(\tau, x) \rangle_\beta = \langle \phi(0) e^{i \op H \tau - \op P_1 x} \phi(0) \rangle_\beta \ ,
\end{equation}
where $\op H$ and $\op P_1$ are the operators previously defined. The thermal mass is then the smallest non-zero eigenvalue of $\op P_1$, and can be extracted from the operator-state correspondence:
\begin{equation}
    m_\text{th} = \frac{1}{\beta} \min_{\Delta, J} \left\langle \mathcal O_{\Delta, J} \left| i \op L_1 \right| \mathcal O_{\Delta, J} \right\rangle = \frac{1}{\beta} \min_{\Delta, J} \lim_{y \to \infty} y^{2\Delta} \langle \mathcal O_{\Delta,J}(y) i \op S_{01} \mathcal O_{\Delta,J}(0) \rangle \ ,
\end{equation}
excluding the zero eigenvalue (corresponding to the identity operator). All correlators here are computed at zero temperature.

\subsubsection{An implicit version of the Cardy formula}

As previously anticipated:
\begin{equation}
    \frac{\partial}{\partial \beta} \langle \mathcal O(0) \rangle_\beta = -\frac{1}{\beta} \int d^{d-1}x \ \langle T^{00}(0, \vec x) \mathcal O(0) \rangle_\beta \ .
\end{equation}
This identity, also found in~\cite{El-Showk:2011yvt}, has a simple and clear interpretation. The inverse temperature $\beta$ is related to the metric component $g_{00}$; since $g_{00}$ acts as a source for $T^{00}$, changing $\beta$ corresponds to inserting $T^{00}$ into correlators.

A notable case is when $\mathcal O = T^{00}$, yielding:
\begin{equation}\label{eq: Cardy formula}
    \frac{\partial}{\partial \beta} \langle T^{00}(0) \rangle_\beta = -\frac{1}{\beta} \int d^{d-1}x \ \langle T^{00}(x) T^{00}(0) \rangle_\beta \ .
\end{equation}
Equation~\eqref{eq: Cardy formula} can be viewed as a differential form of the Cardy formula generalized to arbitrary spacetime dimensions~\cite{El-Showk:2011yvt}. Through thermodynamic identities, one can relate the entropy of the system to the coefficient $b_T$, which plays the role of the central charge $c$ in $d=2$ CFTs\footnote{Indeed, in two dimensions, $b_T \propto c$.}. Since $b_T$ is implicit in~\eqref{eq: Cardy formula}, and not generally known (except in special cases), the generalization of the Cardy formula to higher dimensions remains an open and compelling problem~\cite{DiPietro:2014bca,Assel:2015nca,Mukhametzhanov:2019pzy,Carlip:2000nv,Brevik:2004sd,Wang:2001bf,Benjamin:2023qsc}.

   \subsection{Thermal one-point functions} \label{ssec: Thermal 1-point Functions}

We consider a generic thermal one-point function $\langle \mathcal O(x)\rangle_\beta$, where $\mathcal{O}(x)$ is a local operator. The unbroken Ward identities \eqref{eq: unbroken translations} play a crucial role; in our case, they reduce to:
\begin{equation}
	\frac{\partial}{\partial \tau} \left \langle \mathcal O(\tau,\vec x)\right \rangle_{\beta}=0 \ ,  \qquad \frac{\partial}{\partial x^i} \left \langle \mathcal O(\tau,\vec x)\right \rangle_{\beta}=0 \ .
\end{equation}
From these equations, we conclude that the thermal one-point function of a local operator must be constant throughout the thermal manifold. This allows for significant simplifications of the remaining Ward identities. For instance, the unbroken Ward identity corresponding to spatial rotations becomes:
\begin{equation}\label{eq: 1pt functions unbroken rot}
	\left \langle \op S_{ij} \mathcal O(x) \right \rangle_\beta = 0 \ .
\end{equation}
Dilatations and boosts yield the following simplified broken Ward identities:
\begin{align} \label{eq: 1pt functions dil}
	\Delta_{\co}\langle \mathcal O(x)\rangle_\beta  &=  \beta \int d^{d-1}y \, \Braket{ T^{00}(0,\vec y) \mathcal O(x)}_\beta \ , \\
	i\langle \op S_{i0}  \mathcal O(x)\rangle_\beta &=  \beta \int d^{d-1}y \, \Braket{T^{0i}(0,\vec y)\mathcal O(x)}_\beta \ . \label{eq: 1pt functions rot}
\end{align}

The broken Ward identity for special conformal transformations can also be written explicitly, but it does not provide independent constraints, as can be verified a posteriori.

The unbroken and broken Ward identities of the global conformal group constrain thermal one-point functions to be constant and $SO(d{-}1)$-invariant in their tensor structure. The only dynamical constraints then arise from equations~\eqref{eq: 1pt functions dil} and \eqref{eq: 1pt functions rot}. We now specialize these results to scalar, vector, and rank-two tensor local operators.

\paragraph{Scalar operators.}
Scalar operators are the simplest case, since $\Braket{\op S_{0i} \mathcal O(x) }_\beta = \Braket{ \op S_{ij}\mathcal O(x) }_\beta = 0$ identically. The boost broken Ward identity reduces to:
\begin{equation}
	\int d^{d-1}y \, \Braket{T^{0i}(0,\vec y)\mathcal O(x) }_\beta = 0 \ ,
\end{equation}
which is trivially satisfied by symmetry, as the integrand is odd under spatial inversion. The only nontrivial constraint is given by the dilatation Ward identity:
\begin{equation} \label{eq: 1pt thermal function}
	\Braket{ \co(x) }_{\beta} = \frac{\beta}{\Delta_{\co}} \int d^{d-1}y \Braket{T^{00}(0,\vec y)\co(x)}_{\beta} \ .
\end{equation}
As previously noted, this equation can provide nontrivial constraints on structure constants of the zero-temperature theory.

\paragraph{Vector operators.}
For a local vector operator $\mathcal O^\mu(x)$, the rotational Ward identities are nontrivial. We use the generators of $SO(d)$ in the vector representation:
\begin{equation}
	\tensor{\left(\op S_{ij} \right)}{^\mu_{\nu}}=i\left(\delta^{\mu}_{i}\delta^{\phantom{\mu}}_{j \nu}-\delta^{\mu}_{j}\delta^{\phantom{\mu}}_{i \nu}\right) \ , \qquad \tensor{\left(\op S_{0i} \right)}{^\mu_{\nu}}=i \left(\delta^{\mu}_{0}\delta^{\phantom{\mu}}_{i \nu}-\delta^{\mu}_{i}\delta^{\phantom{\mu}}_{0 \nu}\right) \ .
\end{equation}
The Ward identities become:
\begin{align}
	\left \langle \op S_{ij} \mathcal{O}^{\mu}(x)\right \rangle_\beta &= i \left(\delta^{\mu}_{i}\Braket{\co_{j}(x)}_{\beta}-\delta^{\mu}_{j}\Braket{\co_{i}(x)}_{\beta}\right) = 0 \ , \label{eq: 1pt vec rot 1}\\
	\left \langle \op S_{0i} \mathcal{O}^{\mu}(x)\right \rangle_\beta &= i \left(\delta^{\mu}_{0}\Braket{\co_{i}(x)}_{\beta}-\delta^{\mu}_{i}\Braket{\co_{0}(x)}_{\beta}\right) = -i \beta \int d^{d-1}y \Braket{T^{0i}(0,\vec y)\co^{\mu}(x)}_{\beta} \ . \label{eq: 1pt vec rot 2}
\end{align}

Equation~\eqref{eq: 1pt vec rot 1} implies that $\Braket{\co_i(x)}_\beta = 0$ in $d > 2$ (trivially satisfied for $d=2$). This, in turn, simplifies~\eqref{eq: 1pt vec rot 2}, yielding:
\begin{equation}
	\delta^\mu_i \Braket{\co^0(x)}_\beta = \beta \int d^{d-1}y \Braket{T^{0i}(0,\vec y) \co^\mu(x)}_\beta \ .
\end{equation}
Setting $\mu = j$ and contracting with $\delta^i_j$, we find:
\begin{equation}
	\Braket{\co^0(x)}_\beta = \frac{\beta}{d-1} \int d^{d-1}y \Braket{T^{0i}(0,\vec y) \co^i(x)}_\beta \ .
\end{equation}
From the dilatation Ward identity, we also have:
\begin{equation}
	\Braket{\co^\mu(x)}_\beta = \frac{\beta}{\Delta_{\co}} \int d^{d-1}y \Braket{T^{00}(0,\vec y)\co^\mu(x)}_\beta \ .
\end{equation}
In conclusion, the thermal one-point function of a vector operator satisfies:
\begin{align}
	\Braket{\co^i(x)}_\beta &= 0 \ , \\
	\Braket{\co^0(x)}_\beta &= \frac{\beta}{\Delta_{\co}} \int d^{d-1}y \Braket{T^{00}(0,\vec y)\co^0(x)}_\beta = \frac{\beta}{d-1} \int d^{d-1}y \Braket{T^{0i}(0,\vec y)\co^i(x)}_\beta \ .
\end{align}
In Section~\ref{subsec: constraints in the OPE regime}, we will show that $\left \langle \mathcal O^0(x) \right \rangle_\beta = 0$ if $\mathcal O^\mu$ appears in the OPE of two scalar operators.

\paragraph{Rank-two tensor operators.} \label{sec Two rank operators}
Let us now consider a local symmetric rank-two tensor operator $\co^{\mu \nu}(x)$. The unbroken rotational Ward identity gives:
\begin{equation}
	i\left(\delta_{\mu i}\Braket{\co_{j \rho}(x)}_{\beta}-\delta_{\mu j}\Braket{\co_{i \rho}(x)}_{\beta}+\delta_{i \rho}\Braket{\co_{\mu j}(x)}_{\beta}-\delta_{j\rho}\Braket{\co_{\mu i}(x)}_{\beta}\right) = 0 \ .
\end{equation}
Setting $\mu = \rho = 0$ implies:
\begin{equation}
	\Braket{\co_{00}(x)}_\beta = a \ ,
\end{equation}
for some constant $a$. Similarly, setting $\mu = k$, $\rho = 0$, one finds:
\begin{equation}
	(d-2)\Braket{\co_{j0}(x)}_\beta = 0 \ \Rightarrow\ \Braket{\co_{j0}(x)}_\beta = 0 \quad \text{for } d > 2 \ .
\end{equation}
A similar identity gives $\Braket{\co_{0j}(x)}_\beta = 0$. The remaining condition $\Braket{\co_{ij}(x)}_\beta = b \delta_{ij}$ follows from $SO(d{-}1)$ invariance:
\begin{equation}\label{eq: b definition}
	\Braket{\co_{ij}(x)}_\beta = b \delta_{ij} \ .
\end{equation}
We can now use the broken Ward identities to constrain $a$ and $b$:
\begin{align}
	a &= \frac{\beta}{\Delta_{\co}} \int d^{d-1}y \Braket{T^{00}(0,\vec y)\co^{00}(x)}_\beta \ , \\
	b &= \frac{\beta}{(d-1)\Delta_{\co}} \delta_{ij} \int d^{d-1}y \Braket{T^{00}(0,\vec y)\co^{ij}(x)}_\beta \ .
\end{align}
The broken boost Ward identity imposes:
\begin{equation} \label{eq: a and b constraint}
	b = a - \frac{1}{d-1} \int d^{d-1}y \Braket{T^{0i}(0,\vec y)\co^{0i}(x)}_\beta \ .
\end{equation}
In Section~\ref{subsec: constraints in the OPE regime}, we will show that if $\co^{\mu \nu}$ appears in the OPE of two scalar operators, $a$ and $b$ are further constrained to yield a traceless tensor.

Finally, we remark that all the results above are in agreement with the general form obtained in~\cite{Iliesiu:2018fao}:
\begin{equation}\label{eq: one point functions}
	\langle \mathcal O^{\mu_1\ldots \mu_J}(x)\rangle_{\beta} = \frac{b_{\mathcal O}}{\beta^{\Delta_{\mathcal O}}} \left(e^{\mu_1}\cdots e^{\mu_J}-\text{traces}\right) \ ,
\end{equation}
where $e^\mu$ is a unit vector in the compactified direction $S^1_\beta$. In~\cite{Iliesiu:2018fao}, this form was derived purely from symmetry arguments. In our case, the same result emerges from a detailed analysis of broken and unbroken Ward identities. Whether these identities can be used to extract information beyond the OPE regime remains an open question.

\subsection{OPE blocks and one-point functions}\label{subsec: constraints in the OPE regime}
The constraints set up by the broken symmetries of the global conformal group are difficult to solve by themselves, since the structure of the thermal breaking term requires the knowledge of a thermal $(n+1)$-point function in order to study a thermal $n$-point function. However, in this section we show how the breaking term splits into lower-point functions in the OPE regime, allowing for an explicit derivation of the solution to the bootstrap problem and for additional constraints on the thermal one-point functions. We are going to focus on (broken) dilatations and rotations.

 Let us start by focusing on the new constraint coming from the dilatation broken Ward identity. We study the thermal scalar two-point function 
\begin{equation}
    g(\tau, |\Vec{x}|)=\Braket{\phu(x)\phu(0)}_{\beta} \ ,
\end{equation}
where both scalar operators now have conformal dimension $\Delta_{\phu}$. By introducing the radial coordinate $r=| \Vec{x}|$, the dilatation broken Ward identity reads
\begin{equation} \label{eq: dil br wi to solve}
    \left(\tau \frac{\partial}{\partial \tau}+ r \frac{\partial}{\partial r}+2\Delta_\phu  \right) g\left(\tau,  r\right) =\beta \int d^{d-1}y \Braket{T^{00}(0,\vec y)\phu(x)\phu(0)}_{\beta} \ . 
\end{equation}
If we consider the operators $\phu$ in the OPE regime, the broken Ward identity can be rewritten as
\begin{multline}
    \left(\tau \frac{\partial}{\partial \tau}+ r \frac{\partial}{\partial r}+2\Delta_\varphi  \right) g\left(\tau,  r\right) = \\= \beta \sum_{\co \in \phu \times \phu}\frac{f_{\phu\phu\co}}{c_{\co}} \left(\tau^2+ r^2  \right)^{\frac{1}{2}\left(\Delta_{\co}-2\Delta_\varphi-J \right)} x_{\mu_{1}}\dots x_{\mu_{J}}\int d^{d-1}y \Braket{T^{00}(0,\vec y)\co^{\mu_{1}\dots \mu_{J}}\left(0\right)}_{\beta} \ .
\end{multline}
The right-hand side can be rewritten by plugging in equation \eqref{eq: 1pt thermal function}, which is the dilatation broken Ward identity applied to thermal one-point functions:
\begin{multline}\label{eq: Differential equations}
    \left(\tau \frac{\partial}{\partial \tau}+ r \frac{\partial}{\partial r}+2\Delta_\varphi \right) g\left(\tau,  r\right) = \\= \sum_{\co \in \phu \times \phu}\frac{f_{\phu\phu\co} \Delta_{\co}}{c_{\co}} \left(\tau^2+ r^2  \right)^{\frac{1}{2}\left(\Delta_{\co}-2\Delta_\varphi-J \right)} x_{\mu_{1}}\dots x_{\mu_{J}} \Braket{\co^{\mu_{1}\dots \mu_{J}}}_{\beta} \ .
\end{multline}
The left-hand side of equation \eqref{eq: Differential equations} respects the OPE regime as well, so it can be expanded in a similar fashion. We get
\begin{equation}
     \sum_{\co \in \phu \times \phu}\left(\tau \frac{\partial}{\partial \tau}+ r \frac{\partial}{\partial r}+2\Delta_\varphi- \Delta_{\co}\right)f_\beta^{(\mathcal O)}(\tau, r)=0 \ ,
\end{equation}
where 
\begin{equation}\label{eq: dil bwi sum}
     f_\beta^{(\mathcal O)}(\tau, r)= \frac{f_{\phu\phu\co}}{c_{\co}} \left(\tau^2+ r^2  \right)^{\frac{1}{2}\left(\Delta_{\co}-2\Delta_\varphi-J \right)} x_{\mu_{1}}\dots x_{\mu_{J}} \Braket{\co^{\mu_{1}\dots \mu_{J}}}_{\beta} \ . 
\end{equation}
Since $\langle \mathcal O^{\mu_1\ldots \mu_J}\rangle_\beta \propto \beta^{-\Delta_{\mathcal O}}$ by dimensional analysis, we can interpret the sum over the operators in the OPE appearing in equation \eqref{eq: dil bwi sum} as a sum over powers of $\beta$\footnote{This requires additional care in the case of a degenerate spectrum. In the latter case, we can either distinguish the two operators from the spin or from other global symmetries; in the first case, the contribution of any operator is always captured by the linearity of the differential equation \eqref{eq: dil bwi sum}, in the second case the one-point function is zero because the operator is not in the trivial representation of the global symmetry group.}. Hence, each term of the series constitutes an independent differential equation:
\begin{equation}\label{eq: inhomogeneous equations}
    \left(\tau \frac{\partial}{\partial \tau}+ r \frac{\partial}{\partial r}+2\Delta_\varphi \right) f_\beta^{(\mathcal O)}(r,\tau) = \Delta_{\mathcal O}  f_\beta^{(\mathcal O)} \ .
\end{equation}
If we consider the identity operator $\mathds{1} \in \co_1 \times \co_2$, then $\Delta_{\mathds{1}}=0$ and equation \eqref{eq: inhomogeneous equations} reproduces the zero-temperature result. In all other cases, we can also write the boosts broken Ward identity:
\begin{equation} 
    \left( \tau r \frac{\partial}{\partial r}-r^2 \frac{\partial}{\partial \tau}\right)  g\left(\tau,  r \right)=x_{i}\int d^{d-1}y \Braket{T^{0i}(0,\vec y)\phu(x) \phu(0)}_{\beta} \ .
\end{equation}
This equation can be expanded in the OPE as well; we expand the right-hand side first:
\begin{multline}
    \left( \tau r \frac{\partial}{\partial r}-r^2 \frac{\partial}{\partial \tau}\right)  g\left(\tau,  r \right)=\\=\beta \sum_{\co \in \phu \times \phu}\frac{f_{\phu\phu\co}}{c_{\co}} \left(\tau^2+ r^2  \right)^{\frac{1}{2}\left(\Delta_{\co}-2\Delta_\varphi-J \right)} x_{\mu_{1}}\dots x_{\mu_{J}}x_{i}\int d^{d-1}y \Braket{T^{0i}(0,\vec y)\co^{\mu_{1}\dots \mu_{J}}\left(0\right)}_{\beta} \ ,  
\end{multline}
and then we apply the broken Ward identity \eqref{eq: 1pt functions rot} to rewrite the integral as
\begin{multline} \label{eq: rightHandside}
    \left( \tau r \frac{\partial}{\partial r}-r^2 \frac{\partial}{\partial \tau}\right)  g\left(\tau,  r \right)=\\=i\beta  \sum_{\co \in \phu \times \phu}\frac{f_{\phu\phu\co}}{c_{\co}} \left(\tau^2+ r^2  \right)^{\frac{1}{2}\left(\Delta_{\co}-2\Delta_\varphi-J \right)} x_{\mu_{1}}\dots x_{\mu_{J}}x_{i} \Braket{\op S_{i0}\co^{\mu_{1}\dots \mu_{J}}\left(0\right)}_{\beta} \ . 
\end{multline}
The equation \eqref{eq: inhomogeneous equations}, together with equation \eqref{eq: rightHandside}, can be interpreted as a system of partial differential equations. In general, we find that the first differential equation, i.e., \eqref{eq: inhomogeneous equations}, fixes  
\begin{equation}
    f_\beta^{\mathcal O}(\tau,r) \propto \left (r^2+\tau^2\right )^{\frac{1}{2}\left(\Delta_{\mathcal O}-2\Delta_{\varphi}-J\right )}\tau^J g_{\mathcal O}\left(\frac{r}{\tau}\right) \ ,
\end{equation}
where the function $g_{\mathcal O}(r/\tau)$ remains undetermined. The next step is to plug this solution into the rotation broken Ward identity; order by order in the OPE expansion, equation \eqref{eq: rightHandside} reads 
\begin{equation}
    \frac{r}{\tau^2} \tau^J \left(-J r \tau g_{\mathcal O}\left(\frac{r}{\tau}\right)+\left(r^2+\tau^2\right)g_{\mathcal O}'\left(\frac{r}{\tau}\right)\right) = x_{\mu_1}\ldots x_{\mu_J}x_i \left \langle \op S_{i0}\mathcal O^{\mu_1\ldots \mu_J} \right \rangle_\beta \ .
\end{equation}
For scalars, i.e., $J = 0$, we get $g_{\mathcal O}(r/\tau) = \text{const.}$, meaning that the one-point function of a scalar $\left \langle \mathcal O \right \rangle_\beta $ does not have any kinematical structure, as expected. For vectors, i.e., $J = 1$, by imposing the reality condition on the solution\footnote{The differential equation is also solved by a second function proportional to a hypergeometric function; however, this function is not real for all values of $r$ and $\tau$.}, we get $\left \langle \mathcal O^\mu \right \rangle = 0$ and only a term proportional to an odd power of $\tau$ can contribute to the OPE. However, this contribution is zero because of the symmetry $\tau \to - \tau$, which forbids odd functions of $\tau$ to appear\footnote{The use of $\tau \to -\tau$ is necessary to get the correct result. This corresponds to the use of the full $\mathrm O(d-1)$ symmetry group in \cite{Iliesiu:2018fao}: $\mathrm{SO}(d-1)$ is indeed not sufficient to exclude odd-spin operators from having a non-zero thermal one-point function.}. For rank-2 tensors, i.e., $J=2$, we can recall the results of Section \eqref{sec Two rank operators} to see that the solution to the two differential equations is given by 
\begin{equation}
    f_\beta^{(\mathcal O)}(\tau,r) \propto \left(\tau^2+ r^2  \right)^{\frac{1}{2}\left(\Delta_{\co}-2\Delta_\varphi \right)} C_{2}^{(\nu)}\left(\frac{\tau}{\sqrt{\tau^2+r^2}}\right ) \ .\label{eq: eq1 comp}
\end{equation}
This result can be extended to all values of $J$ by observing that the first differential equation is solved by the ansatz
\begin{equation}
    f_\beta^{(\mathcal O)}(\tau,r) = \frac{\mathcal{C}_{J,\nu}^{\mathcal{O}}}{\beta^{\Delta_{\co}}}\left(\tau^2+ r^2  \right)^{\frac{1}{2}\left(\Delta_{\co}-2\Delta_\varphi \right)} C_{J}^{(\nu)}\left(\frac{\tau}{\sqrt{\tau^2+r^2}}\right ) \ , \label{eq: eq1 comp1}
\end{equation}
where the $\beta^{-\Delta_{\co}}$ factor is introduced by dimensional analysis and $\mathcal{C}_{J,\nu}^{\mathcal{O}}$ is an overall coefficient to be fixed. We can fix it by imposing consistency of the thermal two-point function with the OPE. By comparing expressions \eqref{eq: dil bwi sum} and \eqref{eq: eq1 comp1}, we identify
\begin{equation}
    \frac{\mathcal{C}_{J,\nu}^{\mathcal{O}}}{\beta^{\Delta_{\co}}} C_{J}^{(\nu)}\left(\frac{\tau}{\sqrt{\tau^2+r^2}}\right )=\frac{f_{\phu\phu\co}}{c_{\co}} \left(\tau^2+ r^2  \right)^{-\frac{J}{2}} x_{\mu_{1}}\dots x_{\mu_{J}} \Braket{\co^{\mu_{1}\dots \mu_{J}}}_{\beta} \ .
\end{equation}
If $\nu=\frac{d-2}{2}$, the following identity holds \cite{Iliesiu:2018fao}:
\begin{equation}
    C_J^{(\nu)}\left(\frac{\tau}{\sqrt{\tau^2+r^2}}\right) = \frac{2^J(\nu)_J}{J!}\left(\tau^2+ r^2  \right)^{-\frac{J}{2}} x_{\mu_1}\ldots x_{\mu_J}\left(e^{\mu_1}\ldots e^{\mu_J}-\text{traces}\right) \ ,
\end{equation}
leading to 
\begin{equation}
    \frac{\mathcal{C}_{J,\nu}^{\mathcal{O}}}{b_{\co}}\frac{2^J(\nu)_J}{J!} \frac{b_{\co}}{\beta^{\Delta_{\co}}}\left(e^{\mu_1}\ldots e^{\mu_J}-\text{traces}\right)=\frac{f_{\phu\phu\co}}{c_{\co}}  \Braket{\co^{\mu_{1}\dots \mu_{J}}}_{\beta} \ ,
\end{equation}
from which we can both recover formula \eqref{eq: one point functions} and fix the overall coefficient:
\begin{equation}
    f_\beta^{(\mathcal O)}(\tau,r) = \frac{J!}{2^J (\nu)_J}\frac{1}{\beta^{\Delta_{\co}}}\frac{f_{\phu\phu\co} b_{\co}}{c_{\co}} \left(\tau^2+ r^2  \right)^{\frac{1}{2}\left(\Delta_{\co}-2\Delta_\varphi \right)} C_{J}^{(\nu)}\left(\frac{\tau}{\sqrt{\tau^2+r^2}}\right ) \ ,
\end{equation}
which immediately leads to the final solution\footnote{The procedure highlighted in this short section shows how it is possible to recover the results already presented in \cite{Iliesiu:2018fao} by employing only the broken Ward identities and the OPE. It must be noted that, strictly speaking, formula \eqref{eq: one point functions} was not recovered in full generality, but only for those operators appearing in the OPE between two identical scalar operators. Moreover, the second differential equation has to be solved spin by spin.}:
\begin{equation}\label{eq: OPE decomposition}
    g(\tau,r) = \sum_{\mathcal{O}\in \phu \times \phu}\frac{J!}{2^J (\nu)_J}\frac{1}{\beta^{\Delta_{\co}}}\frac{f_{\phu\phu\co} b_{\co}}{c_{\co}} \left(\tau^2+ r^2  \right)^{\frac{1}{2}\left(\Delta_{\co}-2\Delta_\varphi \right)} C_{J}^{(\nu)}\left(\frac{\tau}{\sqrt{\tau^2+r^2}}\right ) \ .
\end{equation}
Further, observe that the OPE decomposition of the two-point function implies that the one-point functions of odd-spin operators must vanish. This can be easily checked by considering that $C_{J}^{\nu}(\tau/\sqrt{r^2+\tau^2}) = C_{J}^{\nu}(-\tau/\sqrt{r^2+\tau^2})$ only for even-spin operators, and therefore $\tau \to - \tau$ is a symmetry only if the one-point functions of odd-spin operators vanish.

 In conclusion, working with the broken Ward identities in the OPE regime allowed us to obtain already known results about the structure of one-point functions and the OPE decomposition of two-point functions. This was expected, since broken Ward identities contain information about both broken and unbroken symmetries. These results also suggest that broken Ward identities might contain more information, possibly also beyond the OPE regime.

\subsection{Superconformal broken Ward identities}\label{sec: Superconformal BWI}
Whether any trace of supersymmetry is preserved at finite temperature has been a longstanding question \cite{Fuchs:1984ed,Aoyama:1984bk,Das:1978rx}, and significant progress has been made recently, with important developments and applications \cite{Caron-Huot:2008vbk}. The discussion on the broken Ward identities (at finite temperature) presented before can be specialized to the case of superconformal field theories.
In the following, we will focus on supersymmetric and superconformal theories in four dimensions; however, the arguments and procedures can be adapted to other dimensions.

 The Superconformal Group extends the Conformal Group with supersymmetry generators $\op Q_\alpha^{I}$ (and $\op {\overline Q}^{\dot \alpha I}$), superconformal generators $\op S_\alpha^I$ (and $\op {\overline S}^{\dot \alpha I}$), and R-symmetry generators $\op R^{IJ}$, satisfying precise commutation relations that produce a specific decomposition of the Hilbert space of the theory.

 Due to the anti-periodicity of fermionic operators over the thermal circle $S^1_{\beta}$, we expect supersymmetry to be broken; however, it is not obvious \emph{a priori} whether the R-symmetry is broken or unbroken. In this section, we derive explicit broken Ward identities for all the superconformal generators.

\paragraph{Supersymmetry and superconformal symmetry}
The procedure outlined in the previous sections can be followed in every detail, with a few remarks. In particular, the first difference concerns the variation of the action with respect to a symmetry with fermionic generators\footnote{In the following, the spinor indices $\alpha, \beta, \gamma$ will run over $\lbrace 1,2\rbrace$; the same applies to the dotted spinor indices $\dot{\alpha}, \dot{\beta}, \dot{\gamma}$. The R-symmetry indices will be denoted by $I,J,K$ and run over $\lbrace 1, \dots, \mathcal{N} \rbrace$, with $\mathcal{N}$ representing the number of supersymmetries.}:
\begin{equation}
    \delta S = \int_{\mathcal M} d^d x \sqrt{g} \ \nabla_\mu \xi^\alpha(x) \ G_{\alpha}^{\mu}(x) \ ,
\end{equation}
where $G_{\alpha}^{\mu}$ is the fermionic current, $g_{\mu \nu}$ is the metric on the manifold $\mathcal{M}$, and $\xi^{\alpha}$ is an infinitesimal spinor. From the above equation it is clear that, on a generic non-flat manifold, all fermionic generators are broken, since there is no covariantly constant spinor. At finite temperature, i.e. when $\mathcal M_\beta =\mathbb R^{d-1}\times  S^1_{\beta}$, the spacetime is locally flat and this argument does not prevent fermionic generators from realizing quantum symmetries. We can then proceed straightforwardly as in the bosonic case. A second remark concerns the evaluation of the following spatial boundary integral:
\begin{equation}
    \lim_{R\to \infty} \left \langle J_a^{i}(\tau,R,\Omega) \mathcal O_1(x_1) \ldots \mathcal O_n(x_n) \right \rangle_\beta = 0 \ .
\end{equation}
Differently from the bosonic case, this contribution does not lead to an infrared divergence because of clustering and the fact that the one-point function of the fermionic current is necessarily zero: this can follow from the solution of the rotation broken Ward identity, or from symmetry arguments \cite{Iliesiu:2018fao}. The rest of the procedure to compute broken Ward identities is unchanged. To discuss the breaking terms, it is useful to write down the \textit{superconformal current}:
\begin{equation}\label{eq: superconformal currents}
    J_{SC}^\mu(x) = \psi_I^\alpha(x) G_\alpha^{\mu I }(x)\ ,
\end{equation}
where $\psi_I^\alpha(x)$ is the conformal Killing spinor, 
\begin{equation}
    \psi_{I}^\alpha (x) = \lambda_I^\alpha + x_\nu \left(\overline \sigma^{\nu }\right)^{\, \alpha \dot \alpha} \mu_{I \dot \alpha} \ ,
\end{equation}
and $\mu_{I \dot \alpha}$, $\lambda_I^\alpha$ are constant spinors. 

At zero temperature, we define the supercharges and the superconformal charges\footnote{As done in the example \eqref{eq: dil example}, the correct definition of the quantum superconformal generators must take into account the proper spacetime dependence:
$$\op {\overline S^{\dot \alpha} } \, \mathcal O(x) = \int_{\mathbb R^{d-1}}d^{d-1} y \ \left(\overline \sigma^{\nu }\right)_{\, \alpha \dot \alpha}(y_\nu-x_\nu) \ G^{0\alpha}(y) \mathcal O(x) \ .$$}:
\begin{equation}
    Q^\alpha = \int_{\mathbb R^{d-1}}d^{d-1} x \ G^{0\alpha}(x) \ , \hspace{1 cm} \overline S^{\dot \alpha} = \int_{\mathbb R^{d-1}}d^{d-1} x \  \left(\overline \sigma^{\nu }\right)_{\, \alpha \dot \alpha}x_\nu \ G^{0\alpha}(x) \ ,
\end{equation}
When $\lambda_I^{\alpha} = 0$, $\mu_{I \dot \alpha} \neq 0$, the current \eqref{eq: superconformal currents} corresponds to the symmetry generated by the superconformal generators; instead, when $\mu_{I \dot \alpha} = 0$, $\lambda_I^{\alpha} \neq 0$, the current is associated with supersymmetry generators.

 Due to the fermionic nature of $G_{\alpha}^{\mu I}(\tau,\vec x)$, and the anti-periodicity of fermionic operators on the thermal circle, it is natural to write down the most general breaking term—defined in \eqref{eq: breaking term definition}—for the superconformal current \eqref{eq: superconformal currents} as:
\begin{equation}
    \left[2 \lambda_I^{\alpha}+\left((\beta-2\tau) \left(\overline \sigma_0\right)^{\alpha \dot \alpha}+2(x_i-y_i)\left(\overline \sigma^i\right)^{\alpha \dot \alpha}\right)\mu_{I \dot \alpha}\right] G_{\alpha }^{0 I}(\beta,\vec y) \ ,
\end{equation}
where $(\tau,\vec y)$ are the coordinates of the operator to which the generator is applied.

The expression above agrees with the physical expectation that the fermionic nature of $\op Q_\alpha^I$ and $\op S_\alpha^I$ implies an explicit breaking of supersymmetry at finite temperature. Leaving spinor and R-symmetry indices implicit for clarity, the broken Ward identities read:
\begin{multline}\label{eq: SBWI}
    \sum_{r=1}^{n} \left \langle \mathcal O_1(x_1) \ldots \left[\op Q,\mathcal O_r\right ] (x_r)\ldots \mathcal O_n(x_n) \right \rangle_\beta =\\ =  2 \int d^{d-1} y \ \left \langle G^{0}(\beta ,\vec y)\mathcal O_1(x_1) \ldots \mathcal O_n(x_n) \right \rangle_\beta \ ,
\end{multline}
\begin{multline}
    \sum_{r=1}^{n} \left \langle \mathcal O_1(x_1) \ldots \left [\Bar {\op S},\mathcal O_r\right ](x_r)\ldots \mathcal O_n(x_n) \right \rangle_\beta =\\= \sum_{r=1}^{n}(\beta-2 \tau_r)\overline \sigma^0 \int d^{d-1}y \left \langle G^{0}(\beta ,\vec y)\mathcal O_1(x_1) \ldots \mathcal O_n(x_n) \right \rangle_\beta + \\+  \sum_{r=1}^{n}\overline \sigma^j \int d^{d-1}y \ (y_j-x_j)\left \langle G^{0}(\beta ,\vec y)\mathcal O_1(x_1) \ldots \mathcal O_n(x_n) \right \rangle_\beta \  .
\end{multline}
These two equations encode the superconformal symmetry breaking at finite temperature. Equation \eqref{eq: SBWI} governs the behaviour of correlation functions upon acting with supersymmetry generators. In particular, the breaking term contains a correlation function in which the supercurrent appears. From the derivation, it becomes clear that the breaking of supersymmetry at finite temperature is due to the anti-periodic boundary conditions imposed on fermions.

For completeness, we also write down the broken Ward identities for the right-chiral supersymmetry generators:

 \begin{multline}\label{eq: SBWI2}
    \sum_{r=1}^{n} \left \langle \mathcal O_1(x_1) \ldots \left [\Bar{\op Q},\mathcal O_r \right](x_r)\ldots \mathcal O_n(x_n) \right \rangle_\beta =\\ =  2 \int d^{d-1} y \ \left \langle \Bar{G}^{0}(\beta ,\vec y)\mathcal O_1(x_1) \ldots \mathcal O_n(x_n) \right \rangle_\beta \ ,
\end{multline}
\begin{multline}
    \sum_{r=1}^{n} \left \langle \mathcal O_1(x_1) \ldots [ {\op S},\mathcal O_r](x_r)\ldots \mathcal O_n(x_n) \right \rangle_\beta =\\ =\sum_{r=1}^{n}(\beta-2 \tau_r) \sigma^{0}\int d^{d-1}y \left \langle \Bar{G}^{0}(\beta ,\vec y)\mathcal O_1(x_1) \ldots \mathcal O_n(x_n) \right \rangle_\beta + \\+ \sum_{r=1}^{n} \sigma^j\int d^{d-1}y \ (y_j-x_j) \left \langle \Bar{G}^{0}(\beta ,\vec y)\mathcal O_1(x_1) \ldots \mathcal O_n(x_n) \right \rangle_\beta \ .
\end{multline}

\paragraph{The fate of R-symmetry at finite temperature}
The Superconformal Group includes the R-symmetry, which can be understood as a rotation of the supercharges. Even though supersymmetry is broken at finite temperature, it is not immediately clear what happens to the R-symmetry. To address this question, let us explicitly compute the breaking term that appears in the (broken) Ward identities. To do this, we recall that the R-symmetry generators can be defined as
\begin{equation}
    R^{IJ} = \int_{\mathbb R^{d-1}} d^{d-1}x \ J^{0 , IJ}(x) \ ,
\end{equation}
where $J^{\mu, IJ}$ is a bosonic current. Therefore, we must consider the breaking term defined by equation \eqref{eq: breaking term definition} in the case of the R-symmetry current. Since there is no additional coordinate dependence (as in the dilatation broken Ward identity, for example), and since the current is bosonic (hence periodic over the thermal circle), we conclude that the breaking term is simply zero: the Ward identity is unbroken and R-symmetry is preserved.

 This result is noteworthy because it shows that in a supersymmetric/superconformal field theory at finite temperature, although supersymmetry and superconformal symmetry are broken, R-symmetry remains intact as a global symmetry acting on the operators of the theory. This has important consequences, as it implies that every operator charged under the Abelian $\mathfrak{u}(1)$ subalgebra of the R-symmetry at zero temperature must have vanishing one-point function. In formula:
\begin{equation}
    \left \langle \mathcal O \right \rangle_\beta = 0 \ , \hspace{1 cm}\text{charged} \ .
\end{equation}
For instance, in the case of $\mathcal N = 4$ SYM, this means that in the stress-tensor multiplet there are only two operators allowed to have a non-zero one-point function: the Lagrangian operator and the stress-energy tensor. All other operators are charged under R-symmetry and therefore their thermal expectation values vanish.

 Another way to see that the R-symmetry is preserved is to consider the real-time formalism, where an $n$-point function at finite temperature can be seen as a weighted sum of an infinite number of $(n+2)$-point functions at zero temperature. All the zero-temperature $(n+2)$-point functions are invariant under R-symmetry. Hence, it is expected that the R-symmetry remains preserved even at finite temperature. This holds for any global symmetry of the theory generated by a bosonic charge.

 The preservation of R-symmetry can also be understood holographically by considering the duality between type IIB superstring theory on $ \text{AdS}_{5} \times S^5 $ and $\mathcal N=4$ SYM in four dimensions. In this case, the R-symmetry algebra $\mathfrak{so}(6)$ is realized as the isometry algebra of $S^5$, and these isometries remain unaltered even when the non-compact part of the ten-dimensional spacetime is replaced by a black hole geometry. This happens heuristically because the black hole solution does not affect the geometry of $S^5$. Therefore, R-symmetry is preserved in the holographic dual description, which, in the case of a black hole background, corresponds to a finite-temperature CFT.

 All the currents and breaking terms corresponding to supersymmetry, superconformal symmetry, and R-symmetry are summarized in Table \ref{tab: Explicit can susy currents}.

\begin{table}[]
    \centering
    \renewcommand{\arraystretch}{1.5}
    \begin{tabular}{|c|c|c|}
        \hline
        Charge   & Current:  $\omega_{a} J^{\mu}_{a}(x)$ & Breaking term: $\Gamma_a^\beta(\beta, \vec x)$ \\
        \hline
        $\op Q_\alpha^I$  & $\xi G^{\mu}(x)$ & $2 \, G_{\alpha}^{\, 0}(\beta,\vec x)$ \\
        \hline
        $\overline {\op Q}_{\dot \alpha, I}$ & $\xi \overline G^{\mu}(x)$ &  $2 \, \overline G_{\dot \alpha}^{\, 0}(\beta,\vec x)$\\ 
        \hline 
        $\op S ^{\alpha}_{I}$ &  $\xi \sigma^{\nu} x_\nu \overline G^{\mu}(x)$ & $\left [(\beta-2 \tau)\left(\sigma^0\right)^{\alpha }+2(x_i-y_i)\left(\sigma^i\right)^{\alpha }\right ] \overline G^{\, 0}(\beta,\vec x) $ \\ 
        \hline 
        $\op {\overline{S}}_{\dot \alpha}^{I}$  & $\xi \, \overline \sigma^{\nu } x_\nu G^{\mu}(x)$ & $\left [(\beta-2 \tau)\left(\overline \sigma^0\right)_{\dot \alpha }+2(x_i-y_i)\left(\overline \sigma^i\right)_{\dot \alpha}\right ] G^{\, 0}(\beta,\vec x) $ \\ 
        \hline 
        $\op R^{IJ}$ & $J_R^{\mu , I J}(x)$ & $0$ \\ 
        \hline 
    \end{tabular}
    \caption{Explicit currents and breaking terms for the symmetries of the global superconformal group. The coordinates $(\tau ,\vec y)$ locate the operator to which the breaking term is applied. To unclutter the notation, contracted spinor and R-symmetry indices have been left implicit.}
    \label{tab: Explicit can susy currents}
\end{table}

 \subsection{Two dimensions and explicit examples}

Another consistency check is provided by the fact that in two dimensions, the dilatation broken Ward identity yields the correct expression for the one-point function of the stress-energy tensor. The broken Ward identities derived in this work cannot, in general, be directly checked, since in order to say something about an $n$-point thermal correlator, one needs knowledge of a $(n+1)$-point thermal correlator. In higher dimensions, the number of points in the correlation functions has been reduced by studying the OPE regime. However, in two dimensions the equations can be checked directly, since the thermal correlation functions are known. This follows from the equivalence between the two-dimensional cylinder and the thermal manifold, and the existence of a conformal map between the plane $\mathbb R^2 \sim \mathbb C$, where all functions are known, and the cylinder $\mathbb R \times S^1_\beta$. We recall that the conformal map is
\begin{equation}
    \mathbb C \ni \omega = \sigma + i \tau = \frac{\beta}{\pi} \ln z \ ,
\end{equation}
where we use the convention $0 \le \tau \le \beta$ and $-\infty < \sigma < \infty$.

\paragraph{Dilatation broken Ward identity}
Let us test, as an example, the one-point function of the stress-energy tensor. The dilatation broken Ward identity reads
\begin{equation}
    \Delta_T \left \langle T^{00}\right \rangle_\beta = \beta \int d y \ \left \langle T^{00}(0,y)T^{00}(0) \right \rangle_\beta \ . \label{eq: appA int}
\end{equation}
The stress-energy tensor on the complex plane can be written as $2 \pi T^{00}(z,\overline z) = T(z)+\overline T(\overline z)$, and it is known that
\begin{equation}\label{eq:2dStressTensor}
    \left  \langle T(z) T(0) \right \rangle = \frac{c/2}{z^4} \ , \hspace{1 cm} \left  \langle \overline T(\overline z) \overline T(0) \right \rangle = \frac{c/2}{\overline z^4} \ .
\end{equation}
The application of the conformal map yields the two-point function on the cylinder \cite{Datta:2019jeo,DiFrancesco:1997nk,Mussardo:2020rxh}\footnote{Here we are using the standard convention in two-dimensional CFT: in this convention $b_T > 0$ for unitary theories.}
\begin{equation}
    \left \langle T^{00}(\omega,\overline{\omega}) T^{00}(0,0) \right \rangle_\beta = \left(\frac{\pi}{\beta}\right)^4 \frac{c^2}{144 \pi^2}+\frac{c}{8 \pi^2}\left(\frac{\pi}{\beta}\right)^4 \left[\text{csch}^4\left( \frac{\pi \omega }{\beta }\right)+\text{csch}^4\left(\frac{\pi \overline{\omega} }{\beta }\right)\right] \ ,
\end{equation}
which must be interpreted as the thermal two-point function to be inserted into the integral \eqref{eq: appA int}. Up to IR divergence renormalization, the integration gives
\begin{equation}\label{eq:1ptfunctionStress}
    \left \langle T^{00} \right \rangle_\beta = \frac{\pi}{3\beta^2} c \ ,
\end{equation}
which is correct for a free bosonic theory (i.e. $c=1$) \cite{Iliesiu:2018fao}, and also serves as the starting point for deriving the Cardy formula in two dimensions \cite{Mukhametzhanov:2019pzy}.

 Moreover, observe that for a one-point function, it is trivial to check that the breaking term of the dilatation is simply $- \beta \partial_\beta$.

\paragraph{Rotation broken Ward identity}
The next check concerns the rotation broken Ward identity. In this case, we have
\begin{equation}
   -2 \left \langle T^{00}(0) \right \rangle_\beta =  i \left \langle \op S_{01} T^{01}(0) \right  \rangle_\beta = \beta \int d y \ \langle T^{01}(0,y) T^{01}(0)\rangle_\beta \ . 
\end{equation}
All these correlators can be computed explicitly using equation \eqref{eq:2dStressTensor}. Furthermore, observe that
\begin{equation}
    \langle T^{01}(x) T^{01}(0) \rangle_\beta  = - \langle T^{00}(x) T^{00}(0)\rangle_\beta \ ,
\end{equation}
and therefore the rotation broken Ward identity is satisfied with the same computation used for the dilatation broken Ward identity. The rotation broken Ward identity can also be recast as a dilatation broken Ward identity in the case of the two-point function. This is not a generic feature, but a consequence of the chirality of correlation functions (including the stress tensor) in two dimensions. We do not expect this to hold in higher dimensions.

\paragraph{Special conformal transformations broken Ward identities}
The final check concerns special conformal transformations. In this case, we verify that
\begin{equation}
    2 \tau \Delta_T \left \langle T^{00}(\tau,x) \right  \rangle_\beta  = (\beta-2 \tau) \beta \int d y \ \left \langle T^{00}(y) T^{00}(0) \right \rangle_\beta = \Delta_T (\beta-2 \tau) \left \langle T^{00}(\tau,x) \right \rangle_\beta \ ,
\end{equation}
which corresponds to the broken Ward identity for $\op K_0$; the broken Ward identity for $\op K_1$ follows similarly.

In this case, we did not need to compute anything explicitly; we simply applied the dilatation broken Ward identity. To see that the equation above is correct, note that one-point functions are invariant under time translations, so we can compute them at $2\tau$, and by defining $\tilde \tau = 2\tau$ and using the periodicity $\tau \equiv \tau + \beta$, we see that the broken Ward identity becomes a tautology\footnote{Note that the periodicity of time coordinates is a property of the coordinates of local operators in correlation functions. One might be tempted to conclude that every factor of $\beta$ in the broken Ward identities is equivalent to zero, leading to inconsistencies; this is not correct, as $\beta$ is a fixed physical scale in the theory.}.

 When testing higher-point functions, the simplest example is the two-point function of scalar primaries of weights $(h, h)$, such that $2h = \Delta$. These two-point functions read
\begin{equation}\label{eq:phiphi2d2ptfun}
    \left \langle \phi(\tau,\sigma) \phi(0,0) \right \rangle_\beta = \left(\frac{\pi}{\beta}\right)^\Delta  \operatorname{csch}^\Delta \left(\frac{\pi}{\beta}(\tau+i \sigma)\right) \operatorname{csch}^\Delta \left(\frac{\pi}{\beta}(\tau-i \sigma)\right) \ .
\end{equation}
Straightforward computations show that the breaking term of the Ward identity corresponding to $\op K_0$ is given by
\begin{equation}
    \beta \tau \int d \tilde \sigma \  \left \langle T^{00}(0,\tilde \sigma )\phi(\tau,\sigma)\phi(0,0) \right \rangle_\beta + \beta \sigma  \int d \tilde \sigma \  \left \langle T^{01}(0,\tilde \sigma )\phi(\tau,\sigma)\phi(0,0)\right \rangle_\beta \ ,
\end{equation}
whereas the breaking term corresponding to $\op K_1$ is
\begin{equation}
    \beta \sigma \int d \tilde \sigma \  \left \langle T^{00}(0,\tilde \sigma )\phi(\tau,\sigma)\phi(0,0)\right \rangle_\beta - \beta \tau  \int d \tilde \sigma \  \left \langle T^{01}(0,\tilde \sigma )\phi(\tau,\sigma)\phi(0,0)\right \rangle_\beta \ .
\end{equation}
These breaking terms can be written as differential operators acting on the two-point function in equation \eqref{eq:phiphi2d2ptfun}. In this way, the broken Ward identities become trivial, indicating that the dilatation, rotation, and momentum Ward identities automatically imply the special conformal transformation broken Ward identity.

 Similar checks can be performed for the two-point function of the stress-energy tensor \eqref{eq:2dStressTensor}. All results are in agreement with our expectations.

\paragraph{Supersymmetry and superconformal broken Ward identities}
To test the supersymmetry broken Ward identities, we can use the OPE\footnote{We focus on the holomorphic sector; analogous expressions hold in the anti-holomorphic sector.}
\begin{equation}
    G(z_1) G(z_2)  = \frac{2 c}{3 (z_1-z_2)^3}+ \frac{2}{z_1-z_2} T(z)+\ldots 
\end{equation}
to conclude that the superconformal descendant of the supercurrent is the stress-energy tensor $T$. This is also evident from the super-Virasoro algebra anticommutator
\begin{equation}
    \{G_n,G_m\} = 2 L_{n+m}+ \frac{c}{3}\left(n^2-\frac{1}{4}\right) \delta_{n+m,0} \ ,
\end{equation}
recalling that
\begin{equation}
    L_n = \oint \frac{dz}{2 \pi i} z^{n+1} T(z) \ , \hspace{1 cm} G_m = \oint \frac{dz}{2 \pi i} z^{m+1/2} G(z) \ .
\end{equation}
We must therefore check that
\begin{equation}
    \left \langle T^{00} \right \rangle_\beta = 4\int d\sigma \ \left \langle G^{0}(\beta,\sigma) G^{0}(0,0) \right \rangle_\beta \ .
\end{equation}
It is easy to verify that this relation holds in any two-dimensional superconformal theory; in particular, the coefficient of the one-point function of the stress-energy tensor can be obtained from the two-point function of the supercurrent, and it is given by equation \eqref{eq:1ptfunctionStress}.

\section{Analytic structure, inversion formula and dispersion relation}\label{sec:analyticprop}

In the seminal paper \cite{Iliesiu:2018fao}, an inversion formula for OPE coefficients was also proposed. Here, we review this derivation, as it will be useful in the following sections of the thesis. The idea of an inversion formula, originally introduced in a different context in \cite{Caron-Huot:2017vep}, is conceptually simple, yet the result is highly non-trivial. 

Furthermore, we also present a dispersion relation for thermal two-point functions of scalar operators.

\subsection{The inversion formula}

We first focus on the inversion formula, starting with a toy model and then moving on to the thermal two-point function.

\begin{figure}
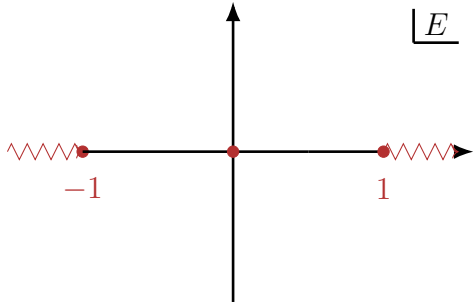

    \centering
    \AnalyticStructureToyModel
    \caption{Analytic structure of the toy model described in the main text.}
    \label{fig:AnalyticStructureToy}
\end{figure}

\paragraph{Inversion formula in a toy model.} Let us first derive an \textit{inversion formula} in a simplified toy model. Consider a function $f$ depending on a single complex variable $E$, and assume it admits an expansion of the form
\begin{equation}
    f(E) = \sum_{J = 0}^\infty  c_J E^J \ .
\end{equation}
Let us further assume that, even though we do not know the full expression of the function $f(E)$, we know its analytic properties. Following \cite{Caron-Huot:2017vep}, we assume for this toy model that the function is analytic except for a branch cut along the real axis for $|E| > 1$, as shown in Fig.~\ref{fig:AnalyticStructureToy}. We also assume that $|f(E)/E|$ is bounded at infinity.

Under these assumptions, it is quite straightforward to compute the coefficients $c_J$ as
\begin{multline}
    c_J = \frac{1}{2 \pi i} \oint_{|E|<1} dE \ \frac{f(E)}{E^{J+1}} = \frac{1}{2 \pi} \int_1^\infty \frac{dE}{E^{J+1}} \left[ \Disc(f(E)) + (-1)^J \Disc(f(-E)) \right] \ ,
\end{multline}
where in the first step we applied Cauchy's theorem, and in the second step we opened the contour (assuming $J > 1$), capturing only the contributions from the branch cuts, given by the discontinuity
\begin{equation}
    \Disc f(E) = i \left(f(E(1+i 0)) - f(E(1-i 0))\right) \ .
\end{equation}

This simple example shows that knowledge of the analytic structure, together with some technical assumptions, allows one to compute the coefficients of an expansion purely in terms of discontinuities across branch cuts.

\paragraph{Inversion formula for CFT at finite temperature.} 
At finite temperature, the idea is essentially the same, but complications arise from the fact that the expansion we would like to invert is the OPE expansion given in equation \eqref{eq: OPE decomposition}. In particular, it is useful to define the \textit{thermal OPE coefficient}
\begin{equation}
    a_{\mathcal O} = \frac{J!}{2^J( \nu)_J} \frac{f_{\varphi \varphi \mathcal O} b_{\mathcal O}}{c_{\mathcal O}} \ ,
\end{equation}
so that
\begin{equation}
    g(\tau,r) = \sum_{\mathcal O \in \varphi \times \varphi} \frac{a_{\mathcal O}}{\beta^{\Delta_{\mathcal O}}} \left(\tau^2+r^2\right)^{\frac{1}{2}(\Delta_{\mathcal O}-2\Delta_\varphi)} C_J^{(\nu)}\left(\frac{\tau}{\sqrt{\tau^2+r^2}}\right)\ .
\end{equation}
From this point on, we set $\beta = 1$: since $\beta$ is the only scale in the theory, the $\beta$-dependence can always be restored by dimensional analysis.

Following \cite{Iliesiu:2018fao}, it is straightforward to derive an analogue of Cauchy's theorem for this specific case. In fact, by using the orthogonality of the Gegenbauer polynomials,
\begin{equation}
    \int_{S^{d-1}}d\Omega \ C_{J}^{(\nu)}(\eta) C_{J'}^{(\nu)}(\eta) = N_J \delta_{J J'} \ ,
\end{equation}
where
\begin{equation}
    N_J = \frac{4^{1-\nu} \pi^{\nu+\frac{3}{2}}\Gamma(J+2\nu)}{J! \Gamma(\nu)^2 (J+\nu)\Gamma \left(\nu+\frac{1}{2}\right)}\ , \qquad \eta = \frac{\tau}{\sqrt{\tau^2+r^2}} \ ,
\end{equation}
it is natural to define the function
\begin{equation}\label{eq:EuclideanIF}
    a(\Delta,J) = \frac{1}{N_J} \int_{|x|<1} d^{d}x \ C_J^{(\nu)}(\eta)|x|^{2\Delta_\varphi-\Delta-d} g(\tau,r) \ .
\end{equation}
The function $a(\Delta,J)$ encodes the thermal dynamics of the theory, and in particular
\begin{equation}
    a(\Delta,J) \sim - \sum_{\Delta_{\mathcal O}} \frac{a_{\mathcal O}}{\Delta-\Delta_{\mathcal O}} \ .
\end{equation}
The definition above is not complete, as we are only interested in the poles (and in particular the residues) of the function $a(\Delta,J)$, and we neglect everything else that will not be relevant for our discussion. Up to this point, we have only used some basic analyticity assumptions\footnote{It is interesting to understand whether these analyticity assumptions can be justified from a physical point of view. For the conformal bootstrap at zero temperature, some results in this direction can be found in \cite{vanRees:2024xkb}. We will not discuss this further in the present thesis.}.

\begin{figure}
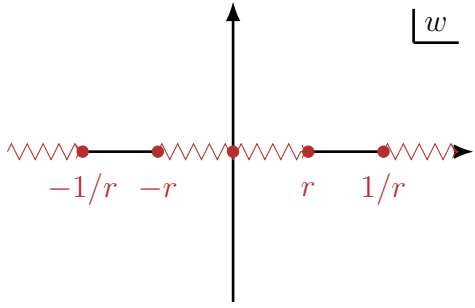

    \centering
    \AnalyticStructureThermal
    \caption{Analytic structure of a thermal correlator in the complex $\omega$ plane.}
    \label{fig:AnalyticStructure}
\end{figure}

As in the toy model, it is necessary to make further assumptions if we want to modify the formula above and compute the function $a(\Delta,J)$ solely in terms of the discontinuities of the two-point function. In particular, it is useful to study the analytic properties in the variables $(\tilde r,\omega)$, defined as
\begin{equation}
    z = \tau + i r = \tilde r \omega \ , \qquad \overline z = \tau - i r = \tilde r \omega^{-1} \ .
\end{equation}
As we will justify in Section \ref{sec:analyticstr}, in the complex $\omega$-plane, the two-point function is analytic except for a branch cut in $(-\tilde r,\tilde r)$, a branch cut in $(1/\tilde r,\infty)$, and a branch cut in $(-\infty,-1/\tilde r)$, as illustrated in Fig.~\ref{fig:AnalyticStructure}. Since $g(z,\overline z)= g(-z,-\overline z)$, the contribution from the discontinuity across the cut $(-\infty, -1/\tilde r)$ is equal to that of the cut along $(1/\tilde r,\infty)$, up to a factor $(-1)^J$. Furthermore, the symmetry $\omega \to \omega^{-1}$ (equivalently, $g(z,\overline z) = g(\overline z,z)$) implies that the contributions from the two cuts extending to $\pm\infty$ are equivalent to that from the cut along $(-\tilde r,\tilde r)$.

 The second technical assumption we need to make concerns the behavior of the two-point function as $|\omega| \to \infty$. We will assume that the correlator is bounded at large $\omega$, and in particular grows more slowly than $\omega^{J_\star}$ for some integer $J_\star$. In the context of the S-matrix bootstrap, this condition is known as the \textit{Regge bound} \cite{Regge:1959mz,Collins:1977jy}, and for this reason, the same terminology is sometimes used in the conformal bootstrap literature. We will also adopt it in the thermal case.

 While the first assumption can be justified by general arguments (to be reviewed in the following subsection), we are not aware of a general derivation of the Regge bound for two-point functions at finite temperature. Nevertheless, no counterexamples are known, and the bound appears to hold in all exactly solvable models (e.g., free theories, large $N$ expansions, two-dimensional correlators).

Under the assumptions above, and considering that
\begin{equation}
    C_J^{(\nu)}(\eta) = \frac{\Gamma(J+2 \nu)}{\Gamma\left(\nu-\frac{1}{2}\right)\Gamma\left(J+\nu+\frac{1}{2}\right)}\left(F_J(\omega^{-1})e^{i \pi \nu}+F_J(\omega)e^{-i \pi \nu}\right) \ ,
\end{equation}
where
\begin{equation}
    F_J(\omega) = \omega^{J+2 \nu}\  {}_2F_1 \left(J+2 \nu,\nu-\frac{1}{2},J+\nu+\frac{1}{2},\omega^2\right) \ ,
\end{equation}
we can easily conclude that equation~\eqref{eq:EuclideanIF} becomes
\begin{equation}\label{eq:INversionFormula}
    a(\Delta,J) = (1+(-1)^J)K_J\int_0^1 \frac{d \overline z}{\overline z}\int_1^{1/\overline z}\frac{d z}{z}(z \overline z)^{\Delta_\varphi-\frac{\Delta}{2}-\nu} (z-\overline z)^{2 \nu} F_J\left(\sqrt{\frac{\overline z}{z}}\right) \Disc g(z,\overline z) \ ,
\end{equation}
where $g(z,\overline z) = \langle \phi(z,\overline z) \phi(0,0)\rangle_\beta$. We have neglected contributions from arc integrals, which only contribute when $J \leq J_\star$.

The proof of the inversion formula above is completed once the analytic structure depicted in Fig.~\ref{fig:AnalyticStructure} is established — we will return to this point in the next subsection.

 In the next section of the thesis, we will make use of the inversion formula to extract dynamical information about the theory at finite temperature.

\subsection{The analytic structure of a thermal two-point function}\label{sec:analyticstr}

To recover the analytic properties of the thermal correlator, we begin by noting that the OPE (in the $s$-channel) converges when $\tilde r < |\omega| < \tilde r^{-1}$. However, OPE convergence near $z, \overline z \sim 1$ and $z, \overline z \sim -1$ extends the analyticity to a larger region. Still, this does not cover the entire complex $\omega$-plane. To achieve full analytic control, we make use of the Kaluza-Klein representation of the correlator.

In particular, it is useful to define the Hamiltonian and momentum operators on $S^1_\beta \times \mathbb{R}^{d-1}$, namely $\op H_{KK}$ and $\op P_{KK}$, generating translations along the non-compact and compact directions, respectively. We can then write:
\begin{equation}
    g(z,\overline z) = \langle \varphi(z,\overline z) \varphi(0,0)\rangle_\beta = \langle \Psi | e^{\frac{i}{2}\left(\op H_{KK}+\op P_{KK}\right)z -\frac{i}{2}\left(\op H_{KK}-\op P_{KK}\right)\overline z}| \Psi\rangle \ ,
\end{equation}
where $\ket{\Psi} = \varphi(0) |0 \rangle_{\beta}$.

The first claim is that $g(z,\overline z)$ is bounded when $\operatorname{Im}z > 0$ and $\operatorname{Im}\overline z < 0$. To show this, it is useful to rewrite:
\begin{equation}
    e^{\frac{i}{2}(\op H_{KK}+\op P_{KK})z - \frac{i}{2}(\op H_{KK}-\op P_{KK})\overline z} = V^{\frac{1}{2}} U V^{\frac{1}{2}} \ ,
\end{equation}
where
\begin{equation}
    V = e^{-\frac{1}{2}(\op H_{KK}+\op P_{KK})\operatorname{Im}z + \frac{1}{2}(\op H_{KK}-\op P_{KK})\operatorname{Im}\overline z} \ ,
\end{equation}
\begin{equation}
    U = e^{\frac{i}{2}(\op H_{KK}+\op P_{KK})\operatorname{Re}z - \frac{i}{2}(\op H_{KK}-\op P_{KK})\operatorname{Re} \overline z} \ .
\end{equation}
From this decomposition and by applying the Cauchy–Schwarz inequality, we find:
\begin{equation}
    |g(z,\overline z)| \le |\langle \Psi | V | \Psi \rangle|^2 \ .
\end{equation}

Positivity of $\op H_{KK} \pm \op P_{KK}$ is, on $\mathbb{R}^d$, a simple consequence of energy positivity. On the thermal geometry, however, this is subtler, since the eigenvalues of $\op P_{KK}$ are quantized according to the Kaluza-Klein prescription, and are associated with quantum numbers $n$. For sufficiently large $|n|$, the compactness of the $\tau$ direction becomes negligible, and the energies approach their flat-space limit. This implies that $\op H_{KK} \pm \op P_{KK}$ are bounded from below.

Let us assume $\op H_{KK} \pm \op P_{KK} > \lambda$ and define $\zeta = \min(\operatorname{Im}z, -\operatorname{Im} \overline z)$. Then:
\begin{equation}
    |g(\tau, r)| \le g(0, \zeta) \, e^{ -\frac{\lambda}{2}(\operatorname{Im}z - \zeta) - \frac{\lambda}{2}(-\operatorname{Im}\overline z - \zeta)} \ .
\end{equation}

Since $g(0, \zeta)$ corresponds to a configuration at $\tau = 0$, it is non-singular, and we conclude that $g(\tau, \vec{x})$ is analytic when $\zeta \ge 0$, i.e., when $\operatorname{Im}z > 0$ and $\operatorname{Im}\overline z < 0$, which corresponds to the upper half-plane in the $\omega$ variable. The symmetry $\omega \to -\omega$ allows us to extend analyticity also to the lower half-plane.

\subsection{A dispersion relation}\label{Sec:DispRel}

The inversion formula relies solely on the analytic structure of the correlation function. By using it, we are—at least in principle—able to write down all the coefficients of the OPE expansion in a single compact formula. This is equivalent to reconstructing the full correlator\footnote{As we will explain later, this statement is not entirely accurate; for now, we neglect certain details.}.

However, to recover the full correlator explicitly, one would need to re-sum all the conformal blocks weighted by the OPE data obtained through the inversion formula. A more efficient method to compute the two-point function directly is to make use of a dispersion relation.

The term \textit{dispersion relation} originates from physical contexts such as optics, where it is used in the derivation of, for example, the \textit{Kramers–Kronig relations}, which describe how a system absorbs and reflects light. In our context, the use of the term is somewhat of an abuse of nomenclature, but it is standard in the literature.
 
 As for the case of the inversion formula it is instructive to derive the dispersion relation first in a simplified examples.
 
\paragraph{Dispersion relation in a toy model.} 
Let us illustrate the dispersion relation in the toy model discussed above, consisting of a function $f(E)$ which admits an expansion of the form
\begin{equation}
    f(E) = \sum_{J = 0}^\infty c_J E^J \ ,
\end{equation}
and has analytic properties summarized in Fig.~\ref{fig:AnalyticStructureToy}.

The first observation is that, if $|f(E)/E| \to f_\infty$ as $|E| \to \infty$, we can apply Cauchy's theorem to conclude that
\begin{equation}
    f(E) - f_\infty = \frac{1}{2 \pi i} \oint_{|E|<1} dE' \, \frac{f(E')}{E - E'} \ .
\end{equation}

We now introduce an additional assumption in our toy model, which will be mirrored in the thermal case: namely, we assume that the function $f$ is symmetric in $E$, i.e., $f(E) = f(-E)$.

By opening up the contour, it is easy to express the right-hand side as an integral over the discontinuities of the function $f$. In fact,
\begin{equation}
    f(E) - f_\infty = \int_1^\infty \frac{\Disc f(E')}{E - E'} \, dE' + \int_{-\infty}^{-1} \frac{\Disc f(E')}{E - E'} \, dE' = - \int_1^\infty dE' \, \frac{2 E' \Disc f(E')}{(E - E')(E + E')} \ .
\end{equation}

In this way, we have written the function $f(E)$ entirely in terms of its discontinuities (plus a boundary term at infinity),
\begin{equation}
    f(E) = f_\infty + \int_1^\infty dE' \, \mathcal{K}(E, E') \Disc f(E') \ , \qquad \mathcal{K}(E, E') = \frac{2 E'}{(E - E')(E + E')} \ ,
\end{equation}
where the kernel $\mathcal{K}(E, E')$ encodes information about the symmetries of $f(E)$—in this case, the presence of two symmetric branch cuts.

This strategy is very general and physically meaningful. For instance, in the context of the S-matrix bootstrap, the \textit{Titchmarsh theorem} relates dispersion relations to causality conditions (see e.g.~\cite{Mizera:2023tfe} for a review).

\paragraph{A dispersion relation for thermal two-point functions.} 
The procedure outlined for the toy model can be easily adapted to the thermal case, in the complex $\omega$-plane, whose analytic properties are summarized in Fig.~\ref{fig:AnalyticStructure}. One important difference is that we do not know the precise behavior of the two-point function at infinity: we can only rely on the Regge assumption, i.e., that $g(\omega, r) \to \omega^{J_\star}$ as $|\omega| \to \infty$.

Therefore, the constant term $f_\infty$ appearing in the toy model must be replaced by the contribution from all conformal blocks associated with operators of spin $J \leq J_\star$. We will refer to this contribution as the arc terms.

The remaining part, which we denote as $g_{\text{dr}}$, is given by:
\begin{equation}
    g_{\text{dr}}(\tilde r, \omega) = \frac{1}{2 \pi i} \int_{0}^{\tilde r} d\omega' \, \mathcal{K}(\omega, \omega') \, \Disc g(\tilde r, \omega') \ ,
\end{equation}
where the integration kernel is given by
\begin{equation}\label{eq:Kernel}
    \mathcal{K}(\omega, \omega') = \frac{\omega (1 - \omega')(1 + \omega')}{\omega' (\omega' - \omega)(1 - \omega \omega')} \ .
\end{equation}

The derivation of the kernel is straightforward and relies only on the following symmetries of the thermal correlator:
\begin{equation}
    g(\tilde r, \omega) = g(\tilde r, \omega^{-1}) \ , \qquad g(\tilde r, \omega) = g(\tilde r, -\omega) \ .
\end{equation}
The first relation simply reflects the reality property of the correlator (it can be written as $g(z, \overline z) = g(\overline z, z)$), while the second is a consequence of parity symmetry (equivalently, $g(z, \overline z) = g(-z, -\overline z)$).

This dispersion relation was first proposed in \cite{Alday:2020eua} and later used more extensively in \cite{NewAnalytic}. In particular those formulae can be used to bootstrap the dynamics at finite temperature as we will see explicitly in \ref{sec:analyticbootstrap}

\subsection{Discontinuities and OPE}\label{sec:disc}

It is possible to combine the OPE expansion with either the inversion formula or the dispersion relation, and compute the discontinuities block by block in the OPE. This procedure is not rigorous, since the discontinuity of a function does not necessarily commute with its expansion. In fact, we will comment on this subtlety and discuss its physical implications in the next chapter.

For the moment, let us neglect these potential issues and assume that we can compute the discontinuity order by order in the OPE expansion. Let us focus on computing the discontinuity in $\overline z$. For polynomial functions, we have:
\begin{multline}\label{eq:disc}
    \Disc\left[(z \overline z)^\alpha (a - z)^\gamma (b - \overline z)^\delta\right] = (z \overline z)^\alpha (a - z)^\gamma \Disc(b - \overline z)^\delta \\
    = 2 i \sin(\pi \delta) (z \overline z)^\alpha (a - z)^\gamma (\overline z - b)^\delta \Theta(\overline z - b) \ ,
\end{multline}
and note that the discontinuity is non-zero if one of the following two conditions holds:
\begin{itemize}
    \item $\delta$ is non-integer;
    \item $\delta$ is integer but negative.
\end{itemize}

The expression above is sufficient to compute discontinuities for the thermal two-point function, following a simple strategy: first expand the correlator (or the conformal blocks we want to invert) around $z, \overline z \sim 1$, then use equation~\eqref{eq:disc} to compute the discontinuity.

To make things concrete and provide a first simple result, we now discuss the discontinuity in the free scalar theory for a generic scaling dimension $\Delta_\phi$.

\paragraph{Discontinuities and free scalars.}
In a free scalar theory, the OPE between fundamental scalar fields can be written as \cite{Osborn:1993cr}
\begin{equation}
    \phi \times \phi = \mathds{1} + [\phi \phi]_{0,J} \ ,
\end{equation}
where $[\phi \phi]_{0,J}$ are higher-spin conserved currents with scaling dimensions $\Delta = 2\Delta_\phi + J$. Due to the equations of motion, operators of the form $\phi \Box^n \partial^J \phi$ are absent in a free theory.

The discontinuity of the identity contribution is non-zero in general. Indeed,
\begin{equation}
    \Disc\left((1 - z)(1 - \overline z)\right)^{-2\Delta_\phi} = -2 i \sin(2 \pi \Delta_\phi) \left((1 - z)(\overline z - 1)\right)^{-2\Delta_\phi} \Theta(\overline z - 1) \ ,
\end{equation}
for any free scalar theory in $d > 2$ dimensions.\footnote{Recall that the dimension of a free scalar in $d$ dimensions is $\Delta_\phi = (d - 2)/2$.}

On the other hand, the conformal block of a generic higher-spin current around $z, \overline z \sim 1$ takes the form
\begin{equation}
    (1 - z)^J (1 - \overline z)^J C_J^{(\nu)}\left( \frac{-z - \overline z + 2}{2 \sqrt{(1 - z)(1 - \overline z)}} \right) \ .
\end{equation}
Since $J$ is a positive integer, the discontinuity of these blocks is identically zero.
Therefore, the entire discontinuity of the correlator in a free scalar theory is fully captured by the identity operator, as already noted in~\cite{Manenti:2019wxs}.

Note that this implies that the correlator can be completely determined solely in terms of the discontinuity, provided that arcs are neglected. The discontinuity can be computed by expanding the correlator in the OPE and analyzing it block by block, as just explained. In the case of the free theory, this means that the dispersion relation only takes as input the identity operator and produces as output the full set of OPE coefficients.

It is worth pointing out that the result of the dispersion relation, as derived in this thesis, is not KMS invariant. In Section~\ref{sec:analyticbootstrap}, we will exploit this tension between the dispersion relation and KMS invariance to make predictions in free and interacting theories, reproducing results from perturbation theory in the $\mathrm{O}(N)$ model in the $\varepsilon$-expansion in Section~\ref{chap:ONmodel}.

\subsection{OPE in momentum space}\label{Sec:OPEmomspace}

One may ask whether there is a notion of OPE in momentum space. The main obstacle lies in the fact that, in position space, the two-point function has poles when the (imaginary) time separation between operators is a multiple of $\beta$. This reflects the fact that the OPE is not convergent everywhere, but only within a specific region $x^2 + \tau^2 < \beta^2$. 

For this reason, it is not possible to simply assert the existence of an OPE expansion in momentum space: the Fourier transform does not commute with the OPE expansion. Nonetheless, one might physically expect a meaningful notion of OPE in the large momentum limit, which corresponds to short distances. This expectation is indeed justified: a well-defined OPE in momentum space exists in the limit where $\omega \to \infty$ with $\xi = |\vec{k}|/\omega$ fixed, as shown in \cite{Caron-Huot:2009ypo}.

Only in this limit can the two-point function in momentum space be approximated by its OPE expansion:
\begin{equation}
    g(\omega, k) = \langle \phi(\omega, k) \phi(-\omega, -k) \rangle_\beta \sim \sum_{\mathcal O} a_{\mathcal O} \tilde f_{\mathcal O}(\omega, k) \ ,
\end{equation}
where $\tilde f_{\mathcal O}$ denotes the Fourier transform of the thermal OPE block. This was computed in \cite{Manenti:2019wxs}:
\begin{equation}
    \tilde f_{\mathcal O}(\omega, k) = \sum_{j = 0}^{J/2} c_{J,j} \frac{2^{\beta_-} \pi^{\frac{d}{2}} \Gamma(\beta_+)}{\Gamma(\alpha)} \frac{{}_2F_1\left(\beta_-, 1 - \beta_-; 1 - \beta_+; \frac{k + i \omega_n}{2k}\right)}{k^{\beta_-} (k + i \omega_n)^{\beta_+}} + \mathcal{O}(e^{-k}) \ ,
\end{equation}
where, with a slight abuse of notation, we define $k = |\vec{k}|$, and
\begin{equation}
    \tilde j = \frac{J}{2} - j \ , \qquad \tilde \Delta = \frac{\Delta}{2} - \Delta_\phi \ , \qquad \alpha = \tilde j - \tilde \Delta \ , \qquad \beta_\pm = \frac{d}{2} + \tilde \Delta \pm \tilde j \ .
\end{equation}

Note that all blocks corresponding to operators $[\phi \phi]_{n,J}$ with dimensions $\Delta = 2\Delta_\phi + 2n + J$ do not appear in the momentum-space OPE. In fact, in a free theory, only the identity contribution is present, yielding
\begin{equation}
    g(\omega, k) \sim \frac{1}{\omega^2 + k^2} \ ,
\end{equation}
which, in this case, is not only an approximation for $\omega, k \to \infty$, but the exact result. 

The fact that all higher-spin currents vanish both from the momentum-space OPE and from the discontinuity of the two-point function in position space suggests a deep connection between these two quantities. This relation can be made precise, as shown in \cite{Manenti:2019wxs}.

\begin{figure}[htb]
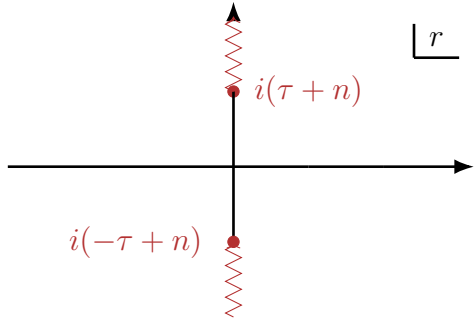

\centering
\rdiscontinuity
\caption{Schematic representation of the analytic structure of the integral in equation \eqref{eq:gg}.}
\label{fig:momdisc}
\end{figure}

The idea is simply to rewrite the momentum-space two-point function as
\begin{equation}
    g(\omega, k) = (2\pi)^{\nu + \frac{1}{2}} \int_0^\infty dr \int_0^1 d\tau \, r^{2\nu} (kr)^{\frac{1}{2} - \nu} J_{\nu - \frac{1}{2}}(kr) e^{-i \omega_n \tau} g(r, \tau) \ .
\end{equation}

Let us assume for the moment that $\nu$ is an integer (equivalently, $d$ even), to simplify the analysis. (An equivalent formula holds also for odd $d$.) In the complex $r$-plane, we encounter two branch cuts: one starting from $r = i(\tau + n)$ and one from $r = i(-\tau + n)$ for $n \in \mathbb{N}$, as illustrated in Fig.~\ref{fig:momdisc}.

Choosing the orientation of the branch cuts so that they do not cross the real axis, one can show that
\begin{equation}\label{eq:gg}
    g(\omega, k) = (2\pi)^{\nu + \frac{1}{2}} \frac{1}{2} \int_0^{i\infty} dr \int_0^1 d\tau \, r^{2\nu} (kr)^{\frac{1}{2} - \nu} J_{\nu - \frac{1}{2}}(kr) e^{-i \omega_n \tau} \Disc g(r, \tau) \ .
\end{equation}

Observe that this formula correctly reproduces the free propagator in the case of a free scalar theory. Since the discontinuities of the blocks associated with $[\phi \phi]_{0,J}$ vanish, only the identity contribution remains, and $g(\omega, k)$ simplifies to
\begin{equation}
    g(\omega, k) \propto \frac{1}{\omega^2 + k^2} \ ,
\end{equation}
where the proportionality symbol indicates that we are neglecting normalization factors, as they are irrelevant for the present discussion.

We will return to this formula in Section \ref{sec:analyticbootstrap} and we will compare it with our analytical bootstrap approach.

\section{Summary of the chapter}

In this chapter, we introduced finite temperature effects in conformal field theories and explained how thermal correlation functions can be expressed in terms of conformal data. As first proposed in the seminal work \cite{Iliesiu:2018fao}, the conformal data at finite temperature is extended by including thermal one-point functions. These coefficients are the key quantities that characterize thermal effects in conformal field theories.

\begin{table}[h!]
    \centering
    \renewcommand{\arraystretch}{1.5}
    \begin{tabular}{|c|c|}
        \hline
        Translations & \cmark \\
        \hline
        Dilatations & \xmark \\
        \hline
        (Spatial) Rotations &  \cmark \\
        \hline
        Boosts & \xmark \\
        \hline
        Special conf. tr. & \xmark \\
        \hline
        Supersymmetry & \xmark \\\hline
        $R$-symmetry & \cmark \\ \hline
    \end{tabular}
    \caption{Preserved (\cmark) and broken (\xmark) (Super)-conformal generators at finite temperature.}
    \label{tab:symmetrypreserved?}
\end{table}

We discussed the broken and unbroken symmetries of the theory (see Tab \ref{tab:symmetrypreserved?} ) and derived the associated broken Ward identities, which capture the symmetry breaking induced by finite temperature. By using these identities, we identified the kinematical structure of thermal one-point functions, and consequently of the OPE decomposition of thermal two-point functions, given by
\begin{equation}\label{eq:thermalOPEreview}
    g(\tau,r) = \sum_{\mathcal{O}\in \phu \times \phu}\frac{J!}{2^J (\nu)_J}\frac{1}{\beta^{\Delta_{\co}}}\frac{f_{\phi\phi\co} b_{\co}}{c_{\co}} \left(\tau^2+ r^2  \right)^{\frac{1}{2}\left(\Delta_{\co}-2\Delta_\varphi \right)} C_{J}^{(\nu)}\left(\frac{\tau}{\sqrt{\tau^2+r^2}}\right ) \ .
\end{equation}

Broken Ward identities can also be derived for superconformal generators, showing that supersymmetry is explicitly broken, while the $R$-symmetry is globally preserved. All the identities have been tested in simple settings where analytical results are available, such as free theories and two-dimensional models.

We also explored the analytic structure of thermal two-point functions and its implications. Under reasonable and physically motivated assumptions, one can show that thermal correlation functions share universal analytic properties. From these, it is possible to derive both an inversion formula and a dispersion relation. The former allows for the extraction of thermal OPE coefficients from the discontinuity of the correlator, while the latter reconstructs the full two-point function from the same discontinuity. These two expressions are conceptually similar and physically equivalent, as they rely on the same input.

Finally, we commented on the analysis of thermal two-point functions in momentum space and reviewed a notion of OPE directly formulated in that setting. We highlighted the relation between the momentum-space correlator and the discontinuity (in position space) of the thermal two-point function.

All results presented in this chapter pertain to the kinematical structure of thermal one- and two-point functions, with some additional insights into higher-point functions via the broken Ward identities. In the following chapters, we will apply these findings to construct a bootstrap problem for thermal dynamics, namely for the thermal one-point functions.

\chapter{Finite Temperature Effects in CFT: Dynamics}\label{ref:dynamics}
In this Section we will provide the tools to formulate a bootstrap problem and we show how this problem can be solved. In some situations a numerical methods, we first proposed in \cite{Barrat:2024fwq}, are necessary while in some other cases analytical bootstrap method, we proposed in \cite{NewAnalytic}, may be more efficient. In this chapter we only apply and test the methods in free and two-dimensional models; in Chapter \ref{chap:ONmodel} we will present non-trivial applications for the case of the $\mathrm O(N)$ models.
\section{Thermodynamics}
Before continuing and exploring the dynamics of the theory, it is useful to understand how the thermal OPE data are related to physical quantities such as the free energy and the entropy. In local theories, a very important operator that always appears is the stress-energy tensor. As we already showed in the previous paragraph, this operator is related to the currents of the conformal group and indeed appears in all the Broken Ward identities. We already commented above that the energy density is defined as
\begin{equation}\label{eq:energydensity}
    \mathcal{E} = - \langle T^{00} \rangle_\beta = - \frac{d-1}{d} \frac{b_T}{\beta^d} > 0 \ .
\end{equation}
The minus sign is conventional and is due to working in Euclidean coordinates. In fact, this sign ensures that \( \langle T_\text{Mink.}^{00} \rangle_\beta = i^2 \langle T_\text{Eucl.}^{00} \rangle_\beta \ge 0 \).

Once the energy density is defined, it is also possible to define the free energy density of the system. To do so, we recall the \textit{thermodynamic equation}:
\begin{equation}
    F = E - TS = E + T \frac{dF}{dT} \ ,
\end{equation}
where free energy, energy, and entropy can be written in terms of their densities:
\begin{equation}
    F = \int_{\mathbb{R}^{d-1}} d^{d-1}x\   f \ , \hspace{1cm} S = \int_{\mathbb{R}^{d-1}} d^{d-1}x\  s \ , \hspace{1cm} E = \int_{\mathbb{R}^{d-1}} d^{d-1}x\  \mathcal{E} \ .
\end{equation}
As an effect of symmetries and dimensional analysis, \( f, \mathcal{E} \propto T^d \) and \( s \propto T^{d-1} \). In particular, \( f, \mathcal{E} \) and \( s \) do not depend on the spatial coordinates \( x \), meaning that the thermodynamic equation can be solved for the densities:
\begin{equation}
    f = \mathcal{E} + T \frac{d f}{d T} \ .
\end{equation}
By plugging into this differential equation the definition of the energy density \eqref{eq:energydensity}, we find
\begin{equation}
    f = \frac{1}{d} \frac{b_T}{\beta^d} \le 0 \ , \hspace{1cm} s = \frac{b_T}{\beta^{d-1}} \le 0 \ .
\end{equation}
All thermodynamic quantities are therefore functions of a single thermal datum: the one-point function of the stress-energy tensor, \( b_T \). The latter is thus a natural target for computation in thermal CFT. Luckily, the stress tensor enters the two-point function of any pair of identical scalar operators in a generic CFT.

In the following, we develop new methods to compute thermal OPE data from the two-point function of two identical scalars $\phi$. These data will include \( a_T \), defined as
\begin{equation}
    a_T = \frac{b_T f_{\phi \phi T}}{c_T} \frac{1}{2\nu} \ .
\end{equation}
As a result of Ward identities,
\begin{equation}
    f_{\phi \phi T} = - \frac{d}{d-1}  \frac{\Delta_\phi}{S_d} \frac{c_{T\, \text{free}}}{c_T} \ , \hspace{1cm} S_d = \text{vol}(S^{d-1}) = \frac{2 \pi^{\frac{d}{2}}}{\Gamma\left(\frac{d}{2}\right)} \ ,
\end{equation}
and therefore
\begin{equation}\label{eq:freeEnergy}
    f = - a_T  \frac{S_d}{\beta^d (d-1) \Delta_\phi}  \frac{c_T}{c_{T\, \text{free}}} \ .
\end{equation}

\section{A bootstrap problem}\label{eq:bootstrapporoblem}
The goal of this chapter is to solve for the dynamics of the thermal CFT. As explained earlier, the thermal CFT data—i.e. the minimal amount of data required to compute correlation functions at finite temperature—consist of the zero-temperature CFT data (given by the conformal dimensions of the operators and the structure constants governing the OPEs), supplemented by the thermal one-point function coefficients. The idea behind the thermal bootstrap is to input the zero-temperature CFT data in order to compute thermal one-point functions. Schematically, we have:
\begin{equation}
    \{\{\Delta_{\mathcal O},J_{\mathcal O}\},f_{\mathcal O \mathcal O' \mathcal O''}\} \hspace{0.5 cm}\overset{\text{thermal bootstrap}}{\longrightarrow} \hspace{0.5 cm}\{b_{\mathcal O}\} \ .
\end{equation}

In particular, the zero-temperature data are treated, for the purposes of this thesis, as input to the program. Nonetheless, consistency conditions of the CFT on non-flat geometries may in principle constrain the zero-temperature conformal data. The most famous and successful example is that of two-dimensional CFTs on the torus, where modular invariance strongly constrains the theory \cite{Cardy:1986ie}. In spacetime dimensions higher than two, the situation is more complex. Still, it is expected that placing the theory on non-trivial manifolds—such as the thermal manifold—imposes constraints on the zero-temperature CFT data \cite{El-Showk:2011yvt}, as illustrated in recent works \cite{Benjamin:2023qsc,Allameh:2024qqp,Benjamin:2024kdg}. Constraining the zero-temperature data is not the goal of this thesis, although we will return to this point later.

Coming back to the thermal bootstrap, the key idea is that the OPE decomposition of the thermal two-point function
\begin{equation}\label{eq:OPEdecomp}
    g(\tau,r) = \langle \varphi(\tau,r)\varphi(0,0)\rangle_\beta =\sum_{\mathcal O} \frac{a_{\mathcal O}}{\beta^{\Delta}} \left(\tau^2+r^2\right)^{\frac{\Delta_{\mathcal O}}{2}-\Delta_\varphi} C_J^{(\nu)}\left(\frac{\tau}{\sqrt{\tau^2+r^2}}\right)\ ,
\end{equation}
is not explicitly periodic in \( \tau \in S_\beta^1 \). Therefore, it is possible to combine the OPE decomposition with the periodicity on the thermal circle to constrain the thermal one-point functions—or more precisely, the thermal OPE coefficients \( a_{\mathcal O} \).

In order to pose a well-defined bootstrap problem, we now enumerate the \textit{bootstrap axioms}, i.e. the set of consistency conditions imposed on the two-point function \cite{Iliesiu:2018fao, Alday:2020eua}.

   \paragraph{KMS condition} 
The periodicity of correlation functions at finite temperature, along the thermal circle \( S_\beta^1 \), is known in the literature as the Kubo-Martin-Schwinger (KMS) condition \cite{Kubo:1957mj,Martin:1959jp,Iliesiu:2018fao,Alday:2020eua}. This condition reads
\begin{equation}\label{KSM condition}
    g(\tau,r) = g(\tau+\beta,r) \ .
\end{equation}
For reasons that will become clear later, it is often useful to combine the KMS condition with the symmetry \( \tau \to -\tau \). The latter is clearly a symmetry in parity-invariant theories. If the CFT at zero temperature is not parity-invariant under this transformation, then it must be accompanied by a reflection of one (or more) directions in the spatial component \( \mathbb{R}^{d-1} \). Combining this symmetry with the KMS condition, we obtain
\begin{equation}
    g(\tau,r) = g(\beta - \tau, r) \ .
\end{equation}

\paragraph{Analyticity} 
In the complex \( \omega \)-plane, where \( z = \tau + i r = \tilde r \omega \) and \( \overline z = \tau - i r = \tilde r \omega^{-1} \), the analyticity condition states that the two-point function may only have branch cuts starting at \( \omega = \pm \tilde r \) and \( \omega = \pm \tilde r^{-1} \); moreover, it has poles at \( \omega = 0, \pm \tilde \omega \), where \( \tilde \omega = \sqrt{(\tau + i|\vec x|)/(\tau - i|\vec x|)} \). Outside of these poles and cuts, the two-point function must be analytic \cite{Iliesiu:2018fao}. This structure is illustrated in Figure~\ref{fig:AnalyticStructure}, and its derivation was discussed in the previous chapter.

\paragraph{Regge limit} 
Define the complex variables \( z = \tau + i|\vec x| \) and \( \overline z = \tau - i|\vec x| \), and introduce \( \rho = \sqrt{z \overline z} \) and \( \eta = \frac{z + \overline z}{2\sqrt{z \overline z}} \). In these coordinates, the two-point function reads \( f_\beta(\rho, \eta) \). The Regge limit is defined as
\begin{equation}
    \lim_{|\eta| \rightarrow \infty} f_\beta(\rho, \eta) \ , \quad \text{with } \rho \text{ fixed} \ .
\end{equation}
The two-point function must be polynomially bounded in the Regge limit \cite{Iliesiu:2018fao,Alday:2020eua}.\footnote{Analyticity in spin is a consequence of polynomial boundedness as \( |\eta| \to \infty \), since this implies that the two-point function does not grow faster than \( \eta^{J_\text{Regge}} \). This is the analog of the Regge limit at zero temperature.}

\paragraph{Large distance limit} 
It is commonly believed that compactification on the thermal manifold induces a mass gap \( m_{\text{th}} > 0 \).\footnote{By dimensional analysis, this mass must be proportional to \( \beta^{-1} \). However, the existence of a thermal mass is not always guaranteed: for instance, in some cases (e.g. free bosonic theories) symmetries prevent the theory from developing a gap. Similarly, spontaneous symmetry breaking at finite temperature, as in \cite{Chai:2020onq,Chai:2020zgq}, could lead to gapless behavior—though we are not aware of any such example in integer spacetime dimensions. Nonetheless, in a generic interacting CFT at finite temperature, the existence of a thermal mass is expected due to the suppression of correlations at large distances.} Assuming this folk-theorem, the two-point function behaves as \cite{Iliesiu:2018fao,Caron-Huot:2022akb}
\begin{equation}\label{eq: clustering}
    g(\tau,r) \overset{r \gg \beta}{\sim} \langle \varphi \rangle_\beta^2 + \mathcal{O}\left(e^{-m_{\text{th}} r} \right) \ .
\end{equation}
Note that, independently of the exponential behavior, the large-distance limit follows from cluster decomposition.

\paragraph{Zero temperature limit} 
Since the thermal two-point function is defined on a thermal manifold for arbitrary values of \( \beta \), we expect \( g(\tau, r) \) to be a continuous function of \( \beta \). Moreover, we expect it to be smooth and admit a well-defined limit as \( \beta \rightarrow \infty \):
\begin{equation}\label{LargeBetaLimit}
   \lim_{\beta \rightarrow \infty} \langle \varphi(\tau,r) \varphi(0,0) \rangle_{\beta} = \langle \varphi(\tau,r) \varphi(0,0) \rangle_{\mathbb{R}^d} = \frac{1}{\left(\tau^2 + r^2 \right)^{\Delta_\varphi}} \ .
\end{equation}

\paragraph{OPE} 
When two local scalar operators \( \varphi \), with conformal dimension \( \Delta_\varphi \), are brought close together, the OPE is exact and reads \cite{Iliesiu:2018fao}
\begin{equation}
    \varphi(x_1) \times \varphi(x_2) \sim \sum_{\mathcal O \in \mathcal{O}_1 \times \mathcal{O}_2} \frac{f_{\mathcal{O}_1 \mathcal{O}_2 \mathcal O}}{c_{\mathcal O}} \left( \tau^2 + |\vec x|^2 \right)^{\frac{1}{2}(\Delta_{\mathcal O} - 2\Delta_\varphi - J)} x_{\mu_1} \cdots x_{\mu_J} \, \mathcal O^{\mu_1 \ldots \mu_J}(x_2) \ ,
\end{equation}
where \( x^\mu = x_1^\mu - x_2^\mu \), \( \Delta_{\mathcal O} \) is the conformal dimension of \( \mathcal{O} \), \( f_{\mathcal{O}_1 \mathcal{O}_2 \mathcal O} \) is the zero-temperature OPE coefficient, and \( c_{\mathcal O} \) is the normalization factor in the two-point function:
\begin{equation}
    \langle \mathcal O^{\mu_1 \ldots \mu_J}(x) \mathcal O_{\nu_1 \ldots \nu_J}(0) \rangle = c_{\mathcal O} \frac{I_{(\nu_1 \ldots \nu_J)}^{(\mu_1 \ldots \mu_J)}}{|x|^{2\Delta_{\mathcal O}}} \ , \quad I^\mu_{\, \nu} = \delta^\mu_{\, \nu} - \frac{2x^\mu x_\nu}{x^2} \ .
\end{equation}
The geometry of the thermal manifold imposes a minimal radius of convergence for the OPE, since the largest sphere with flat interior has radius \( \beta \). Consistency with the OPE regime requires the thermal two-point function to be compatible with the expansion \eqref{eq:OPEdecomp} within the convergence radius.

The bootstrap problem for the thermal scalar two-point function consists in finding a function of the coordinates \( x_1 \), \( x_2 \), and the inverse temperature \( \beta \), such that it satisfies all the conditions listed above.

  \subsection{Uniqueness of the solution and free scalar theory in $d = 4$}

Whether the bootstrap problem defined above admits a unique solution remains an open question. In particular, the minimal set of conditions required to uniquely determine the two-point function is not known. Suppose there exist two distinct solutions, \( g_1 \) and \( g_2 \), to the problem. First, consider the case where the two solutions differ by the contribution of a finite number of local operators in the OPE regime. Then, the large-distance limit of the difference \( \Delta g = g_1 - g_2 \) is\footnote{The limit holds only for even-spin operators; however, this is sufficient since the one-point function of an odd-spin operator vanishes, and so does their contribution to the two-point function.}
\begin{equation}
    \lim_{r \to \infty} \left| \Delta g(\tau, r) \right| = \infty \ ,
\end{equation}
and therefore, if \( g_1 \) satisfies the clustering property expressed in equation \eqref{eq: clustering}, then \( g_2 \) does not (and \emph{vice versa}). Hence, only one of \( g_1 \) or \( g_2 \) can be a solution to the bootstrap problem.

If instead \( g_1 \) and \( g_2 \) differ by an overall factor, then their limits in \eqref{LargeBetaLimit} also differ by the same factor. This discrepancy can be traced back to the normalization of the two-point function at zero temperature. If we fix this normalization before solving the bootstrap problem, then it also fixes the residues of the poles of the finite-temperature two-point function.

Two solutions \( g_1 \) and \( g_2 \) that do not differ by an overall factor must have the same poles with the same residues, since the poles are determined by the zero-temperature data, and the \( \beta \to \infty \) limit determines the coefficients in front of them.

We were not able to find a method to discriminate between two solutions that comply with all the above reasoning but differ in their branch cut structure. This ambiguity was already noted in \cite{Alday:2020eua}, where the authors proved uniqueness under specific assumptions\footnote{The necessary condition for uniqueness proposed in \cite{Alday:2020eua} is \( \text{Disc}(z, \overline z) = 0 \) for \( 0 \le \overline z = \tau - i r \le 1 \) and \( 1 \le z = \tau + i r \).}, using Regge boundedness and subtracted thermal dispersion relations.

However, in the case of a free scalar theory in four spacetime dimensions, it is possible to reconstruct the thermal two-point function using only the bootstrap conditions outlined above, providing evidence in favor of uniqueness in this case. This approach appears to work only for the free scalar theory, and the underlying reason is the simplicity of the OPE between two fundamental scalars, which allows the two-point function to factorize into \emph{holomorphic} components. In general, such factorization is not evident from the OPE, and we do not expect that all two-point functions admit such a decomposition. Nonetheless, we do not exclude the possibility that a more general uniqueness result could be obtained with more sophisticated tools.

   \paragraph{Free scalar theory in $d = 4$} \label{sec: Exact computation of phiphi for a free scalar theory}

We will explicitly solve the bootstrap problem posed above for the two-point function of two fundamental scalar fields $\phu(x)$. We will also use the resulting two-point function to extract several one-point functions, as done in \cite{Petkou:2021zhg,Iliesiu:2018fao}, focusing in particular on the one-point function of the stress-energy tensor $T^{\mu \nu}$.

In Section \ref{subsec: constraints in the OPE regime}, we computed the explicit expression of the two-point function in the OPE regime directly from the dilatation broken Ward identity:
\begin{equation}
    g(\tau,r) = \sum_{\mathcal{O}\in \phu \times \phu}\frac{J!}{2^J (\nu)_J}\frac{1}{\beta^{\Delta_{\co}}}\frac{f_{\phu\phu\co} b_{\co}}{c_{\co}} \left(\tau^2+ r^2  \right)^{\frac{1}{2}\left(\Delta_{\co}-2\Delta_\varphi \right)} C_{J}^{(\nu)}\left(\frac{\tau}{\sqrt{\tau^2+r^2}}\right) \ .
\end{equation}

Specializing to a free scalar theory, we can use the explicit OPE expansion:
\begin{equation}
	\phu(x)\times \phu(0) \sim 1+a\, \phu^2(0) + \underbrace{b_{\mu}\phu(0)\partial^{\mu}\phu(0)}_{\sim V^{\mu}(0)}+\underbrace{c_{\mu \nu}\phu(0)\partial^{\mu}\partial^{\nu}\phu(0)}_{\sim T^{\mu \nu}(0)}+ \dots \ , \label{eq: free scal 2pt}
\end{equation}
where the explicit structure of the operators is not important for our purposes. The crucial information is that the conformal dimension of any operator $\mathcal{O} \in \phu \times \phu$ (excluding the identity operator $\mathds{1}$) is determined by its spin $J$:
\begin{equation}\label{eq: Dimensions of operator FST}
	\Delta_{\co} = d - 2 + J \ ,
\end{equation}

Using this, we can rewrite the OPE sum as a sum over the spin $J$:
\begin{equation} \label{eq: thermal 2pts function}
	g(\tau, r) = \frac{1}{\left(\tau^2 + r^2 \right)^{\frac{d-2}{2}}} + \sum_{J=0}^{\infty}k_{J}\frac{1}{\beta^{d-2+J}}\frac{J!}{2^{J}(\nu)_{J}} \left(\tau^2+ r^2  \right)^{\frac{J}{2}} C_{J}^{(\nu)}\left(\frac{\tau}{\sqrt{\tau^2+r^2}} \right) \ ,
\end{equation}
where all dynamical information is grouped into the coefficients
\begin{equation}\label{eq: kJ definition}
	k_{J} = \frac{f_{\phu\phu\co} b_{\co}}{c_{\co}} \ .
\end{equation}

The expression \eqref{eq: thermal 2pts function} lacks manifest periodicity over the thermal circle, which should be enforced by properly constraining the coefficients $k_J$. Since $k_J$ are functions of the one-point function data $b_{\mathcal{O}}$, imposing periodicity on the scalar two-point function provides information about the one-point functions in the free scalar theory itself \cite{Iliesiu:2018fao}.

From now on, we set $d = 4$, so that \eqref{eq: thermal 2pts function} becomes:
\begin{equation} 
	f_{\beta}^{d=4}(\tau, r) = \frac{1}{\tau^2 + r^2} + \sum_{J=0}^{\infty} \frac{1}{\beta^{2 + J}} \frac{k_J}{2^J} \left(\tau^2 + r^2\right)^{J/2} U_J\left(\frac{\tau}{\sqrt{\tau^2 + r^2}}\right),
\end{equation}
where we used the identity:
\begin{equation}
	C_J^{(1)}(x) = U_J(x),
\end{equation}
with $U_J(x)$ the $J^\text{th}$ Chebyshev polynomial of the second kind. These polynomials admit a closed-form expression:
\begin{equation}
	U_J(x) = \frac{(x + \sqrt{x^2 - 1})^{J+1} - (x - \sqrt{x^2 - 1})^{J+1}}{2 \sqrt{x^2 - 1}}.
\end{equation}

This can be plugged into the $d=4$ solution:
\begin{equation}
	g(\tau, r) = \frac{1}{\tau^2 + r^2} + \frac{1}{2 \beta^2 r} \sum_{J=0}^\infty k_J \left( \frac{i}{2 \beta} \right)^J \left[ (r - i \tau)^{J+1} + (-1)^J (r + i \tau)^{J+1} \right].
\end{equation}

We further rewrite the flat-space term as:
\begin{equation}
	\frac{1}{\tau^2 + r^2} = \frac{1}{2 \beta^2 r} \left(-\frac{1}{4} \right)\left(\frac{i}{2 \beta} \right)^{-2} \left[ (r - i \tau)^{-1} + (-1)^{-2} (r + i \tau)^{-1} \right],
\end{equation}
and add the following term, identically zero:
\begin{equation}
	0 = \frac{1}{2 \beta^2 r} k_{-1} \left(\frac{i}{2 \beta} \right)^{-1} \left[ (r - i \tau)^0 + (-1)^{-1}(r + i \tau)^0 \right].
\end{equation}

Defining $k_{-2} = -1/4$, the full solution reads:
\begin{equation}
	g(\tau, r) = \frac{1}{2 \beta^2 r} \sum_{J = -2}^{\infty} k_J \left( \frac{i}{2 \beta} \right)^J \left[ (r - i \tau)^{J+1} + (-1)^J (r + i \tau)^{J+1} \right].
\end{equation}

Now relabel \( J = \ell - 2 \) and enforce invariance under \( \tau \to -\tau \), forcing the sum to run only over even \( \ell = 2n \). This is achieved by setting \( k_{\text{odd}} = 0 \):
\begin{equation}
	g(\tau, r) = \frac{\pi}{2 \beta r} \sum_{n = 0}^\infty \frac{k_{2n - 2}}{\pi^{2n}} \left( -\frac{1}{4} \right)^{n - 1} \left\{ \left[ \frac{\pi}{\beta}(r - i \tau) \right]^{2n - 1} + \left[ \frac{\pi}{\beta}(r + i \tau) \right]^{2n - 1} \right\}. \label{eq: boot start pt}
\end{equation}

Introducing the dimensionless variables:
\begin{equation}
	w = \frac{\pi}{\beta} (r - i \tau), \quad \overline{w} = \frac{\pi}{\beta} (r + i \tau), \label{eq: w coord}
\end{equation}
the two-point function becomes:
\begin{equation}
	g(w, \overline{w}) = \left( \frac{\pi}{\beta} \right)^2 \frac{\coth(w) + \coth(\overline{w})}{w + \overline{w}}. \label{eq: 2pt free}
\end{equation}

The sums converge for \( 0 < |w| < \pi \). The function is periodic, with simple poles at \( w = 0 \) and its periodic images, and the residue is fixed by \( k_{-2} = -1/4 \).

We now test whether this is the unique solution. Suppose a second solution exists:
\begin{equation}
	\widehat g(w, \overline{w}) = \left( \frac{\pi}{\beta} \right)^2 \frac{[\coth(w) + \delta g(w)] + [\coth(\overline{w}) + \delta g(\overline{w})]}{w + \overline{w}},
\end{equation}
where \( \delta g(w) \) is holomorphic, regular, and periodic. Its general form is:
\begin{equation}
	\delta g(w) = \sum_n \left( a_n \sinh(nw) + b_n \cosh(nw) \right).
\end{equation}

The solution must be odd in \( w \), so we set \( b_n = 0 \):
\begin{equation}
	\widehat g(w, \overline{w}) = g(w, \overline{w}) + \sum_n a_n \left( \frac{\sinh(nw)}{w + \overline{w}} + \frac{\sinh(n\overline{w})}{w + \overline{w}} \right). \label{eq: test regge}
\end{equation}

Now take the Regge limit:
\begin{equation}
	w = \frac{\rho}{\eta - \sqrt{\eta^2 - 1}}, \quad \overline{w} = \rho (\eta - \sqrt{\eta^2 - 1}).
\end{equation}

We find:
\begin{align}
	\frac{\coth(w)}{w + \overline{w}} &\to \frac{1}{2 \eta \rho}, & \frac{\coth(\overline{w})}{w + \overline{w}} &\to \frac{1}{\rho^2}, \\
	\frac{\sinh(nw)}{w + \overline{w}} &\to \frac{e^{2n \eta \rho}}{4 \eta \rho}, & \frac{\sinh(n\overline{w})}{w + \overline{w}} &\to \frac{n}{4 \eta^2}.
\end{align}

Since \( \delta g(w) \) grows exponentially in the Regge limit, it is excluded. Thus \( \delta g(w) = 0 \) and
\begin{equation}\label{eq: final two point function}
	g(\tau,r) = \frac{\pi}{2 \beta r} \left\{ \coth\left( \frac{\pi}{\beta}(r + i\tau) \right) + \coth\left( \frac{\pi}{\beta}(r - i\tau) \right) \right\},
\end{equation}
which agrees with the known result in \cite{Rodriguez-Gomez:2021pfh}.

\section{Channel duality and consequences}
We now focus on two conditions from the list of bootstrap requirements that become particularly interesting when combined. These are the OPE expansion and the periodicity along the thermal circle, i.e., the KMS condition. In fact, observe that a single block in \eqref{eq:OPEdecomp}, i.e., the contribution of a single operator in the OPE, is not invariant under KMS. Therefore, we expect that imposing the KMS condition within the OPE regime should allow us to constrain the OPE coefficients $a_{\mathcal O}$.

However, note that not both sides of equation \eqref{KSM condition} can be expanded using the OPE. For example, suppose $0 < \tau < \beta$: then the left-hand side can be expanded in an OPE if $\tau^2 + r^2 < \beta^2$. But if this is the case, then $(\tau + \beta)^2 + r^2 > \beta^2$, and therefore the right-hand side cannot be expanded in OPE. 

To be able to expand both sides simultaneously, we can combine the KMS condition with the symmetry $\tau \to -\tau$, which leads to a more tractable constraint.
 \begin{equation}\label{eq:KMS2}
    g(\tau,r) = g(\beta-\tau,r) \ .
\end{equation}
Observe that now there are infinitely many values of $\tau$ and $r$ for which we can expand both sides of the equation, creating an infinite number of non-trivial constraints:
\begin{multline}
   \sum_{\mathcal O} \frac{a_{\mathcal O}}{\beta^{\Delta}} \left(\tau^2+r^2\right)^{\frac{\Delta_{\mathcal O}}{2}-\Delta_\varphi} C_J^{(\nu)}\left(\frac{\tau}{\sqrt{\tau^2+r^2}}\right) =\\ 
   = \sum_{\mathcal O} \frac{a_{\mathcal O}}{\beta^{\Delta}} \left((\beta-\tau)^2+r^2\right)^{\frac{\Delta_{\mathcal O}}{2}-\Delta_\varphi} C_J^{(\nu)}\left(\frac{\beta-\tau}{\sqrt{(\beta-\tau)^2+r^2}}\right) \ .
\end{multline}
We will sometimes refer to this equation as \textit{channel duality}, since one can think of the left-hand side as analogous to the $s$-channel in conformal bootstrap, and the right-hand side as the $t$-channel\footnote{The terminology $s$- and $t$-channel originates from an analogy with the crossing equation in $S$-matrix theory.}.

As already noted, not only is a single block not KMS invariant, but it is easy to show that a single block in one of the two channels must be compensated by an infinite sum of contributions from the opposite channel. One way to see this is by using the inversion formula: if we input a single block in the $t$-channel, we find that it generates an infinite number of operators in the $s$-channel.

A simpler illustration is obtained by considering, for simplicity, the case $r = 0$ and taking the limit $\tau \to \beta$. On the right-hand side, the leading contribution comes from the most relevant operator: in a unitary theory, this is the identity. We thus obtain:
\begin{equation}
    \sum_{\mathcal O} \frac{a_{\mathcal O }}{\beta^{\Delta_{\mathcal O}}} C_{J}^{(\nu)}(1) \tau^{\Delta-2\Delta_\varphi} = \frac{1}{(\beta-\tau)^{2\Delta_\varphi}}+\ldots \ .
\end{equation}
Observe that the identity contribution on the right-hand side leads to a divergence at $\tau = \beta$, which is nothing but the image (by periodicity) of the UV divergence caused by the two operators colliding. However, no single operator on the left-hand side produces such a divergence, meaning that only the resummation of the OPE can reproduce the identity block of the opposite channel.

Physically, this highlights that not only is a single block not KMS invariant, but compensating the identity requires the collective contribution of an infinite number of operators, in close analogy with conformal bootstrap and $S$-matrix theory.

Having channel duality is thus crucial for any bootstrap problem. In the following, we will organize the resulting equations into an infinite set of sum rules for thermal OPE coefficients, and derive some asymptotic properties for heavy operators, which will be essential for solving the problem.

\subsection{Sum rules for one-point functions}

The equation \eqref{eq:KMS2} was first identified by El-Showk and Papadodimas \cite{El-Showk:2011yvt} as the analog of the crossing equation for thermal CFTs. In this section, we explore this idea explicitly by extracting \emph{explicit sum rules} from the condition \eqref{eq:KMS2}, making the analogy proposed by the aforementioned authors more concrete.

It is useful to notice that there exists a specific point in kinematic space, corresponding to $\tau = \beta/2$ and $r = 0$, where the KMS condition is trivially satisfied for any choice of $a_{\mathcal O}$. This motivates expanding around this point, which we will refer to as the \textit{KMS fixed point}, so that KMS can be written as
\begin{equation}
    g\left(\frac{\beta}{2}+\tau, r\right) = g\left(\frac{\beta}{2}-\tau, r\right) \ .
\end{equation}

Schematically, the idea is the following: we expand both sides of \eqref{eq:KMS2} using the OPE, and then further expand the result in powers of $\tau$ and $r$. From a perspective close to the zero-temperature analytical (and numerical) conformal bootstrap (see \cite{Fitzpatrick:2012yx,Komargodski:2012ek,Alday:2013cwa,Alday:2013opa,Hartman:2022zik,Poland:2018epd} and references therein), this procedure corresponds to setting to zero the $(2\ell+1)$-th derivative in $\tau$ and the $n$-th derivative in $r$ of $f(\tau, r)$, evaluated at $\tau = \beta/2$ and $r = 0$ \cite{Iliesiu:2018fao}:
\begin{equation}\label{KMSaroundfixedpoint}
    \left.\frac{\partial^{2 \ell+1}}{\partial \tau^{2 \ell+1}} \frac{\partial^{n}}{\partial r^{n}}\left( g\left(\frac{\beta}{2}+\tau, r\right) - g\left(\frac{\beta}{2}-\tau, r\right)\right) \right|_{\tau = \frac{\beta}{2},\, r = 0} = 0 \ .
\end{equation}

This can be achieved using the definition of the Gegenbauer polynomials and the binomial theorem, applied twice. It is easy to see from equation \eqref{KMSaroundfixedpoint} that the even powers in $\tau$ cancel identically, while the odd powers yield non-trivial constraints. Imposing these conditions order by order in $r$, we obtain explicit sum rules for the combination $b_{\mathcal O} f_{\mathcal O \varphi\varphi}$.

As a first step, we expand the Gegenbauer polynomials as
\begin{equation} \label{eq: gegdef}
   C_{J}^{(\nu)}\left(\frac{\tau}{\sqrt{\tau^2+r^2}} \right) = \sum_{k = 0}^{[J/2]} 2^{J-2k}  (-1)^k \frac{\Gamma\left(J-k+\nu\right)}{\Gamma(\nu) k! (J-2k)!} \left(\frac{\tau}{\sqrt{r^2+\tau^2}}\right)^{J-2k} \ .
\end{equation}

Plugging this decomposition into the OPE expansion gives:
\begin{equation}
    g(\tau,r)  =  \sum_{\mathcal O \in \varphi \times \varphi} \sum_{k = 0}^{[J/2]} \frac{a_{\mathcal O}}{\beta^{\Delta}} (-1)^k \frac{\Gamma\left(J-k+\nu\right)}{\Gamma(\nu) k! (J-2k)!} 2^{J-2k} |\tau^2+r^2|^{\frac{h-2 \Delta_{\varphi}+2k}{2}} \tau^{J-2k} \ ,
\end{equation}
where we have adopted the twist variable \eqref{eq: twist}. Using the generalized binomial theorem, we obtain:
\begin{equation}
    g(\tau,r)  =  \sum_{\mathcal O \in \varphi \times \varphi} \sum_{k = 0}^{[J/2]} \sum_{n = 0}^\infty \frac{a_{\mathcal O}}{\beta^{\Delta}} (-1)^k \frac{\Gamma\left(J-k+\nu\right)}{\Gamma(\nu) k! (J-2k)!} 2^{J-2k} \binom{\frac{h-2\Delta_{\varphi}+2k}{2}}{n} \tau^{\Delta-2\Delta_\varphi - 2n} r^{2n} \ . \label{eq: r=0}
\end{equation}

We now compute the function at $\frac{\beta}{2} \pm \tau$ and apply the binomial theorem again:
\begin{multline}
    g\left(\frac{\beta}{2} \pm \tau, r\right)  = \sum_{\mathcal O \in \varphi \times \varphi} \sum_{k = 0}^{[J/2]} \sum_{n, \ell = 0}^\infty \frac{a_{\mathcal O}}{\beta^{2\Delta_{\varphi}+2n+\ell}} (-1)^k \frac{\Gamma\left(J-k+\nu\right)}{\Gamma(\nu) k! (J-2k)!} \times \\ 
    \times 2^{-h-2k+2 \Delta_{\varphi}+2n+\ell} \binom{\Delta - 2\Delta_\varphi - 2n}{\ell} \binom{\frac{h - 2\Delta_{\varphi} + 2k}{2}}{n} (\pm \tau)^{\ell} r^{2n} \ .
\end{multline}

It is crucial to note that the expression is sensitive to the sign difference in $\frac{\beta}{2} \pm \tau$ if and only if $\ell$ is an odd positive integer. Imposing the KMS condition \eqref{KMSaroundfixedpoint} leads to:
\begin{multline}
    \sum_{\mathcal O \in \varphi \times \varphi} \sum_{k = 0}^{[J/2]} \sum_{n = 0}^\infty \frac{a_{\mathcal O}}{\beta^{2n}} (-1)^k \frac{\Gamma\left(J-k+\nu\right)}{\Gamma(\nu) k! (J-2k)!} 2^{-h-2k+2n} \times \\
    \times \binom{\Delta - 2\Delta_\varphi - 2n}{\ell} \binom{\frac{h - 2\Delta_{\varphi} + 2k}{2}}{n} r^{2n} = 0 \ ,
\end{multline}
for fixed $\ell \in 2\mathbb{N} + 1$. Since these equations must hold for any $n \in \mathbb{N}$ and any such $\ell$, they reduce to:
\begin{equation}
    \sum_{\mathcal O \in \varphi \times \varphi} \sum_{k = 0}^{[J/2]} a_{\mathcal O} (-1)^k \frac{\Gamma\left(J-k+\nu\right)}{\Gamma(\nu) k! (J-2k)!} 2^{-h-2k} \binom{\Delta - 2\Delta_\varphi - 2n}{\ell} \binom{\frac{h - 2\Delta_{\varphi} + 2k}{2}}{n} = 0 \ . \label{eq: sumrulesappe}
\end{equation}

 Once we re-sum over $k$, the equation can be recast as
\begin{equation}
    \sum_{\mathcal O \in \varphi \times \varphi} b_{\mathcal O} f_{\mathcal O \varphi \varphi} F_{\ell, n}(h,J) = 0 \ ,
\end{equation}
where we used the definition of the $a_{\mathcal O}$ coefficients (with normalization constants $c_{\mathcal O}$ set identically to 1):
\begin{equation}
    a_{\mathcal O}= \frac{f_{\varphi\varphi\mathcal O}b_{\mathcal O}}{c_{\mathcal O}} \frac{J!}{2^J (\nu)_J} \ , 
\end{equation}
and we introduced the function
\begin{equation} \label{eq: vector}
    F_{\ell, n}(h,J)=\frac{1}{2^{h+J}} \binom{\frac{h-2 \Delta_{\varphi}}{2}}{n} \binom{h+J-2 \Delta_{\varphi} -2 n }{\ell} \, {}_{3}F_{2}{\left[\left.\genfrac..{0pt}{}{\frac{1-J}{2},-\frac{J}{2}, \frac{h}{2}-\Delta_{\varphi} +1}{\frac{h}{2}-\Delta_{\varphi}-n +1,-J-\nu +1}\right| 1\right]} \ .
\end{equation}

This concludes the derivation, and the final result is:
\begin{equation}\label{eq:sumrule}
    \sum_{\mathcal O \in \varphi \times \varphi} b_{\mathcal O} f_{\mathcal O \varphi \varphi} \,  F_{\ell, n}(h,J) = 0 \ , \qquad  n \in \mathbb{N}, \ \ell \in 2 \mathbb{N}+1 \ ,
\end{equation}
where we introduced the twist variable
\begin{equation}
    h = \Delta - J \ , \label{eq: twist}
\end{equation}
and defined the function
\begin{equation}
    F_{\ell, n}(h,J)=\frac{1}{2^{h+J}} \binom{\frac{h-2 \Delta_{\varphi}}{2}}{n} \binom{h+J-2 \Delta_{\varphi} -2 n }{\ell} \, {}_{3}F_{2}{\left[\left.\genfrac..{0pt}{}{\frac{1-J}{2},-\frac{J}{2}, \frac{h}{2}-\Delta_{\varphi} +1}{\frac{h}{2}-\Delta_{\varphi}-n +1,-J-\nu +1}\right| 1\right]} \ , \label{eq: sumrules}
\end{equation}
setting $c_{\mathcal O} = 1$.

As expected, the function $F_{\ell, n}(h,J)$ depends both on the conformal dimensions $\Delta$ (through the twist variable $h$) and the spins $J$ of the operators appearing in the OPE. It also depends on two positive integers, $\ell$ and $n$.

In the following, we will verify the sum rules in the four-dimensional free scalar theory. Then, we will set $r=0$ and study configurations where the two operators in the correlation function lie on the same thermal circle.

   \paragraph{A comment on free scalar theory}
Let us comment on the simplest example, where $\varphi$ is a fundamental free scalar field. In this particular case, the upper entry of the first binomial coefficient in equation \eqref{eq: sumrules} is fixed to zero. Indeed, apart from the identity operator $\mathds{1}$, the conformal dimensions of the operators $\mathcal{O} \in \varphi \times \varphi$ are given by $\Delta = 2\Delta_\varphi + J$, which implies $h - 2\Delta_\varphi = 0$. This in turn means that all the equations with $n \neq 0$ are trivially satisfied.

This behavior is to be expected, since the operators $\mathcal{O} \in \varphi \times \varphi$ are labeled solely by their spin, with a one-to-one correspondence between their conformal dimensions and their spin. It is therefore natural to expect that setting the spatial coordinates to zero from the beginning (which is equivalent to setting $n = 0$) is sufficient to determine the OPE coefficients.

This theory can be solved explicitly \cite{Iliesiu:2018fao, Marchetto:2023fcw}, and the thermal OPE coefficients can be computed in closed form. Since only the sum rules for $n = 0$ are non-trivial, a more general discussion of the correlator at $r = 0$ will be relevant. We anticipate here that the sum rules can be numerically verified, and we observe that the larger the value of $\ell$, the larger the conformal dimensions of the operators involved must be for the sum rule to yield accurate results.

   \subsubsection{Reduction to zero spatial coordinates} \label{ssec: zero}

In this section, we consider only a specific subset of the equations coming from \eqref{eq:sumrule}, namely those with $n = 0$. These correspond to the case in which the spatial coordinates in the two-point function are set to zero. In this setting, the general sum rules \eqref{eq:sumrule} reduce to the simpler form
\begin{equation}\label{eq:BootstrapFor1D}
    \frac{\Gamma \left(2 \Delta_\varphi+ \ell \right)}{\Gamma \left(2 \Delta_\varphi\right)} = \sum_{\Delta} \frac{a_{\Delta}}{2^{\Delta}} \frac{\Gamma\left(\Delta-2\Delta_\varphi+1\right)}{\Gamma \left(\Delta-2\Delta_\varphi-\ell+1\right)} \ ,
\end{equation}
where $\ell \in 2\mathbb N+1$. To show this, we note from equation \eqref{eq: r=0} that the only non-trivial equation is obtained by setting $n = 0$. This significantly simplifies the expression \eqref{eq: vector}:
\begin{equation} 
      F_{\ell, 0}(h,J)=\frac{1}{2^{h+J}}  \binom{h+J-2 \Delta_{\varphi} }{\ell} \, {}_{2}F_{1}{\left[\left.\genfrac..{0pt}{}{\frac{1-J}{2},-\frac{J}{2}}{-J-\nu +1}\right| 1\right]} \ .
\end{equation}

This form of $F_{\ell,0}$ makes it clear that the sum rule \eqref{eq:sumrule} is equivalent in the $n = 0$ reduction. To see this, we recall that by symmetry, $J \in 2 \mathbb N$, and by definition, $\nu \in \mathbb N/2$. It can then be shown that
\begin{equation}
      C_{J}^{(\nu)}(1)=\frac{2^J (\nu)_{J}}{J!} \, {}_{2}F_{1}{\left[\left.\genfrac..{0pt}{}{\frac{1-J}{2},-\frac{J}{2}}{-J-\nu +1}\right| 1\right]} \ .
\end{equation}

The sum rules \eqref{eq: sumrulesappe} then become, for fixed odd integer $\ell$,
\begin{equation}
     \sum_{\mathcal O \in \varphi \times \varphi} a_{\mathcal O} \, C_J^{(\nu)}(1) \frac{\Gamma(\Delta-2\Delta_\varphi+1)}{2^\Delta \Gamma(\Delta-2\Delta_\varphi-\ell+1)}=0   \ .
\end{equation}

Since the only spin-dependent term is $a_{\mathcal O} \, C_J^{(\nu)}(1)$, we define a weighted sum of thermal OPE coefficients as
\begin{equation}
      a_{\Delta} \equiv \sum_{\mathcal{O} \in \varphi \times \varphi}^{\Delta \text{ fixed}}  a_{\mathcal O} \,  C_J^{(\nu)}(1) \ .
\end{equation}

Then, by separating the contribution of the identity operator $\mathds{1}$,
\begin{equation}
      -\frac{\Gamma(2\Delta_\varphi+\ell)}{\Gamma(2\Delta_\varphi)} \ ,
\end{equation}
we recover equation \eqref{eq:BootstrapFor1D}.\footnote{Alternatively, one can directly start from the correlator at zero spatial coordinates and apply the binomial theorem, as done in the proof of equation \eqref{eq:sumrule}.}

The sum runs over all possible conformal dimensions appearing in the OPE $\varphi \times \varphi$; due to the reduction, we are no longer sensitive to the spins of the operators. Indeed, when a $d$-dimensional two-point function is restricted to zero spatial separation, the thermal OPE takes the form
\begin{equation}\label{eq:twopintOPEZeror}
    \langle \varphi(\tau,0) \varphi(0,0)\rangle_\beta = \sum_{\Delta} \frac{a_{\Delta}}{\beta^{\Delta}} \tau^{\Delta-2\Delta_\varphi} \ , \qquad a_{\Delta} \equiv \sum_{\mathcal{O} \in \varphi \times \varphi}^{\Delta \text{ fixed}}  a_{\mathcal O}  \, C_J^{(\nu)}(1) \ ,
\end{equation}
where the sum is over operators of dimension $\Delta$ (possibly with different spin $J$) in the OPE.

These equations provide, in principle, non-trivial constraints on the OPE coefficients, and thus on the one-point functions of the operators appearing in the OPE $\varphi \times \varphi$. In deriving equation \eqref{eq:BootstrapFor1D}, we used the fact that the identity operator $\mathds{1}$ appears in the OPE. Its conformal dimension and spin are both zero, and its one-point function and structure constant are conventionally set to $b_{\mathds{1}} = f_{\varphi \varphi \mathds{1}} = 1$. We can then isolate the identity contribution, which gives the left-hand side of equation \eqref{eq:BootstrapFor1D}. This information is crucial to fix the normalization; otherwise, the coefficients $a_\Delta$ would only be defined up to an overall scaling (as is clear from equation \eqref{eq:sumrule}).

We now present some examples and consequences of equation \eqref{eq:BootstrapFor1D}. Let us also remark that equations \eqref{eq:sumrule} and \eqref{eq:BootstrapFor1D} can be straightforwardly generalized to the case of unequal external operators. The structure of the equations remains unchanged, but in that case, the identity does not appear in the OPE.

\begin{figure}[h!]
\begin{subfigure}[t]{.48\textwidth}
  \centering
  \includegraphics[width=\textwidth]{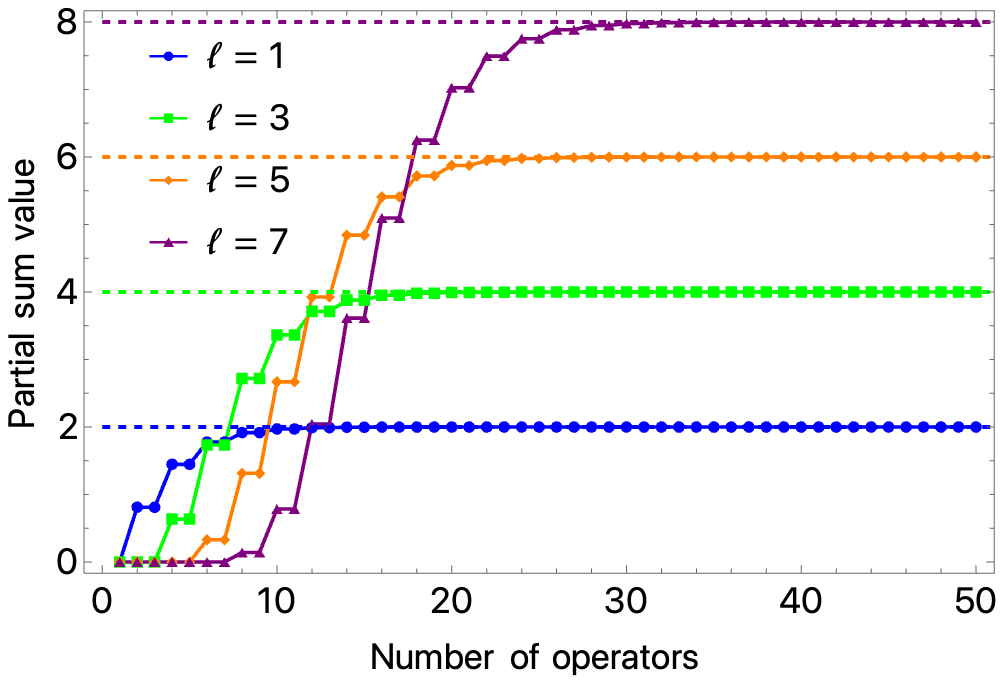}%.png}
  \caption{}
  \label{fig:D1GFF}
\end{subfigure}%
\hfill
\begin{subfigure}[t]{.49\textwidth}
  \centering
  \includegraphics[width=\linewidth]{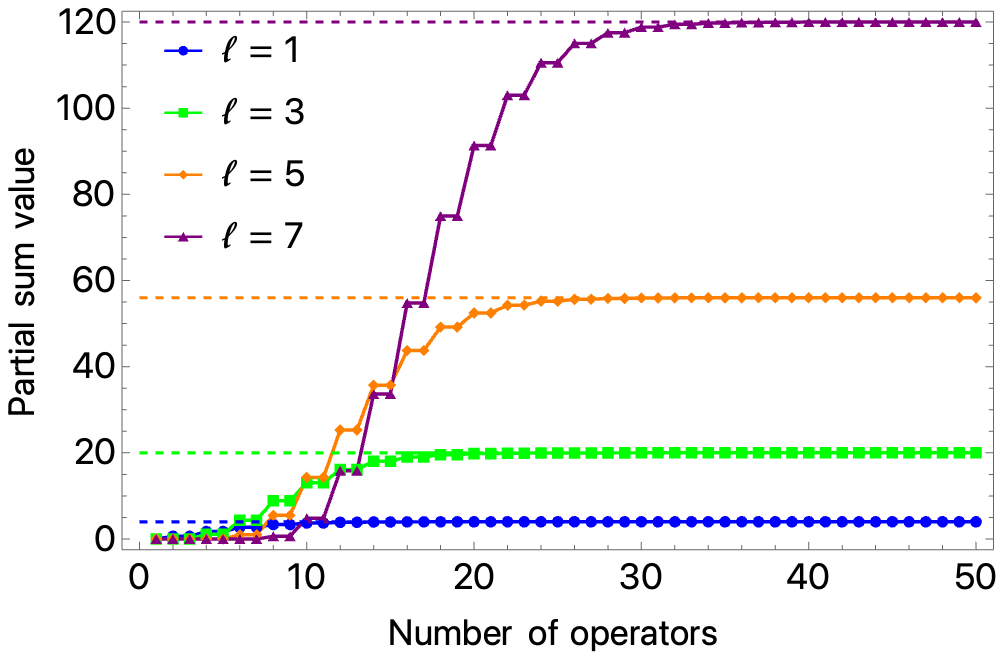}%png}
  \caption{}
  \label{fig:D3GFF}
\end{subfigure}
\label{fig:GFFCheck}
\caption{Comparing the left (straight, dashed lines) and right (continuous lines) sides of equation \eqref{eq:BootstrapFor1D} for different values of $\ell$. The bigger $\ell$ is, the more operators need to be summed for the right-hand side to converge to the correct value, namely the left-hand side. The \textbf{left panel (a)} corresponds to a numerical test for a \underline{GFF with $\Delta_\varphi = 1$}, or equivalently for a four-dimensional scalar free theory. The \textbf{right panel (b)} corresponds to a numerical test for a \underline{GFF with $\Delta_\varphi = 2$}, or equivalently for a six-dimensional scalar free theory. In both cases, it is convenient to multiply both sides of the equation \eqref{eq:BootstrapFor1D} by a factor $1/\ell!$.}
\end{figure}

\paragraph{First test: generalized free fields in different dimensions}\label{GFF1d}
Equation \eqref{eq:BootstrapFor1D} is the only set of equations for thermal CFTs correlators restricted to zero spatial coordinates. Among all possible thermal CFTs, some can be solved with other methods. One class of such theories is represented by the generalized free fields (GFF) defined by the OPE 
\begin{equation}
    \varphi \times \varphi = \mathds{1} + [\varphi \varphi]_{p, q}\ ,
\end{equation}
where the conformal dimensions of the operators in the OPE are $\Delta= 2 \Delta_\varphi+p+2 q$, $p, q \in \mathbb N$ (up to the identity $\mathds{1}$, whose dimension is zero). These theories are well studied and they are solved for any spacetime dimensions \cite{Iliesiu:2018fao, Alday:2020eua} thanks to the possibility to employ the method of images, which allows writing the thermal propagator as an infinite sum over zero-temperature propagators. When restricted to zero spatial coordinates this infinite sum can be recognized as a linear combination of Hurwitz $\zeta$-functions\footnote{The Hurwitz $\zeta$-function is defined as \begin{equation}
    \zeta_{H}(s,a) = \sum_{n = 0}^\infty \frac{1}{(n+a)^s} \ .
\end{equation}}
\begin{equation}\label{eq:GFFsolution}
    \langle \varphi(\tau) \varphi(0) \rangle_\beta = \sum_{m = -\infty}^\infty \frac{1}{(\tau+ m \beta)^{2\Delta_\varphi}} = \frac{1}{\beta^{2 \Delta_\varphi}} \left[\zeta_H\left(2 \Delta_\varphi, \frac{\tau}{\beta}\right)+\zeta_H\left(2 \Delta_\varphi, 1-\frac{\tau}{\beta}\right)\right] \ .
\end{equation}
There are cases in which the above solution simplifies even more. If we consider the GFF theory with $\Delta_\varphi = 1$,
\begin{equation}\label{D=1GFF}
    \langle \varphi(\tau) \varphi(0) \rangle_\beta = \frac{\pi^2}{\beta^2} \csc^2\left( \frac{\pi \tau}{\beta}\right) \ .
\end{equation}
This two-point function is particularly interesting since it also corresponds to the two-point functions of a four-dimensional fundamental free scalar $\phu_{\text{4d}}$: recalling
\begin{equation}
    \langle \phu_{\text{4d}}(\tau,r)\phu_{\text{4d}}(0,0) \rangle_\beta = \frac{\pi}{2 \beta r } \left[\coth\left(\frac{\pi}{\beta} (r+ i\tau )\right)+\coth\left(\frac{\pi}{\beta} (r- i\tau )\right)\right]\  ,
\end{equation}
then 
\begin{equation}
    \langle \varphi(\tau) \varphi(0) \rangle_\beta=\lim_{r\rightarrow 0} \langle \phu_{\text{4d}}(\tau,r)\phu_{\text{4d}}(0,0) \rangle_\beta \ .
\end{equation}
The two-point function \eqref{D=1GFF} also corresponds to the restriction to $r=0$ of the energy-energy two-point function $\Braket{\epsilon(\tau, r) \epsilon(0,0)}_{\beta}$ in the critical two-dimensional Ising model
\begin{equation}
    \langle \varphi(\tau) \varphi(0) \rangle_\beta=\lim_{r\rightarrow 0} \Braket{\epsilon(\tau, r) \epsilon(0,0)}_{\beta} \ .
\end{equation}
Another special case is given by the GFF with $\Delta_\varphi = 2$. In this scenario, the solution \eqref{eq:GFFsolution} reduces to 
\begin{equation}\label{eq:GFFD=2}
    \langle \varphi(\tau) \varphi(0) \rangle_\beta = \frac{\pi^4}{3\beta^4} \left(2+\cos\left(\frac{2\pi \tau}{\beta}\right)\right)\csc^4 \left(\frac{\pi}{\beta} \tau\right) \ .
\end{equation}
Similarly to the previous one, this case can also be interpreted as the restriction of a higher dimensional free theory to one dimension by setting $r=0$. This is indeed the case for the two-point function of a six-dimensional fundamental free scalar $\phu_{\text{6d}}$
\begin{multline}\label{eq:6dfreetheory}
     \langle \phu_{\text{6d}}(\tau,r) \phu_{\text{6d}}(0) \rangle_\beta =  \frac{\pi}{4 \beta^2 r^3} \left[\beta \coth \left(\frac{\pi  (r-i \tau )}{\beta }\right)+\beta \coth \left(\frac{\pi  (r+i
   \tau )}{\beta }\right)+\right. \\ \left . \pi  r \ \text{csch}^2\left(\frac{\pi  (r-i \tau )}{\beta
   }\right)+\pi r\ \text{csch}^2\left(\frac{\pi  (r+i \tau )}{\beta }\right)\right] \ .
\end{multline}
Equation \eqref{eq:6dfreetheory} reduces to equation \eqref{eq:GFFD=2} in the limit $r \to 0$. 
GFF theories represent an optimal playground to test the equation \eqref{eq:BootstrapFor1D} since the OPE coefficients $a_{\Delta}$ can be easily extracted from the exact solution \eqref{eq:GFFsolution}. The numerical tests for GFF theories with $\Delta_\varphi = 1$ and $\Delta_\varphi = 2$ are presented in Figs. \ref{fig:D1GFF} and \ref{fig:D3GFF}.
What we immediately learn from these simple checks is that, even if the equations \eqref{eq:BootstrapFor1D} are satisfied for every odd value of $\ell$, the bigger $\ell$ is, the more operators we need to sum on the right-hand side of equation \eqref{eq:BootstrapFor1D} to converge to the correct value, or better to the \emph{truncated sum} on the right-hand side to be a good approximation.  This makes the sum rules \eqref{eq:BootstrapFor1D} extremely difficult to solve by truncation of the right-hand side,  since the more OPE coefficients $a_{\Delta}$ we want to extract, the more operators we need to take into consideration. At this stage this is an observation, nonetheless in the next Section we will show a general result about heavy operators.
\newline

\paragraph{Second test: two-point functions of Virasoro primaries}

Another test we can easily perform concerns two-dimensional correlators of Virasoro primary fields. These two-point functions can be computed using the conformal map between the plane and the cylinder \cite{DiFrancesco:1997nk,Mussardo:2020rxh,Datta:2019jeo}:
\begin{equation}
    \langle \varphi (\tau,\sigma) \varphi(0,0)\rangle_\beta = \left(\frac{\pi}{\beta}\right)^{2 h_\varphi+2\overline h_\varphi} \operatorname{csch}^{2h_\varphi}\left[\frac{\pi}{\beta}(\sigma+i \tau)\right]\operatorname{csch}^{2\overline h_\varphi}\left[\frac{\pi}{\beta}(\sigma-i \tau)\right]\ ,
\end{equation}
where $(h_\varphi,\overline h_\varphi)$ are the conformal weights of the Virasoro primary $\varphi$ with conformal dimension $\Delta = h_\varphi + \overline h_\varphi$.

\begin{figure}[h!]
\begin{subfigure}[t]{.49\textwidth}
  \centering
  \includegraphics[width=\textwidth]{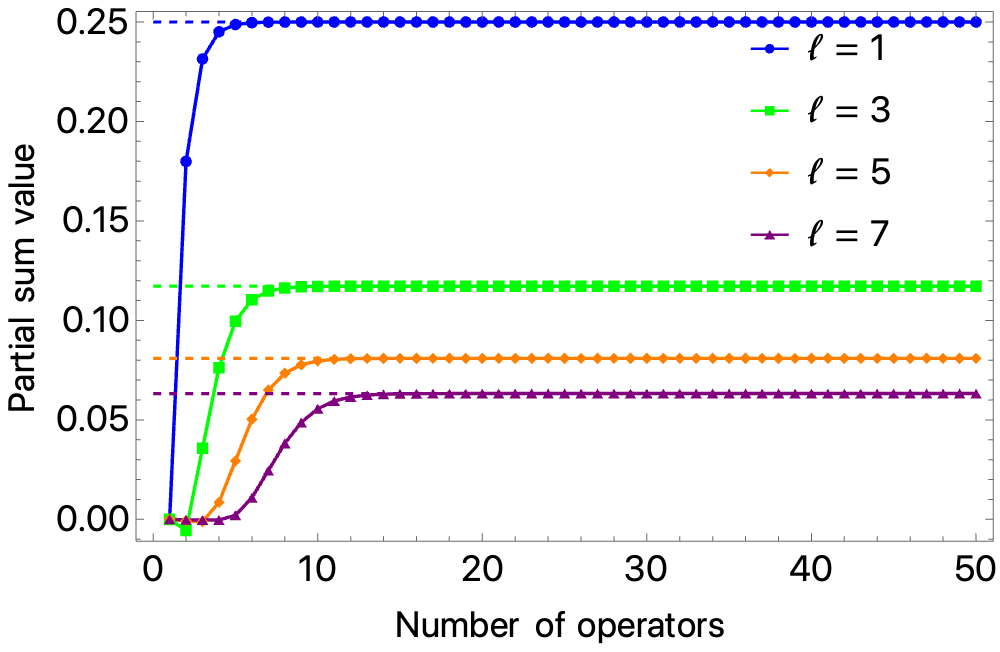}
  \caption{}
  \label{fig:Ising2dcheck}
\end{subfigure}%
\hfill
\begin{subfigure}[t]{.49\textwidth}
  \centering
  \includegraphics[width=\linewidth]{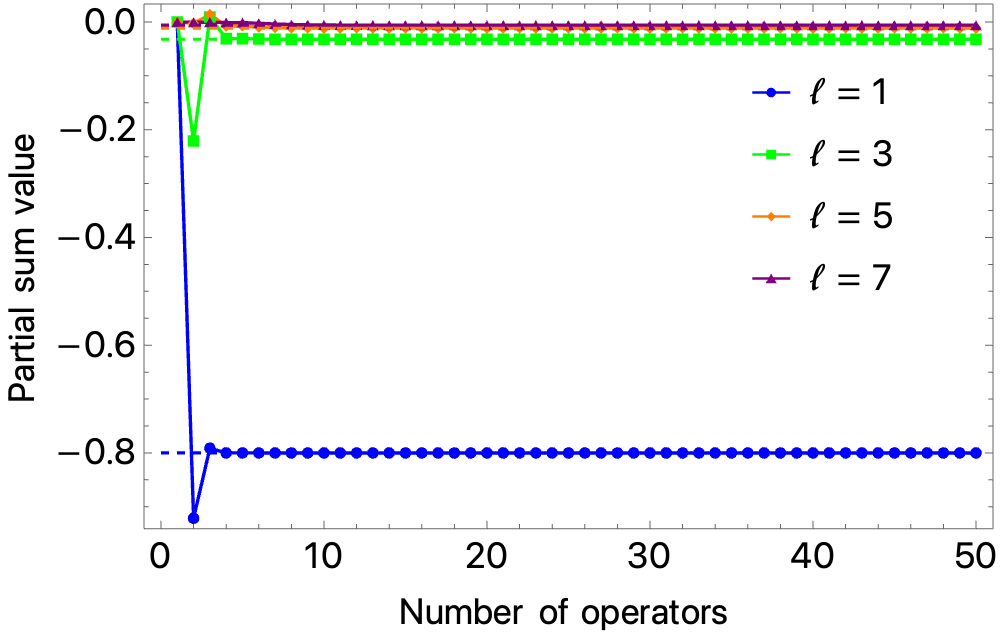}
  \caption{}
  \label{fig:YL2dCheck}
\end{subfigure}
\label{fig:2dCheck}
\caption{Comparison between the left-hand side (dashed lines) and right-hand side (solid lines) of equation \eqref{eq:BootstrapFor1D} for various values of $\ell$. As in Figs. \ref{fig:D1GFF} and \ref{fig:D3GFF}, larger $\ell$ requires summing more operators for convergence. \textbf{Left (a)}: test for the correlator $\langle \sigma(\tau)\sigma(0)\rangle_\beta$ in the \underline{two-dimensional critical Ising model}. \textbf{Right (b)}: test for the correlator $\langle \varphi(\tau)\varphi(0)\rangle_\beta$ in the \underline{Lee–Yang model}, i.e., the non-unitary minimal model $\mathcal{M}_{2,5}$. In both panels, a rescaling by $1/\ell!$ has been applied for convenience.}
\end{figure}

When restricted to zero spatial coordinate, the correlator simplifies to:
\begin{equation}
    \langle \varphi (\tau,0) \varphi (0,0)\rangle_\beta = \left(\frac{\pi}{\beta}\right)^{2\Delta} \csc^{2\Delta}\left(\frac{\pi \tau}{\beta}\right) \ .
\end{equation}
We can therefore test equation \eqref{eq:BootstrapFor1D} for arbitrary values of $\Delta$. Numerical tests are shown in Figs. \ref{fig:Ising2dcheck} and \ref{fig:YL2dCheck}: similarly to Figs. \ref{fig:D1GFF} and \ref{fig:D3GFF}, they compare the left- and right-hand sides of equation \eqref{eq:BootstrapFor1D} for various values of $\ell$ and different levels of truncation in the sum.

The considered examples are:
- $\langle \sigma(\tau) \sigma(0)\rangle_\beta$, where $\sigma$ is the Virasoro primary of conformal weights $(1/16, 1/16)$ in the two-dimensional critical Ising model;
- $\langle \varphi(\tau)\varphi(0)\rangle_\beta$, where $\varphi$ is the only Virasoro primary in the Lee–Yang minimal model.

The Lee–Yang model is the simplest conformal field theory: it is a non-unitary minimal model with a single primary field of conformal weights $(-1/5, -1/5)$. Despite its simplicity, it provides a non-trivial test of the sum rules. In particular, the non-unitarity of the theory highlights that the derivation of equation \eqref{eq:BootstrapFor1D} only assumes the KMS condition, and thus the sum rules also hold for non-unitary theories, such as Lee–Yang.

\paragraph{A first attempt at solutions}
The simplest way to use equations \eqref{eq:sumrule} is to truncate the OPE. This is analogous to the \textit{Gliozzi method} in conformal bootstrap, originally proposed in \cite{Gliozzi:2013ysa}. In general, it is not guaranteed that a simple truncation of the infinite sum over operators will give reliable results: we will discuss this point in more detail below. In fact, we will later show analytically that this approach does not work, and we will provide a complete explanation of why this is the case. 

\medskip

Finally, let us comment that in our approach the equations \eqref{eq:sumrule} are used to constrain the thermal OPE coefficients. However, it would also be interesting to use them to constrain the zero-temperature CFT data, as discussed for example in \cite{Marchetto:2023xap}.

\subsection{Tauberian theorems and heavy operators}
In this section, we explore how the divergences of the two-point correlation function, together with the KMS condition, can be used to extract information about the thermal OPE coefficients when the spatial coordinates are fixed to zero. This result can be derived by invoking \emph{Tauberian theory} (see \cite{tauberian} for an extensive review). Tauberian theory is a branch of mathematics that found early applications in physics in \cite{Vladimirov:1978xx}, and more recently in \cite{Pappadopulo:2012jk,Das:2017vej,Mukhametzhanov:2019pzy,Pal:2019zzr,Ganguly:2019ksp,Das:2020uax,Mukhametzhanov:2020swe,Pal:2020wwd,Kusuki:2023bsp,Pal:2023cgk}. Notably, its application to one-dimensional CFTs is rigorously treated in \cite{Qiao:2017xif}. 

\noindent Motivated by these works, we aim to derive the thermal OPE density for heavy operators by combining a Tauberian theorem with the KMS condition and the parity transformation $\tau \to -\tau$ on the thermal circle. 

\noindent We begin with a heuristic derivation of the theorem to provide intuition about its meaning and implications (see Section \ref{ssec: deriv}). We will then outline the steps necessary for a rigorous proof. A crucial hypothesis of the theorem—namely, the boundedness of the OPE coefficients—will be justified by showing that the OPE coefficients corresponding to heavy operators all share the same sign (beyond a certain cutoff).

\subsubsection{Derivation of the theorem} \label{ssec: deriv}
Here we present the derivation of the thermal Tauberian theorem. We begin with a heuristic derivation to clarify the physical intuition, and later provide a justification of the result.

\subsubsection{Heuristic derivation}  \label{ssec: heur}
Let us consider a scalar, local operator $\varphi(\tau, \Vec{x})$ and its thermal two-point function 
\begin{equation}
    \langle \varphi(\tau,r) \varphi(0,0)\rangle_\beta \ ,
\end{equation}
and study the same two-point function in the limit of vanishing spatial separation:
\begin{equation}
       \langle \varphi(\tau) \varphi(0)\rangle_\beta \equiv  \langle \varphi(\tau,0) \varphi(0,0)\rangle_\beta \ .
\end{equation}
In this limit, the thermal OPE expansion of the two-point function reads
\begin{equation}\label{eq:TwoPointFunctionInt}
    \langle \varphi(\tau) \varphi(0)\rangle_\beta = \tau^{-2 \Delta_\varphi}\sum_{\mathcal O \in \varphi \times \varphi} \frac{a_{\mathcal O}}{\beta^{\Delta_{\mathcal O}}} \tau^{\Delta_{\mathcal O}} = \tau^{-2\Delta_\varphi} \int_0^\infty d\Delta \ \rho(\Delta) x^{\Delta} \ ,
\end{equation}
where we introduced the \emph{spectral density} $\rho(\Delta)$:
\begin{equation}\label{eq:densitydef}
    \rho(\Delta) = \sum_{\mathcal O \in \varphi \times \varphi} a_{\mathcal O} \ \delta(\Delta-\Delta_{\mathcal O}) \ ,
\end{equation} 
and the dimensionless variable $x = \tau/\beta$. The thermal OPE converges only within the open interval $0< x < 1$ \cite{Iliesiu:2018fao}, due to the presence of an infinite number of poles on the real $\tau$ axis.

This can be seen by noting that the KMS condition implies
\begin{equation}
    \langle \varphi(\tau) \varphi(0)\rangle_\beta \overset{\tau \to k \beta}{\sim} (k\beta-\tau)^{-2\Delta_\varphi} \ ,
\end{equation}
for any $k \in \mathbb Z$. Therefore, from equation \eqref{eq:TwoPointFunctionInt}, we deduce 
\begin{equation}
    \int_0^\infty d\Delta \  \rho(\Delta) x^{\Delta}\  \overset{x \to 1}{\sim}\  (1-x)^{-2\Delta_\varphi} \ .
\end{equation}

Now, we make the \emph{ansatz} that the large-$\Delta$ behavior of $\rho(\Delta)$ follows a power law:
\begin{equation}\label{eq:assumption}
    \rho(\Delta) \overset{\Delta \rightarrow \infty}{\sim} A \Delta^{-\alpha+1} \ .
\end{equation} 
This ansatz will be justified in Section \ref{ssec: precise}. Plugging it into the two-point function \eqref{eq:TwoPointFunctionInt}, we obtain
\begin{equation} \label{eq: taubeta}
    \langle \varphi(\tau) \varphi(0) \rangle_\beta \overset{\tau \to \beta}{\sim}  A \, \Gamma(2-\alpha ) \tau^{-2 \Delta_\varphi } \left(1-\frac{\tau}{\beta}\right)^{\alpha-2} \ ,
\end{equation}
from which we read:
\begin{equation}
   \alpha = -2\Delta_\varphi+2 \ , \qquad A = \frac{1}{\Gamma(2\Delta_\varphi)} \ .
\end{equation}
Therefore, the asymptotic behavior of $\rho(\Delta)$ becomes  
\begin{equation}\label{eq:Tauberian}
   \rho(\Delta) \overset{\Delta \rightarrow \infty}{\sim}   \frac{1}{\Gamma(2 \Delta_\varphi)} \Delta^{2 \Delta_\varphi-1}\ .
\end{equation}

The physical interpretation of this result is that it captures the average density of heavy operators in the $\varphi \times \varphi$ OPE (in the limit $\Delta \rightarrow \infty$). Although the spectrum is generally discrete and the density $\rho(\Delta)$ is a sum of $\delta$-functions, the expression \eqref{eq:Tauberian} makes sense when interpreted in an averaged sense:
\begin{equation}\label{eq:TauberianMath}
    \int_0^{\Delta} \rho(\widetilde \Delta) \ d\widetilde \Delta \ \overset{\Delta \to \infty}{\sim} \frac{\Delta^{2 \Delta_\varphi}}{\Gamma(2 \Delta_\varphi+1)} \ .
\end{equation}

\subsubsection{The precise derivation} \label{ssec: precise}
The derivation above is valid once the ansatz in equation \eqref{eq:assumption} is justified—namely, that the density of states grows as a power law in the conformal dimension $\Delta$ as $\Delta \rightarrow \infty$. As previously discussed, this must be understood in an ``integrated'' sense, as shown in equation \eqref{eq:TauberianMath}. The proof is non-trivial and relies on the theory of \emph{Tauberian theorems}.

First, we stress that the assumption is not universally valid. A well-known counterexample is the series \cite{tauberian2,Qiao:2017xif}
\begin{equation}\label{eq:contr}
    \frac{1}{(x+1)^2(1-x)} = 1 - 2x + 2x^2 - 3x^3 + 3x^4 + \ldots \ .
\end{equation}
While this series has the expected divergence near $x \to 1$, behaving as $(1-x)^{-1}$, the partial sums of its coefficients oscillate: $1, 0, 2, 0, 3, 0, \ldots$, alternating between 0 and $n$ as $n \to \infty$. This prevents a clean power-law behavior in the partial sums. A more general problematic case was analyzed in \cite{Qiao:2017xif}, where it was shown that, in absence of further assumptions, partial sums can oscillate between two different power-law regimes.

This issue was rigorously studied by mathematicians in the last century, and the corresponding results are known as \textit{Tauberian theorems}. In \cite{Pappadopulo:2012jk,Qiao:2017xif}, the key hypothesis is the non-negativity of the weighted spectral density $\rho(\Delta)$. However, this does not hold in our case: from the definition \eqref{eq:densitydef}, the coefficients $a_{\mathcal O}$ can take negative values. The lack of positivity in thermal OPE coefficients is a well-known obstruction in the thermal conformal bootstrap program.

To overcome this issue, we relax the positivity requirement and instead consider the following three sufficient conditions:
\begin{itemize}
    \item[$\star$] The OPE coefficients $a_{\mathcal O}$ are real;
    \item[$\star$] The OPE coefficients $a_{\mathcal O}$ are bounded from below;
    \item[$\star$] The external scaling dimension satisfies $\Delta_\varphi > \frac{1}{2}$.
\end{itemize}
For a rigorous derivation of the Tauberian theorem under these assumptions, see \cite{Marchetto:2023xap} or Theorems 95 and 97 in \cite{Hardy}. 

The last condition ensures that $\alpha > 1$, which controls the divergence $(1-x)^{-\alpha}$ as $x \to 1$. This is crucial for the validity of the theorem in cases where positivity fails but the density is bounded from below. Physically, this condition is motivated by the unitarity bound for scalar fields:
\begin{equation}
    \Delta_\varphi \ge \frac{d-2}{2} \ ,
\end{equation}
which implies that the only unitary, physically relevant theory violating this bound (for $d > 2$) is the free scalar theory in $d = 3$, where the bound is saturated. Since this theory can be solved analytically, we do not consider it problematic. Similarly, in two dimensions, scalar operators may have $\Delta < 1/2$, but those cases can also be solved analytically using the conformal map from the plane to the cylinder.

The reality of the OPE coefficients $a_{\mathcal O}$ is ensured by unitarity, and follows from clustering and reflection positivity:
\begin{equation}
   \lim_{x \to \infty} \langle \varphi(x) \varphi(0) \rangle_\beta = \langle \varphi^2 \rangle_\beta \ge 0 \ .
\end{equation}

The remaining task is to prove that the $a_{\mathcal O}$ are bounded from below. To do so, we show that there exists a cutoff dimension $\widehat{\Delta}$ such that for all $\Delta > \widehat{\Delta}$, the thermal OPE coefficients share the same sign (say, positive\footnote{If the sign is negative—as in the Lee-Yang model—the same result holds by flipping the overall sign of the density.}). Since each coefficient is finite, boundedness from below follows directly.

This behavior is not universal (as illustrated by the counterexample \eqref{eq:contr}), but is instead enforced by the KMS condition. Physically, since individual OPE terms are regular as $\tau \to \beta$, the divergence $(\beta - \tau)^{-2\Delta_\varphi}$ must arise from the cumulative contribution of heavy operators with $\Delta \gg 1$. This channel duality enforces sign stability.

The proof of this stability is presented in Appendix C of \cite{Marchetto:2023xap}. The idea is to use the inversion formula and compute $a_\Delta$ in terms of other OPE coefficients. By comparing $a_\Delta$ and $a_{\Delta + \delta \Delta}$ (where $\delta \Delta$ is the smallest gap in dimension), one finds:
\begin{equation}\label{eq:samesign}
    a_{\Delta + \delta \Delta} = a_\Delta \left[1 + \mathcal{O}\left(\frac{1}{\Delta}\right)\right] \ , \quad \Delta \gg 1 \ .
\end{equation}
If $a_\Delta$ does not vanish asymptotically, then beyond some $\widehat{\Delta}$ the correction becomes negligible, and the signs stabilize.

This is supported by all known examples. In addition to free and 2D theories, holographic calculations confirm the behavior using the geodesic approximation \cite{Rodriguez-Gomez:2021pfh}. As another check, we analyze the 3D $\mathrm{O}(N)$ model at large $N$ in the next section.

Note that this does not imply the positivity of the $a_\Delta$ coefficients, but rather that they all have the same sign in the heavy sector. This is sufficient to establish the Tauberian result by redefining the sign of the density, if needed.

As an alternative to the analysis in \cite{Marchetto:2023xap}, we take a complementary route: we invert the identity operator using the inversion formula. This amounts to applying the thermal inversion formula to $1/(r^2+\tau^2)^{\Delta_\varphi}$ in the KMS-dual channel. The computation yields poles in $a(\Delta)$ located at
\begin{equation}\label{eq:GFFidinv}
    \Delta = 2\Delta_\varphi + 2n + J \ .
\end{equation}
These correspond to the spectrum of a generalized free field, and match results from conformal bootstrap \cite{Fitzpatrick:2012yx,Komargodski:2012ek,Kravchuk:2021kwe,Pal:2022vqc,vanRees:2024xkb}. Notably, this inversion does not include anomalous dimensions, which appear only upon inverting other (non-identity) operators.

The thermal inversion formula also encodes information about the thermal OPE coefficients. By taking residues at \eqref{eq:GFFidinv}, one finds:
{\small
\begin{equation}
    a_{2\Delta_\varphi + 2n + J} = \sum_{r = 0}^n \frac{\left((-1)^J + 1\right) (-1)^n (J+\nu) (J+2r) \Gamma(\nu) \Gamma(\nu +1) \Gamma(J + r) \binom{-\Delta_\varphi}{n - r} \Gamma(n + r + \Delta_\varphi)}{\Gamma(\Delta_\varphi) \Gamma(r+1) \Gamma(n+r+1) \Gamma(-r + \nu +1) \Gamma(J + r + \nu +1)} \ .
\end{equation}
}
These coefficients are all positive, and dominate the asymptotic spectrum. Inverting other light operators produces only subleading corrections. As a concrete case, inverting $\varphi^2$ (spin zero, $n = 0$) shows that the resulting anomalous dimension corrects the positions \eqref{eq:GFFidinv}.

The inversion formula computes contributions proportional to $a_{\mathcal O}$, regardless of their sign. However, the heavy sector remains dominated by the identity inversion. Thus, in the discrete case, while the coefficients $a_\Delta$ are not necessarily positive, they are bounded from below.

In conclusion, under mild assumptions of unitarity and analyticity (as discussed in the previous section), the thermal Tauberian result is rigorously valid for $d > 3$. A complete proof can be found in Appendix C of \cite{Marchetto:2023xap}.

\paragraph{Heavy operators' asymptotic}
The most interesting application of the Tauberian theorem is the approximation of the OPE coefficients $a_{\Delta}$ in the limit $\Delta \to \infty$, i.e., for heavy operators. Recalling the definition of the thermal OPE density \eqref{eq:densitydef}, we can isolate the OPE coefficient of a heavy operator of dimension $\Delta_{H}$ by integrating over the interval $\left[\Delta_{H}-\delta \Delta, \Delta_{H} \right]$, where $\delta \Delta$ parametrizes the gap between $\Delta_{H}$ and the closest smaller-dimension operator appearing in the OPE:
\begin{equation}
    a_{\Delta_{H}}=\int_{\Delta_{H}-\delta \Delta}^{\Delta_{H}} \rho(\Delta) \, d \Delta \ , \qquad \rho(\Delta)=\sum_{\mathcal O \in \varphi \times \varphi} a_{\mathcal O} \ \delta(\Delta-\Delta_{\mathcal O}) \ .
\end{equation}
By employing the Tauberian theorem, in the heavy operator regime we can invoke its asymptotic form \eqref{eq:TauberianMath}\footnote{Making this statement mathematically rigorous is highly non-trivial: deriving a Tauberian theorem for short intervals from one of the form \eqref{eq:TauberianMath} is a notoriously difficult problem.}
\begin{equation}
    a_{\Delta_{H}} = \left(\int_{0}^{\Delta_{H}} d\Delta - \int_{0}^{\Delta_{H}-\delta \Delta} d\Delta \right)\rho(\Delta) \overset{\Delta_{H} \to \infty}{\sim}  \frac{\Delta_{H}^{2\Delta_\varphi-1}}{\Gamma(2\Delta_\varphi)}\delta \Delta \ ,
\end{equation}
in the limit $\delta \Delta = 2 \ll \Delta_{H}$. Therefore, we obtain:
\begin{equation}\label{eq:largeDcoeff}
    a_{\Delta_{H}} \overset{\Delta_{H} \to \infty }{\sim}   2 \frac{\Delta_{H}^{2\Delta_\varphi-1}}{\Gamma(2\Delta_\varphi)}\ .
\end{equation}
Let us emphasize that in this derivation, $\Delta_H$ refers to the conformal dimension of a heavy operator, and $\delta \Delta = 2$ corresponds to the spacing with the next lower operator in the generalized free field (GFF) asymptotic spectrum. We do not claim that this result holds for arbitrary values of $\Delta_H$ and $\delta \Delta$. To make such a claim, one would need to prove a finite-size version of the Tauberian theorem, which is generally stronger than the version employed here.

\subsubsection{Tauberian theorem and the tail of heavy operators}\label{sec:error}
Previously, we used the KMS condition to predict thermal one-point function coefficients of light operators, introducing the hypothesis of a gap between the light and heavy sectors. In this Section, we employ the Tauberian theorem to estimate the contribution of the heavy operator tail.

\noindent Let us consider a theory with $N$ light operators. Suppose that the first heavy operator has conformal dimension $\Delta_H \gg 1$: the sum rule \eqref{eq:BootstrapFor1D} can then be split as
\begin{equation}
    \sum_{\Delta <\Delta_H} a_{\Delta} F(\Delta,2p-1)+\sum_{\Delta \ge \Delta_H} a_{\Delta} F(\Delta,2p-1) = 0 \ , \quad p=1, \dots, N \ , \label{eq: gapsum}
\end{equation}
where $F(\Delta,2p-1)$ is defined in equation \eqref{eq:BootstrapFor1D}. Plugging the asymptotic behavior from equation \eqref{eq:largeDcoeff} into the second sum gives
\begin{equation}
  \sum_{\Delta \ge \Delta_H} a_{\Delta} F(\Delta,2p-1) \sim \sum_{\Delta \ge \Delta_H} 2 \frac{\Delta^{2\Delta_\varphi-1}}{\Gamma(2\Delta_\varphi)}  F(\Delta,2p-1) \ .  \label{eq: sumheav}
\end{equation}
Assuming $\Delta_H \ge 2\Delta_\varphi + 2N - 2$, the sum \eqref{eq: sumheav} contains only positive terms. We can estimate the right-hand side by replacing the sum with an integral:
\begin{equation}\label{eq:predictionerr}
  2 \int_{\Delta_H}^\infty d \Delta \frac{\Delta^{2\Delta_\varphi-1}}{\Gamma(2\Delta_\varphi)} F(\Delta,2p-1)\overset{\Delta_H \gg 1}{\sim} \frac{1}{2^{\Delta_H}} \frac{\Delta_H^{2(\Delta_\varphi+p-1)}}{\Gamma(2\Delta_\varphi)\log 2} \ .
\end{equation}
We conclude that the sum rule \eqref{eq: gapsum} can be rewritten by bounding the heavy sector contribution as follows:
\begin{equation}
    \sum_{\Delta <\Delta_H} a_{\Delta} F(\Delta,2p-1)=\mathcal{O}\left( \frac{\Delta_H^{2(\Delta_\varphi+p-1)}}{2^{\Delta_H}}\right)  \ , \quad p=1, \dots, N \ . \label{eq: gapppsum}
\end{equation}

The behavior above can be explicitly checked in theories for which the two-point function admits an analytic expression. In particular, in Figs. \ref{fig:D1GFF}, \ref{fig:D3GFF}, \ref{fig:Ising2dcheck}, \ref{fig:YL2dCheck}, we plot the right-hand side of equation \eqref{eq:BootstrapFor1D} as a function of the cutoff dimension and compare it with the identity contribution (the left-hand side of \eqref{eq:BootstrapFor1D}) in 2d CFTs and generalized free field theories. It is also possible to include the identity contribution in the right-hand side of \eqref{eq:BootstrapFor1D} and compare the result with the bound \eqref{eq: gapppsum}.

The interpretation is the following: the appearance of double-twist operators makes it impossible to solve the sum rules by a simple truncation, as in the \textit{Gliozzi method} in the conformal bootstrap. In fact, a naive truncation would require
\begin{equation}
    \frac{\Delta_H^{2(\Delta_\varphi+p-1)}}{2^{\Delta_H}} \ll 1  \ ,
\end{equation}
but if we want as many equations as unknowns, then $p$ must scale with $\Delta_H$, and the condition above is never satisfied. Nevertheless, thanks to the Tauberian theorem, we obtain a good estimate of the tail of heavy operators. This suggests a promising direction: solving the bootstrap problem by implementing the Tauberian asymptotic contribution directly into the sum rules.

\section{Numerical bootstrap}\label{sec:numericalProblem}
We now construct a numerical method to solve for thermal OPE data, based on the sum rules given in equation \eqref{eq:BootstrapFor1D} and the improved Tauberian estimate \eqref{eq:largeDcoeff}.

The starting point of our analysis is the KMS condition. The sum rules can be written as
\begin{equation}
    \sum_{\Delta} a_\Delta \, \mathtt{F}(\Delta,\Delta_\phi,m) = 0 \,,
    \label{eq:SumRules}
\end{equation}
where the sum runs over all operators in the OPE of $\phi \times \phi$. The kernel $\mathtt{F}$, defined in equation \eqref{eq:BootstrapFor1D}, depends solely on the zero-temperature CFT data, which we treat as input. On the other hand, the coefficients $a_\Delta$ encode the thermal dynamical information,
\begin{equation}
    a_\Delta = \sum_{\mathcal O_\Delta} \frac{b_{\mathcal O} f_{\phi \phi \mathcal O}}{c_{\mathcal O}} \frac{J! }{2^J(\nu)_J} C_J^\nu(1) \,,
    \label{eq:aDelta}
\end{equation}
where the sum is over operators of fixed dimension $\Delta$ but varying spin $J$.

The ultimate goal of the thermal bootstrap program is to compute these coefficients and thus complete the thermal CFT data.

A naive attempt to solve equation \eqref{eq:SumRules} would involve truncating the sum at some cutoff dimension $\Delta_\text{max}$. However, this approach fails because the contribution of heavy operators cannot be neglected.\footnote{The error from neglecting the heavy tail is analyzed in \cite{Marchetto:2023xap} and the previous Section. This is in contrast to the zero-temperature case, where a naive truncation can still yield useful approximations \cite{Gliozzi:2013ysa,Gliozzi:2014jsa,Gliozzi:2016cmg}. Nonetheless, even there, accounting for the heavy tail improves precision, as discussed in \cite{Su:2022xnj,Li:2023tic,Poland:2023bny}. In particular, \cite{Li:2023tic,Poland:2023bny} adopt a similar strategy to estimate the error.}

To overcome this issue, we approximate the contribution from the heavy operators using their asymptotic behavior as predicted by the Tauberian theorem:
\begin{equation}
    a_\Delta^{\text{heavy}} = \frac{\Delta^{2\Delta_\phi-1}}{\Gamma(2\Delta_\phi+1)}\delta \Delta \left(1+\frac{c_1}{\Delta} + \ldots \right)\,,
    \label{eq:Tauberianpred}
\end{equation}
where $\delta\Delta$ is the gap between the given scaling dimension $\Delta$ and the one immediately below it in the OPE spectrum. The coefficient $c_1$ encodes the first subleading correction, which is theory-dependent. 

Note that to derive equation \eqref{eq:Tauberianpred}, additional analyticity assumptions on $a_\Delta$ are needed, since the Tauberian theorem controls only the leading term. While the power of $\Delta$ in the first correction is universal, higher-order corrections are theory-dependent.

We can now split the constraints \eqref{eq:SumRules} into two parts:
\begin{equation}
    \mathtt f(m) =
    \sum_{\Delta \leq \Delta_\text{max}} \hspace{-0.75em} a_\Delta \, \mathtt {F}(\Delta,\Delta_\phi,m)
    +
    \sum_{\Delta > \Delta_\text{max}} \hspace{-0.75em} a_\Delta^{\text{heavy}} \, \mathtt {F}(\Delta,\Delta_\phi,m)\,.
    \label{eq:ConstraintsWithTauberian}
\end{equation}

Since the spectrum in the heavy sector is not fully known, further approximation is necessary. In this work, we include only the operators $[\phi\phi]_{n,\ell}$ in the second sum in equation \eqref{eq:ConstraintsWithTauberian}, which—by channel duality—are dual to the identity.\footnote{The identity block inverts to double-twist operators in both zero- and finite-temperature Lorentzian inversion formulas \cite{Iliesiu:2018fao}.} These operators take the schematic form $\phi \partial^\ell \Box^n \phi$, and their dimensions are approximated by their mean-field value: $\Delta_{n,\ell} = 2 \Delta_\phi + 2n + \ell$. The main sources of error are the anomalous dimensions of these operators and the neglect of additional trajectories, which are estimated in \cite{Barrat:2024fwq}.

With this approximation, we are left with a finite number of unknowns: the coefficients $a_\Delta$ for $\Delta \leq \Delta_\text{max}$, and the subleading corrections in \eqref{eq:Tauberianpred}, i.e., $c_1$, etc.

The constraints \eqref{eq:SumRules} can now be imposed by minimizing the cost function:
\begin{equation}
    \eta(\{\omega_i\}) =
    \sum_{m \leq m_\text{max}} \omega_m \, \mathtt f(m)^2\,,
    \label{eq:CostFunction}
\end{equation}
where $m_\text{max}$ is the maximum number of derivatives considered, and $\omega_i \in (0,1)$ are random weights used to test the numerical stability of the procedure, similarly to \cite{Poland:2023bny}.

\vspace{1em}
\noindent To summarize, we propose the following pseudo-code for the algorithm:
\begin{figure}[H]
  \centering
  \begin{minipage}{.7\linewidth}
    \begin{algorithmic}[1]
\Require $\Delta<\Delta_\text{max}, m_\text{max}, \text{iter}$
\State Define the heavy tail approximant;
\State Compute $\mathtt f(m)$ for $m<m_\text{max}$;
\State Initialize $a_\Delta = \{\}$;
\For{$n \le \text{iter}$}
    \State Sample random weights $\omega_1,...,\omega_{m_{\text{max}}}$;
    \State Compute cost $\eta = \sum_{m = 1}^{m_\text{max}} \omega_m \mathtt f(m)^2$;
    \State Minimize $\eta$ $\rightarrow$ Add result to $a_\Delta$;
\EndFor
\State Output: $\text{Mean}(a_\Delta)$
\end{algorithmic}
  \end{minipage}
\end{figure}

\paragraph{Estimation of the error}
The minimization procedure yields estimates for the unknown parameters, which are affected by numerical errors stemming from two sources:
\begin{itemize}
    \item A \textit{statistical} error, estimated via the square root of the variance across multiple runs of the minimization of equation \eqref{eq:CostFunction};
    \item A \textit{systematic} error, due to the approximation of the heavy operator contribution using equation \eqref{eq:Tauberianpred}, estimated below.
\end{itemize}
The errors quoted in this work should be understood as estimates and do not correspond to rigorous bounds.

\noindent The statistical error is straightforward to evaluate. Since the cost function $\eta(\{\mathtt \omega_i\})$ depends on the random weights $\{\mathtt \omega_i\}$, different random initializations will yield slightly different values for the OPE coefficients. This motivates computing the final result as the average over several runs, with the (square root of the) variance serving as a measure of the statistical error. This procedure was also adopted in the context of the zero-temperature bootstrap in \cite{Poland:2023bny}.

\noindent The systematic error is more subtle to estimate. We explain the general strategy below and refer concretely to the case of the 3d Ising model, discussed in detail in Chapter \ref{chap:ONmodel}.

This error originates from the subleading corrections to the Tauberian approximation for the heavy tail. The outcome of the minimization procedure depends on three key parameters:
\begin{itemize}
    \item \textbf{Number of derivatives} $m_\text{max}$: As shown in \cite{Marchetto:2023xap}, the contribution of the heavy tail increases with $m_\text{max}$. The error can thus be minimized by choosing the smallest $m_\text{max}$ compatible with having as many equations as unknowns;
    
    \item \textbf{Number of Tauberian corrections} $c_1, \ldots$: As discussed, the asymptotic form can be refined with corrections of the type $c_i / \Delta^{\alpha_i}$, where $c_i$ and $\alpha_i$ are theory-dependent. For the $\mathrm{O}(N)$ model, where $\Delta_\phi \in [1/2,1]$, we retain only one correction (i.e., $c_1$), as higher-order terms are suppressed. In Section \ref{...}, we explicitly show the impact of subleading corrections in the 4d free scalar theory. A more systematic treatment is left for future work;

    \item \textbf{Cutoff dimension} $\Delta_\text{max}$: This parameter sets the separation between light and heavy sectors. There is no universal prescription for choosing it. We tested values in the range $\Delta_\text{max} \in (3,10)$ and selected the one that gave the most stable minimization, i.e., the smallest variance across runs. Table \ref{tab:resultsvarying} shows the results for the OPE coefficient $a_T$ for various choices of $m_\text{max}$ and $\Delta_\text{max}$. For $\Delta_\text{max} = 6$, the coefficient $a_{\epsilon'}$ was provided as input to ensure convergence. As expected, increasing $m_\text{max}$ increases instability, due to the greater influence of heavy operators. The case $\Delta_\text{max}=3$ corresponds to retaining only $\epsilon$ and $T^{\mu \nu}$ in the light sector. The inclusion of $\epsilon'$ significantly improves convergence. In conclusion, choosing three light operators optimizes numerical stability. An alternative would be to compute a variance-weighted average over all $\{\Delta_\text{max}, m_\text{max}\}$ combinations, but the selected value would still dominate the outcome. We also note that current zero-temperature bootstrap results naturally limit how large $\Delta_\text{max}$ can be taken.
\end{itemize}

\begin{table}[b]
    \centering
    \caption{OPE coefficient $a_T$ in the $3d$ Ising model for $m_\text{max} = \{5,7,9,11\}$ and $\Delta_\text{max} =\{3,4,6\}$. Results are shown as \texttt{value(variance)}. For $\Delta_\text{max}=6$, $m_\text{max}=7$, the number of unknowns exceeds the number of equations and the minimization fails. The highlighted value corresponds to the choice adopted in this work.}
    \renewcommand{\arraystretch}{1.5}
    \begin{tabular}{|c|c|c|c|c|}
    \hline 
    \multicolumn{5}{|c|}{$a_T$ (variance)} \\
    \hline
    \diagbox{$\Delta_\text{max}$}{$m_\text{max}$} & 5 &  7 & 9 & 11\\  
    \hline
    3 & $1.492(1 \cdot 10^{-10})$ & $1.599(3 \cdot 10^{-2})$ & $1.66(6 \cdot 10^{-2})$ & $1.69(4 \cdot 10^{-2})$ \\
    \hline
    4  & no solution & $\bm{1.973(2 \cdot 10^{-18})}$ &  $2.038(4 \cdot 10^{-4})$ &  $2.068(2 \cdot 10^{-3})$ \\ 
    \hline 
    6 & no solution & no solution & $2.053(5 \cdot 10^{-4})$ & $2.105(1 \cdot 10^{-3})$\\ 
    \hline 
    \end{tabular}
    \label{tab:resultsvarying}
\end{table}

\noindent Once these parameters are set, the systematic error—dominant over the statistical one—can be estimated. We distinguish three main contributions:
\begin{enumerate}
    \item[I.] \textbf{Neglecting anomalous dimensions}: In the heavy sector, we approximate operators as double-twist composites with $\Delta = 2\Delta_\phi + 2n + \ell$. The dominant contributions come from large-spin operators. For these, subleading corrections due to anomalous dimensions are of order $O(1/\ell)$ \cite{Fitzpatrick:2012yx}. In the $3d$ Ising model, this introduces a relative error of about $\sim 2\%$ in $a_T$.
    
    \item[II.] \textbf{Neglecting higher Tauberian corrections}: We include only the first correction term in equation \eqref{eq:Tauberianpred}. To estimate the effect of the next correction, we assume $c_2 \sim c_1$ and $\alpha_2 \approx \alpha_1 + 1$, consistent with expectations from the inversion formula. The resulting relative error in $a_T$ is about $\sim 3\%$.
    
    \item[III.] \textbf{Neglecting subdominant trajectories}: In the heavy sector, we consider only $[\phi \phi]_{n,\ell}$ operators. Other contributions, such as $\epsilon^m$ in the $3d$ Ising model, are neglected. To estimate their effect, we use the large-$N$ approximation where these correspond to $\sigma^m$ composites. Their thermal OPE coefficients are estimated as
    \begin{equation}
        a_{\sigma^m} = \frac{m_{\text{th}}^{2m}}{\Gamma(2m+1)} \ ,
    \end{equation}
    where $m_{\text{th}}$ is the thermal mass from Eq. \eqref{eq:TM}. These corrections are subdominant, contributing a relative error of $\sim 0.5\%$ in $a_T$.
\end{enumerate}

\noindent The total error is estimated by combining the three effects above in quadrature. This is conservative, since some of the contributions are not fully independent—for instance, correction II already incorporates anomalous dimensions via pole shifting, as predicted by the inversion formula \cite{Iliesiu:2018fao}.

A more refined analysis of the heavy tail and its systematic contributions would require further developments, possibly via improved inversion techniques or numerical bootstrap at finite temperature.

\subsection{Tests in free scalar theory in $d = 4$}

To test the validity of the method proposed above, we apply it to simple models where analytical results are available.
The simplest case is the free scalar field theory in four dimensions.
The two-point function can be computed explicitly using the method of images, and the thermal OPE data can be directly extracted. As shown previously, the bootstrap problem can also be solved analytically in this case, with the solution given in equation \eqref{eq: final two point function}.

When reduced to zero spatial separation, the two-point function of fundamental scalars reads
\begin{equation}
   g(\tau,0) = \frac{\pi^2}{\beta^2} \csc^2\left(\frac{\pi \tau}{\beta} \right) \,.
\end{equation}
Only double-twist operators appear in the OPE between two fundamental scalars.
Furthermore, the equation of motion $\Box \phi = 0$ implies that only conserved currents are allowed, i.e., operators of the schematic form $\phi \partial^{\mu_1} \ldots \partial^{\mu_J} \phi$, with spin $J$ and conformal dimensions $\Delta = 2 + J$.

To obtain predictions for the OPE coefficients via the numerical method, we include a varying number of corrections to the Tauberian approximation.
This is a useful exercise, as increasing the number of corrections demonstrates that the only source of error is due to approximating the tail of heavy operators.
In the free theory, the spectrum is exactly given by double-twist operators with integer scaling dimensions.
The Tauberian asymptotic takes the form
\begin{equation}
    a_{\Delta}^{\text{heavy}} = \Delta \left(1+\frac{c_1}{\Delta}+\frac{c_2}{\Delta^2}+\frac{c_3}{\Delta^3}+\ldots \right)\,.
\end{equation} 
Recall that only the first correction is universal, but in this case further corrections can be added since anomalous dimensions vanish.

\begin{figure*}[h!]
\centering
\begin{subfigure}[t]{.50\textwidth}
   \includegraphics[width=\textwidth]{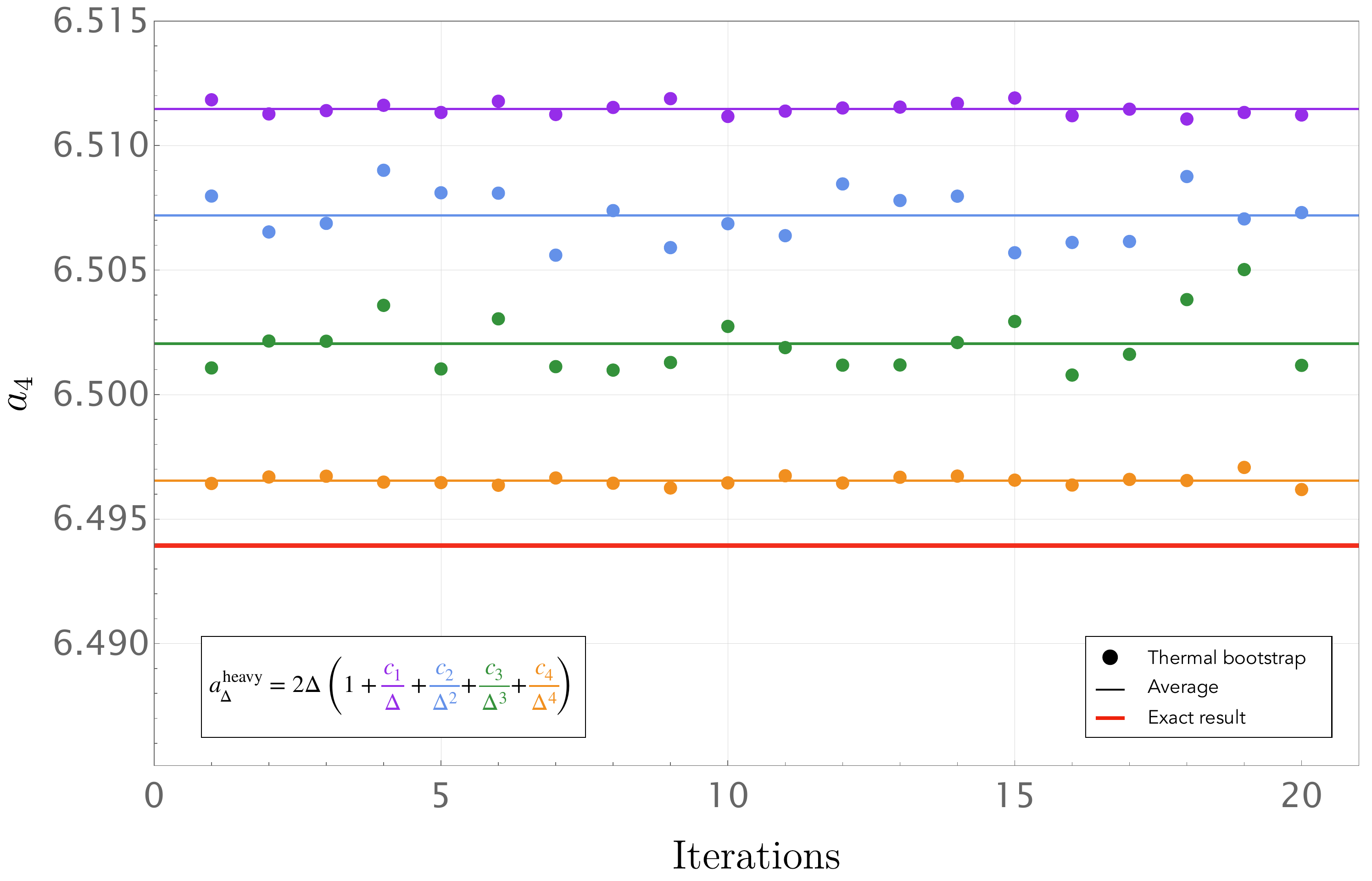}
   \caption{}\label{fig:TaubcorrFree}
\end{subfigure}%
\hfill
\begin{subfigure}[t]{.48\textwidth}
    \includegraphics[width=\textwidth]{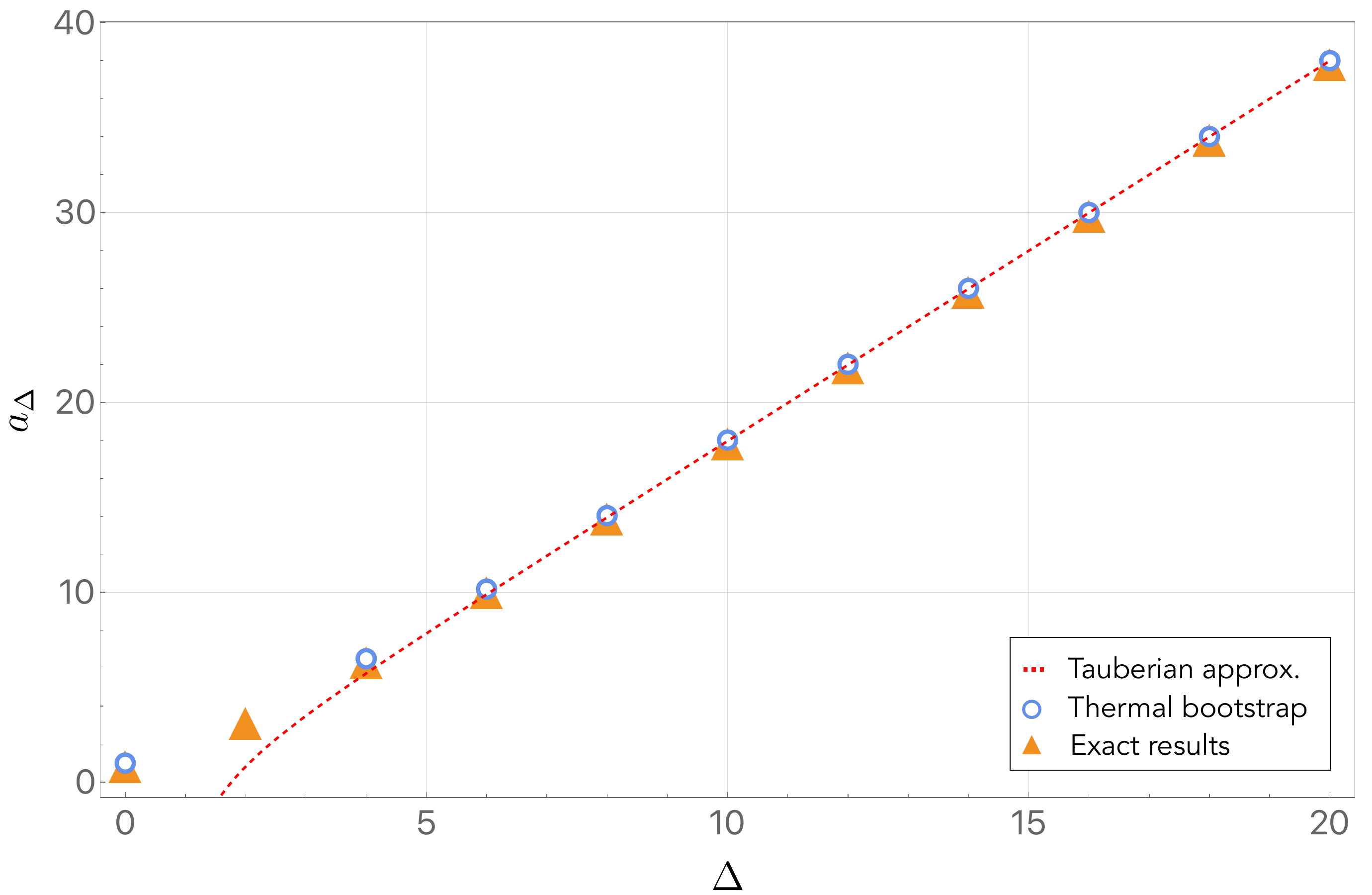}
    \caption{}\label{fig:FreeTops}
\end{subfigure}
  \caption{\textbf{Left panel (a):} Contribution of the stress-energy tensor to the two-point function of fundamental scalars in the free theory, for different approximations of the heavy operator tail. \textbf{Right panel (b):} Numerical versus analytical predictions in the free scalar theory. The operator $\phi^2$ ($\Delta = 2$) contributes to the two-point function as a constant and is therefore not constrained by the KMS condition.}
  \label{Fig:FreeTheory}
\end{figure*}

The results of this analysis are presented in Fig. \ref{Fig:FreeTheory}, where we compare the exact results to the numerical predictions.
Note that the discrepancy decreases as more Tauberian corrections are included.
In these tests we neglect the error bars, as the precision can be systematically improved by adding further corrections to the asymptotic expansion.

\subsection{Tests in the two-dimensional Ising model}

Another useful model for testing the numerical method is the two-dimensional Ising model, where analytical results are also available.
In this case, all one-point functions vanish, except for those of operators in the vacuum module.
This follows from the existence of an anomalous conformal map between the plane and the cylinder.

The two-point function of the Virasoro primary field $\sigma$, with conformal dimension $\Delta_\sigma = 1/8$, is given by
\begin{equation}
    \langle \sigma(\tau) \sigma(0) \rangle_\beta
    =
     \left|\frac{\pi}{\beta}\csc \left(\frac{\pi}{\beta} \tau\right)\right |^{1/4}\,.
\end{equation}
As noted above, the non-vanishing thermal one-point functions correspond to the operators $1, T^{\mu \nu}, T^{\mu \nu} T^{\rho \sigma}, \ldots$ in the vacuum module.
These operators have conformal dimensions $\Delta \in 2\mathbb{N}$, with $\Delta \le J$, and their one-point functions are proportional to the central charge.

\begin{figure*}[h!]
\centering
\begin{subfigure}[t]{.50\textwidth}
   \includegraphics[width=\textwidth]{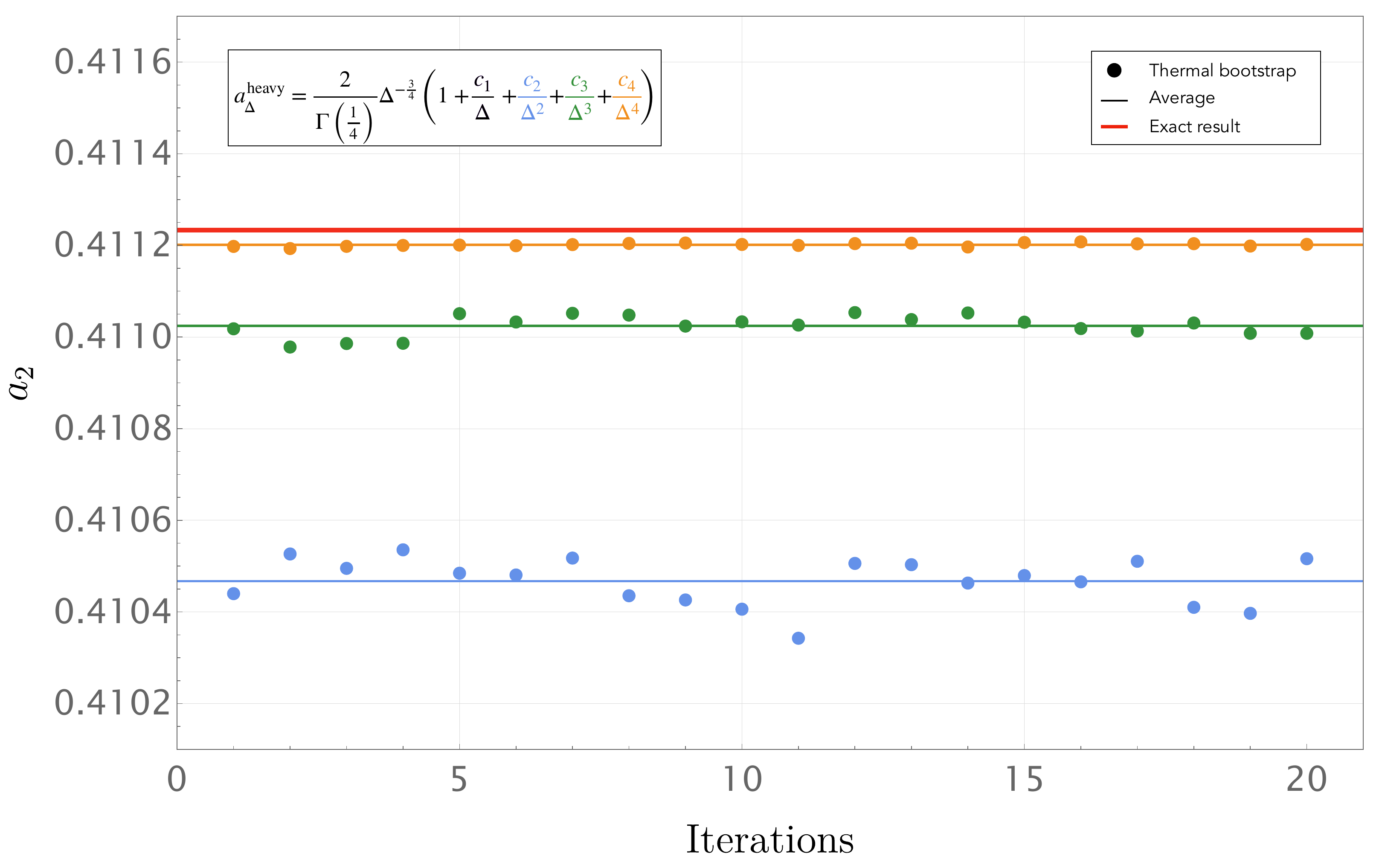}
   \caption{}\label{fig:Taubcorris}
\end{subfigure}%
\hfill
\begin{subfigure}[t]{.48\textwidth}
    \includegraphics[width=\textwidth]{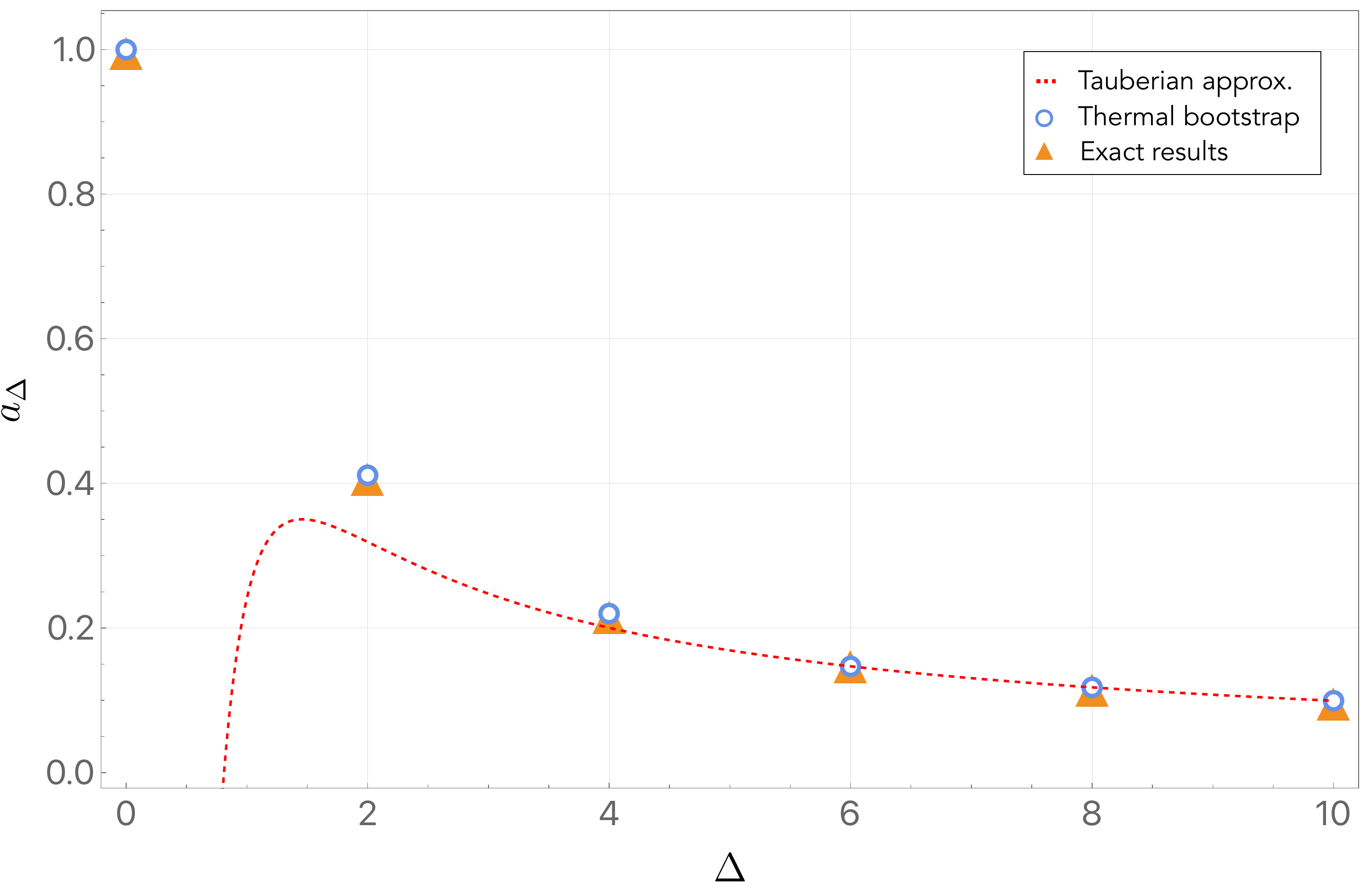}
    \caption{}\label{fig:FreeIsings}
\end{subfigure}
  \caption{\textbf{Left panel (a):} Contribution of the stress-energy tensor to the two-point function of the lightest scalar in the 2d Ising model, for different approximations of the heavy operator tail. \textbf{Right panel (b):} Numerical versus analytical predictions in the 2d Ising model.}
  \label{Fig:2dIsing}
\end{figure*}

The corrections to the Tauberian approximation can also be predicted, since the conformal dimensions are integer-valued.
We perform the same analysis as in the free scalar case, and present the comparison between analytical and numerical results in Fig. \ref{Fig:2dIsing}.

\noindent This analysis can in principle be extended to any two-point function of primary fields in two-dimensional conformal field theories. 
We focus here on the 2d Ising model to provide a concrete and physically relevant example.

\section{Analytic bootstrap}\label{sec:analyticbootstrap}

At zero temperature, considerable effort has been devoted to analytical approaches to the bootstrap of correlation functions in CFTs (see for instance \cite{Alday:2013cwa,Alday:2014tsa,Fitzpatrick:2012yx,Dey:2016zbg,Bianchi:2020hsz,Harris:2024nmr}). 
In the same spirit, we now explore possible analytical strategies for the bootstrap at finite temperature. In this Section we will make use of analytical properties of thermal two-point functions encoded in the inversion formula and the dispersion relation presented in \ref{sec:analyticprop} to construct an analytical bootstrap problem at finite temperature.

\section{A formula for thermal correlators}

We begin by considering the special kinematical configuration in which the two operators lie on the same thermal circle.\footnote{I thank Dalimil Mazac for interesting comments and suggestions regarding this section.}

\subsection{The complex-time plane}

Let us consider a thermal two-point function in the zero spatial direction limit. It becomes a function $g(\tau)$ of the coordinate $\tau$ along the thermal circle. By complexifying the $\tau$ coordinate, we promote the correlator to a function of $\xi=\tau+i t$ in the complex plane. The analytic structure of $f(\xi)$ is strongly constrained by the KMS condition: the function must have simple poles on the real axis at
\begin{equation}
    \xi= n \beta \ , \quad n \in \mathbb{Z} \ .
\end{equation}
We cannot exclude \textit{a priori} the presence of branch cuts connected to each of these poles. If present, KMS invariance requires them to be vertical and all aligned with respect to the real axis.

In the $\xi$ plane, the two-point function can be written using a dispersion relation, i.e., via the Cauchy integral formula:
\begin{equation} \label{eq: disp1}
    g(\xi)=\frac{1}{2 \pi i}\oint_{\mathcal{C}} d \xi' \frac{f(\xi')}{\xi'-\xi}
\end{equation}
Using KMS symmetry, the contour can be decomposed into an infinite sum of contours, each surrounding a different image of the branch cut:
\begin{equation} \label{eq: disp2}
    g(\xi)=\sum_{m=-\infty}^{\infty} \frac{1}{2 \pi i}\oint_{\mathcal{C}_m} d \xi' \frac{g(\xi'+m)}{(\xi'+m)-\xi}
\end{equation}
Neglecting arc contributions, and assuming $g(\xi)$ is polynomially bounded at infinity (as verified in explicit examples), we arrive at:
\begin{equation}
    g(\xi) = \sum_{m=-\infty}^{\infty} \frac{1}{2 \pi i} \int_0^{i \infty} \frac{d\xi'}{\xi'+m-\xi} \Disc f(\xi') \ .
\end{equation}

\paragraph{On the arcs}

Until now, we have neglected potential arc contributions. Here we show that they must vanish. Define the dispersion part as
\begin{equation}
    g_\text{dr}(\xi) = \sum_{m=-\infty}^{\infty} \frac{1}{2 \pi i} \int_0^{i \infty} \frac{d\xi'}{\xi'+m-\xi} \Disc g(\xi') \ ,
\end{equation}
so that
\begin{equation}
    g(\xi) = g_\text{dr}(\xi) + g_\text{arc}(\xi) \ .
\end{equation}
Since $g_\text{dr}(\xi)$ accounts for all non-analytic structure, $g_\text{arc}(\xi)$ must be entire. Moreover, periodicity of $g(\xi)$ and $g_\text{dr}(\xi)$ implies that $g_\text{arc}(\xi)$ is periodic on the real axis. \begin{figure}
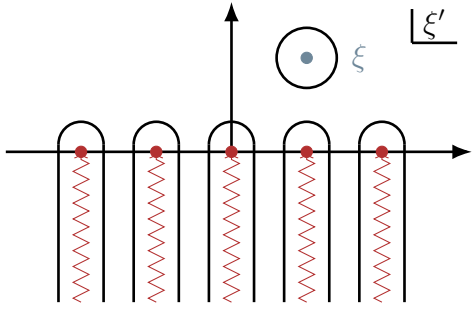

    \centering
    \ContourThermaltau
    \caption{Analytic structure in the complex $\xi=\tau+i t$ coordinate. In the simplest cases, the branch cuts are absent and the contour surrounds only the real poles. In the case of the $\mathrm{O}(N)$ model at first order in $\varepsilon$, the branch cuts extend to infinity.}
    \label{fig:collanalytictau}
\end{figure}

We now show that $g_\text{arc}(\xi)$ is bounded at large $|\xi|$. Following the argument in \cite{Iliesiu:2018fao}, we write
\begin{equation}
    g(\xi) = \langle \Psi | e^{i \xi \op P_{KK} } |\Psi \rangle \ .
\end{equation}

This can be decomposed as
\begin{equation}
    e^{i \xi \op P_{KK}} = V^{\frac{1}{2}} U V^{\frac{1}{2}} \ , \quad
    V = e^{-\frac{1}{2} t \op P_{KK}}, \quad U = e^{i \tau \op P_{KK}} \ ,
\end{equation}
where $V$ is Hermitian and positive, and $U$ is unitary. By the Cauchy–Schwarz inequality,
\begin{equation}
    |g(\xi)| \le \langle \Psi | V | \Psi \rangle = |g(i t/2)| \ .
\end{equation}
We use the real-time representation
\begin{equation}
    g(\xi) = \sum_{i,j} e^{-\beta E_i} e^{i \xi E_j} |\langle E_i | \phi | E_j \rangle|^2 \ ,
\end{equation}
and, assuming $E_i > \lambda$, we get
\begin{equation}
    g(i t) \le e^{-t \lambda} \cdot \text{const.}
\end{equation}
In a thermal CFT on $S^1 \times \mathbb{R}^{d-1}$, the spectrum is $E_i = \Delta_i / R$, and in unitary theories, the minimal $\Delta$ is zero, hence $\lambda = 0$, and
\begin{equation}
    |g(\xi)| \le \text{const.} \cdot e^{|t|} \ .
\end{equation}
Thus $g_\text{arc}(\xi)$ is an entire function, periodic on the real axis, and exponentially bounded in the complex plane, implying it must be constant.

\paragraph{The zero spatial direction limit in the $\omega$-plane}

In the $\omega$ coordinate, and in the limit of vanishing spatial separation, the analytic structure includes:
\begin{itemize}
    \item two cuts from $\infty$ to $\omega = \pm 1/\sqrt{\tau}$;
    \item two cuts from $\omega = \pm \tau$ to $\omega = 0$;
    \item poles at $\omega = \pm 1$.
\end{itemize}
This structure is meaningful only if $|\tau| < 1$, suggesting that the analysis in the $\omega$ plane is valid only within a fundamental domain of the $\xi$ plane.

\begin{figure}
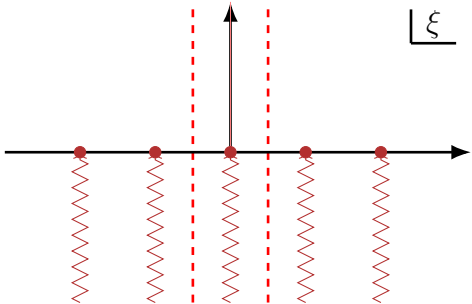

    \centering
    \ContourThermaltauandw
    \caption{Fundamental domain in the $\xi$ plane, corresponding to the domain for the variable $w$ used to study the analytic properties of thermal correlators. Each domain represents a different image.}
    \label{fig:collanalytic}
\end{figure}

The logic is then as follows:
\begin{itemize}
    \item Reduce to a fundamental domain in the $\xi$ plane;
    \item Use dispersion relations to compute $g(\xi)$ in that domain;
    \item Extend the result via image summation to cover the full $\xi$ plane;
    \item Generalize to the case with non-zero spatial separation.
\end{itemize}

\subsection{Combining KMS and dispersion relations}

We already observed that the result from the dispersion relation in Section \ref{Sec:DispRel} is not KMS invariant. A possible strategy to restore KMS invariance is to sum over images. This procedure naturally gives a KMS-invariant function with the correct analytic structure, though arc terms might still be needed.

\paragraph{The KMS compensator}

One strategy is to define a \emph{KMS compensator}:
\begin{equation}
    g(z,\overline z) = g_\text{dr}(z,\overline z) + g_\text{comp.}(z,\overline z) \ .
\end{equation}
While $g_\text{comp.}(z,\overline z)$ is difficult to guess exactly, the inversion formula suggests that contributions from outside the OPE regime are exponentially suppressed in spin, of order $2^{-J}$ \cite{Iliesiu:2018fao}. Thus, a reasonable ansatz for perturbative theories is to model non-perturbative spin corrections via
\begin{equation}\label{eq:corre}
    a_{\mathcal O} = a_{\mathcal O}^{\text{(dr)}} \left(1+\sum_{k}\frac{c_k}{k^J}\right) \ ,
\end{equation}
with unknown coefficients $c_k$. Truncating this sum and imposing KMS symmetry in the OPE regime allows one to solve for $c_k$. In simple theories, this yields closed-form results. However, this approach is heuristic, and the general validity of \eqref{eq:corre} is unclear.

\paragraph{Images over DR}

A second approach is
\begin{equation}\label{eq:formula}
   g(z,\overline z) = \frac{1}{2} \sum_{m = -\infty}^\infty g_\text{dr}(z-m,\overline z - m) + g_\text{arcs}(z,\overline z) \ ,
\end{equation}
where summing over images restores KMS invariance, and $g_\text{arcs}$ includes only operators of spin $J < J_\star$.
The prefactor $1/2$ avoids double-counting due to $\tau \to -\tau$ symmetry already implemented in $g_\text{dr}$.

Importantly, $g_\text{arcs}$ must reduce to a constant in the limit $z \to \overline z$.
For example, in free theory, the identity operator might seem to contribute to arcs, but its conformal block is not constant in this limit.

The motivation for \eqref{eq:formula} comes from the structure of the complex-time plane. While a general proof is lacking, the formula has been shown to reproduce correct results in various cases. It satisfies all bootstrap axioms (KMS, analyticity, OPE, zero-temperature limit), with arcs possibly needed to ensure the correct Regge and large-distance behavior.

In cases where the image sum does not converge, one must treat $z, \overline z$ as complex variables and interpret the result via analytic continuation.

\subsection{Tests in 4d free theory}
    
The simplest possible application is the free scalar theory in $d = 4$: we already encountered this example in the thesis. Recall that the two-point function of two fundamental scalars can be decomposed into an OPE, where only higher spin currents and the identity contribute. Schematically:
\begin{equation}
    \phi \times \phi = [\op{1}]+ [\phi \phi]_{0,J} \ , 
\end{equation} 
where the higher spin currents can be written as $[\phi \phi]_{0,J} = \phi \partial^{\mu_1}\ldots \partial^{\mu_J} \phi$, and their dimensions are given by $\Delta = 2 + J$ (in $d = 4$). The two-point function can be computed directly using the method of images, yielding:
\begin{equation}\label{eq:gfreetheoryexact}
   g(z,\overline z)  = \sum_{m = -\infty}^\infty (z-m)^{-2} (\overline z-m)^{-2} = \frac{ \pi }{ \beta (z-\overline z)} \left(-\cot\left(\frac{\pi}{\beta} z \right)+\cot\left(\frac{\pi}{\beta}\overline z \right)\right)\ .
\end{equation}
The block expansion of this equation automatically gives the thermal OPE data: $a_{[\phi \phi]_{0,J}} = 2 \zeta(2+J)$ and $a_{\op{1}} = 1$ \cite{Iliesiu:2018fao,Marchetto:2023fcw}.

\paragraph{The dispersion relation (DR):}
As recalled in Section~\ref{sec:disc}, the discontinuity over individual blocks in the OPE vanishes for all higher spin currents. Therefore, the only contribution to the discontinuity in the OPE regime comes from the identity operator, whose discontinuity is given by:
\begin{equation}
    \Disc [\op{1}] = -2 i \sin (\pi \Delta_\phi ) r^{2\Delta_\phi }  \left(\frac{r}{\omega'} - 1 \right)^{-\Delta_\phi} (1 - r \omega')^{-\Delta_\phi} \Theta \left( \frac{r}{\omega'} - 1 \right) \ .
\end{equation}
This must be integrated against the appropriate kernel in equation \eqref{eq:Kernel}. While direct integration is difficult, a power series expansion in $r$ allows us to reconstruct the result order by order. In $(z, \overline z)$ coordinates, we obtain:
\begin{equation}\label{eq:gdrfree}
    g_{dr}(z,\overline z) = \frac{\xi^{-\Delta_\phi}}{z \overline z} \left( \left( \frac{1 - z}{z + 1} \right)^{\Delta_\phi} \left( \frac{1 - \overline z}{\overline z + 1} \right)^{\Delta_\phi} + 1 \right) \ ,
\end{equation}
where
\begin{equation}
    \xi = \frac{(1 - z)(1 - \overline z)}{z \overline z}\ .
\end{equation}
As expected, the result from the dispersion relation is not KMS invariant. The OPE coefficients extracted from this correlator differ from those of the exact free scalar theory. In fact, expanding \eqref{eq:gdrfree} in thermal blocks gives exactly the same result as the inversion formula, as both rely on the same input.

\paragraph{Lorentzian Inversion Formula (LIF):}
Using the inversion formula, it is well known that the contribution of an operator $\mathcal O$ to the OPE always contains a pole at $\Delta = 2\Delta_\phi + n + J$. Specifically, the contribution to $[\phi \phi]_{0,J}$ is given by \cite{Iliesiu:2018fao}:
\begin{equation}
    a_{[\phi \phi]_{n = 0}}^{\mathcal O}(J) = a_{\mathcal O} (1 + (-1)^J) \frac{K_J}{K_{J_{\mathcal O}}} S_{h_{\mathcal O} - \Delta_\phi, \Delta_\phi}(\overline h) \ ,
\end{equation}
with
\begin{equation}
    K_j = \frac{\Gamma(j + 1) \Gamma(\nu)}{4 \pi \Gamma(j + \nu)} \ , \quad
    S_{c,\Delta}(\overline h) = \frac{1}{\Gamma(-c)} \frac{\Gamma(\overline h - \Delta - c)}{\Gamma(\overline h - \Delta + 1)} - \frac{B_{1/z_\text{max}}(\overline h - \Delta - c, 1 + c)}{\Gamma(-c) \Gamma(1 + c)} \ .
\end{equation}
The residues at these poles yield:
\begin{equation}
    a_{[\phi \phi]_{n,J}} = 1 + (-1)^J \ .
\end{equation}
This matches the expansion of \eqref{eq:gdrfree}, confirming that the inversion formula and the dispersion relation provide the same answer. We compare the exact results with those from DR and LIF in Table~\ref{tab:drFree}.

\begin{table}[h]
    \begin{center}
        \renewcommand{\arraystretch}{1.5}
        \begin{tabular}{|c|c|c|c|}
        \hline
        $\mathcal O$ & $a_{\mathcal O}$ exact & from DR & from LIF \\ \hline
        $\op{1}$ & 1 & 0 & 0 \\ \hline
        $[\phi \phi]_{0,J}$ & $2 \zeta(2 + J)$ & 2 & 2 \\ \hline
        \end{tabular}
        \caption{Thermal OPE coefficients in the $\phi \times \phi$ OPE in free theory, computed from the exact correlator, DR, and LIF. Only even-spin operators contribute.}
        \label{tab:drFree}
    \end{center}
\end{table}

\paragraph{Imposing KMS:}
In the free theory, we know that the only operator contributing to the arcs is the identity operator: its one-point function is simply the normalization of the scalar field $\phi$, and we can choose to unit-normalize the zero-temperature two-point function so that $a_{\op{1}} = 1$. This contribution must be added by hand, yet the resulting function is still not KMS invariant. According to the discussion in the previous section, the ansatz for the correlator is:
\begin{equation}\label{eq:freeansaz}
    g(z,\overline z) = g_{dr}(z,\overline z) + \frac{1}{z \overline z} + \sum_{n = 1}^\infty c_n g_\text{comp.}^{(n)} \ ,  
\end{equation}
where the first term is the outcome of the dispersion relation, the second is the identity contribution (i.e., the only arc term in this case), and the last term represents corrections from operators outside the OPE regime. The ansatz for these corrections is:
\begin{equation}
    a_{\mathcal O} = a_{\mathcal O}^{(dr)} \left(1 + \sum_{n = 1}^\infty \frac{c_n}{n^J} \right)\ .
\end{equation}
In the free scalar case, the only operators are the higher-spin currents $[\phi \phi]_{0,J}$: the compensators $g_\text{comp.}^{(n)}$ can be computed directly. We obtain:
\begin{equation}
    g_\text{comp.}^{(n)} = \frac{n^2 (n^2 + z \overline z)}{(n^2 - z^2)(n^2 - \overline z^2)} \ .
\end{equation}
To impose KMS, we truncate the ansatz \eqref{eq:freeansaz} at $n_\text{max}$ and numerically fit the coefficients $c_n$ by expanding the correlator around both $z, \overline z \sim 0$ and $z, \overline z \sim 1$. Varying $n_\text{max}$ allows us to verify the stability of the solution. For sufficiently large $n_\text{max}$, all coefficients turn out to be rational (as expected from the exact result), and can be identified exactly using \texttt{Rationalize} in \texttt{Mathematica}.

\begin{table}[h!]
    \caption{Values of the first correction coefficients $c_n$ for the case $\Delta_\phi = 1$, extracted by imposing KMS.}
    \centering
    \begin{tabular}{ccccccl}
        \hline
        $n_\text{max}$ & 2 & 3 & 4 & 5 & 10 & Exact result \\ \hline
        $c_2$ & 0.614 & 0.486 & 0.501 & 0.500 & 0.500 & $1/2$ \\
        $c_3$ & – & 0.384 & 0.172 & 0.232 & 0.222 & $2/9$ \\
        $c_4$ & – & – & 0.323 & 0.017 & 0.125 & $1/8$ \\ \hline
    \end{tabular}
    \label{tab:KMS_Free}
\end{table}

\begin{figure}[h]
\centering
 \includegraphics[width=96mm]{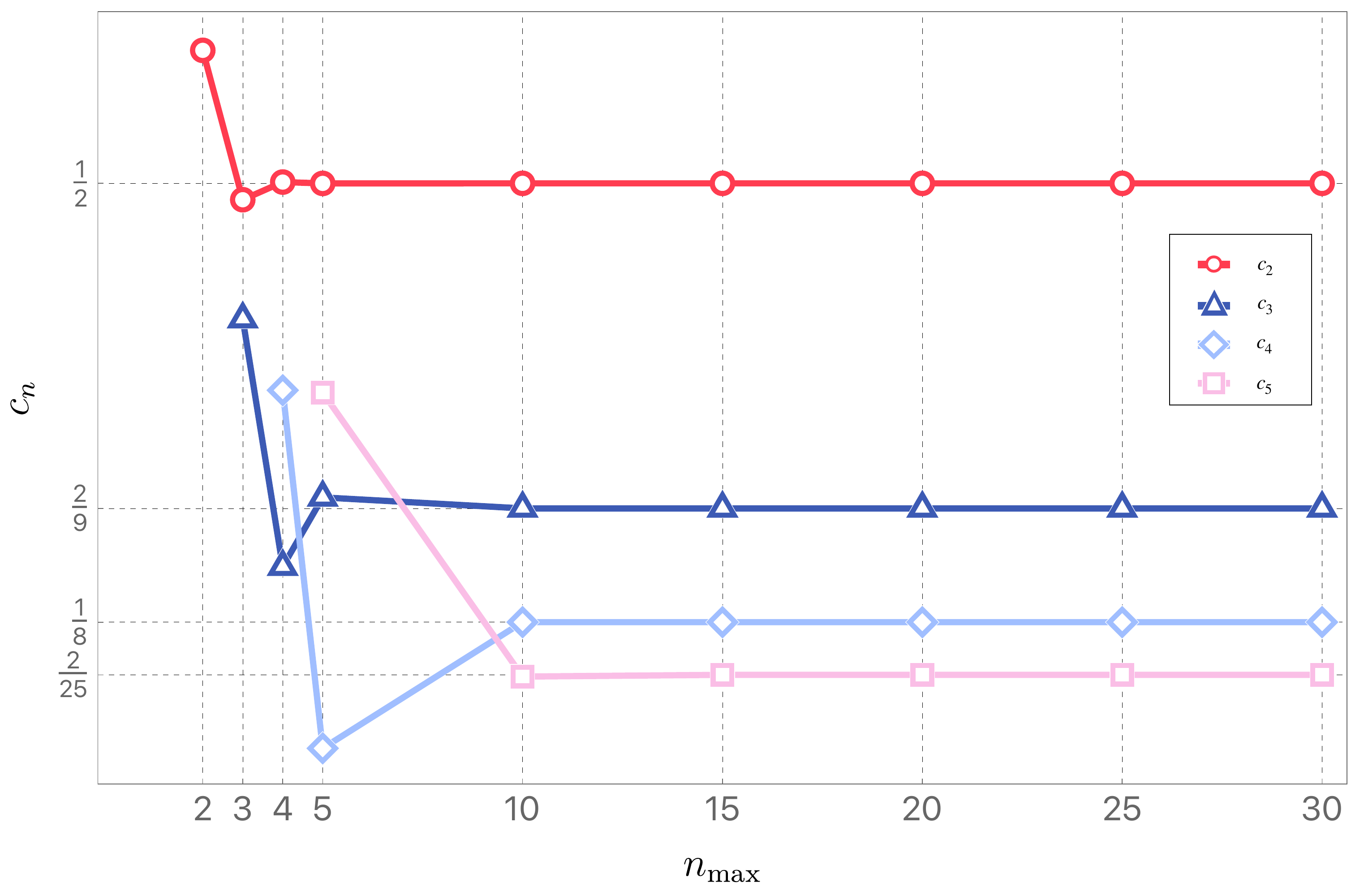}
    \caption{First five corrections in equation \eqref{eq:freeansaz}, computed by imposing KMS and varying the number of included corrections $n_\text{max}$.}
    \label{fig:cnvsmmax}
\end{figure}

Table~\ref{tab:KMS_Free} and Figure~\ref{fig:cnvsmmax} show the convergence of the coefficients $c_n$ as $n_\text{max}$ increases. We observe very good agreement, especially for the first few coefficients. A clear pattern emerges:
\begin{equation}
    c_n = \frac{2}{n^2} \ ,
\end{equation}
which leads to the correct OPE coefficients: $a_{[\phi \phi]_{0,J}} = 2 \zeta(2+J)$, perfectly matching the exact result.

\paragraph{Using images over DR:}
We now consider the KMS-symmetrized version of the dispersion relation:
\begin{equation}
    g(z,\overline z) = \sum_{m = -\infty}^\infty \left( g_{dr}(z - m, \overline z - m) + \frac{a_{\op{1},\text{arc}}}{(z - m)(\overline z - m)} \right) \ .
\end{equation}
A direct computation shows that choosing $a_{\op{1},\text{arc}} = -1$ yields:
\begin{equation}
    g(z,\overline z) = \frac{ \pi }{ \beta (z - \overline z)} \left(-\cot\left(\frac{\pi}{\beta} z \right) + \cot\left(\frac{\pi}{\beta} \overline z \right) \right) \ ,
\end{equation}
which is the exact result. Note that this is not a naive method of images acting on the identity block: the contribution is nontrivial and results from the full analytic structure.

\paragraph{The momentum space interpretation:}
As discussed in Section~\ref{Sec:OPEmomspace}, in free theory the thermal OPE in momentum space yields:
\begin{equation}\label{eq:propfreet}
    g(k,\omega_n) \propto \frac{1}{k^2 + \omega_n^2} \ .
\end{equation}
As observed in \cite{Manenti:2019wxs}, operators with $\Delta = 2\Delta_\phi + 2n + J$ do not appear in the momentum-space OPE. In free theory, this means only the identity contributes. This matches the prediction from the dispersion relation, see also equation \eqref{eq:gg}. While $e^{-k}$ corrections are in principle allowed, they originate from non-local interactions and are absent in free theories.

Thus, equation \eqref{eq:propfreet} is exact. Performing the Fourier transform and using Poisson resummation gives:
\begin{equation}
    g(\tau,\vec x) = \sum_{m = -\infty }^\infty \frac{1}{x^2 + (\tau - m)^2} \ ,
\end{equation}
which can be computed explicitly and reproduces the correlator in \eqref{eq:gfreetheoryexact}.

   \subsection{Tests in 2d CFTs}

It is interesting to apply the same analytic bootstrap methodology to the case of Virasoro primaries in two spacetime dimensions. As already explained earlier in this thesis, the two-point functions of identical Virasoro primaries can be written exactly thanks to a conformal map from the plane to the cylinder. This map is anomalous: in fact, the only non-vanishing one-point functions are those of the Virasoro vacuum module, as they are the ones affected by conformal anomalies. For a primary of conformal weights $(h, \overline h)$, we have:
\begin{equation}\label{eq:2dcorr}
    \langle \phi(z, \overline z) \phi(0,0) \rangle_\beta = \left(\frac{\pi}{\beta}\right)^{h + \overline h} \csc^h\left(\frac{\pi z}{\beta} \right)\csc^{\overline h}\left(\frac{\pi \overline z}{\beta} \right) \ .
\end{equation}
We focus on scalar primaries, so that $h = \overline h = \Delta_\phi/2$. The operators appearing in the OPE belong to the vacuum module, meaning that they satisfy $\Delta \leq J$, with $\Delta \in 2\mathbb{N}$ and $J \in 2\mathbb{N}$. In the following, we study the cases $\Delta_\phi = 1$ and $\Delta_\phi = 2$.

We note that the block expansion defined in general dimensions is not well-defined in $d = 2$, because the OPE coefficients contain $\Gamma(\nu)$, which diverges when $\nu = 0$ ($d=2$). On the other hand, the blocks involve Gegenbauer polynomials $C_J^{(0)}(\eta)$, which vanish. The divergence in the OPE coefficients exactly cancels the zero of the Gegenbauer polynomials in the limit $\nu \to 0$. To handle this, we redefine the blocks as:
\begin{equation}
    f_{\Delta,J}(z,\overline z) = \lim_{\nu \to 0} \frac{J!}{(\nu)_J} z^{\frac{\Delta}{2} - \Delta_\phi} \overline z^{\frac{\Delta}{2} - \Delta_\phi} C_J^{(\nu)}\left(\frac{z + \overline z}{2 \sqrt{z \overline z}} \right)\ ,
\end{equation}
so that the OPE coefficients become:
\begin{equation}
    a_{\mathcal O} = \frac{f_{\phi \phi \mathcal O} b_{\mathcal O}}{2^J c_{\mathcal O}} \ .
\end{equation}
With slight abuse of notation, we will use the same symbols for blocks and coefficients in $d=2$ and $d>2$, keeping in mind this redefinition.

\subsubsection{The case $\Delta_\phi = 1$}

We study here the case where $\Delta_\phi = 1$.

\paragraph{Dispersion relation (DR):}
If $\Delta_\phi$ is an integer, e.g., $\Delta_\phi = 1$ or $2$, one might expect that all block discontinuities vanish except for the identity, as happens for the operators $[\phi \phi]_{n,J}$ in the GFF case. This is not so in two dimensions: twist is not bounded below by a positive number. In particular, operators with zero twist (i.e., $\Delta = J$) contribute to the discontinuity. The contribution of a single such block to the dispersion relation is:
\begin{equation}
    g_\text{dr}^{(J)}(z,\overline z) = a_J \frac{2^{1-J}(z \overline z + 1)(z \overline z - 1)^J}{z \overline z (z^2 - 1)(\overline z^2 - 1)} \ .
\end{equation}
We can compute the coefficients $a_J$ using the exact result:
\begin{equation}
    a_J = -2^{J + 1} \Li_J(-1) \ .
\end{equation}
Summing over all spins:
\begin{equation}
    g_\text{dr}(z,\overline z) = -\frac{2 \pi z \overline z (z^2 \overline z^2 - 1) \csc(\pi z \overline z)}{(z^2 - 1)(\overline z^2 - 1)} \ .
\end{equation}
Expanding in blocks, we find $a_{\op{1}} = 0$, while the remaining coefficients arrange into:
\begin{equation}
    a_{\Delta,J} = -4 \Li_{(\Delta - J)/2}(-1) \ ,
\end{equation}
which correctly reproduces the large-spin behaviour.

\paragraph{Imposing KMS:}
It is possible to use compensators as before, but in this case, we need infinitely many $1/n^J$ corrections for each twist. Thus, the final ansatz becomes:
\begin{equation}
    g(z,\overline z) = g_\text{dr}(z,\overline z) + g_\text{arc}(z,\overline z) + \sum_{n=2}^\infty \sum_{\tau = 0}^\infty c_{n,\tau} g_\text{comp.}^{(\tau,n)}(z,\overline z) \ ,
\end{equation}
with
\begin{equation}
    g_\text{comp.}^{(\tau,n)}(z,\overline z) = \frac{z^{\tau/2} \overline z^{\tau/2} (n^4 - z^2 \overline z^2)}{z \overline z (n^2 - z^2)(n^2 - \overline z^2)} \ .
\end{equation}
We must truncate both the number of corrections ($n_\text{max}$) and the maximum twist ($\tau_\text{max}$). This procedure is systematic but numerically costly. For instance, with $n_\text{max} = 12$ and $\tau_\text{max} = 16$, the first few $c_{n,0}$ converge well:
\begin{equation}
    \frac{c_{8,0}}{c_{8,0}^\text{exact}} = 0.9976\ , \quad \frac{c_{9,0}}{c_{9,0}^\text{exact}} = 0.9652 \ .
\end{equation}
However, generalization becomes hard. The image method is more efficient.

\paragraph{Using images over DR:}
We can test the image-summed DR in special kinematic limits. First, $z = \overline z$, i.e., operators on the same thermal circle:
\begin{equation}
    g(z,z) = \pi^2 \csc^2(\pi z) \ ,
\end{equation}
matching \eqref{eq:2dcorr} with $h = \overline h = 1/2$. The second limit is $z = -\overline z$, i.e., equal times and spatial separation:
\begin{multline}
    g(ix,-ix) = \sum_{m=-\infty}^\infty g_\text{dr}(ix - m, -ix - m) \\
    = \frac{1}{2} \sum_{m} \left( \frac{1}{x^2 + (m - 1)^2} + \frac{1}{x^2 + (m + 1)^2} \right)
    - \sum_{\Delta \geq 2} 2^\Delta \Li_\Delta(-1) \sum_m \frac{(m \pm 1)^\Delta}{x^2 + (m \pm 1)^2} \ .
\end{multline}
After simplification:
\begin{equation}
    g(ix, -ix) = \pi^2 \text{csch}^2(\pi x) \ ,
\end{equation}
which again matches the exact result.

\subsubsection{The case $\Delta_\phi = 2$}

\paragraph{Dispersion relation (DR):}
Also in this case, only zero-twist operators ($\Delta = J$) contribute. One might expect twist-2 operators to appear, but they vanish in the thermal OPE. The contribution of each zero-twist operator is:
\begin{multline}
    g_\text{dr}^{(J)}(z,\overline z) = a_J \frac{2(1 - z \overline z)^{J - 2}}{(1 - z^2)^2 (1 - \overline z^2)^2}
    \bigg[1 + \overline z^2 - (J + 2) z \overline z (1 - z^2)(1 - \overline z^2) \\
    + z^4 \overline z^4 \left(1 + \overline z^2 + J(1 - \overline z^2) + z^2(1 - (J + 6)\overline z^2 + (J + 1)\overline z^4)\right) \bigg] \ .
\end{multline}
This can be resummed to match the large-spin limit of the exact result:
\begin{equation}
    a_{\tau,J} = (\tau - 2)(2J + \tau - 2) \Li_{\tau/2}(1) \Li_{J + \tau/2}(1) \ , \quad
    a_{\tau,J} = 2J(\tau - 2)\Li_{\tau/2}(1) + \mathcal{O}(2^{-J}) \ .
\end{equation}
Since the KMS correction method is inefficient here, we focus directly on the image-summed DR.

\paragraph{Using images over DR:}
Again, two kinematic limits simplify the computation. First, $z = \overline z$ (same thermal circle), only the identity and stress tensor ($\Delta = J = 2$) contribute. Resumming:
\begin{equation}
    g(z, z) = \pi^4 \csc^4(\pi z) \ ,
\end{equation}
which matches the exact result. In the collinear limit ($z = i x = -\overline z$), after analytic continuation:
\begin{equation}
    g(i x, -i x) = \pi^4 \text{csch}^4(\pi x) \ ,
\end{equation}
again perfectly agreeing with \eqref{eq:2dcorr}.

   \section{Summary of the chapter}

This chapter represents the conceptual core of the thesis. After reviewing how thermal CFT data are connected to thermodynamic quantities, we proposed a bootstrap problem for thermal one-point functions, based on the KMS condition applied to thermal two-point functions and its compatibility with the operator product expansion (OPE).

We began by enumerating the general bootstrap conditions expected to hold for thermal correlators and discussed the issue of uniqueness. The KMS condition was then recast as a channel duality: the two-point function at $\tau$ must equal the two-point function at $\beta - \tau$. This duality has a fixed point at $\tau = \beta/2$, around which we can expand to derive sum rules for the thermal OPE coefficients, which are proportional to the thermal one-point functions. We derived these sum rules in full generality and then specialized them to the case $\vec{x} = 0$. We also provided explicit checks in free theories and in the two-dimensional Ising model.

While the sum rules arise from an expansion around $\tau = \beta/2$, one can also consider the limit $\tau \to \beta$. In this limit, the identity operator dominates one of the two channels, while reproducing the full correlator in the other requires an infinite sum over heavy operators. This observation allows us to extract the asymptotic behavior of thermal OPE coefficients for heavy operators in the limit $\vec{x} = 0$. This result can be heuristically obtained via a simple Legendre transform, but can also be rigorously derived using Tauberian theorems under appropriate assumptions.

By combining the heavy-operator asymptotics with the sum rules derived earlier in the chapter, we constructed a numerical method to bootstrap thermal one-point functions. This method was tested in the cases of the four-dimensional free scalar theory and the two-dimensional Ising model. We also discussed how to estimate the numerical error. The approach relies on a truncated OPE expansion in which the contribution from heavy operators is not discarded, but instead approximated using an analytic continuation of the asymptotic formula obtained from the KMS condition.

Schematically, the logic of the method can be summarized as follows:
\begin{align}
    \tau \sim \frac{\beta}{2} &\quad \Rightarrow \quad \text{sum rules for thermal OPE coefficients} \ , \\
    \tau \to \beta &\quad \Rightarrow \quad \text{asymptotics for heavy operators} \ ,
\end{align}
and by combining these two limits, we are able to obtain numerical predictions also for the one-point functions of light operators.

In addition to the numerical approach, we also proposed an analytic method based on the dispersion relation constructed in the previous chapter. This method was tested in simple theories, and we introduced a hybrid analytic-numerical strategy that provides highly accurate results. In particular, we proposed a closed-form expression—up to arcs—in which the two-point function can be interpreted as a \textit{KMS-invariant version} of the dispersion relation. Concretely, we suggest applying the method of images to the dispersion relation output in order to enforce KMS invariance. In the complex $\tau$-plane (with $\vec{x} = 0$), this procedure can be shown to be well-defined, and the resulting correlator is exact up to an additive constant.

Applications of both the numerical and analytical methods to the $\mathrm{O}(N)$ model will be presented in Chapter~\ref{chap:ONmodel}.

\def\Oh{{\widehat{\mathcal{O}}}}
\def\veps{\varepsilon}
\newcommand{\Db}{\bar{\Delta}}

\chapter{Conformal line defects at finite temperature}\label{sec:defectsBO}

In this chapter, we discuss the symmetries of a CFT with a defect at finite temperature. We present possible configurations and analyze which symmetries are broken or preserved when the defect wraps the thermal circle.
We then examine the correlation functions associated with the thermal defect CFT and identify the OPE data required for a complete characterization.

\section{Line defects on the thermal manifold}
\label{subsec:PolyakovLoopsOnTheThermalManifold}

\subsubsection{Configurations}
\label{subsubsec:Configurations}

Our starting point is a Euclidean CFT at zero temperature, i.e., a CFT on the flat manifold $\mathbb{R}^{d}$.
We label the representations of the conformal group by the scaling dimensions $\Delta$ and the spin $J$.
A \textit{defect CFT} can be defined by inserting an extended operator of codimension $q$ (called a \textit{conformal defect}) that preserves part of the original symmetry.
Specifically, for a conformal line defect, the codimension is $q=d-1$ and the symmetry is broken in the following way:
\begin{equation}
    \mathrm{SO}(d+1,1)
    \longrightarrow
    \mathrm{SO}(2,1) \times \mathrm{SO}(d-1)\,.
    \label{eq:DefectCFT_ZeroT}
\end{equation}
The group $\mathrm{SO}(2,1)$ can be interpreted as the one-dimensional conformal symmetry preserved along the defect, while $\mathrm{SO}(d-1)$ corresponds to the unbroken rotations around the defect.
From the $1d$ perspective, the latter can be viewed as a \textit{global} symmetry \cite{Kapustin:2005py, Drukker:2005af}.
Defect representations are then labeled by the quantum numbers $\Dh$, the $1d$ scaling dimension, and $s$, the transverse spin, associated with the group $\mathrm{SO}(d-1)$.

\begin{figure}
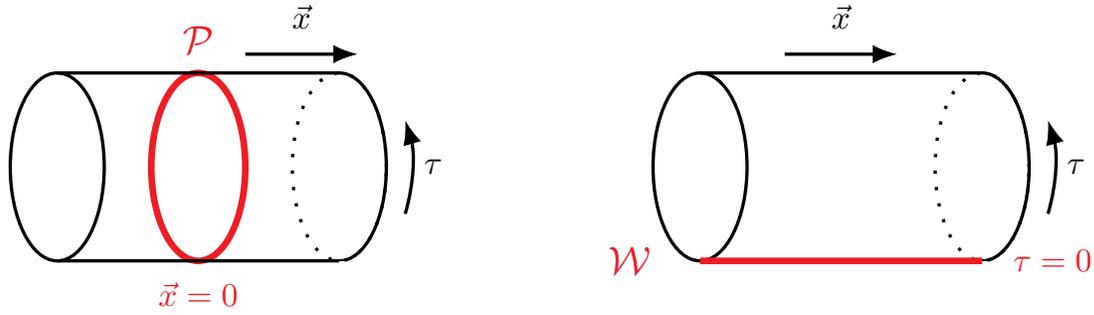

    \begin{subfigure}[b]{0.48\textwidth}
        \centering
        \scalebox{1.25}{\PolyakovLoop}
    \end{subfigure}
    \hfill
    \begin{subfigure}[b]{0.48\textwidth}
        \centering
        \scalebox{1.25}{\WilsonLine}
    \end{subfigure}
\caption{
\emph{The left figure shows a temporal line defect wrapping the thermal circle at the spatial position $\vec{x}=0$.
This configuration corresponds to case (a) in the main text, and is the focus of this paper.
On the right side, the setup (b) is illustrated, in which a spatial line defect is placed at $\tau=0$ and extends along a spatial direction.}
}
\label{fig:Configurations} 
\end{figure}

Our goal is to study this system at finite temperature.
More precisely, we consider the defect CFT not on flat space, but on the \textit{thermal manifold}
\begin{equation}
    \Mm_\beta
    =
    S^1_\beta
    \times
    \mathbb{R}^{d-1} \ ,
    \label{eq:ThermalManifold}
\end{equation}
where the original time direction is compactified along a circle of length $\beta = 1/T$, with $T$ being the temperature.
In the absence of a defect, this compactification explicitly breaks the conformal symmetry of the bulk theory \cite{Iliesiu:2018fao,Marchetto:2023fcw}.
Contrary to the zero-temperature case, the orientation of the line defect plays a significant role in how the symmetries are broken or preserved.
We can consider \textit{three} different configurations:
\begin{enumerate}
    \item[(a)] The line defect wraps the thermal circle;
    \item[(b)] The line defect is placed along one of the (non-compactified) spatial directions;
    \item[(c)] The line defect is placed in a direction that includes both time and space components. In two dimensions, this defect may take the shape of a \textit{spiral} wrapping the thermal cylinder.
\end{enumerate}
Graphical representations of cases (a) and (b) are shown in Fig.~\ref{fig:Configurations}.
At zero temperature, all configurations are equivalent in Euclidean space.
However, at finite temperature, these setups describe \textit{distinct physical systems}.
For instance, in setup (a), the defect preserves translations along the time direction and maintains an $\mathrm{SO}(d-1)$ symmetry. In contrast, in setup (b), only rotations orthogonal to the defect remain unbroken, reducing the symmetry group to $\mathrm{SO}(d-2)$.

Although all configurations have interesting physical applications, this work focuses on case (a), which maximizes the residual conformal symmetry of the defect CFT.
Importantly, in the context of gauge theories, this configuration corresponds to \textit{Polyakov loops}, whose one-point function acts as the order parameter for the confinement/deconfinement transition.

\subsubsection{Broken and unbroken symmetries}
\label{subsubsec:BrokenAndUnbrokenSymmetries}

Let us summarize the effect of turning on the temperature on the symmetries of the conformal group in the case where the defect wraps the thermal circle.
Building upon the discussion of the previous section, we have:
\begin{itemize}
    \item[$\star$] \textbf{Translations}: Time translations around the thermal circle are unbroken, while spatial translations are broken by the presence of the defect.
    \item[$\star$] \textbf{Rotations}: Spatial rotations are unbroken, while boosts are broken by both the defect and finite-temperature effects.
    \item[$\star$] \textbf{Dilatations and SCTs}: Dilatations and special conformal transformations (SCTs) are broken only by finite-temperature effects.
    \item[$\star$] \textbf{Global symmetries}: Any global symmetry is preserved at finite temperature, but can be broken by the defect. The potential breaking of external symmetries by the defect depends on the details of the theory. 
\end{itemize}
This set of symmetries defines the \textit{thermal defect CFT} that we study in the rest of this work.
It is possible to derive (broken) Ward identities in the spirit of \cite{Marchetto:2023fcw}.
These result in non-trivial constraints on the correlation functions of the theory. The (broken) Ward identities are derived and presented below.

\subsubsection{Broken Ward identities}
\label{sec:BrokenWardIdentities}

We derive (broken) Ward identities for both broken and unbroken generators, following the procedure adapted in \cite{Marchetto:2023fcw} for the case without defects, where many details are given.
The main idea is that, since conformal symmetry is expected to be non-anomalous, the action of the theory must be non-invariant under the broken generators. Using the path integral formulation, one can derive how a symmetry generator acts on a correlation function and relate it to the variation of the action.

Concretely, consider the following formal action as a starting point:
\begin{equation}
    \label{eq: start}
    S = S_{\text{bulk}} + h \int_0^\beta d\tau \, \phi(\tau, \vec{0}) \ ,
\end{equation}
where we include a term describing a line wrapping the thermal circle located at the spatial origin.\footnote{The field $\phi$ is not necessarily a fundamental field of the theory, but rather an operator of conformal dimension one.} The term $S_{\text{bulk}}$ denotes the action in the absence of the defect.

For any non-anomalous, infinitesimal symmetry transformation $\delta \mathcal{O} = i \omega^a \op G_a \mathcal{O}$, the following Ward identity holds:
\begin{equation}
    \delta \langle \mathcal O_1(x_1) \ldots \mathcal O_n(x_n) \Pm \rangle_\beta = \langle \delta S_{\text{bulk}} \, \mathcal O_1(x_1) \ldots \mathcal O_n(x_n) \Pm \rangle_\beta \ .
\end{equation}
Isolating the variation of the Polyakov loop, we obtain:
\begin{equation}
    \langle \delta\left( \mathcal O_1(x_1) \ldots \mathcal O_n(x_n) \right) \Pm \rangle_\beta 
    + \langle \mathcal O_1(x_1) \ldots \mathcal O_n(x_n) \delta \Pm \rangle_\beta
    = \langle \delta S_{\text{bulk}} \, \mathcal O_1(x_1) \ldots \mathcal O_n(x_n) \Pm \rangle_\beta \ .
\end{equation}
Expanding the defect term using \eqref{eq: start}:
\begin{multline}
    \langle \delta\left( \mathcal O_1(x_1) \ldots \mathcal O_n(x_n) \right) \Pm \rangle_\beta
    - h\, \delta \int_0^{\beta} d\tau \, 
    \langle \mathcal O_1(x_1) \ldots \mathcal O_n(x_n) \phi(\tau, \vec{0}) \Pm \rangle_\beta \\
    = \langle \delta S_{\text{bulk}} \, \mathcal O_1(x_1) \ldots \mathcal O_n(x_n) \Pm \rangle_\beta \ .
\end{multline}
Following \cite{Marchetto:2023fcw}, we obtain:
\begin{equation}
    i \sum_i \langle \mathcal O_1(x_1) \ldots \op G_a \mathcal{O}_i(x_i) \ldots \mathcal O_n(x_n) \Pm \rangle_\beta = \int d^{d-1}y \, \langle \Gamma_a^{\Pm,\beta}(\vec{y}) \, \mathcal O_1(x_1) \ldots \mathcal O_n(x_n) \Pm \rangle_\beta \ ,
\end{equation}
where the breaking term $\Gamma_a^{\Pm,\beta}$ is a defect-induced correction to the thermal breaking term $\Gamma_a^\beta$:
\begin{equation}
    \Gamma_a^{\Pm,\beta}(\vec{y}) = \Gamma_a^\beta(\vec{y}) + h\, \delta(\vec{y}) \, \op G_a \int_0^\beta d\tau\, \phi(\tau, \vec{y}) \ .
\end{equation}

\paragraph{Broken translations.} Let us begin with translations. Since the purely thermal breaking term is $\Gamma^\beta_\mu = 0$, we have:
\begin{equation}
    \Gamma_\mu^{\Pm,\beta}(\vec{y}) = h\, \delta(\vec{y}) \int_0^\beta d\tau\, (\partial_\mu \phi)(\tau, \vec{y}) \ .
\end{equation}
If $\phi$ is periodic on the thermal circle, this simplifies to:
\begin{equation}
    \Gamma_0^{\Pm,\beta} = 0 \ , \qquad
    \Gamma_i^{\Pm,\beta}(\vec{y}) = h\, \delta(\vec{y}) \int_0^\beta d\tau\, D_i(\tau, \vec{y}) \ ,
\end{equation}
where $D_i$ is the displacement operator. Hence, as expected, time translations are preserved, while spatial translations are broken, leading to the Ward identity:
\begin{equation} \label{eq:sys}
    \sum_i \langle \mathcal O_1(x_1) \ldots \partial_i \mathcal{O}_i(x_i) \ldots \mathcal O_n(x_n) \Pm \rangle_\beta
    = h \int_0^\beta d\tau\, \langle D_i(\tau, \vec{0}) \mathcal O_1(x_1) \ldots \mathcal O_n(x_n) \Pm \rangle_\beta \ .
\end{equation}
The unintegrated form can be written as an operator equation:
\begin{equation}
    \partial_\mu \op T^{\mu}_{\;\; i}(\tau, \vec{y}) = \delta(\vec{y}) \, \op D_i(\tau, \vec{y}) \ ,
\end{equation}
which is well known from early studies of defects in CFTs \cite{Billo:2016cpy}. Finite-temperature effects do not modify the breaking term.

\paragraph{Broken dilatations.} We now consider dilatations, which are broken both by thermal effects and by the defect. The thermal breaking term is $\Gamma^\beta = \beta T^{00}$, so:
\begin{equation}
    \Gamma^{\Pm,\beta}(\vec{y}) = \beta T^{00}(0,\vec{y}) + h\, \delta(\vec{y})\, \op D \int_0^\beta d\tau\, \phi(\tau, \vec{y}) \ .
\end{equation}
The action of the dilatation operator is:
\begin{align}
    \op D \int_0^\beta d\tau\, \phi(\tau, \vec{y}) 
    &= \int_0^\beta d\tau\, \left(\tau \partial_\tau + y^i \partial_i + \Delta_\phi \right) \phi(\tau, \vec{y}) \nonumber \\
    &= (\Delta_\phi - 1) \int_0^\beta d\tau\, \phi(\tau, \vec{y})
    + y^i \int_0^\beta d\tau\, D_i(\tau, \vec{y})
    + \beta \phi(0, \vec{y}) \ .
\end{align}
The corresponding Ward identity is:
\begin{align}
    \op D \langle \mathcal O_1 \ldots \mathcal O_n \Pm \rangle_\beta 
    &= \beta \int d^{d-1}y \, \langle \left(T^{00}(0,\vec{y}) + h\, \delta(\vec{y}) \phi(0, \vec{y}) \right) \mathcal O_1 \ldots \mathcal O_n \Pm \rangle_\beta \nonumber \\
    &\quad + h(\Delta_\phi - 1) \int_0^\beta d\tau\, \langle \phi(\tau, \vec{y}) \mathcal O_1 \ldots \mathcal O_n \Pm \rangle_\beta \ .
\end{align}
If the defect is conformal, i.e., $\Delta_\phi = 1$, the expression simplifies:
\begin{equation}
    \op D \langle \mathcal O_1 \ldots \mathcal O_n \Pm \rangle_\beta
    = \beta \int d^{d-1}y \, \langle \left(T^{00}(0,\vec{y}) + h\, \delta(\vec{y}) \phi(0, \vec{y}) \right) \mathcal O_1 \ldots \mathcal O_n \Pm \rangle_\beta \ .
\end{equation}
This can be interpreted operatorially as:
\begin{equation} \label{eq: D defect}
    \op D = -\beta \left( \op H + E_0 + h \op \phi \right) \ ,
\end{equation}
where $\op H$ is the thermal Hamiltonian and $E_0$ is the vacuum energy \cite{Marchetto:2023fcw}:
\begin{equation}
    \op H = -\int d^{d-1}y \left( T^{00}(\tau, \vec{y}) + E_0 \right) \ , \qquad
    E_0 = \frac{d-1}{d} \frac{b_T}{\beta^d} \ .
\end{equation}
In this case, dilatations receive corrections from both the thermal background and the defect. Equation~\eqref{eq: D defect} is strictly valid for conformal defects admitting a Lagrangian realization as in \eqref{eq: start}.

Finally, note that the definition \eqref{eq: start} is not the most general way to define a conformal line defect. For example, disorder-type defects are often implemented via boundary conditions. Nevertheless, identities based purely on symmetry (such as \eqref{eq:sys}) are expected to remain valid.

\subsection{Correlation functions}
\label{sec:CorrelationFunctions}

We now turn our attention to the implications of these symmetries: our aim is to provide the \textit{OPE data} necessary to (in principle) fully solve the system through repeated use of the bulk and defect OPEs.

\subsubsection{Bulk and defect OPEs}
\label{subsubsec:BulkAndDefectOPEs}

One crucial aspect of this setup is that the thermal geometry described by \eqref{eq:ThermalManifold} is \textit{conformally flat}.\footnote{A manifold is \textit{conformally flat} if each point has a neighborhood that can be mapped to an open subset of flat space via a conformal transformation. This does not imply the existence of a conformal map between the full thermal manifold and flat space: this only happens in $d = 2$.}
This implies that \textit{local} properties of the zero-temperature theory are preserved when the time dimension is compactified.
In other words, the scaling dimensions of the operators remain unaffected, and products of two (or more) operators can be expanded using the same OPE as at zero temperature.
For instance, the \emph{bulk OPE} for two identical scalar operators $\phi$ of dimension $\Delta_\phi$ is given by
\begin{equation}
\phi (x_1) \phi (x_2)
=
\frac{1}{x_{12}^{2 \Delta_\phi}}
\sum_{\Op \in \phi \times \phi}
\mu_{\phi \phi \Op} |x_{12}|^{\Delta - J} x_{12\, \mu_1} \ldots x_{12 \, \mu_J} \Op^{\mu_1 \ldots \mu_J} (x_2) \,,
\label{eq:BulkOPE}
\end{equation}
where the conformal data (i.e., the scaling dimensions $\Delta$ and the structure constants $\mu_{\phi\phi\Op}$) appearing in this equation are zero-temperature data.
Before proceeding, let us note that the OPE convergence, discussed in similar CFT contexts in \cite{Pappadopulo:2012jk,Antunes:2021qpy,Cuomo:2024vfk}, is now limited to a \textit{finite} region.
Concretely, the convergent region of the bulk OPE \eqref{eq:BulkOPE} in the presence of the defect is given by
\begin{equation}
    \tau_{12}^2+\vec x_{12}^2
    \le \min
    \left\lbrace
    \beta^2, |\vec x_1|^2,|\vec x_2|^2
    \right\rbrace\,,
    \label{eq:ConvergenceBulkOPE}
\end{equation}
with $|\vec{x}_i|$ the transverse positions of the two operators with respect to the defect.

In defect CFT, it is well known that bulk operators can be expanded in terms of the defect operators introduced in Section \ref{subsubsec:Configurations}, by bringing the bulk operator close to the defect.
For a scalar operator $\phi$, this results in the \emph{defect OPE}
\begin{equation}
\phi (\tau, \vec x)
=
\frac{1}{|\vec{x}|^\Delta}
\sum_{\Oh}
\lambda_{\phi \Oh}
|\vec{x}|^{\Dh-s} x_{i_1} \ldots x_{i_s} \Oh^{i_1 \ldots i_s} (\tau)\,,
\label{eq:DefectOPE}
\end{equation}
where $|\vec{x}|$ is the transverse distance between the operator and the defect. The OPE above is expected to converge in the region $|\vec x|<\beta$, unless other operators are closer to the defect.
Once again, the OPE data appearing in this equation correspond to the zero-temperature case.
Here, the coefficients $\lambda_{\phi \Oh}$ describe bulk-defect two-point functions, which are fixed kinematically at $T=0$ \cite{Billo:2016cpy}.

\subsubsection{Bulk and defect one-point functions}
\label{subsubsec:BulkAndDefectOnePointFunctions}

Non-local properties of the theory can however be affected by the compactification of the time dimension.
Since dilatations are not preserved, one-point functions of local operators (without defects) can in general be non-zero.
Translational invariance along the thermal circle fixes them to be constant and dependent on a single coefficient \cite{Iliesiu:2018fao,Marchetto:2023fcw}
\begin{equation}
    \vev{\Op_{\Delta}^{\mu_1 \ldots \mu_J}}_\beta
    =
   \frac{ b_\Op }{\beta^\Delta} \left(e^{\mu_1}\ldots e^{\mu_J}-\text{traces}\right)\,,
    \label{eq:ThermalOnePointFunctions}
\end{equation}
where $e^\mu$ is non-zero only in the time direction.
We refer to these as \emph{bulk thermal one-point functions}.

To understand the OPE data required to characterize the thermal CFT in the presence of a defect, we study the first kinematically non-trivial correlator: the one-point function of a bulk operator in presence of the defect $\Pm$.
At zero temperature, it is kinematically fixed, but since translations orthogonal to the defect are broken, the correlator becomes
\begin{equation}
\vev{\phi (\tau,\vec{x}) \Pm}_\beta
=
\frac{F_{\phi}(z)}{|\vec{x}|^{\Delta_\phi}}\,,
\label{eq:OnePoint_Correlator}
\end{equation}
for a scalar operator $\phi$, where $z$ is the dimensionless ratio
\begin{equation}
    z = \frac{|\vec{x}|}{\beta}\,.
    \label{eq:OnePoint_CrossRatio}
\end{equation}
Using the defect OPE \eqref{eq:DefectOPE}, we can rewrite \eqref{eq:OnePoint_Correlator} as
\begin{equation}
    \vev{\phi (\tau, \vec x)\Pm}_\beta 
    =
    \frac{1}{|\vec{x}|^{\Delta_\phi}}
    \sum_{\Oh}
    \lambda_{\phi \Oh}
    |\vec{x}|^{\Dh-s} x_{i_1} \ldots x_{i_s} \langle \Oh^{i_1 \ldots i_s} (\tau)\Pm\rangle_\beta \,,
\label{eq:OnePoint_DefectOPE}
\end{equation}
which converges for $|\vec{x}| \leq \beta$.
As mentioned in Section \ref{subsubsec:Configurations}, the defect operators $\Oh$ are labeled by their quantum numbers $\Dh$ and $s$, corresponding respectively to $\mathrm{SO}(2,1)$ and $\mathrm{SO}(d-1)$.
At zero temperature, the correlator on the right-hand side vanishes for all operators except for the identity $\widehat{\mathds{1}}$.
From time-translation invariance, defect one-point functions take the form
\begin{equation}
    \vev{ \Oh_{\Dh}^{i_1\ldots i_s} }_\beta
    =
    \frac{\widehat{b}_{\Oh}}{\beta^{\Dh}}\Pi^{i_1\ldots i_s}\,,
    \label{eq:DefectOnePoint}
\end{equation}
where $\Pi^{i_1 \ldots i_s}$ is an $\mathrm{SO}(d-1)$-invariant tensor structure.\footnote{We choose the normalization such that $x_{i_1}\ldots x_{i_s} \Pi^{i_1\ldots i_s} =  |\vec x|^s$ to simplify later expressions.} We refer to these as \textit{defect one-point functions} $\widehat{b}_{\widehat{\mathcal O}}$.

Importantly, these are \textit{not} one-dimensional thermal one-point functions.
Indeed, in a 1d CFT on $S_\beta^1$, conformal invariance and absence of anomalies imply that thermal one-point functions vanish.
The apparent contradiction with \eqref{eq:DefectOnePoint} is resolved by noting that \textit{thermal bulk effects} induce non-trivial one-point functions on the 1d defect theory.

Contracting the tensor structure with $x_{i_1}\ldots x_{i_s}$ gives
\begin{equation}
    x_{i_1}\ldots x_{i_s} \vev{\Oh_{\Dh}^{i_1\ldots i_s}}_\beta 
    =
    \bh_{\Oh}
    \frac{|\vec{x}|^{s} }{\beta^{\Dh}}\,,
\end{equation}
so that \eqref{eq:OnePoint_Correlator} becomes
\begin{equation}
      F_{\phi} (z)
      =
      \sum_{\Oh}
      \lambda_{\phi \Oh} \, \bh_{\Oh} \, 
      z^{\Dh}\,.
      \label{eq:OnePoint_DefectOPE2}
\end{equation}
Operators with the same $\Dh$ but different $s$ contribute at the same order in $z$.
In interacting theories, degeneracies are rare and can be lifted by considering several bulk insertions.

The function $F_{\phi}(z)$ must satisfy the limits
\begin{equation}
    F_{\phi} (z)
    \overset{z \to 0}{\sim}
    \lambda_{\phi \hat{\mathds{1}}}\,,
    \qquad
    F_{\phi} (z)
    \overset{z \to \infty}{\sim}
    b_{\phi} z^{\Delta_\phi}\,.
    \label{eq:Limit1pt}
\end{equation}
The first corresponds to the zero-temperature limit, where only the identity contributes.\footnote{We assume the normalization $\vev{ \widehat{\mathds{1}}} = \vev{\widehat{\mathds{1}}}_\beta = 1$.} The second corresponds to the high-temperature limit, and implies that $\widehat{b}_{\widehat{\mathcal O}}$ can be non-zero in general.

In summary, any correlator in this \textit{thermal defect CFT} can be expanded via the bulk and defect OPEs, using the zero-temperature data (scaling dimensions and OPE coefficients) together with the thermal one-point functions $b_{\Op}$ and $\widehat{b}_{\Oh}$.
For example, two-point functions of defect operators $\widehat{\phi}_1$ and $\widehat{\phi}_2$ are controlled by $\widehat{\mu}_{\widehat{\phi}_1 \widehat{\phi}_2 \Oh}$ and $\bh_\Oh$.
We summarize the necessary data in Table \ref{table:CFTData}.\footnote{At zero temperature, certain observables (e.g., the Bremsstrahlung function) depend on the normalization of defect two-point functions. This normalization can be viewed as a function of the bulk-to-defect OPE coefficients, so we omit them here.}
While thermal correlators can, in principle, be reconstructed from zero-temperature data, doing so is difficult in practice \cite{Iliesiu:2018fao}. Instead, we will see that a bootstrap approach for the $\widehat{b}_{\Oh}$ is more tractable and closely parallels the standard thermal bootstrap.

\begin{table}
    \centering
    \setlength{\tabcolsep}{12pt}
    \begin{tabular}{lrlrlrl}
        \hline \\[-.75em]
        \, & \multicolumn{2}{c}{\hspace{-2.3em} bulk data} & \multicolumn{2}{c}{\hspace{-1.1em} bulk-defect data} & \multicolumn{2}{c}{\hspace{-1.8em}defect data} \\[.5em] 
        \hline \\[-.75em]
        \multirow{2}{*}{$T=0$} & $ \Delta_{\mathcal{O}} \hspace{-1.7em}  $ & $\sim\vev{\Op \Op}$ & $\lambda_{\Op \Oh} \hspace{-1.7em} $ & $ \sim\vev{\Op \Oh}$&  $\Dh_{\Oh}\hspace{-1.7em}  $ &$\sim\vev{\Oh \Oh}$ \\[.5em]
         & $ \mu_{ijk}\hspace{-1.7em}  $ & $\sim \vev{\Op_i \Op_j \Op_k}$&  & & $\widehat{\mu}_{ijk}\hspace{-1.7em}  $ &$\sim\vev{\Oh_i \Oh_j \Oh_k}$ \\[.5em] \hline \\[-.75em]
        $T=1/\beta$ & $ b_{\Op}\hspace{-1.7em}$ & $\sim\vev{\Op}_\beta$ & & &  $ \widehat{b}_{\Oh}\hspace{-1.7em} $ & $\sim\vev{\Oh}_\beta$\\[.5em] \hline
    \end{tabular}
    \caption{
    \emph{Summary of the OPE data needed to characterize the thermal defect CFT, and how the coefficients are related to correlation functions at zero and finite temperature.
    Notice that the only two new sets of coefficients appearing at $T \neq 0$ are the thermal bulk one-point functions $b_\Op$ and the defect one-point functions $\bh_\Oh$.}
    }
    \label{table:CFTData}
\end{table}

\subsubsection{An inversion formula for defect one-point functions}
\label{subsubsec:AnInversionFormulaForDefectOnePointFunctions}

We have seen in the previous Section that defect one-point functions are the fundamental quantities to determine, and that they are induced by the bulk theory as thermal excitations.
We now show how these coefficients can be computed through a bulk calculation.
Indeed, it is possible to write the OPE expansion of $F_\phi(z)$ (defined in \eqref{eq:OnePoint_Correlator}) as
\begin{equation}
    F_{\phi} (z)
    =
    \int_{-\veps-i \infty}^{-\veps+i \infty}d \Delta\
    \frac{a(\Delta)}{2 \pi i} z^{\Delta}\,,
    \label{eq:inversion1}
\end{equation}
where the function $a(\Delta)$ has poles at the scaling dimensions of physical operators, with residues fixed to be
\begin{equation}
    a(\Delta)
    \sim
    -\frac{\lambda_{\phi \Oh}\, \widehat{b}_{\Oh}}{\Delta-\Dh}\,,
    \label{eq:aDeltaDef}
\end{equation}
and, as a technical condition, $a(\Delta)$ is required not to grow exponentially in the positive-$\Delta$ half (complex) plane.
Here, $\veps$ parametrizes the vertical contour in the complex plane, which can be chosen arbitrarily as long as the integral converges.
A Mellin transform can be applied to \eqref{eq:inversion1} to extract the function $a(\Delta)$:
\begin{equation}
    a(\Delta)
    =
    \int_0^1 d z\
    z^{-\Delta-1}\ F_{\phi}(z)\,.
\end{equation}
Using \eqref{eq:aDeltaDef}, we obtain the inversion formula for the OPE coefficients:
\begin{equation}
    \lambda_{\phi \Oh}\, \widehat{b}_{\Oh}
    =
    \operatorname{Res}_{\Delta \to \Dh}
    \int_0^1 d z\
    z^{-\Delta-1} \ F_{\phi} (z)\,.
    \label{eq:inversion}
\end{equation} 
This equation provides a way to extract defect one-point functions from a scalar bulk one-point function and the known zero-temperature structure constants $\lambda_{\phi \Oh}$.

To conclude, note that in \eqref{eq:inversion} operators are identified solely by their scaling dimensions.
As mentioned in the previous Section, degeneracy in the transverse spin cannot be resolved using bulk one-point functions alone.
However, it is possible that two degenerate operators appear in different bulk correlators with different structure constants; in such cases, the degeneracy can in principle be resolved.

\section{Setting up a bootstrap problem}
\label{sec:SettingUpABootstrapProblem}

This section is dedicated to formulating a bootstrap problem for defect one-point functions at finite temperature.
We present novel sum rules derived from the KMS condition, which impose strong constraints on the coefficients $\widehat{b}_\Oh$.
With this in mind, we also discuss the behavior of heavy operators.

\subsection{From the KMS condition to sum rules}
\label{sec:FromTheKMSConditionToSumRules}

\begin{figure}
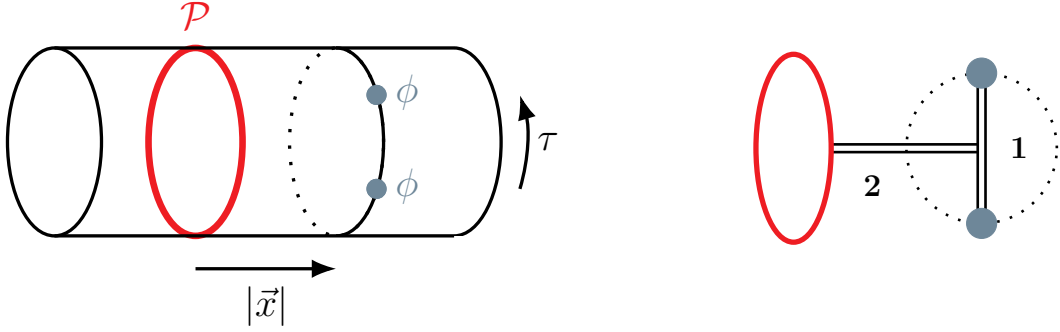

\centering
\begin{subfigure}{.45\textwidth}
  \raggedright
  \scalebox{1.25}{\KMSCollinear}
\end{subfigure}%
\begin{subfigure}{.45\textwidth}
  \raggedleft
  \scalebox{1}{\OrderOfOPE}
\end{subfigure}
\caption{
\emph{The configuration for the KMS condition in the collinear limit is depicted on the left.
Here, the two local operators are located at a distance $|\vec{x}|$ from the Polyakov loop, and separated in the time direction by $\tau$.
The figure on the right shows the order in which the OPEs must be taken to obtain the sum rules: (1) the bulk OPE $\phi \times \phi \to \sum \Op$, (2) the defect OPE $\Op \times \Pm \to \sum \Oh$.
The dotted circle emphasizes the condition $\tau < |\vec{x}| < \beta$, necessary to be in the appropriate OPE regime.}
}
\label{fig:KMS}
\end{figure}

Our starting point is the KMS condition for two identical scalar operators $\phi (\tau, \vec{x})$. In order to write an equation where both sides can be expanded in OPEs, we additionally consider $\tau \to -\tau$.\footnote{For parity-invariant theories, $\tau \to -\tau$ is a symmetry. In general, we can compose it with a parity action on one spatial direction to preserve the same $\mathbb Z_2$ symmetry \cite{Iliesiu:2018fao}.}
Explicitly,
\begin{equation}
0
=
\vev{\phi (\beta/2+\tau,\vec{x}_1) \phi (0, \vec{x}_2) \Pm}_\beta
-
\vev{\phi (\beta/2-\tau,\vec{x}_1) \phi (0, \vec{x}_2) \Pm}_\beta\,.
\label{eq:KMSCondition}
\end{equation}
We now describe how to use the bulk and defect OPEs to derive sum rules for the one-point function coefficients $\widehat{b}_\Oh$.

We begin by applying the bulk OPE \eqref{eq:BulkOPE} to both terms in \eqref{eq:KMSCondition}.
For the expansion to be convergent, the operators must satisfy $|x_{12}| < \min(|\vec{x}_1|, |\vec{x}_2|) < \beta$.
We obtain:
\begin{multline}\label{eq:KMSBef}
\sum_{\Op \in \phi \times \phi}
\mu_{\phi \phi \Op} \vev{ \Op^{\mu_1 \ldots \mu_J} (\vec{x}_2) \Pm}_\beta
|v_+|^{\Delta - 2 \Delta_\phi - J} v_{+\mu_1} \ldots v_{+\mu_J}
=\\=
\sum_{\Op \in \phi \times \phi}
\mu_{\phi \phi \Op} \vev{ \Op^{\mu_1 \ldots \mu_J} (\vec{x}_2) \Pm}_\beta
|v_-|^{\Delta - 2 \Delta_\phi - J} v_{-\mu_1} \ldots v_{-\mu_J}\,,
\end{multline}
with the vectors $v_\pm$ defined as
\begin{equation}
v_\pm^\mu
:=
(\beta/2 \pm \tau, \vec{x}_{12})\,.
\label{eq:vpm_Definition}
\end{equation}

In the \textit{collinear limit}, where the operators are placed on the same thermal circle, we have:
\begin{equation}
v_\pm^\mu
=
\delta^{0\mu}
\left(
\frac{\beta}{2} \pm \tau
\right)\,.
\label{eq:CollinearLimit}
\end{equation}
This simplifies the expression considerably. The resulting equation can be expanded in $\tau$, and each term must vanish separately:
\begin{equation}
0
=
\sum_{\Op}
\mu_{\phi \phi \Op} \vev{ \Op^{0 \ldots 0} (\vec{x}) \Pm}_\beta
\frac{\beta^\Delta}{2^\Delta} \binom{\Delta - 2 \Delta_{\phi}}{n}\,,
\label{eq:KMS_CollinearLimit}
\end{equation}
with $n \in 2 \mathbb{N} + 1$, since even powers cancel.

We now perform the defect OPE \eqref{eq:DefectOPE} for $\Op^{0 \ldots 0} (\vec{x})$.
Thanks to the collinear configuration, spinning blocks are avoided and only primaries contribute due to time translation invariance.\footnote{Note that in the OPE $\phi \times \phi$ (Equation \eqref{eq:KMSBef}), not only bulk primaries contribute. However, in the collinear setup, descendants are obtained by acting with time derivatives, and their one-point functions vanish by time translation invariance.}
We arrive at the sum rule:
\begin{equation}
0
=
\sum_{\Op, \Oh}
\mu_{\phi \phi \Op} \lambda_{\Op \Oh} \widehat{b}_{\Oh} \binom{\Delta - 2\Delta_\phi}{n} \frac{z^{\Dh - \Delta}}{2^\Delta}\,,
\label{eq:KMS_DefectOPE}
\end{equation}
or, relabeling as $\Bar{\Delta} = \Dh - \Delta$,
\begin{equation}
0
=
\sum_{\Delta, \Db}
\mu_{\phi \phi \Op} \lambda_{\Op \Oh} \widehat{b}_{\Oh} \binom{\Delta - 2\Delta_\phi}{n} \frac{z^{\Db}}{2^\Delta}\,.
\label{eq:BootstrapEquations}
\end{equation}

These equations are more constraining than in the bulk thermal case (e.g. \cite{Marchetto:2023xap}), as they depend on two indices $(\Bar \Delta, n)$ instead of just one.
Despite the added complexity due to spin, this system of equations offers a promising route to constrain the defect data $\widehat{b}_\Oh$.

A few comments:
- These sum rules only explore a subset of constraints, namely those in the collinear configuration.
- The binomial coefficients can take negative values, and there's no positivity requirement on the OPE coefficients—making standard bootstrap techniques more subtle.

\subsection{Heavy operators}
\label{sec:HeavyOperators}

The sum rules above relate light and heavy operators, via the KMS condition.
This mirrors known results at zero temperature, where relations among light operators and heavy ones arise from crossing symmetry \cite{Qiao:2017xif,Fitzpatrick:2012yx,Komargodski:2012ek}, and similar relations also appear at finite temperature \cite{Marchetto:2023xap}.
Here, we generalize these results to the defect setting.

In the collinear limit, we can write:
\begin{equation}
    \vev{\phi(\tau, \vec x) \phi(0, \vec x) \Pm}_\beta
    =
    \sum_{\Op ,\Oh} \mu_{\Op \phi \phi} \lambda_{\Op \Oh} \widehat{b}_{\Oh }
    \frac{\tau^{\Delta-2\Delta_\phi}}{|\vec x|^\Delta } z^{\Dh}\,.
\end{equation}
The identity defect operator $\widehat{\mathds{1}}$ certainly contributes. Projecting to $\Delta = \Dh$ (i.e., $|\vec x|^0$), we isolate:
\begin{equation}
    f_\beta(\tau)=
    \left.
    \vev{\phi(\tau,\vec x) \phi(0,\vec x) \Pm}_\beta
    \right|_{|\vec{x}|^0}
    =
    \sum_{\Delta = \Dh} \mu_{\Op \phi \phi} \lambda_{\Op \Oh} \frac{\tau^{\Delta-2\Delta_\phi}}{\beta^{\Delta}}\,.
\end{equation}

KMS symmetry implies:
\begin{equation}
    f_\beta (\tau)
    =
    f_\beta (\beta - \tau)\,.
\end{equation}

We define the OPE spectral density:
\begin{equation}
    \rho(\widetilde \Delta )
    =
    \sum_{\Delta} a_{\Delta} \delta(\widetilde \Delta-\Delta)\,,
\end{equation}
with $a_\Delta = \mu_{\Op \phi \phi} \lambda_{\Op \Oh}$. Then:
\begin{equation}
    \int_0^\infty d \Delta \, \rho(\Delta) \frac{\tau^{\Delta-2\Delta_\phi}}{\beta^\Delta}
    \sim
    \frac{1}{(\beta-\tau)^{2\Delta_\phi}}\,.
\end{equation}

Using inverse Laplace methods or Tauberian arguments, we find:
\begin{equation}
    \rho(\Delta)
    \overset{\Delta \to \infty}{\sim}
    \frac{\Delta^{2\Delta_\phi-1}}{\Gamma(2\Delta_\phi)}
    \left[1+\Om \left(\frac{1}{\Delta}\right)\right]\,.
    \label{eq:Tauberian1}
\end{equation}

This result is standard in channel duality literature (e.g. \cite{Cardy:1986ie,Fitzpatrick:2012yx}), though not always rigorous.
If $a_\Delta > 0$ at large $\Delta$, then Tauberian theorems ensure \eqref{eq:Tauberian1} holds.

Finally, since the spectrum is discrete, the statement is best understood as:
\begin{equation}
    \int_0^\Delta d \widetilde \Delta\ \rho(\widetilde \Delta)
    \overset{\Delta \to \infty }{\sim}
    \frac{\Delta^{2\Delta_\phi}}{\Gamma(2\Delta_\phi+1)}
    \left[1+\Om \left(\frac{1}{\Delta}\right)\right]\,.
\end{equation}

\section{Defect thermodynamics}
\label{sec:DefectThermodynamics}

We now discuss the thermodynamics of defects at finite temperature.
We determine the free energy and entropy density of the system in terms of OPE data, starting from the one-point function of the stress-energy tensor in the presence of a line defect.
To conclude, we discuss the free energy of moving quarks in gauge theories.

\subsection{Free energy density and entropy}
\label{sec:FreeEnergy}

Among all bulk operators, a special role is played by the stress-energy tensor. It is directly related to standard thermodynamic quantities, such as the free energy and entropy of the system. As a starting point, let us consider the general constraints imposed by spatial rotation symmetry. At zero temperature, its one-point function in the presence of a conformal line defect is fixed up to a single function of the marginal couplings \cite{Kapustin:2005py}. At finite temperature, the one-point function depends on \textit{two} functions of the dimensionless ratio $z$ defined in \eqref{eq:OnePoint_CrossRatio} and the marginal couplings.

By combining $\mathrm{SO}(d-1)$ invariance with the tracelessness condition, we can write the following Ansatz for $d > 2$:
\begin{equation}\label{eq:asz}
    \vev{T^{ij} (\tau,\vec x) \Pm}_\beta
    =
    \frac{f_{1}(z)}{|\vec x|^d}  \delta^{ij}-2 \frac{x^i x^j}{|\vec x|^{d+2}} f_2(z)\,,
\end{equation}
\begin{equation}\label{eq:asz2}
    \vev{T^{00} (\tau,\vec x) \Pm}_\beta
    =
    -\frac{(d-1) f_1(z) -2 f_2(z)}{|\vec x|^d}\,,
    \quad
    \vev{T^{i0}(\tau,\vec x)\Pm}_\beta
    =
    0\,.
\end{equation}
The case $d=2$ will be discussed separately below. There are no further constraints from symmetry alone, but we can expand the stress tensor in the OPE regime. For instance, the time-time component yields:
\begin{equation}
   - (d-1) f_1(z) + 2 f_2(z)
   =
   \sum_{\Oh}
   \lambda_{T \Oh} \,  \bh_\Oh \, z^{\Dh}\,.
   \label{eq:TensorExpansion}
\end{equation}
It is important to notice that $f_1(z)$ and $f_2(z)$ are not independent. The (broken) Ward identities associated with spatial translations imply:
\begin{equation}
    \partial_\mu \langle T^{\mu j}(\tau, \vec x)\Pm \rangle_\beta = \delta^{(d-1)}(\vec x) D^j(\tau)\,,
\end{equation}
which, away from the defect ($|\vec{x}| \ne 0$), leads to a differential equation relating $f_1(z)$ and $f_2(z)$:
\begin{equation}
    -d\, f_1(z) + 4 f_2(z) + z (f_1'(z) - 2 f_2'(z))
    =
    0 \,.
    \label{eq:thesecondforf}
\end{equation}
This equation admits the analytical solution
\begin{equation}
    f_2(z)
    =
    c_1 z^2 + z^2 \int_0^z d y \ \frac{-d\, f_1(y) + y f_1'(y)}{2 y^3}
    \label{eq:f2andf1}
\end{equation}
with $c_1$ an integration constant.

Using \eqref{eq:TensorExpansion} and \eqref{eq:thesecondforf}, we obtain the following expression for $f_1$ and $f_2$:
\begin{equation}
\begin{split}
f_1(z) &= \sum_{\Oh}
   \lambda_{T \Oh} \,  \bh_\Oh \, \frac{2-\Dh}{(d-2)(\Dh-1)}z^{\Dh}\,, \\
f_2(z) &= \sum_{\Oh} \frac{\lambda_{T \Oh} \,  \bh_\Oh}{2} \left(
1+\frac{d-1}{d-2}\frac{2-\Dh}{\Dh-1}\right)z^{\Dh} \,.
\end{split}
\end{equation}
The asymptotic behavior is:
\begin{equation}
    f_1(z) \overset{z \to 0}{\sim}
    \frac{2}{d-2}\lambda_{T\widehat{\mathds{1}}},\quad
    f_2(z) \overset{z \to 0}{\sim}
    -\frac{d/2}{d-2}\lambda_{T\widehat{\mathds{1}}},\quad
    f_1(z) \overset{z \to \infty}{\sim} \frac{b_T}{d}z^d,\quad
    f_2(z) \overset{z \to \infty}{\sim} 0\,,
    \label{eq:BoundaryConditions}
\end{equation}
where $\lambda_{T \widehat{\mathds{1}}}$ is the zero-temperature one-point function of the stress tensor, while $b_T$ is the bulk thermal coefficient in absence of the defect.

The stress tensor determines the energy density of the system:
\begin{equation}
    E(\vec{x})
    =
    \frac{\Em (z)}{|\vec x|^d}  =- \vev{T^{00}(\tau, \vec x) \Pm }_\beta \,.
\end{equation}
Using the thermodynamic identity
\begin{equation}
    F(\vec x)
    =
    E(\vec x)- T S(\vec x)
    =
    E(\vec x)+ T \frac{d F(\vec x)}{d T}\,,
    \label{eq:ThermodynamicEquation}
\end{equation}
and defining $F(\vec{x}) = |\vec{x}|^{-d} \mathfrak f(z)$, this becomes
\begin{equation}\label{eq:thermoequation}
    \mathfrak f(z)
    =
    \Em(z) + z \mathfrak f'(z)\,.
\end{equation}
Its solution is
\begin{equation}
     \mathfrak f(z)
    =
    c_2 z  - z \int_1^z \frac{d y}{y^{2}} \ \Em (y)\,,
    \label{eq:freeneergy}
\end{equation}
and the entropy density reads
\begin{equation}
    S(\vec{x})
    =
    -\frac{d F}{d T}
    =
    -\frac{1}{|\vec{x}|^{d-1}}
    \left(
    c_2-\int_1^z\frac{d y}{y^2}\Em(y)-\frac{\Em(z)}{z}
    \right)\,.
    \label{eq:entropy}
\end{equation}

In terms of CFT data,
\begin{equation} \label{eq:FreeEn}
F(\vec{x})=\frac{1}{|\vec x|^d}\left(c_3 z+ \sum_{\Oh}
    \lambda_{T \Oh} \,  \bh_\Oh \, \frac{1}{\Dh-1}z^{\Dh}\,\right)
\end{equation}
and
\begin{equation}
S(\vec{x})=-\frac{1}{|\vec x|^{d-1}}\left(c_3+ \sum_{\Oh}
   \lambda_{T \Oh} \,  \bh_\Oh \, \frac{\Dh}{\Dh-1}z^{\Dh-1}\right)
   \label{eq:entropydens}
\end{equation}
The constant $c_3$ is fixed by requiring that the entropy vanishes at zero temperature.
This imposes that no defect operators with $\Dh \le 1$ and $\lambda_{T \Oh} \bh_\Oh \neq 0$ contribute.
This condition is verified explicitly in Appendix D of \cite{Barrat:2024aoa}. When satisfied, we can set $c_3 = 0$.

\paragraph{Defect thermodynamics in $d = 2$}

In $d=2$, equations \eqref{eq:asz} and \eqref{eq:asz2} simplify. There is only one symmetric invariant tensor structure:
\begin{equation}
    \langle T^{11}\Pm \rangle_\beta =-\langle T^{00}\Pm \rangle_\beta  =\frac{f(z)}{|x_1|^2}, \quad \langle T^{10}\rangle_\beta = \langle T^{01}\rangle_\beta = 0 \ .
\end{equation}
Stress tensor conservation $\partial_1 \langle T^{11}\Pm \rangle_\beta = 0$ implies
\begin{equation}
    z f'(z) -2 f(z) = 0 \Rightarrow f(z) = c_1 z^2\,.
\end{equation}
Thus, the energy density is constant. Only defect operators with $\widehat{\Delta}=2$ contribute.

Integrating over space, the entropy is
\begin{equation}\label{eq:entropy1}
   S = -2\frac{c_1}{\beta} \text{Vol}(\mathbb R)+ \text{const.} \ .
\end{equation}
This agrees with the result from modular invariance:
\begin{equation}\label{eq:entropy2}
    S = \frac{\pi c}{3 \beta} \text{Vol}(\mathbb R) + \text{const.} \ ,
\end{equation}
where $c$ is the central charge. Comparing \eqref{eq:entropy1} and \eqref{eq:entropy2}, we fix:
\begin{equation}\label{eq:eqc1}
    c_1 =\sum_{\widehat{\Delta}=2}\lambda_{T \widehat{O}}\,\widehat{b}_{\widehat{O}}=- \frac{\pi c}{3} \ .
\end{equation}
This matches the known thermal one-point function of $T$ on the cylinder \cite{Marchetto:2023fcw,Datta:2019jeo} and reflects the fact that only the vacuum module contributes.

\subsection{The free energy of a quark}
\label{sec:FreeEnergyOfAMovingQuark}

In gauge theory, the expectation value of the Polyakov loop is an order parameter for the confinement/deconfinement transition \cite{Polyakov:1975rs}.
Although we do not study this transition here, our setup applies to Polyakov loops.

We define the loop as
\begin{equation} \label{eq: loop1}
    P = \frac{1}{N}\operatorname{Tr}\mathrm P\exp \left(\int_0^\beta d \tau \ A_0 (\tau) \right) \,,
\end{equation}
where $\mathrm P$ denotes path ordering. This quantity requires regularization.

We normalize correlation functions by $\langle P \rangle$:
\begin{equation}
    \langle \mathcal O_1(x_1) \ldots \mathcal O_{n}(x_n) \Pm  \rangle = \frac{\langle \mathcal O_1(x_1) \ldots \mathcal O_{n}(x_n)  P  \rangle}{\langle P\rangle} \,.
\end{equation}
With this convention, $\vev{\Pm}_\beta = \vev{\Pm} = 1$.\footnote{The divergences at finite temperature coincide with those at $T=0$, so $\langle P \rangle$ is finite.}

To match the standard definition of Polyakov loops as measuring the quark free energy, we define instead
\begin{equation}\label{eq:regularization}
    \Pm =  \frac{\operatorname{Tr}\mathrm {P}\exp \left(\int_0^\beta d \tau \ A_0(\tau) \right)}{N\langle  P_\text{free} \rangle_\beta} \,,
\end{equation}
so that
\begin{equation} \label{eq: free en}
    \langle  \Pm\rangle_\beta = e^{-\beta \left(F_\text{quark}-F_\text{free quark}\right)}\,.
\end{equation}
In this definition, the Polyakov loop becomes a well-defined order parameter: finite in the deconfined phase and vanishing in the confining phase.
This definition assumes a Lagrangian description.

In our work, we focus on the deconfined phase of CFTs at infinite volume, and use $\Pm$ to denote the regularized defect insertion.

Finally, the free energy difference is given by
\begin{equation}
    F_\text{quark}-F_\text{free quark} = \int d^{d-1} x \left[F(\vec x)-F_\text{free}(\vec x)-\frac{b_T}{d}\right] \ ,
\end{equation}
where $F(\vec x)$ is determined via the defect OPE as in \eqref{eq:FreeEn}.
This highlights the usefulness of studying line defects at finite temperature via bootstrap techniques.

\section{Examples}
\label{sec:Applications}

We present here applications of the concepts developed in the previous sections to simple models, where computations can be carried out analytically.
The simplest example is the case of generalized free fields.

\subsection{Magnetic lines in generalized free field theory}
\label{subsec:MagneticLinesInGeneralizedFreeFieldTheory}

We begin with the simple case of a magnetic line in Generalized Free Field (GFF) theory, in any spacetime dimension $d$. The thermal defect CFT is defined by the action
\begin{equation}
    S
    =
    S_\text{GFF}
    +
    h \int_0^\beta d \tau\ \phi(\tau,\vec 0)\,,
    \label{eq:DefectCFT_GFF}
\end{equation}
where $\phi$ denotes the fundamental field of the GFF theory and $h$ is the defect coupling.
The defect in \eqref{eq:DefectCFT_GFF} is conformal when $\Delta_\phi = 1$, corresponding to non-local theories in $2 \leq d < 4$, and to a free scalar field in $d = 4$.

Since the theory is free, we can write the exact scalar propagator at finite temperature using the method of images. We represent it in Feynman diagrams using a solid black line:
\begin{align}
    \ScalarPropagator\
    &=
    \vev{\phi(\tau,\vec x) \phi(0,\vec 0)}_\beta \notag \\
    &= \frac{\pi}{2 \beta |\vec x|} \left[\coth\left(\frac{\pi}{\beta}(|\vec x|-i \tau)\right)+\coth\left(\frac{\pi}{\beta}(|\vec x|+i \tau)\right)\right]\,.
    \label{eq:ScalarPropagator}
\end{align}

The defect itself is defined as
\begin{equation}
    \Pm=\frac{1}{\vev{\Pm}_\beta}\exp\left[-h \int_0^\beta d \tau\ \phi(\tau,\vec 0)\right] \,.
\end{equation}

In Feynman diagrams, the defect is represented by a red line, with Feynman rule:
\begin{equation}
    \DefectFeynmanRuleTwo\
    =
    - h \int_0^\beta d \tau\,.
\end{equation}

\subsubsection{One-point functions of bulk operators}
\label{subsubsec:OnePointFunctionsOfBulkOperators}

We now compute the one-point functions of the bulk operators $\phi$ and $\phi^2$ using standard Wick contractions in the free theory.

\paragraph{One-point function of $\phi$.}
The one-point function of $\phi$ is given by a single connected diagram:\footnote{An additional disconnected term, corresponding to the thermal one-point function of $\phi$, vanishes due to the $\mathbb{Z}_2$ symmetry of GFF in absence of the defect.}
\begin{equation}
    \vev{\phi(\tau,\vec{x}) \Pm}_\beta
    =
    \phiGFF\,.
    \label{eq:GFF_phi_Diagram}
\end{equation}

Using \eqref{eq:ScalarPropagator}, this evaluates to
\begin{align}
    \phiGFF
    &=
   - h \int_0^\beta d \tilde \tau\
    \vev{\phi (\tau,\vec x) \phi (\tilde \tau)}_\beta 
    = -\frac{h \pi}{|\vec x|}\,.
    \label{eq:GFF_phi_Result}
\end{align}

This corresponds to the zero-temperature one-point function in presence of the magnetic line.
By comparing with \eqref{eq:OnePoint_DefectOPE}, we find that only the identity defect operator contributes to this correlator.
This is consistent with the fact that, in free theory, the defect OPE of $\phi$ contains only derivatives of $\widehat{\phi}$, whose expectation values vanish.

\paragraph{One-point function of $\phi^2$.}
For $\phi^2$, the computation includes both the connected and disconnected parts:
\begin{equation}
    \vev{\phi^2 (\tau, \vec{x}) \Pm}_\beta
    =
    \phisquaredGFFOne
    +
    \phisquaredGFFTwo\,.
\end{equation}

After normal ordering, the result reads
\begin{equation}
    \vev{\phi^2 (\tau, \vec{x}) \Pm}_\beta
    =
    \frac{1}{|\vec{x}|^2}\left(
    h^2 \pi^2
    +
    \frac{\pi^2}{6}z^2\right)\,,
    \label{eq:GFF_phisquared_Result}
\end{equation}
where the second term corresponds to the thermal one-point function of $\phi^2$.

This implies that both the defect identity and the operator $\widehat{\phi}^2$ (with $\Dh = 2$) contribute.
Symmetry forbids a contribution from the displacement operator $D^i = \partial^i \phi$, confirming that the thermal piece is due to $\widehat{\phi}^2$.

\subsubsection{The two-point function $\vev{\phi \phi \Pm}_\beta$}
\label{subsubsec:TheTwoPointFunctionphiphi}

The two-point function of scalar fields in presence of the defect reads:
\begin{align}
    \vev{\phi (\tau_1, \vec{x}_1) \phi (\tau_2, \vec{x}_2) \Pm}_\beta
    &=
    \vev{\phi (\tau_1, \vec{x}_1) \phi (\tau_2, \vec{x}_2)}_\beta
    +
    \vev{\phi (\tau_1, \vec{x}_1) \Pm}_\beta
    \vev{\phi (\tau_2, \vec{x}_2) \Pm}_\beta\,.
\end{align}

In the collinear limit ($|\vec x_1| = |\vec x_2|$), the thermal two-point function becomes
\begin{equation}
    \vev{\phi (\tau, \vec{x}) \phi (0, \vec{x})}_\beta
    =
    \frac{\pi^2}{\beta^2} \csc^2
    \left(
    \frac{\pi \tau}{\beta}
    \right)
    = \frac{2}{\beta^2} \sum_{k=0}^\infty  (2k-1) \zeta_{2k}\left(\frac{\tau}{\beta}\right)^{2k-2}\,.
    \label{eq:GFF_Bulk2Point}
\end{equation}

This structure matches the form expected from the OPE expansion. Only operators satisfying $\Delta = \Dh$ contribute to thermal corrections, and parity ensures that only even powers survive.
It is straightforward to check that the sum rules \eqref{eq:KMS_DefectOPE} are satisfied.

\subsubsection{Free energy and entropy}
\label{subsubsec:FreeEnergyAndEntropy}

We now compute the one-point function of the stress-energy tensor in the $d=4$ free scalar theory.

The result is
\begin{equation}
    \vev{T^{00}\Pm }_\beta
    =
    - \frac{1}{|\vec x|^4}
    \left(
    \frac{h^2 \pi^2}{2} + \frac{2 \pi^2}{15 }z^4
    \right)\,.
\end{equation}

The corresponding free energy is given by
\begin{equation}
    F(\vec{x})
    =
    \frac{1}{|\vec{x}|^{4}}
    \left(
    c_2 z + \frac{\pi^2}{90} (1-z) (45 h^2 + 4 z (1+z+z^2))
    \right)\,,
\end{equation}
and the entropy density reads
\begin{equation}
    S(\vec{x})
    =
    \frac{s(\vec{x})}{|\vec{x}|^3} = \frac{\pi ^2 \left(45 h^2+16 z^3-4\right)-90 c_2}{90 |\vec{x}|^3}\,.
\end{equation}

To ensure the entropy vanishes at $T=0$ ($z \to 0$), we fix
\begin{equation}
    c_2 = \frac{\pi^2}{90}(45 h^2 - 4)\,.
\end{equation}

Substituting this back, we find
\begin{equation}
    F(\vec{x})
    =
    \frac{\pi ^2 \left(45 h^2-4 z^4\right)}{90 |\vec{x}|^4}\,,
    \quad
    S(\vec{x})
    =
    \frac{8 \pi ^2 T^3}{45}\,.
\end{equation}

In this free theory, the entropy density is independent of the defect, though this is not expected to hold in interacting theories. As expected, the free energy diverges near the defect core ($|\vec{x}|\to 0$) and asymptotes to the free energy of the bulk theory at large $z$.

\section{Summary of the chapter}

In this chapter, we extended the physical setup discussed previously by adding a conformal line defect to the thermal CFT. These defects preserve a specific subgroup of the original conformal group, including in particular the conformal symmetry along the line. Physically, this implies that the theory living on the defect is a one-dimensional conformal field theory (1D CFT). We considered the defect placed along the (imaginary) time direction, such that, after thermal compactification, it becomes a circular defect wrapping the thermal circle.

We analyzed the pattern of broken and unbroken symmetries and discussed their consequences for correlation functions. In particular, we showed that the data needed to fully characterize the thermal defect CFT includes new one-point functions of operators localized on the defect. These are analogous to thermal one-point functions, but originate from a distinct mechanism: they arise because a circular defect wrapping the thermal circle cannot be mapped to a straight line via a conformal transformation.

We expressed bulk one-point functions in terms of this new defect-CFT data, and from the stress-energy tensor one-point function, we derived expressions for the local energy density, free energy density, and entropy density of the system. We also discussed the special case of gauge theories, where the defect can be identified with a Polyakov loop, and its expectation value serves as an order parameter for confinement.

Moving to bulk two-point functions in the presence of the defect, we formulated a bootstrap problem in analogy with the one proposed in Chapter~\ref{ref:dynamics}. The sum rules derived in this context follow a similar derivation and structure to those in the defect-free case, with the KMS condition again playing a central role. However, providing a rigorous derivation of the heavy-operator asymptotics is more challenging in this setup, since the assumptions required for invoking Tauberian theorems are more difficult to verify.

We concluded the chapter with explicit examples in (generalized) free field theories, where all relevant quantities can be computed analytically. More involved examples, such as magnetic lines in the $\mathrm{O}(N)$ model at large $N$ and in the $\varepsilon$-expansion, will be presented in Chapter~\ref{chap:ONmodel}.

\chapter{The $\mathrm O(N)$ model at finite temperature}\label{chap:ONmodel}
\section{A primer on the \texorpdfstring{$\mathrm O(N)$}{O(N)} model}

We have already discussed the 3D Ising model and, more generally, the $\mathrm{O}(N)$ models throughout this thesis. Before turning to finite temperature effects and applying the tools developed in previous chapters, we review the zero-temperature features of these theories.

In the Landau paradigm \cite{Landau:1937obd}, based on the concept of universality, strongly coupled conformal models can often be described effectively by weakly coupled field theories, provided they share the same symmetries. From a Wilsonian point of view, this connection is made precise by the existence of renormalization group (RG) flows from weakly coupled UV fixed points (often free theories) to the IR CFTs of interest. Among all universality classes, the Ising model and its simplest generalization—the $\mathrm{O}(N)$ model—play a particularly central role.

In fact, the $N=1$ case describes the critical behavior of spin chains or lattice models governed by the Hamiltonian
\begin{equation}
    \op H = -J \sum_{\langle i,j\rangle} \op \sigma_i \op \sigma_j - h \sum_i \op \sigma_i \,,
\end{equation}
where $\langle i,j\rangle$ denotes a sum over nearest-neighbor spin sites, and $\sigma_i = \pm 1$ is a spin variable. This universality class appears in a wide range of physical systems, and its critical exponents have been measured experimentally. Theoretically, the model exhibits no phase transition in $d=1$ \cite{Ising1925BeitragZT}, while in $d=2$ it was famously solved by Onsager \cite{Onsager:1943jn}. Its critical point corresponds to the minimal model $\mathcal{M}(3,4)$, which is the simplest unitary two-dimensional CFT \cite{Belavin:1984vu}. In $d=3$, the model is strongly coupled, and precise information on critical exponents has been obtained only via Monte Carlo simulations and modern numerical bootstrap methods.

The generalizations to $N=2$ and $N=3$ are also of great interest: the former describes the $\lambda$-transition in liquid helium and phase transitions in binary fluid mixtures, while the latter is relevant for statistical Heisenberg models \cite{Pelissetto:2000ek}. The $N=4$ model is noteworthy due to its connection with QED in three dimensions at its IR fixed point \cite{Chester:2024waw}, making it a basic theoretical framework for studying confinement. Motivated by the ubiquity and importance of these models, we now study their behavior using the Landau paradigm.

To identify the Landau-Ginzburg description of a theory in the $\mathrm{O}(N)$ universality class, one writes the simplest $\mathrm{O}(N)$-invariant Lagrangian:
\begin{equation}
    \mathcal{L} = \frac{1}{2}(\partial \phi_i)^2 + \frac{\lambda}{4!} (\phi_i \phi_i)^2 \,.
\end{equation}
A standard perturbative analysis shows that the operator $(\phi_i \phi_i)^2$ is relevant for $d<4$, marginally irrelevant in $d=4$, and irrelevant for $d>4$. Therefore, the theory flows to a nontrivial IR fixed point in dimensions $d < 4$. The fixed point is weakly coupled near $d = 4$, making it accessible via the $\varepsilon$-expansion \cite{Wilson:1971dc}. At leading orders in $\varepsilon = 4-d$, the fixed-point coupling is
\begin{equation}
    \frac{\lambda}{(4 \pi)^2} = \frac{3}{N+8} \varepsilon + \frac{9(3N+14)}{(N+8)^3} \varepsilon^2 + \mathcal{O}(\varepsilon^3) \,.
\end{equation}
Perturbative calculations allow one to extract a great deal of information about the CFT data, including scaling dimensions and OPE coefficients. However, extrapolating these results reliably to the physical dimensions $d=2$ and $d=3$ remains challenging. Fortunately, recent developments in the conformal bootstrap have yielded remarkably precise determinations of physical quantities in these models, matching experimental and Monte Carlo results. A comparison of estimates for the critical exponents $\eta$ and $\nu$ in the three-dimensional Ising model is shown in Table~\ref{tab:critical-exponents}. For more comprehensive reviews, we refer to \cite{Henriksson:2022rnm, Pelissetto:2000ek}.

\begin{table}[h]
    \centering
    \renewcommand{\arraystretch}{1.2}
    \begin{tabular}{|l|c|c|l|}
        \hline
        \textbf{Method} & \(\eta\) & \(\nu\) & \textbf{Reference} \\
        \hline
        Conformal Bootstrap (2024) & 0.036297612(48)  & 0.62997097(12) & \cite{Chang:2024whx} \\
        Monte Carlo (2021) & 0.036284(40) & 0.62998(5) & \cite{Hasenbusch:2021tei} \\
        $\varepsilon$-expansion (2020) & 0.03653(65) & 0.62977(22) & \cite{Shalaby:2020xvv} \\
        Fuzzy Sphere (2020) & 0.036298(20) & 0.629971(4) & \cite{Zhu:2023} \\
        Experiment (1988) & 0.032(13) & 0.629(3) & \cite{Lipa:1996} \\
        Experiment (2004) & 0.041(5) & 0.632(2) & \cite{Ahlers:1983} \\
        \hline
    \end{tabular}
    \caption{Estimates and measurements of the critical exponents \(\eta\) and \(\nu\) in the three-dimensional Ising model from various theoretical and experimental approaches.}
    \label{tab:critical-exponents}
\end{table}

Having introduced the $\mathrm{O}(N)$ models and their physical significance, we now focus on their zero-temperature CFT data and features, which serve as the basis for our study of finite temperature effects. We begin with the simplest tractable limit—the large-$N$ $\mathrm{O}(N)$ model—which is analytically accessible even at finite temperature. We then proceed to investigate the $3d$ $\mathrm{O}(1)$, $\mathrm{O}(2)$, and $\mathrm{O}(3)$ models, corresponding to the Ising, XY, and Heisenberg universality classes, all of which are experimentally relevant. Finally, we explore the evolution of the free energy in the $\varepsilon$-expansion in the range $3 \leq d < 4$.

\section{Large \texorpdfstring{$N$}{N} analysis}

The $\mathrm O(N)$ model simplifies drastically in the limit $N \to \infty$, where exact results can be extracted using the Hubbard–Stratonovich transformation, yielding the effective Lagrangian
\begin{equation}
    \mathcal L = \frac{1}{2}(\partial \phi_i)^2+ \frac{1}{2}\sigma \phi_i \phi_i \,.
\end{equation}
In this formulation, the momentum-space propagator takes the form
\begin{equation}
    G_{ij}(\omega_n, \vec k) = \frac{\delta_{ij}}{\omega_n^2+\vec k^2+m_{\text{th}}^2} \,,
\end{equation}
where $m_{\text{th}}^2 = \langle \sigma \rangle_{\beta}$ is the thermal mass. Fourier transforming to position space, we obtain the two-point function as a sum over free massive propagators in $d$ dimensions:
\begin{equation}\label{eq:2ptlargeN}
   \langle \phi_i(\vec x,\tau) \phi_j(0,0) \rangle_\beta  = \delta_{ij}\left(\frac{m_{\text{th}}}{2 \pi}\right)^{d-2}\sum_{m \in \mathbb Z}  \frac{ \mathrm K_{(d-2)/2}(m_{\text{th}}\sqrt{x^2+(\tau+m\beta)^2})}{ \left(m_{\text{th}}\sqrt{x^2+(\tau+m\beta)^2}\right)^{(d-2)/2}}\,,
\end{equation}
where $\mathrm K$ is the modified Bessel function of the second kind.

\begin{figure*}[t]
\centering
\begin{subfigure}[t]{.49\textwidth}
   \includegraphics[width=\textwidth]{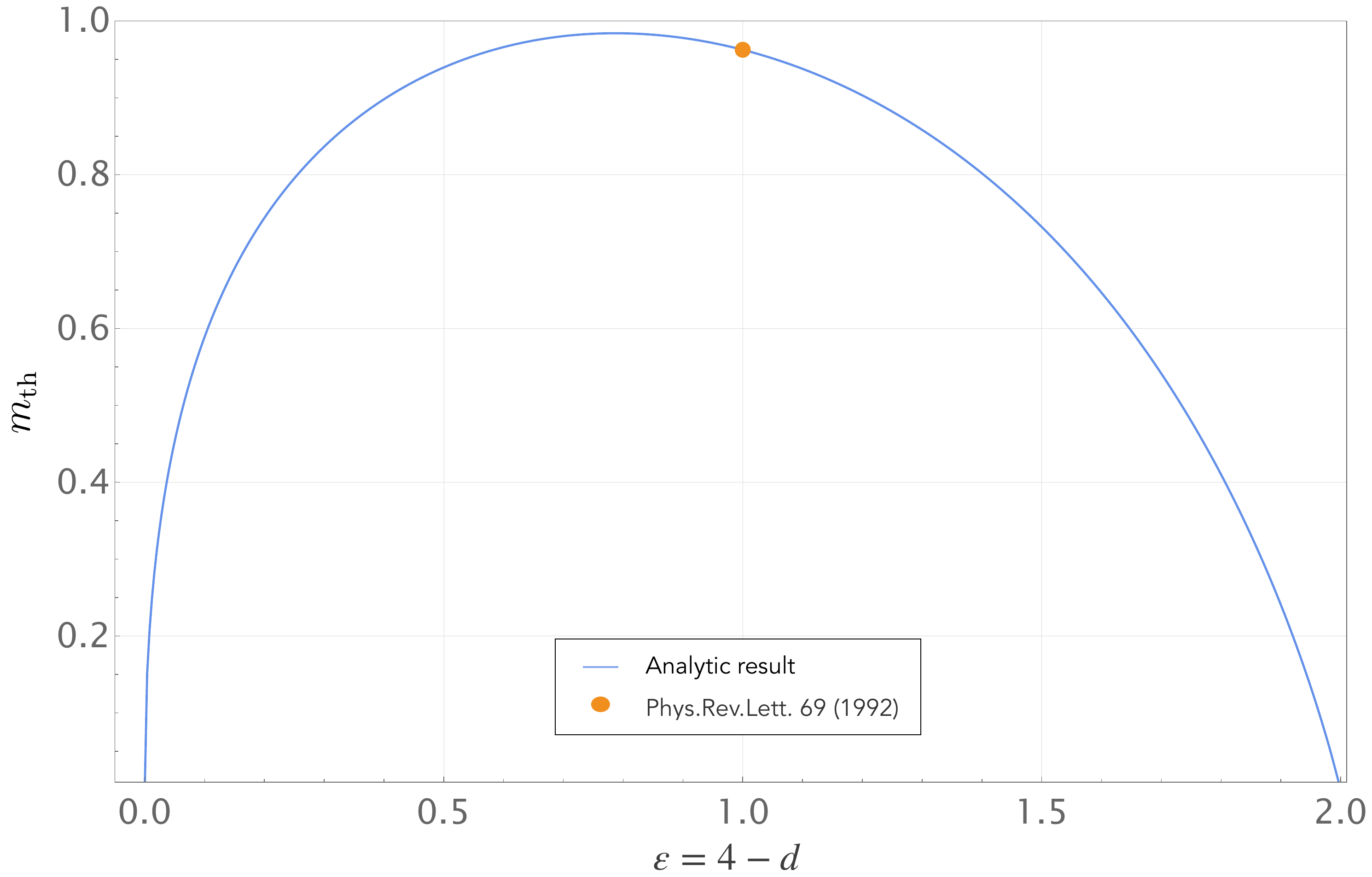}
  \caption{Thermal mass $m_\text{th}$ vs.\ dimension $d$}
  \label{fig:thermalmassesLargeN}
\end{subfigure}%
\hfill
\begin{subfigure}[t]{.49\textwidth}
    \includegraphics[width=\textwidth]{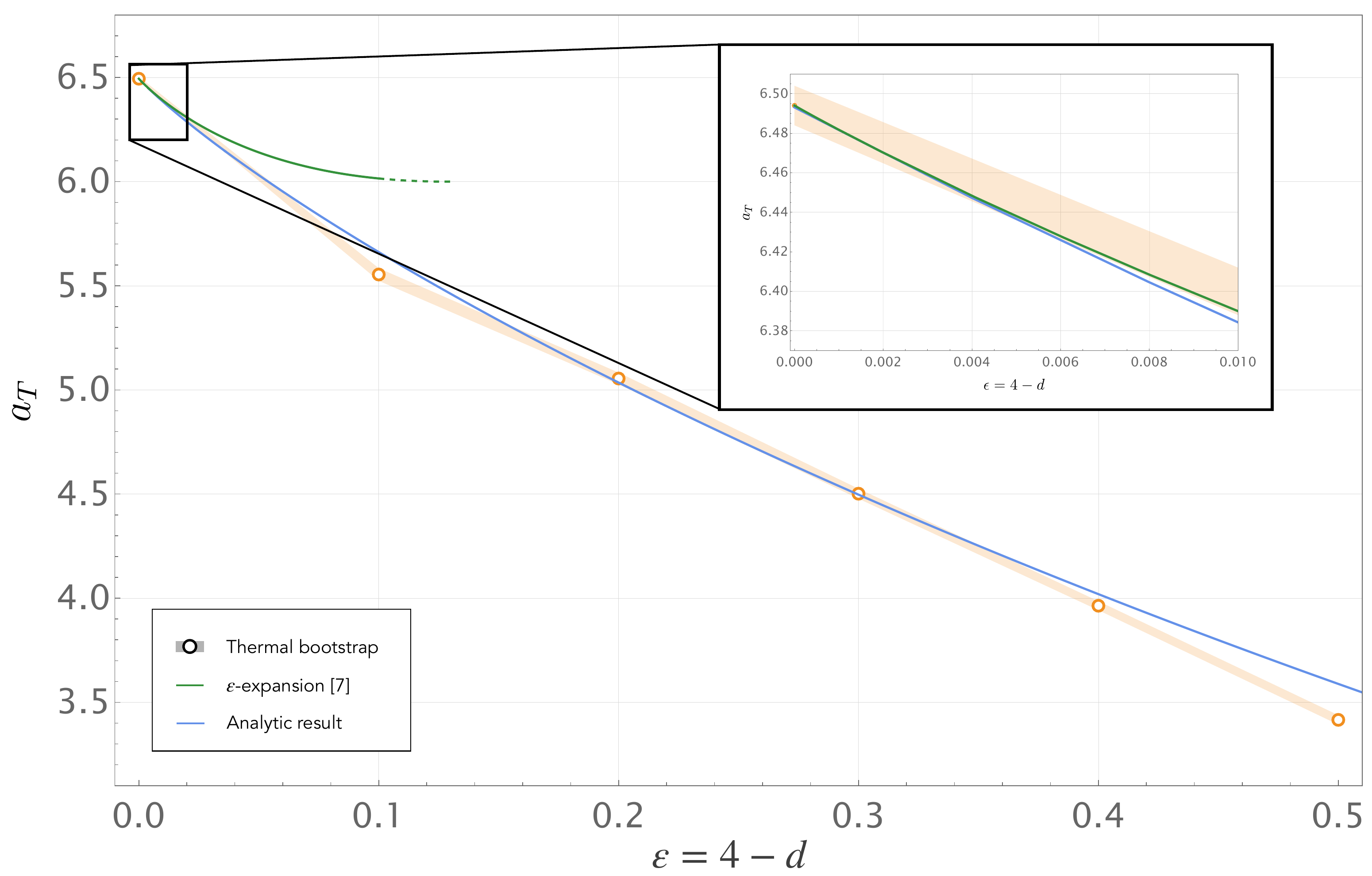}
    \caption{Thermal OPE coefficient $a_T$ vs.\ $d$}
    \label{fig:atlargeN}
\end{subfigure}
  \caption{\textbf{Left (a):} Thermal mass as a function of the spacetime dimension. \textbf{Right (b):} Stress-energy tensor OPE coefficient in the two-point function of fundamental scalars in the $\mathrm O(N)$ model at large $N$. Numerical results are compared with $\varepsilon$-expansion and exact values.}
  \label{Fig:LargeN}
\end{figure*}

To fully determine the two-point function, we must compute the thermal mass. This is obtained by evaluating the saddle point of the effective action,
\begin{equation}
    Z = \int \, D \sigma \ e^{-\frac{N}{2}\text{Tr}\log \left(\Box+\sigma\right)} \,,
\end{equation}
which in flat space yields $\sigma = 0$, but on $S^1_\beta \times \mathbb{R}^{d-1}$ results in a nonzero constant saddle. Imposing the saddle point equation
\begin{equation}
    \frac{\partial}{\partial \sigma} \text{Tr}\log \left(\Box+\sigma\right) = \sum_{n = -\infty}^{\infty} \int \frac{d^{d-1}p}{(2 \pi)^{d-1}} \frac{1}{\omega_n^2 +\vec p^2+\sigma} = 0 \,,
\end{equation}
ensures that $\phi^2$ does not contribute to the two-point function of fundamental scalars at leading order in $1/N$. This sum can be computed analytically in $d = 3$, yielding \cite{Sachdev:1992py,Iliesiu:2018fao,Petkou:2018ynm}
\begin{equation}\label{eq:TM}
    -m_{\text{th}}^{(3d)} = 2 \log \left(1-e^{-m_{\text{th}}^{(3d)}}\right)  \hspace{1 cm}\Rightarrow \hspace{1 cm} m_{\text{th}}^{(3d)} = 2 \log \left(\frac{1+\sqrt 5}{2}\right) \,.
\end{equation}
For $d \neq 3$, the sum can be evaluated numerically to high precision, as shown in Fig.~\ref{fig:thermalmassesLargeN}.

Once $m_\text{th}$ is known, the thermal OPE coefficients can be extracted from the small $\tau$ expansion of the two-point function. In particular, the OPE coefficient for the stress-energy tensor in $d = 3$ is given by \cite{Sachdev:1992py,Iliesiu:2018fao}
\begin{equation}
    a_T^{(3d)} = \frac{8}{5} \zeta(3) \,.
\end{equation}
The numerical results in Fig.~\ref{fig:atlargeN} are in excellent agreement with the analytical predictions derived from the thermal mass data in Fig.~\ref{fig:thermalmassesLargeN}. As expected, the $\varepsilon$-expansion accurately approximates the result in the vicinity of $d=4$, as shown in the inset.

Beyond the expansion, observe that the correlator \eqref{eq:2ptlargeN} does not simplify in general for arbitrary $\tau$ and $r$, nor for arbitrary $d$. However, in the case $r = 0$ and $d = 3$, the summation can be carried out explicitly, giving the closed-form expression
\begin{equation}\label{eq:LargeN2p}
    \langle \phi_i(\tau,0)\phi_j(0,0)\rangle_\beta = \delta_{ij} \left[
    e^{\tau -1} \Phi \left(\frac{1}{e},1,1-\tau \right)
    +e^{-m_\text{th}  \tau } \frac{\Gamma (\tau )}{\Gamma(1+\tau)} \,
   {_2F_1}\left(1,\tau ;\tau +1;e^{-m_\text{th} }\right)\right] \,,
\end{equation}
where $m_\text{th}$ is the $3d$ thermal mass, $\Phi$ is the Hurwitz–Lerch function,\footnote{Defined as $\Phi(z,s,a) = \sum_{k = 0}^\infty z^k (k+a)^{-s}$.} and ${_2F_1}$ is the ordinary hypergeometric function. We have set $\beta = 1$, but the temperature can easily be restored via dimensional analysis.

\begin{figure*}[t]
\centering
\begin{subfigure}[t]{.48\textwidth}
   \includegraphics[width=\textwidth]{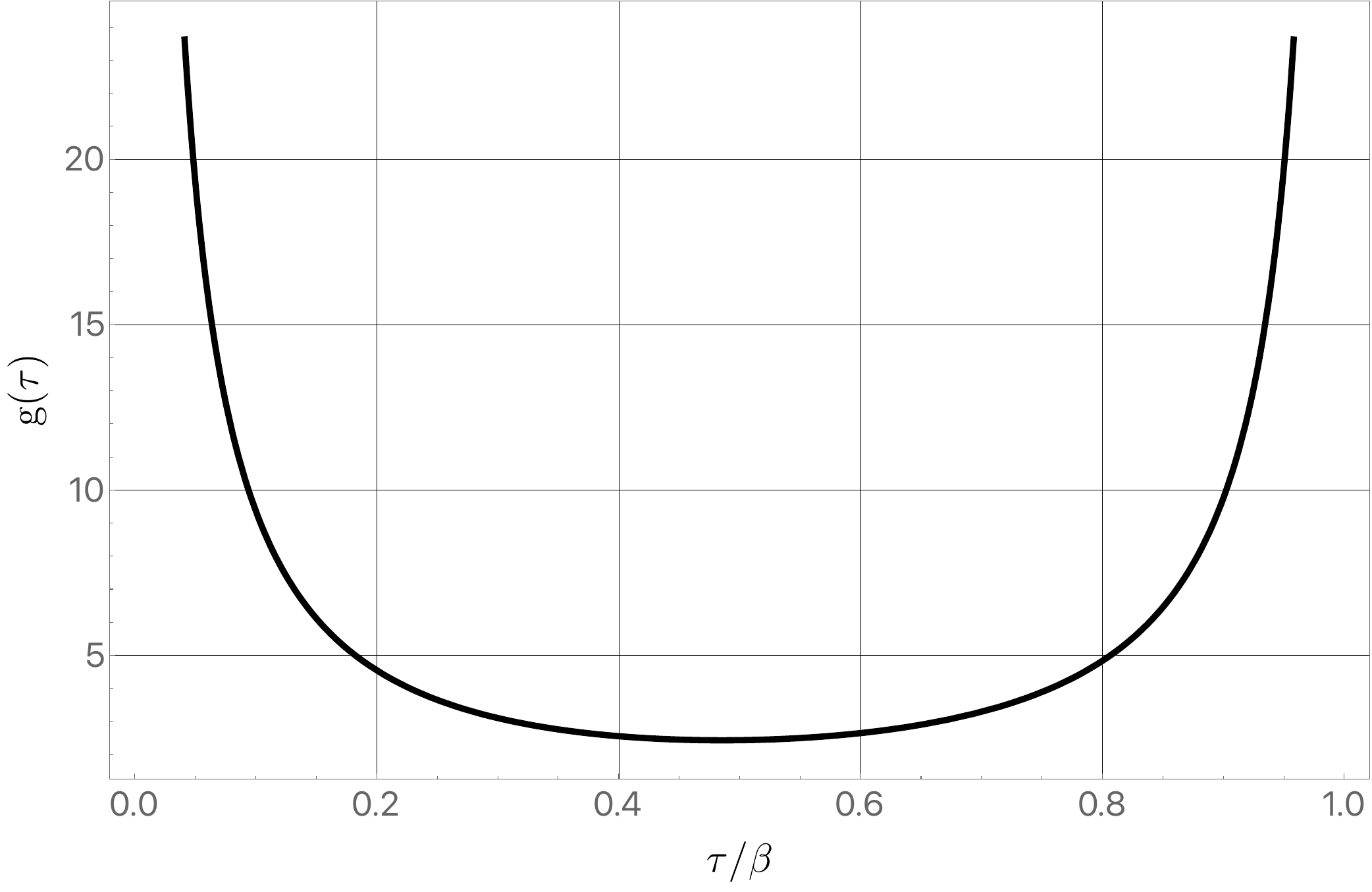}
  \caption{Two-point function \eqref{eq:LargeN2p} vs.\ $\tau$}
\end{subfigure}%
\hfill
\begin{subfigure}[t]{.49\textwidth}
    \includegraphics[width=\textwidth]{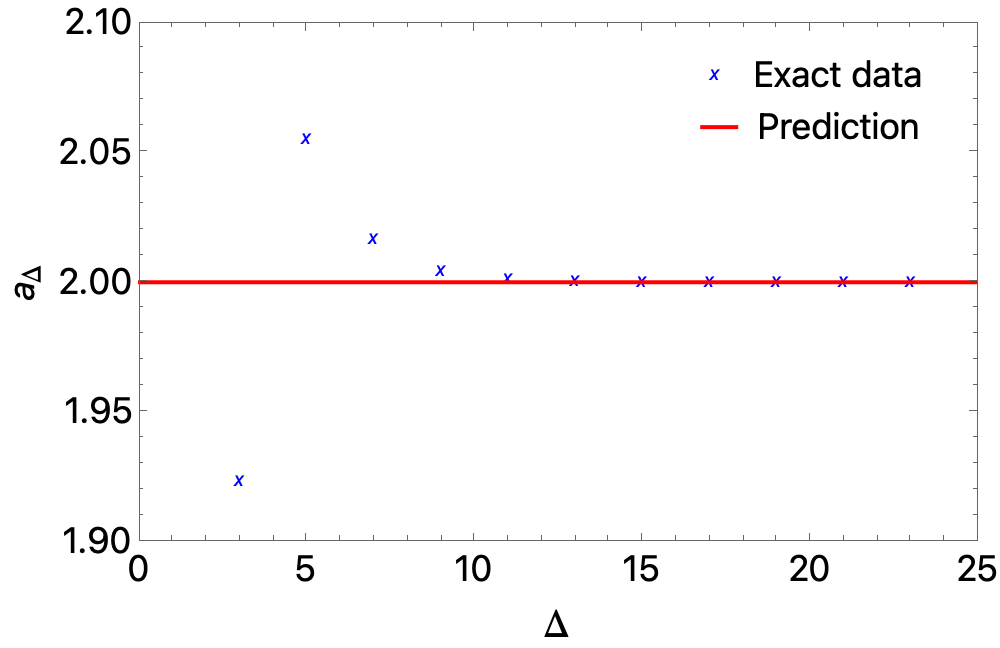}
    \caption{Thermal OPE data vs.\ Tauberian prediction}
\end{subfigure}
  \caption{\textbf{Left (a):} Two-point function $\langle \phi_i(\tau,0)\phi_j(0,0)\rangle_\beta$ in the large-$N$ $\mathrm{O}(N)$ model. \textbf{Right (b):} Thermal OPE coefficients $a_\Delta$ compared with Tauberian predictions at large scaling dimensions.}
  \label{Fig:LargeNtp}
\end{figure*}

Figure~\ref{Fig:LargeNtp} presents the exact thermal two-point function at $r = 0$ in $d = 3$, along with a comparison between the extracted thermal OPE coefficients and the asymptotic prediction obtained via Tauberian theorems. The agreement at large $\Delta$ confirms the consistency of our results with the general bootstrap expectations developed in earlier chapters.

\section{Ising model in \texorpdfstring{$3d$}{3d}}

The opposite limit of the $N \to \infty$ case is $N = 1$. The $\mathrm{O}(1)$ model is the Ising model, which in $2$ and $3$ dimensions describes a variety of physical phenomena, as already discussed in this thesis. Thanks to recent advances in the conformal bootstrap program, we now have access to certain key physical quantities at zero temperature with extremely high precision. We summarize relevant operator data in Table~\ref{tab:3dIsingBotstrap}.

\begin{table}[h]
\begin{center}
\caption{Some operators in the $3d$ Ising model, their classical (Ginzburg–Landau) descriptions ($\mathcal O_{GL}$), and their conformal dimensions obtained via conformal bootstrap. Only the operators relevant for this thesis are listed.}
\renewcommand{\arraystretch}{1.25}
\begin{tabular}{  |c| c | c|}
\hline 
$\mathcal O$ & $\mathcal O_{GL}$ & $\Delta_{\mathcal{O}}$ \cite{Kos:2016ysd,Reehorst:2021hmp,Chang:2024whx} \\  \hline 
$\sigma$ & $\phi$ & 0.518148806(24)  \\ \hline 
$\epsilon$ & $\phi^2$ &1.41262528(29)  \\ \hline 
$T_{\mu \nu}$ & $T_{\mu \nu}$ & 3 \\ \hline 
$\epsilon'$ & $\phi^4$ & 3.82951(61) \\ \hline 
$C_{\mu \nu \rho \sigma}'$ & $\phi \partial^\mu \partial^\nu \partial^\rho \partial^\sigma \phi$ & 5.022665(28)   \\ \hline 
$T_{\mu \nu}'$ & $\phi \partial^\mu \partial^\nu \Box \phi$ & 5.50915(44)  \\ \hline 
\end{tabular}
\label{tab:3dIsingBotstrap}
\end{center}
\end{table}

To numerically determine thermal one-point functions (i.e., thermal OPE coefficients), it is sufficient to know the conformal dimensions of the operators appearing in the relevant OPE. However, to compute the free energy, we also need the central charge. The most recent estimate at the time of writing is given in \cite{Chang:2024whx}:
\begin{equation}
    \frac{c_T}{c_{T\, \text{free}}} = 0.946538675(42) \,.
\end{equation}

With this zero-temperature data, we can bootstrap thermal OPE coefficients entering the expansion of the thermal two-point function
\begin{equation}
    \langle \sigma(\tau)\sigma(0)\rangle_\beta
    = \sum_{\mathcal{O} \in \sigma \times \sigma} a_{\mathcal{O}} \, f_{\mathcal{O}}(\tau, \beta)
    = [\mathds{1}] + [\epsilon] + [T_{\mu \nu}] + [\epsilon'] + \ldots
\end{equation}
For the heavy sector, a good approximation is to include only double-twist operators of the form $[\sigma \sigma]_{n,J}$, which encompass operators like $C_{\mu \nu \rho \sigma}'$, $T_{\mu \nu}'$, and others.

\begin{figure*}[t]
\centering
\begin{subfigure}[t]{.48\textwidth}
   \vspace*{-6.5cm}
   \includegraphics[width=\textwidth]{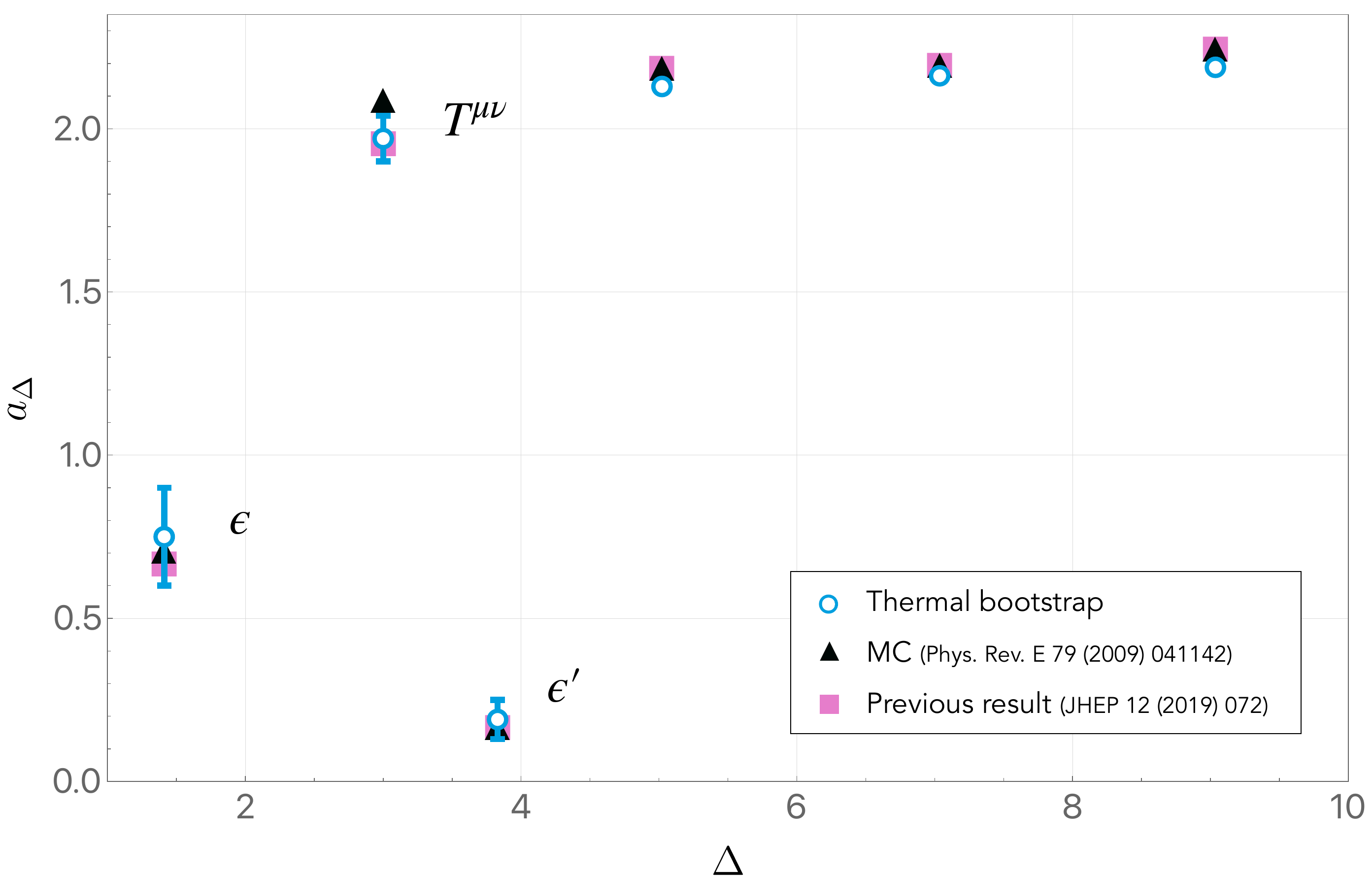}
  \caption{Thermal OPE data for $\langle \sigma\sigma\rangle_\beta$ compared to \cite{Iliesiu:2018zlz} and Monte Carlo results \cite{PhysRevE.79.041142,PhysRevE.53.4414}.}
\end{subfigure}%
\hfill
\begin{subfigure}[t]{.49\textwidth}
    \includegraphics[width=\textwidth]{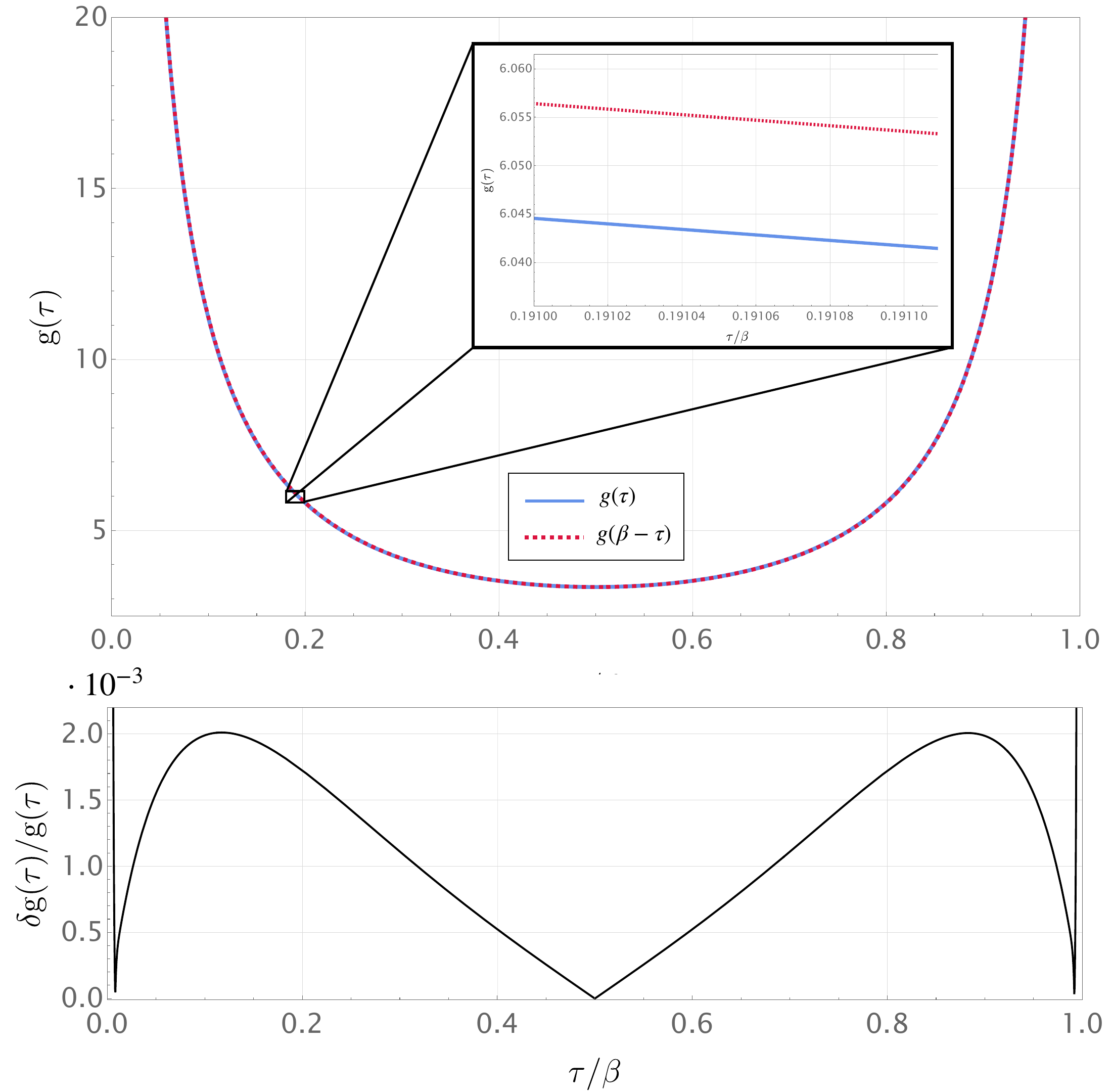}
    \caption{Comparison of $\langle \sigma(\tau,0)\sigma(0,0)\rangle_\beta$ and $\langle \sigma(\beta-\tau,0)\sigma(0,0)\rangle_\beta$.}
\end{subfigure}
\caption{\textbf{Left (a):} Thermal OPE coefficients. \textbf{Right (b):} KMS condition check: the relative discrepancy is $\sim 10^{-3}$.}
\label{Fig:IsingResults}
\end{figure*}

In Fig.~\ref{Fig:IsingResults}, we plot the thermal OPE coefficients and compare them to previous bootstrap \cite{Iliesiu:2018zlz} and Monte Carlo results \cite{PhysRevE.79.041142,PhysRevE.53.4414}. With knowledge of the light spectrum and an approximation for the heavy sector, we reconstruct the thermal two-point function in the OPE regime. We check KMS symmetry by comparing $\langle \sigma(\tau,0)\sigma(0,0)\rangle_\beta$ to $\langle \sigma(\beta-\tau,0)\sigma(0,0)\rangle_\beta$, finding a relative discrepancy of order $10^{-3}$—a strong indication of numerical convergence.

A more quantitative comparison of OPE coefficients is given in Table~\ref{tab:3dIsingResults}.

\begin{table}[h]
\begin{center}
 \renewcommand{\arraystretch}{1.25}
\begin{tabular}{ | c|| c| c| c|}
\hline 
$\mathcal O$ & This work (\cite{Barrat:2024fwq}) & MC \cite{PhysRevE.79.041142,PhysRevE.53.4414} & PR \cite{Iliesiu:2018zlz} \\  \hline 
$\epsilon$ & 0.75(15) & 0.711(3) & 0.672(74) \\ \hline 
$T_{\mu \nu}$ & 1.97(7) & 2.092(13) & 1.96(2) \\ \hline 
$\epsilon'$ & 0.19(6) & 0.17(2) & 0.17(2) \\ \hline 
\end{tabular}
\caption{Thermal OPE coefficients $a_\Delta$ for light operators in the $3d$ Ising model, compared with Monte Carlo (MC) and previous results (PR). The value of the Tauberian correction is $c_1 \sim -0.065$, with negligible error.}
\label{tab:3dIsingResults}
\end{center}
\end{table}

All results are consistent with previous findings, with the exception of the stress-energy tensor coefficient, which shows a mild discrepancy with Monte Carlo results.

Using the relation \eqref{eq:freeEnergy}, we can convert $a_T$ into a free energy density estimate. Our results, compared with MC values, are:
\begin{equation}
    f_{N=2} = -\frac{0.275(12)}{\beta^3} \ , \qquad
    f_{N=3} = -\frac{0.393(17)}{\beta^3} \,.
\end{equation}
Although the two values do not perfectly match, they are within $2\sigma$ agreement.\footnote{After our publication, we were informed of new Monte Carlo results from an independent group \cite{NewLattice}. It would be interesting to clarify the origin of the discrepancy and compare additional observables.}

\paragraph{Comparison with large spin perturbation theory:}

An alternative approach to the $3d$ Ising thermal bootstrap was proposed in \cite{Iliesiu:2018zlz}, based on large spin perturbation theory. The authors also imposed the KMS condition on $\langle \sigma\sigma\rangle_\beta$, but took a different route: instead of truncating the OPE, they modeled the heavy tail using only double-twist operators $[\sigma \sigma]_{0,J}$ and $[\sigma \sigma]_{1,J}$, which dominate the asymptotics.

Using the inversion formula, one can show that the contribution of light operators such as $\mathds{1}$, $\epsilon$, and $T_{\mu \nu}$ generates poles at $\Delta = 2\Delta_\phi + J + n$, corresponding to the classical dimensions of $[\sigma \sigma]_{n,J}$. This allows one to express the heavy operator coefficients in terms of the light data:
\begin{equation}
    a_{[\sigma \sigma]_{0,J}} = a_{[\sigma \sigma]_{0,J}}(a_\epsilon, a_T) \ , \qquad
    a_{[\sigma \sigma]_{1,J}} = a_{[\sigma \sigma]_{1,J}}(a_\epsilon, a_T) \ ,
\end{equation}
and hence the entire correlator becomes a function of just two parameters:
\begin{equation}
    g(r,\tau) = g(r,\tau)(a_\epsilon, a_T) \,.
\end{equation}
One can then numerically optimize KMS to fix $a_\epsilon$ and $a_T$. For a full implementation, see \cite{Iliesiu:2018zlz}.

In contrast, our approach uses a Tauberian tail with open parameters $(c_1, \ldots, c_n)$ that are not explicitly tied to the light operator data. This makes the method more flexible: the Tauberian tail can incorporate multiple subleading trajectories, and the cost function $\eta$ is universal across theories, depending only on known CFT data. On the downside, our method is currently limited to collinear configurations ($\vec x = 0$), while \cite{Iliesiu:2018zlz} has access to full spatial dependence.

\section{XY and Heisenberg models in \texorpdfstring{$3d$}{3d}}

Other crucial models for simulations and experimental applications are the $\mathrm{O}(2)$ and $\mathrm{O}(3)$ models. The case $N = 2$ describes the $\lambda$-line in helium, phase transitions in XY ferromagnets, and the statistical XY model. The $N = 3$ model captures isotropic magnets and the statistical Heisenberg model \cite{Pelissetto:2000ek}.

\begin{table}[h]
\begin{center}
\caption{Conformal dimensions of selected operators in the $3d$ $\mathrm{O}(2)$ and $\mathrm{O}(3)$ models, together with their classical (Ginzburg–Landau) counterparts. Only operators relevant for this thesis are listed.}
\renewcommand{\arraystretch}{1.25}
\begin{tabular}{|c|c|c|}
\hline
$\mathcal{O}_{GL}$ & $\Delta_{\mathcal{O}}$ in $N = 2$ \cite{Kos:2016ysd,Chester:2019ifh,Liu:2020tpf} & $\Delta_{\mathcal{O}}$ in $N = 3$ \cite{Kos:2016ysd,Chester:2020iyt} \\
\hline
$\phi_i$ & 0.519088(22) & 0.518942(51) \\ \hline
$\phi_i \phi_i$ & 1.51136(22) & 1.59489(59) \\ \hline
$T_{\mu \nu}$ & 3 & 3 \\ \hline
$(\phi_i \phi_i)^2$ & 3.794(8) & 3.7668(100) \\ \hline
\end{tabular}
\label{tab:3dO2O3}
\end{center}
\end{table}

As in previous examples, only the conformal dimensions of the operators in the relevant OPE are needed to numerically determine thermal one-point functions (i.e., thermal OPE coefficients). To compute the free energy, the central charge is also needed. The most up-to-date estimates, at the time of writing, are given by \cite{Chester:2019ifh,Chester:2020iyt}:
\begin{equation}
    N = 2 \ :\quad \frac{c_T}{c_{T \, \text{free}}} = 0.944056(15)\,,
\end{equation}
\begin{equation}
    N = 3 \ :\quad \frac{c_T}{c_{T \, \text{free}}} = 0.944524(28)\,.
\end{equation}
This zero-temperature data is sufficient to bootstrap thermal OPE coefficients in the two-point function $\langle \phi_i(\tau) \phi_j(0) \rangle_\beta$, using the OPE expansion of $\phi_i \times \phi_j$. For the heavy operator sector, it is a good approximation to consider only the double-twist operators $[\phi_i \phi_j]_{n,J}$.

\begin{figure*}[t]
\centering
\begin{subfigure}[t]{.48\textwidth}
  \includegraphics[width=75mm]{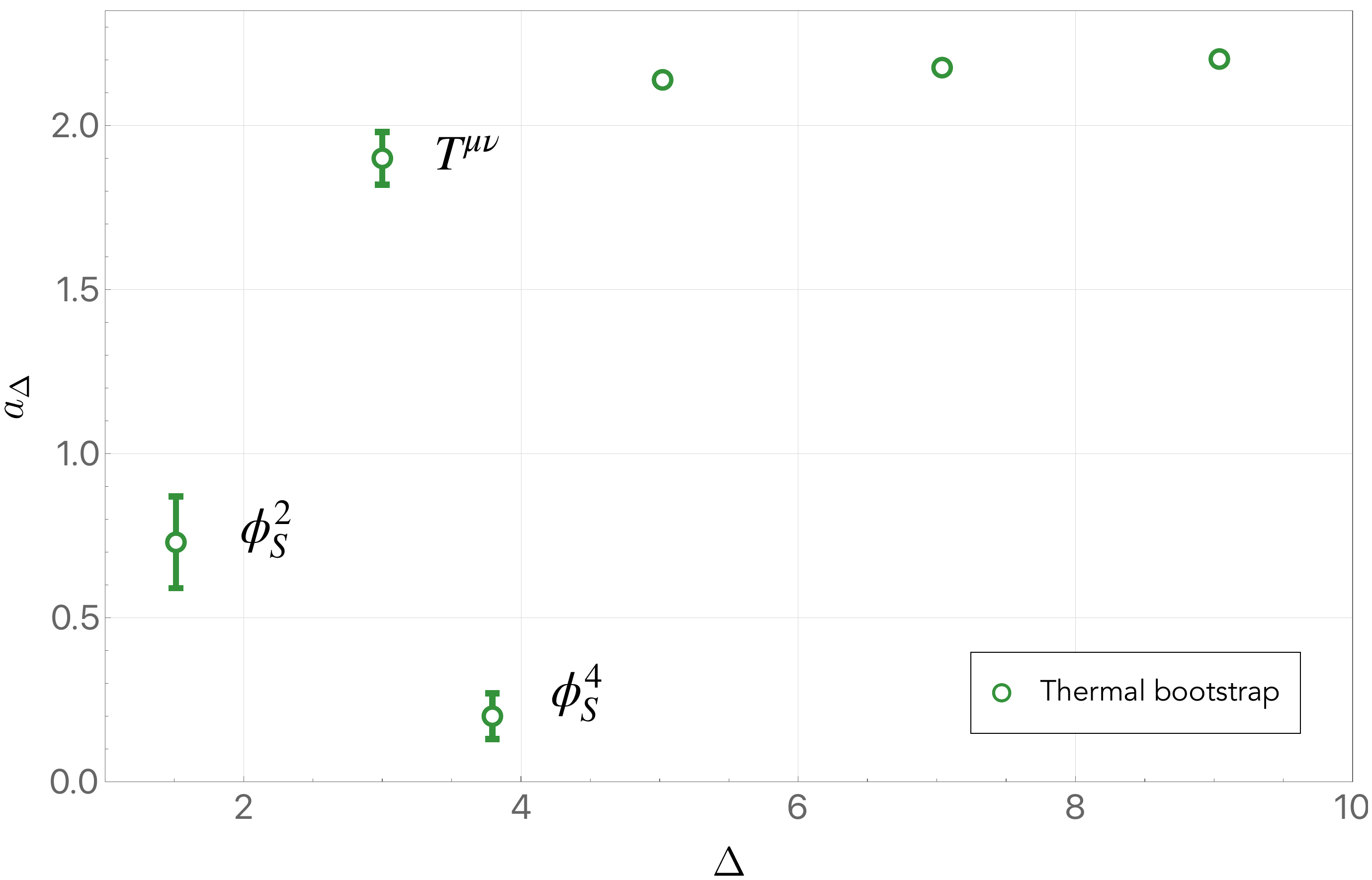}
  \caption{$\mathrm{O}(2)$ model (XY)}
\end{subfigure}%
\hfill
\begin{subfigure}[t]{.48\textwidth}
  \includegraphics[width=75mm]{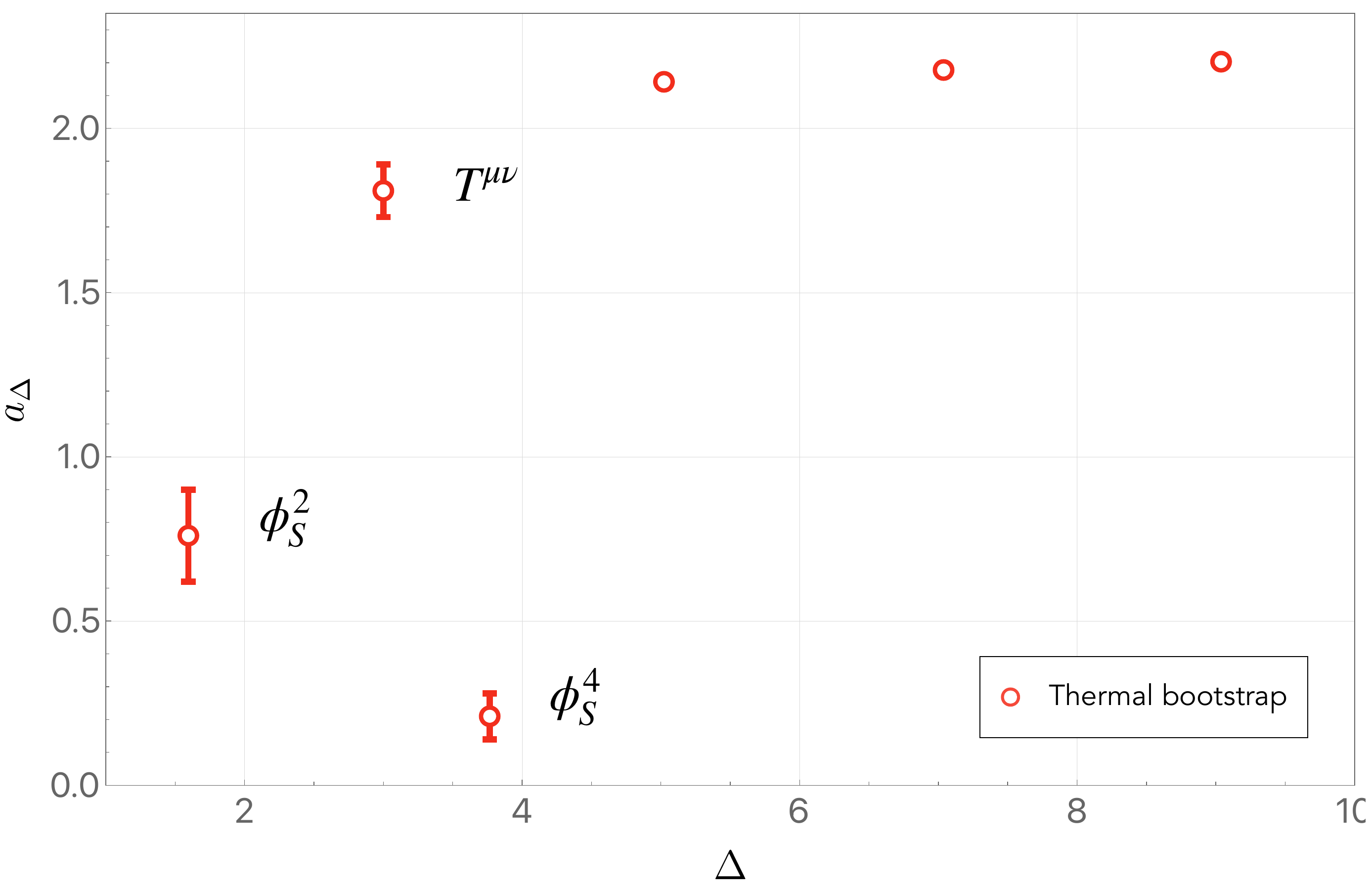}
  \caption{$\mathrm{O}(3)$ model (Heisenberg)}
\end{subfigure}
\caption{Thermal OPE coefficients for the lightest operators in the OPE spectrum of the $\mathrm{O}(2)$ and $\mathrm{O}(3)$ models. Points without error bars correspond to Tauberian predictions.}
\label{fig:OPE_Coefficients}
\end{figure*}

Figure~\ref{fig:OPE_Coefficients} displays the thermal OPE coefficients for the XY ($N = 2$) and Heisenberg ($N = 3$) models. In Table~\ref{tab:3do2o3Results}, we provide the values for the first three light operators.

\begin{table}[h]
\begin{center}
\renewcommand{\arraystretch}{1.25}
\begin{tabular}{|c||c|c|}
\hline
$\mathcal O$ & $N = 2$ & $N = 3$ \\
\hline
$\phi_i \phi_i$ & 0.73(14) & 0.76(14) \\ \hline
$T_{\mu \nu}$ & 1.90(8) & 1.81(8) \\ \hline
$(\phi_i \phi_i)^2$ & 0.20(7) & 0.21(7) \\ \hline
\end{tabular}
\caption{Thermal OPE coefficients $a_\Delta$ for light operators in the $3d$ $\mathrm{O}(2)$ and $\mathrm{O}(3)$ models. The Tauberian tail corrections are $c_1 \approx -0.0539$ ($N=2$) and $c_1 \approx -0.0471$ ($N=3$), with negligible error.}
\label{tab:3do2o3Results}
\end{center}
\end{table}

Using the relation in Eq.~\eqref{eq:freeEnergy}, we convert the values of $a_T$ into estimates for the free energy density. Our results, compared to Monte Carlo simulations, are:
\begin{equation}
    f_\text{Boot.} = -\frac{0.143(5)}{\beta^3} \,, \qquad
    f_\text{MC} = -\frac{0.1526(16)}{\beta^3}\,.
\end{equation}
Interestingly, the large $N$ prediction is given by
\begin{equation}
    \frac{f}{N} = -\frac{2}{5 \pi} \frac{\zeta(3)}{\beta^3} + \mathcal{O}\left(\frac{1}{N}\right)\,.
\end{equation}
Somewhat unexpectedly, this result lies closer to the $N=1$ case than to $N=2$ or $N=3$. This behavior can arise if the $1/N$ corrections are large and do not have a fixed sign. In fact, this is known to be the case \cite{Katz:2014rla}, and has been further verified through Monte Carlo simulations \cite{NewLattice}.

\section{The free energy in \texorpdfstring{$\varepsilon$}{ε}-expansion for \texorpdfstring{$N = 1,2,3$}{N = 1,2,3}}

One of the most physically relevant observables we can compute is the free energy density of a system. As shown in equation~\eqref{eq:freeEnergy}, the free energy density is expressed in terms of the coefficient $a_T$. In this section, we investigate how this quantity evolves in the range $3 \le d < 4$.

In $d = 4$, the $\mathrm{O}(N)$ models describe a set of $N$ free, non-interacting scalar fields. Therefore, the free energy density is simply $N$ times the free energy of a single free scalar field. In $d = 3$, the conformal bootstrap program provides extremely precise zero-temperature data, enabling accurate computations of the free energy density for $N = 1, 2, 3$.

However, it is particularly interesting to understand how the free energy interpolates between the free theory in $d = 4$ and the strongly interacting regime in $d = 3$. To study this interpolation, we employ zero-temperature data obtained from the $\varepsilon$-expansion. Specifically, we use the following expressions for the conformal dimensions of relevant operators:
\begin{align}
    \Delta_{\phi_i} &= 1 - \frac{\varepsilon}{2} + \frac{\varepsilon^2 (N+2)}{4 (N+8)^2} + \mathcal{O}(\varepsilon^3)\,, \\
    \Delta_{\phi_i \phi_i} &= 2 - \frac{6 \varepsilon}{N+8} - \frac{\varepsilon^2 (N+2)(13N + 44)}{2 (N+8)^3} + \mathcal{O}(\varepsilon^3)\,, \\
    \Delta_{T_{\mu \nu}} &= 4 - \varepsilon\,, \\
    \Delta_{[\phi_i\phi_i]_{0,4}} &= 6 - \varepsilon + \frac{7 \varepsilon^2 (N+2)}{20 (N+8)^2} + \mathcal{O}(\varepsilon^3)\,.
\end{align}
Additional operators and their anomalous dimensions in the $\varepsilon$-expansion can be found in \cite{Henriksson:2022rnm}. The central charge is also known perturbatively and is given by
\begin{equation}
    \frac{c_T}{c_{T \, \text{free}}} = 1 - \frac{5 (N+2)}{12 (N+8)} \varepsilon^2 + \mathcal{O}(\varepsilon^3)\,.
\end{equation}

With this input, we impose the KMS condition on the OPE $\phi_i \times \phi_i$ and numerically compute $a_T$ for $N = 1,2,3$ across the interval $3 \le d \le 4$ (or equivalently, $0 \le \varepsilon \le 1$). We then translate these values into free energy densities using equation~\eqref{eq:freeEnergy}.

\begin{figure}[t]
\centering
\includegraphics[width=96mm]{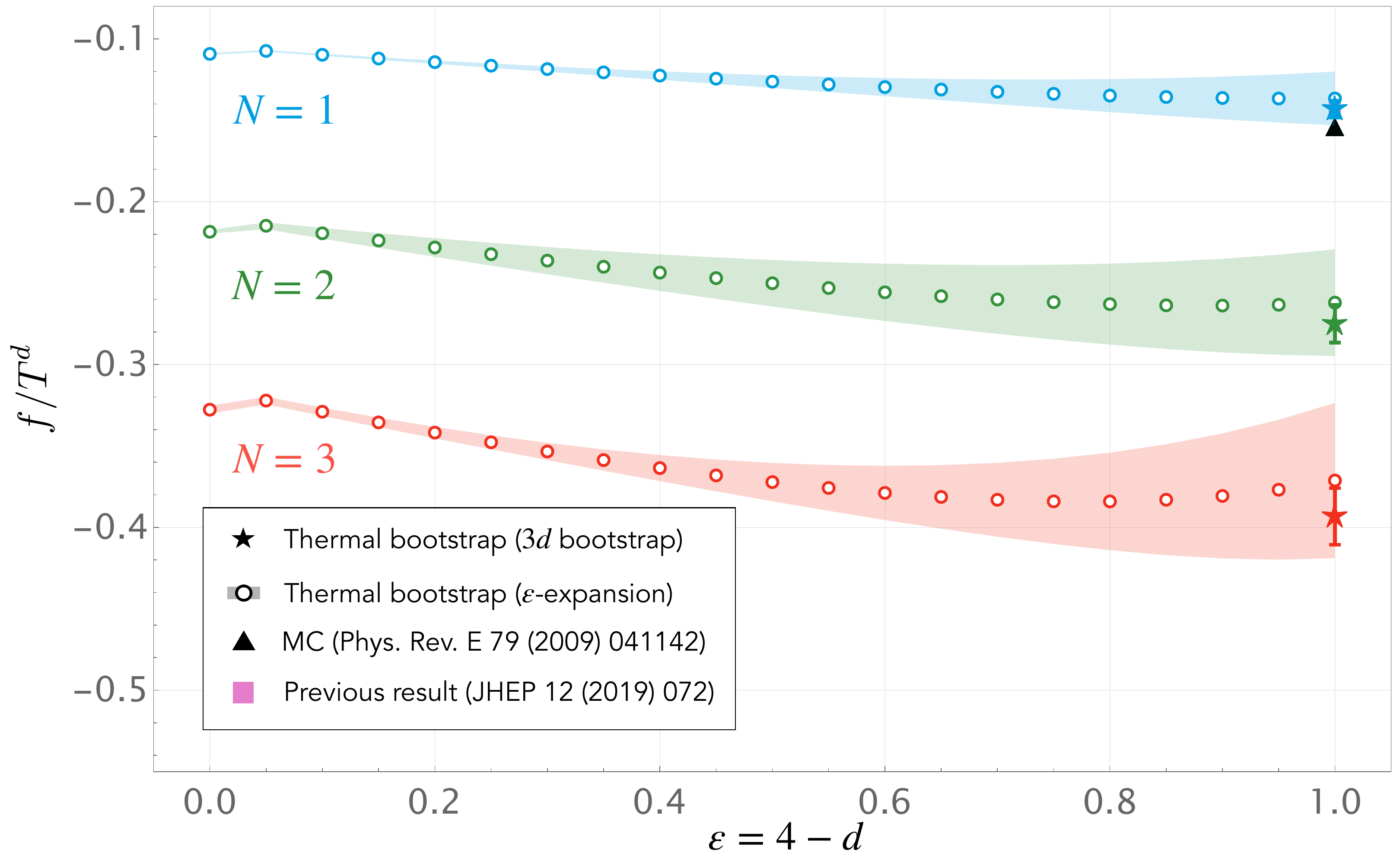}
\caption{Free energy density of the critical $\mathrm{O}(N)$ models for $N = 1, 2, 3$ in $3 \le d \le 4$ (i.e., $0 \le \varepsilon \le 1$), computed using the $\varepsilon$-expansion.}
\label{fig:Free_Energy}
\end{figure}

The results are displayed in Fig.~\ref{fig:Free_Energy}. The error bars shown reflect uncertainties in the zero-temperature input data from the $\varepsilon$-expansion, which propagate through the minimization procedure used to solve the bootstrap problem. In particular, predictions at $d = 3$ derived from $\varepsilon$-expansion are expected to be less precise than those obtained using direct $3d$ bootstrap data.

To estimate the uncertainty, we compare $\varepsilon$-expansion data with the high-precision bootstrap results at $d = 3$.\footnote{Bootstrap errors are negligible compared to the propagated uncertainties from the $\varepsilon$-expansion.} We find that deviations in the conformal dimensions of operators contribute relatively little to the overall error, while variations in the central charge dominate the uncertainty in the computed free energy.

These findings demonstrate the importance of accurately determining the central charge across dimensions. The bootstrap approach in $d = 3$ provides a reliable benchmark against which $\varepsilon$-expansion results can be compared.

\section{Analytical bootstrap: \texorpdfstring{$\mathrm O(N)$}{O(N)} critical model in \texorpdfstring{$\varepsilon$}{ε}-expansion}

In this section, we demonstrate the validity of the procedures outlined in Section \ref{sec:analyticbootstrap} in concrete, physically relevant examples. In particular, we focus on the free theory in $d = 4$ and on the critical $\mathrm{O}(N)$ model at leading order in the $\varepsilon$-expansion.

\subsubsection{\texorpdfstring{$\mathrm O(N)$}{O(N)} model at criticality in \texorpdfstring{$\varepsilon$}{ε}-expansion}

It is known that the operator product expansion (OPE) between fundamental scalars in the $\mathrm{O}(N)$ model, to first order in $\varepsilon$, takes the schematic form
\begin{equation}
    \phi \times \phi = [\mathds{1}] + [\phi \phi]_{0,J} + [\phi \phi]_{1,J} \,.
\end{equation}
The operators $[\phi \phi]_{1,J}$ correspond classically to $\phi\, \partial^J \Box \phi$. At the free point ($d = 4$), these operators are null states due to the equations of motion. However, in the interacting theory at $d = 4 - \varepsilon$, the equation of motion reads
\begin{equation}
    \Box \phi \sim \lambda \phi^3 \sim \varepsilon\, \phi^3\,,
\end{equation}
implying that $[\phi \phi]_{1,J} \sim \varepsilon\, \phi \partial^J \phi^3$ and hence are present at leading order in $\varepsilon$.

Regarding the spectrum, we have $\Delta_\phi = 1 - \frac{\varepsilon}{2} + \mathcal{O}(\varepsilon^2)$, as expected for a free scalar in $d = 4 - \varepsilon$. At this order, the only operator acquiring a non-zero anomalous dimension is $\phi^2$ (i.e., $[\phi \phi]_{0,0}$).

\paragraph{Diagrammatic calculation}

We begin with a direct diagrammatic computation. At first order in perturbation theory, the relevant contribution comes from the tadpole diagram:
\begin{equation}
    \begin{split}
        \DiagramB
        =\ &
        \lambda (N+2) \frac{\Gamma^3(d/2 - 1)}{64 \pi^{3d/2 + 2}}
        \left( \sum_{m = -\infty}^{\infty} \frac{1}{m^{d - 2}} \right) \\
        &\times
        \sum_{n = -\infty}^{\infty}
        \sum_{\ell = -\infty}^{\infty}
        \int_0^1 d\tau'\, \int d^{d-1} x'\,
        G(x,x'+n) G(x',0+\ell)\,,
    \end{split}
\end{equation}
where $G(x,x')$ is the thermal propagator of a massless scalar field in $d$ dimensions,
\begin{equation}
    G(x,x') = \frac{1}{[(\vec{x} - \vec{x}')^2 + (\tau - \tau')^2]^{d/2 - 1}}\,.
\end{equation}

Shifting $\tau'$ and performing the sums and integrals, the diagram reduces to a convolution of massless propagators:
\begin{equation}
    \DiagramB
    =
    \lambda (N+2) \frac{\Gamma(d/2 - 1)\Gamma(d/2 - 2)}{32 \pi^{d+2}} \zeta_{d-2}
    \sum_{n = -\infty}^{\infty}
    \frac{1}{[(n+z)(n+\zb)]^{d/2 - 2}}\,.
\end{equation}
Expanding around $d = 4 - \varepsilon$, we find
\begin{equation}
    \begin{split}
        \DiagramB = -\frac{\lambda (N+2)}{96 \pi^4}
        \sum_{n = -\infty}^\infty
        \left\{
        \frac{1}{\varepsilon}
        + \gamma_E + \log \pi
        + \frac{1}{2} \log[(n+z)(n+\zb)]
        - \frac{6}{\pi^2} \zeta_2'
        \right\}
        + \mathcal{O}(\varepsilon^2)\,.
    \end{split}
\end{equation}

After regularizing the divergent sum (we use the same prescription as in the dispersion relation approach), the renormalized result reads
\begin{equation}\label{eq:perturbativeepislonres}
    \begin{split}
        \tilde{g}(z,\zb)
        =
        - \frac{N+2}{N+8} \frac{\pi^2}{6}\,
        z \zb
        \biggl( &
        \log (z\zb)
        +
        \log \Gamma(z) + \log \Gamma(-z)
        +
        \log \Gamma(\zb) + \log \Gamma(-\zb) \\
        &-
        12 \log A
        -
        \log (2\pi^2)
        +
        2\pi i
        \biggr)\,,
    \end{split}
\end{equation}
where $A = 1.282427\ldots$ is the Glaisher–Kinkelin constant, related to $\zeta_2'$ by
\begin{equation}
    \zeta_2' = \zeta_2 \left(\gamma_E + \log(2\pi) - 12 \log A\right)\,.
\end{equation}

From this expression, we can extract the thermal OPE coefficients and the anomalous dimension of $\phi^2$:
\begin{equation}
    \gamma_{\phi^2} = \varepsilon\, \frac{N+2}{N+8}\,,
\end{equation}
which agrees with well-known results. The corresponding OPE coefficients are presented in Table~\ref{tab:drepsilon}.

Note: the full $\mathcal{O}(\varepsilon)$ corrections also include contributions from the shift in $\Delta_\phi$ in the GFF blocks:
\begin{equation}
    a_{[\phi \phi]_{0,J}} = 2 \zeta(2+J) - 2 \varepsilon \zeta'(2+J) + \mathcal{O}(\varepsilon^2)\,.
\end{equation}

\paragraph{Dispersion relation (DR)}

The conformal dimensions of all operators (except $\phi^2$) match those of the GFF spectrum in $d = 4 - \varepsilon$. Thus, the only new contribution to the discontinuity in the dispersion relation comes from $\phi^2$. Its discontinuity reads
\begin{equation}
    \mathrm{Disc}[\phi^2] = -\gamma_{\phi^2} \frac{i \pi z (\overline z - z \overline z)}{z - 1} \Theta(\overline z - 1)\,.
\end{equation}
The resulting contribution to the correlator is
\begin{equation}
    g_{\text{DR}}^{(\phi^2)} = a_{\phi^2}^{(0)} \gamma_{\phi^2} \left[ \log(1 - z^2) + \log(1 - \overline z^2) \right]\,,
\end{equation}
where $a_{\phi^2}^{(0)}$ is the free theory one-point function.

Interestingly, imposing KMS naively suggests
\begin{equation}
    a_{\phi^2}^{(0)} = \frac{2 \varepsilon}{\varepsilon - \gamma_{\phi^2}}\,,
\end{equation}
but this is inconsistent with the known free theory result. We interpret this as an artifact of the arc contribution: $\phi^2$ sits on the arc, and must be treated separately in the DR formulation.

\paragraph{Lorentzian inversion formula (LIF)}

Using the inversion formula for a scalar exchange, we find that the contribution of $\phi^2$ to the thermal OPE coefficients is
\begin{equation}
    a_{[\phi \phi]_{0,J}}^{(\phi^2)} = -\left(1 + (-1)^J\right) \frac{a_{\phi^2}^{(0)} \gamma_{\phi^2}}{2J}\,, \quad
    a_{[\phi \phi]_{1,J}}^{(\phi^2)} = \left(1 + (-1)^J\right) \frac{a_{\phi^2}^{(0)} \gamma_{\phi^2}}{2(J+2)}\,,
\end{equation}
with all higher $n$ vanishing at this order. Table~\ref{tab:drepsilon} summarizes the results from all approaches.

\begin{table}[h]
\centering
\renewcommand{\arraystretch}{1.5}
\begin{tabular}{|c|c|c|c|}
\hline
Operator $\mathcal{O}$ & $a_{\mathcal{O}}$ (Exact) & DR & LIF \\
\hline
$\mathds{1}$ & 1 & 0 & 0 \\
$[\phi\phi]_{0,J\ge 2}$ & $-2 \zeta'(2+J) - \frac{1}{J} a_{\phi^2}^{(0)} \gamma_{\phi^2} \zeta(J)$ & $-\frac{1}{J} a_{\phi^2}^{(0)} \gamma_{\phi^2}$ & $-\frac{1}{J} a_{\phi^2}^{(0)} \gamma_{\phi^2}$ \\
$[\phi\phi]_{1,J}$ & $\frac{1}{J+2} a_{\phi^2}^{(0)} \gamma_{\phi^2} \zeta(2+J)$ & $\frac{1}{J+2} a_{\phi^2}^{(0)} \gamma_{\phi^2}$ & $\frac{1}{J+2} a_{\phi^2}^{(0)} \gamma_{\phi^2}$ \\
\hline
\end{tabular}
\caption{Thermal OPE coefficients in the $\phi \times \phi$ OPE to leading order in $\varepsilon$. We compare results from exact block decomposition, the dispersion relation (DR), and the Lorentzian inversion formula (LIF). We do not consider the operator $\phi^2$, since inversion formula gives divergent results and its contribution is surely in the arcs of the dispersion relation. Only even spin operators contribute, due to symmetry. }
\label{tab:drepsilon}
\end{table}

   \paragraph{Imposing KMS}

The first method we propose to complete the correlator computed from the dispersion relation is to impose KMS symmetry. For simplicity, we work order-by-order in the $\varepsilon$-expansion. Furthermore, we subtract the free theory contribution:
\begin{equation}
    \tilde{g}(z,\overline{z}) = g(z,\overline{z}) - g_{\text{free}}(z,\overline{z})\,,
\end{equation}
where $g_{\text{free}}$ is the free scalar two-point function in $d = 4 - \varepsilon$. Since both $g$ and $g_{\text{free}}$ are KMS invariant, the difference $\tilde{g}$ also preserves KMS. 

At first order in $\varepsilon$, $\tilde{g}$ is proportional to $\gamma_{\phi^2}$. Thus, we only need to impose KMS on this contribution. The identity operator and $\phi^2$ contribute from the arc term. While the scalar $[\phi\phi]_{1,0}$ would also naively contribute from the arc, we will show its arc contribution vanishes.

We make the ansatz:
\begin{equation}
    \tilde{a}_{[\phi \phi]_{0,J}} = -\frac{1}{J} a_{\phi^2}^{(0)} \gamma_{\phi^2} \left(1 + \sum_{n=1}^\infty \frac{\tilde{c}_n}{n^J} \right) \,, \quad
    \tilde{a}_{[\phi \phi]_{1,J}} = \frac{1}{J+2} a_{\phi^2}^{(0)} \gamma_{\phi^2} \left(1 + \sum_{n=1}^\infty \frac{\tilde{d}_n}{n^J} \right) \,,
\end{equation}
where $\tilde{a}$ refers to the thermal OPE coefficients in $\tilde{g}$. The coefficients $\tilde{c}_n$, $\tilde{d}_n$ are numerically determined by imposing KMS. Resumming the corrections leads to the contribution of the compensator:
\begin{multline}
    \tilde{g}_{\text{comp}}^{(n)}(z, \overline{z}) =
    a_{\phi^2}^{(0)} \gamma_{\phi^2} \tilde{c}_n
    \frac{z \log \left(1 - \frac{z^2}{n^2} \right) - \overline{z} \log \left(1 - \frac{\overline{z}^2}{n^2} \right)}{2(z - \overline{z})} \\
    + a_{\phi^2}^{(0)} \gamma_{\phi^2} \tilde{d}_n
    \frac{
    n^2 \left[
    z \log\left(1 - \frac{\overline{z}^2}{n^2}\right)
    - \overline{z} \log\left(1 - \frac{z^2}{n^2}\right)
    + z \overline{z} (\overline{z} - z)
    \right]
    }{2(z - \overline{z})}\,.
\end{multline}

The scalar operators $\phi^2$ and $[\phi \phi]_{1,0}$ are treated separately as arc contributions, with open coefficients $a_{\phi^2}$ and $a_{[\phi \phi]_{1,0}}$. However, since $\phi^2$ contributes as a constant in the correlator, its value cannot be fixed by KMS.

The numerically determined values with $n_{\text{max}} = 30$ are:

\begin{table}[H]
    \centering
    \renewcommand{\arraystretch}{1.25}
    \begin{tabular}{cccc|cccc}
        \hline
        $\tilde{c}_2$ & $\tilde{c}_3$ & $\tilde{c}_4$ & $\ldots$ & $\tilde{d}_2$ & $\tilde{d}_3$ & $\tilde{d}_4$ & $\ldots$ \\
        \hline
        1.0000 & 1.0000 & 1.0000 & $\ldots$ & 0.2500 & 0.1111 & 0.06250 & $\ldots$  \\
        %\hline\hline
        %$\tilde{d}_2$ & $\tilde{d}_3$ & $\tilde{d}_4$ & $\ldots$ \\
        %\hline
        %0.2500 & 0.1111 & 0.06250 & $\ldots$ \\
        \hline
    \end{tabular}
\end{table}

We also find:
\begin{equation}
    a_{[\phi \phi]_{1,0}} = 0.3073\,.
\end{equation}

We thus conclude:
\begin{equation}
    \tilde{c}_n = 1\,, \qquad \tilde{d}_n = \frac{1}{n^2}\,,
\end{equation}
which reproduces the exact results from Table~\ref{tab:drepsilon}.

\paragraph{Sum over images on the dispersion relation}

We now perform the sum over images on the dispersion relation, following the prescription in Eq.~\eqref{eq:sumoverphi2}. We restrict to $\tilde{g}$, since $g_{\text{free}}$ can be treated exactly in $d = 4 - \varepsilon$.

The relevant sum is:
\begin{equation}
    \sum_{m = -\infty}^{\infty} g_{\text{DR}}^{(\phi^2)}(z - m, \overline{z} - m)\,.
\end{equation}

This sum diverges on the real axis, so we compute it via analytic continuation. The basic building block is:
\begin{equation}
    \sum_{m = -\infty}^\infty \log(1 + (x - m)^2)
    = \sum_{m} \log(1 - (x - m)) + \sum_{m} \log(1 + (x - m))\,.
\end{equation}

To handle each term, consider:
\begin{equation}
    \log(1 + (x - m)) = -\left. \frac{\partial}{\partial s} \left[ \frac{1}{(1 + z - m)^s} \right] \right|_{s \to 0}\,.
\end{equation}

We analytically continue the sum:
\begin{equation}
    \sum_{m = -\infty}^{\infty} \frac{1}{(1 + z + m)^s}
    =
    (-1)^s \left[\zeta_H(s, -z) + (-1)^s \zeta_H(s, 1 + z)\right]\,,
\end{equation}
where $\zeta_H$ is the Hurwitz zeta function. Differentiating at $s = 0$, we obtain:
\begin{equation}
    \sum_{m} \log(1 + (x - m)) =
    -\log \Gamma(-z - 1) + \log \Gamma(z) + \log z + \log(-z - 1) - \log(2\pi)\,.
\end{equation}

Using this, we find that the sum over images reproduces the perturbative result in Eq.~\eqref{eq:perturbativeepislonres}, up to a constant. This constant can be attributed to the arc contribution of the $\phi^2$ block.

\paragraph{Momentum-space interpretation}

We finally verify that the momentum-space OPE matches expectations. As discussed in \cite{Manenti:2019wxs}, only the identity and $\phi^2$ contribute at leading order in large momentum.

The identity contribution is:
\begin{equation}
    \text{Identity} \sim \frac{1}{k^2 + \omega_n^2}\,,
\end{equation}
while the leading correction from $\phi^2$ is:
\begin{equation}
    \phi^2 \sim \frac{2 \gamma_{\phi^2}}{(k^2 + \omega_n^2)^2}\,.
\end{equation}

Thus, the one-loop corrected two-point function in momentum space is:
\begin{equation}
    g(k, \omega_n) =
    \frac{1}{k^2 + \omega_n^2}
    \left(1 + a_{\phi^2}^{(0)} \frac{2 \gamma_{\phi^2}}{k^2 + \omega_n^2} \right)\,.
\end{equation}

Since the expansion is perturbative and local, no exponential corrections (e.g., $\sim e^{-k}$) arise. Therefore, the OPE in momentum space is exact at this order. One can explicitly check that its Fourier transform matches the results from both the dispersion relation and diagrammatic computation.

 \section{Mixing numerical and analytical methods in the $3d$ Ising model}

The analytic techniques proposed in this thesis—such as the dispersion relation in the complex $\xi$-plane—can be tested numerically for any strongly coupled theory, even when analytic control is not available. In this section, we test the validity of our approach in the case of the three-dimensional Ising model, focusing on the two-point function $\langle \sigma(\tau)\sigma(0)\rangle_\beta$ in the limit where $\vec{x}=0$. We make use of thermal one-point functions and operator dimensions extracted in \cite{Barrat:2024fwq} and compare our results with the numerically constructed correlator in that work.

We consider the thermal OPE:
\begin{equation}
    \sigma \times \sigma = [\mathds{1}] + [\epsilon] + [T^{\mu\nu}] + [\epsilon'] + \cdots \,,
\end{equation}
and restrict to operators with conformal dimension $\Delta < 4$. The corresponding conformal data is listed in Table~\ref{tab:3dIsingBotstrapwitht}.

\begin{table}[h]
    \centering
    \renewcommand{\arraystretch}{1.25}
    \begin{tabular}{|c|c|c|c|}
        \hline
        $\mathcal{O}$ & $\mathcal{O}_{\text{GL}}$ & $\Delta_\mathcal{O}$~\cite{Kos:2016ysd,Reehorst:2021hmp,Chang:2024whx} & $a_{\mathcal{O}}^{(\langle \sigma \sigma \rangle)}$~\cite{Barrat:2024fwq} \\
        \hline
        $\sigma$ & $\phi$ & 0.518148806(24) & 0 \\
        \hline
        $\epsilon$ & $\phi^2$ & 1.41262528(29) & 0.75(15) \\
        \hline
        $T_{\mu\nu}$ & $T_{\mu\nu}$ & 3 & 1.97(7) \\
        \hline
        $\epsilon'$ & $\phi^4$ & 3.82951(61) & 0.19(6) \\
        \hline
        $C'_{\mu\nu\rho\sigma}$ & $\phi \partial^\mu \partial^\nu \partial^\rho \partial^\sigma \phi$ & 5.022665(28) & / \\
        \hline
        $T'_{\mu\nu}$ & $\phi \partial^\mu \partial^\nu \Box \phi$ & 5.50915(44) & / \\
        \hline
    \end{tabular}
    \caption{Light operators in the 3D Ising model, their classical (Ginzburg-Landau) interpretation, conformal dimensions from bootstrap, and thermal one-point coefficients from~\cite{Barrat:2024fwq}. Only operators with $\Delta < 4$ are used in this analysis.}
    \label{tab:3dIsingBotstrapwitht}
\end{table}

To reconstruct the correlator, we invert the contribution of the blocks listed above using the dispersion relation in the complex $\xi$-plane. Strictly speaking, this inversion is not formally justified: the OPE may not commute with the discontinuity in a strongly coupled theory. However, since we are truncating the expansion and assuming that the asymptotic spectrum consists of generalized free field (GFF) double-twist operators~\cite{Fitzpatrick:2012yx,Komargodski:2012ek,Kravchuk:2021kwe,Pal:2022vqc,vanRees:2024xkb}, we expect the dominant discontinuity to come from the low-dimension operators included here. The inversion of each block in the complex plane can be performed exactly and is expressible in terms of Hurwitz zeta functions.

\begin{figure*}[h!]
    \centering
    \begin{subfigure}[t]{.48\textwidth}
        \includegraphics[width=\textwidth]{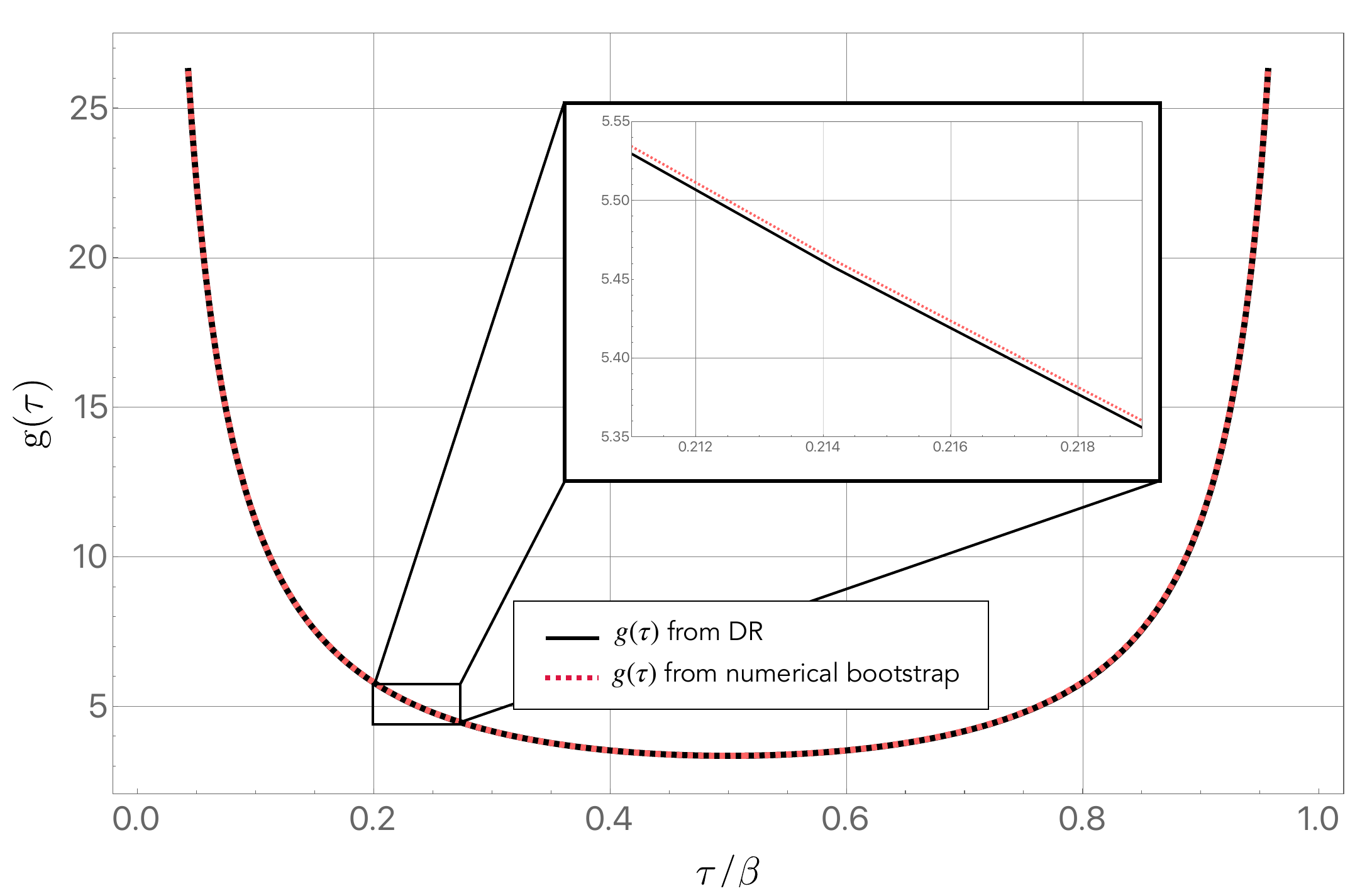}
        \caption{}
        \label{fig:3dIsing}
    \end{subfigure}
    \hfill
    \begin{subfigure}[t]{.48\textwidth}
        \includegraphics[width=\textwidth]{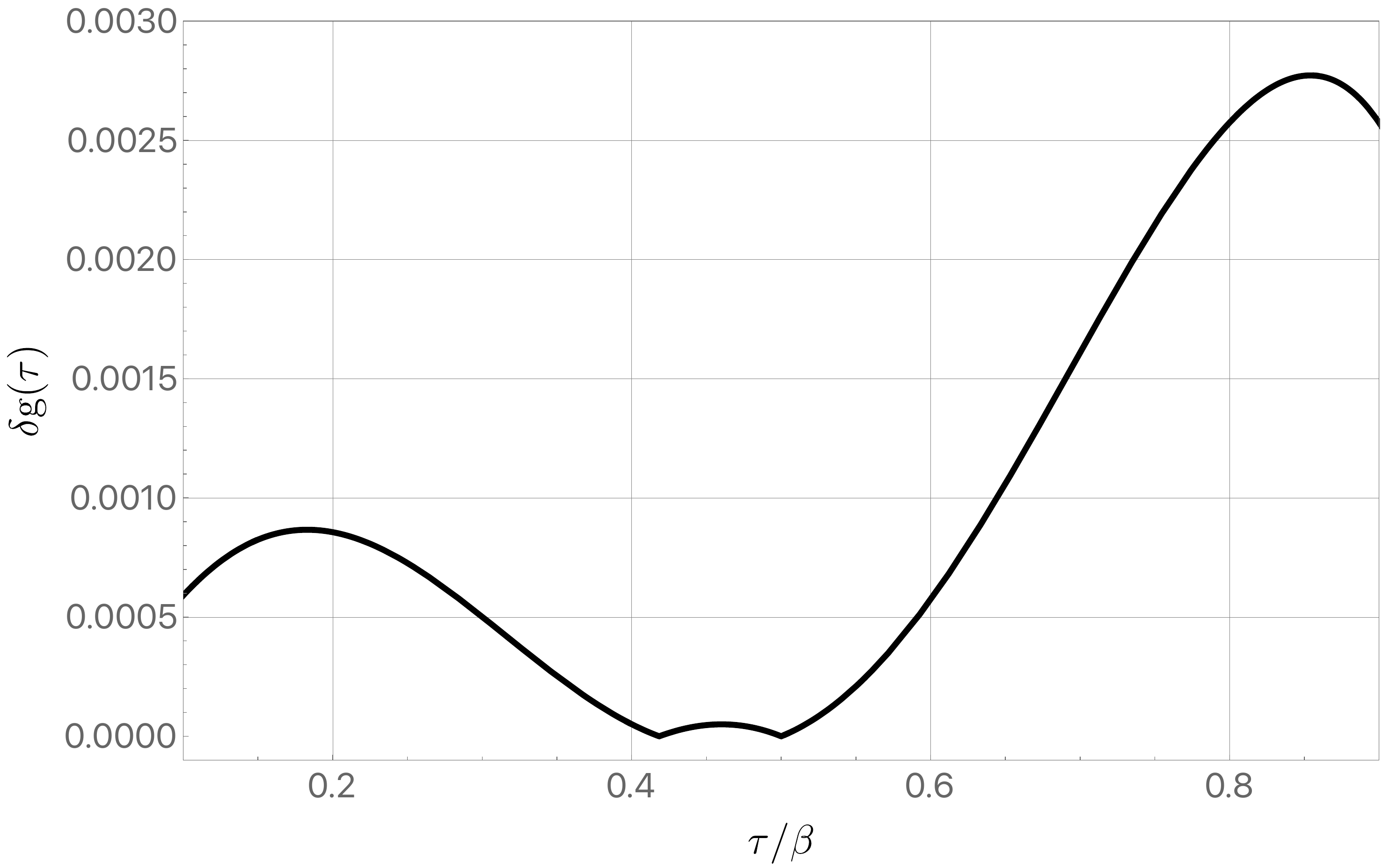}
        \caption{}
        \label{fig:3dIsingdiff}
    \end{subfigure}
    \caption{
    \textbf{(a)} Two-point function $\langle \sigma(\tau)\sigma(0) \rangle_\beta$ at $\vec{x} = 0$, computed from the dispersion relation in the complex $\xi$-plane using thermal OPE coefficients from~\cite{Barrat:2024fwq} (black curve), compared to the direct numerical evaluation from~\cite{Barrat:2024fwq} (dashed curve). 
    \textbf{(b)} Relative discrepancy between the two results. The discrepancy is of order $10^{-3}$.
    }
    \label{Fig:3dIsingDR}
\end{figure*}

To fix the overall additive constant ambiguity in the dispersion relation, we match the result to the numerical data of~\cite{Barrat:2024fwq} at $\tau = \beta/2$. This choice is convenient, as it sits symmetrically between the region dominated by the identity and its KMS-dual.

In Fig.~\ref{Fig:3dIsingDR}, we show the reconstructed two-point function and its relative difference with the numerical result. The discrepancy is $\sim 10^{-3}$—comparable to the numerical precision reported in~\cite{Barrat:2024fwq}. The maximal deviation appears near $\tau/\beta \sim 0.8$, where corrections to the Tauberian approximation used in~\cite{Barrat:2024fwq} become more important, as expected.

\vspace{1em}
\noindent\textbf{Extending the method: the $\langle \epsilon(\tau) \epsilon(0) \rangle_\beta$ correlator}

The dispersion relation method can now be extended beyond the $\langle \sigma \sigma \rangle_\beta$ channel. In particular, the Tauberian approximation in~\cite{Barrat:2024fwq} was not sufficiently accurate to bootstrap $\langle \epsilon(\tau) \epsilon(0)\rangle_\beta$. However, one can use the thermal OPE coefficients extracted in the $\sigma \times \sigma$ channel, combined with zero-temperature OPE coefficients from~\cite{Chang:2024whx}, to estimate the thermal OPE coefficients in the $\epsilon \times \epsilon$ channel:
\begin{equation}
    a_{\mathcal{O}}^{(\langle \epsilon\epsilon \rangle)} = a_{\mathcal{O}}^{(\langle \sigma\sigma \rangle)} \cdot \frac{f_{\mathcal{O}\epsilon\epsilon}}{f_{\mathcal{O}\sigma\sigma}} \,.
\end{equation}

From this, we obtain:
\begin{equation}
    a_{\epsilon}^{(\langle \epsilon\epsilon \rangle)} = 1.09267\,, \qquad a_{T}^{(\langle \epsilon\epsilon \rangle)} = 5.37898\,.
\end{equation}

\begin{figure}[h]
    \centering
    \includegraphics[width=0.5\textwidth]{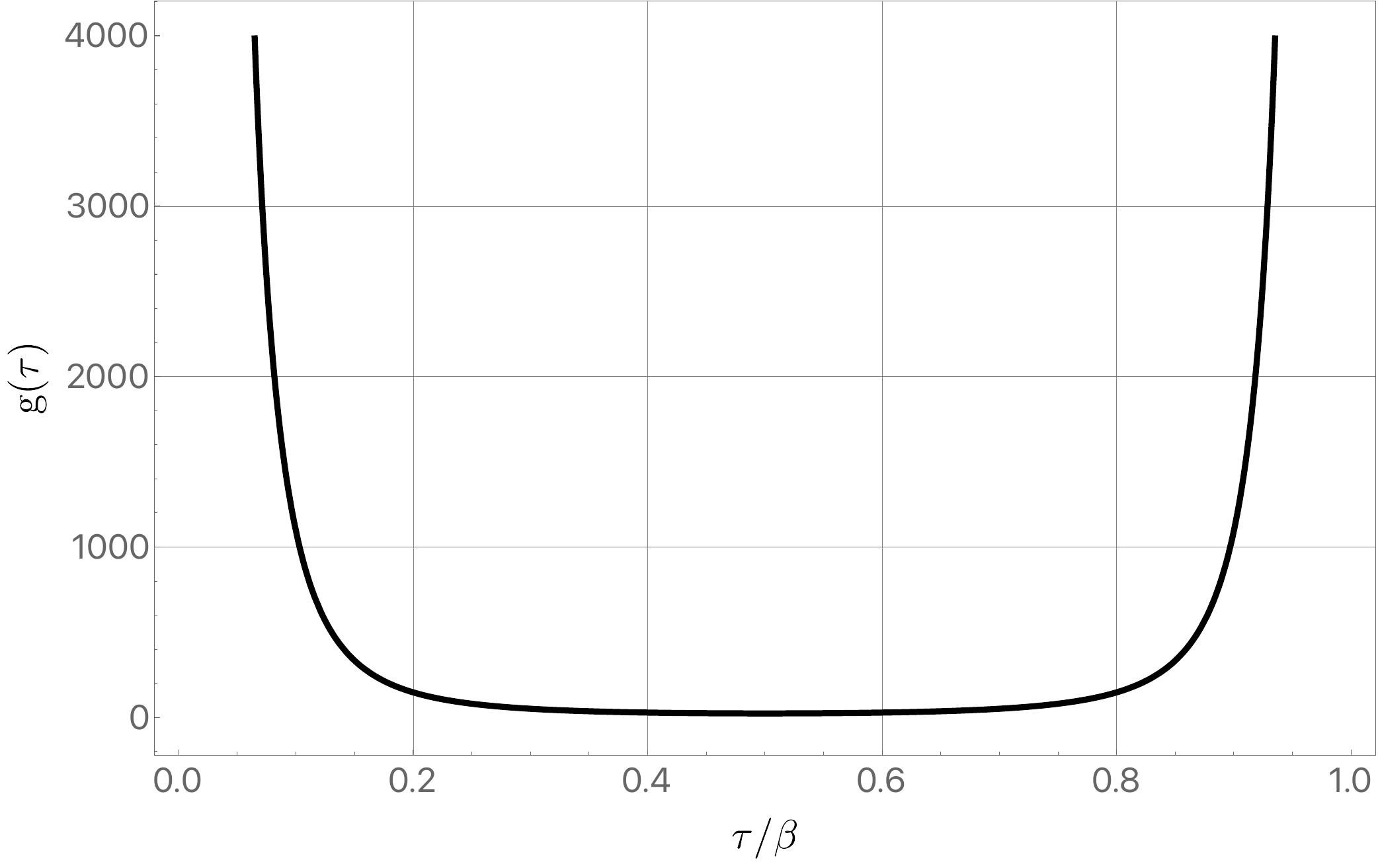}
    \caption{
    Two-point function $\langle \epsilon(\tau)\epsilon(0)\rangle_\beta$ computed from inverting $\mathds{1}$, $\epsilon$, and $T_{\mu\nu}$, using the dispersion relation in the complex $\xi$-plane. The overall additive constant was fixed by requiring agreement with the OPE approximation at $\tau/\beta = 0.03$.
    }
    \label{fig:3dIsingee}
\end{figure}

We then consider the thermal OPE:
\begin{equation}
    \epsilon \times \epsilon = [\mathds{1}] + [\epsilon] + [T^{\mu\nu}] + \cdots\,,
\end{equation}
and invert only the operators explicitly shown. The resulting two-point function is displayed in Fig.~\ref{fig:3dIsingee}.

As before, the result is determined up to an additive constant. We fix this by matching the OPE expansion near $\tau/\beta = 0$, choosing $\tau/\beta = 0.03$. One can verify that including additional operators, even with Tauberian-approximated thermal coefficients, has a negligible effect on the correlator in this regime.

\section{The magnetic line in the $\mathrm O(N)$ model}
\label{subsec:ONModel}

We now turn our attention to the $\mathrm O(N)$ model to provide an example in which interactions are turned on, while computations can still be carried out analytically.
At zero temperature, this theory has been extensively studied in the absence of defect insertions \cite{Kos:2016ysd,Dey:2016zbg,Rychkov:2018vya,Giombi:2019upv,Giombi:2020enj}, while its coupling to a localized magnetic line has also received considerable attention \cite{Cuomo:2021kfm,Bianchi:2022sbz,Gimenez-Grau:2022ebb,Gimenez-Grau:2022czc}.
At finite temperature, the $\mathrm O(N)$ model without defects is one of the most widely studied thermal CFTs \cite{Iliesiu:2018fao,Petkou:2018ynm,David:2023uya,David:2024naf}.
In this section, we present two different yet complementary approaches: the $\veps$-expansion and the large-$N$ analysis.\footnote{We are grateful to Simone Giombi for insightful discussions related to this section.}

\subsubsection{$\veps$-expansion}
\label{subsubsec:epsilonexpansion}

\paragraph{The Wilson-Fisher fixed point.}
The $\mathrm{O}(N)$ model in $d$ dimensions, at finite temperature and in the presence of a magnetic line, is described by the action
\begin{equation}
    S
    =
    \int_0^\beta d \tau \ \int d^{d-1}x \
    \left[
    \frac{1}{2}(\partial_\mu \phi_i)^2+\frac{\lambda}{4!} (\phi_i \phi_i)^2+h \, \delta^{d-1}(\vec x) \phi_1(\tau)
    \right]\,.
    \label{eq:ON_Action}
\end{equation}
This action defines a conformal defect for specific values of the coupling constants $\lambda$ and $h$.
The strategy of the $\veps$-expansion is to set $d=4-\veps$ and determine the fixed point perturbatively in $\veps$.
This fixed point, known as the \textit{Wilson-Fisher} fixed point, corresponds to the following couplings \cite{Cuomo:2021kfm,Gimenez-Grau:2022ebb}:
\begin{align}
    \frac{\lambda_\star}{(4 \pi)^2}
    &=
    \frac{3}{N+8} \veps
    +
    \frac{9(3N+14)}{(N+8)^3} \veps^2
    +
    \mathcal{O}(\veps^3)\,, \\
    h_\star^2
    &=
    N+8
    +
    \frac{4 N^2 + 45 N + 170}{2N+16} \veps
    +
    \mathcal{O}(\veps^2)\,.
\end{align}
Note that, at leading order, $\lambda_\star \sim \veps$ and $h_\star \sim 1$. The Feynman rule associated with the bulk quartic interaction is given by
\begin{equation}
    \FourVertex
    =
    - \lambda \int_0^\beta d\tau \int d^{d-1}x\,,
\end{equation}
while the thermal bulk propagator in arbitrary dimensions reads
\begin{align}
    \ScalarPropagator\
    =
    \frac{\Gamma(d/2-1)}{4 \pi^{d/2}} \sum_{m \in \mathbb Z} \frac{1}{(|\vec x|^2+|\tau+m \beta|^2)^{d/2-1}}\,.
    \label{eq:ScalarPropagator2}
\end{align}

\paragraph{One-point function of $\phi_1$.}
At order $\mathcal{O}(\veps)$, the one-point function of $\phi_1$ is given by the following Feynman diagrams:
\begin{equation}
    \vev{\phi_1 (\tau, \vec{x}) \Pm}_\beta
    =
    \phiGFF
    +
    \phiEpsExpTwo
    +
    \phiEpsExpThree
    +
    \mathcal{O} (\veps^2)\,.
    \label{eq:ON_phi_Diagrams}
\end{equation}
To compute these diagrams, we adopt the conventions for the normalization of vertices and propagators used in \cite{Gimenez-Grau:2022ebb}, and we make use of the following master integrals:
\begin{align}
    &\int_{-\infty}^\infty \frac{d \tau}{(|\vec{x}|^2 + \tau^2)^{1-\veps/2}}
    =
    \frac{\sqrt{\pi} \Gamma \left( \frac{1-\veps}{2} \right)}{\Gamma ( 1 - \veps/2 ) |\vec{x}|^{1-\veps}}\,,\\[2ex]
    &\int \frac{d^{d-1} x_2}{|\vec{x}_{12}|^a |\vec{x}_2|^b}
    =
    \frac{ \Gamma \left( \frac{d-1-a}{2} \right) \Gamma \left( \frac{d-1-b}{2} \right) \Gamma \left( \frac{a+b+1-d}{2} \right) }{\Gamma \left( \frac{a}{2} \right) \Gamma \left( \frac{b}{2} \right) \Gamma \left( \frac{2(d-1)-a-b}{2} \right)}
    \frac{\pi^{(d-1)/2}}{|\vec{x}_1|^{a+b+1-d}}\,.
\end{align}
The first diagram is similar to \eqref{eq:GFF_phi_Result}:
\begin{equation}
    \phiGFF
    =
    -\frac{1}{4} h\pi ^{\frac{\veps -3}{2}}  \Gamma \left(\frac{1-\veps}{2}\right) \frac{1}{|\vec{x}|^{1-\veps}}\,.
\end{equation}
The second diagram in \eqref{eq:ON_phi_Diagrams} does not appear at zero temperature (as massless tadpoles vanish) and can be interpreted as a manifestation of the \textit{thermal mass}.
At order $\mathcal{O}(\veps)$, the thermal mass is given by \cite{Dolan:1973qd,Weinberg:1974hy,Kapusta:2006pm,Chai:2020zgq}
\begin{equation}
    m_\text{th}^2
    =
    \frac{\veps N}{2 \beta^2 (N+8)} + \mathcal{O} (\veps^2)\,,
\end{equation}
and the diagram evaluates to
\begin{equation}
    \phiEpsExpTwo
    =
    -\frac{1}{16} h m_{\text{th}}^2 \pi^{\frac{\veps -3}{2}} \Gamma \left(-\frac{1+\veps }{2}\right) |\vec{x}|^{1+\veps}\,.
\end{equation}
The third diagram also appears at zero temperature and does not receive thermal corrections:
\begin{equation}
    \phiEpsExpThree
    =
    - \frac{1}{768} h^3 \lambda\, \pi^{\frac{3(\veps-3)}{2}} \frac{\Gamma^3 \left( \frac{1-\veps}{2} \right)}{  \veps (1-3 \veps)} \frac{1}{|\vec{x}|^{1-3 \veps}}\,.
    \label{eq:ON_phi_ThirdDiagram}
\end{equation}
This expression \eqref{eq:ON_phi_ThirdDiagram} diverges in the $\veps \to 0$ limit and requires renormalization. We follow the procedure explained in \cite{Cuomo:2021kfm,Gimenez-Grau:2022ebb}.

Putting everything together, we obtain
\begin{align}
    \frac{\vev{\phi_1(\vec x) \Pm}_\beta}{\sqrt{\vev{\phi_1(\infty) \phi_1(0)}}}
    =
    - \frac{\sqrt{N+8}}{2 |\vec{x}|}
    \biggl[
    1
    +
    \left(
    \frac{N^2 - 3N - 22 + 2(N+8)^2 \log (2 |\vec{x}|)}{4 (N+8)^2}
    +\right.\\
   \left. \frac{N}{4(N+8)} z^2
    \right) \veps 
    + \mathcal{O} (\veps^2)
    \biggr]
    \label{eq:finalphi1}
\end{align}
where $\vev{\phi_1(\infty) \phi_1(0)}$ refers to the normalization of the two-point function, and $z$ is the usual temperature-dependent cross-ratio defined in \eqref{eq:OnePoint_CrossRatio}.

\paragraph{OPE data.}
The first and second terms in \eqref{eq:finalphi1} correspond to the zero-temperature result. The logarithmic dependence on $|\vec{x}|$ arises from the expansion of $\Delta_\phi$ at order $\veps$, and matches the results of \cite{Cuomo:2021kfm,Gimenez-Grau:2022ebb}.
The $z$-dependent (hence, $\beta$-dependent) terms are induced by the thermal mass.
The powers of $z$ indicate the presence, at this order, of a defect operator with bare $1d$ scaling dimension $\widehat{\Delta} = 2$.
Among the operators with this dimension, only $\widehat{t}^{\,\,2}$ — the square of the tilt operator $\widehat{t}_{a} = \widehat{\phi}_{a \ne 1}$ — contributes, since the one-point function of the displacement operator $D_i$ vanishes by $\mathrm{SO}(d-1)$ symmetry.
Following the notation of \eqref{eq:OnePoint_DefectOPE2}, we extract the OPE data:
\begin{equation}
    \lambda_{\phi_1 \widehat{t}^{\,\,2}} \,\widehat{b}_{\widehat{t}^{\,\,2}} = -\frac{N}{8 \sqrt{N +8}} \veps + \mathcal{O}(\veps^2)\,.
\end{equation}
This is a new prediction, which we will also use in the next section after performing the large $N$ analysis.

\subsubsection{Large $N$ analysis}
\label{subsubsec:LargeN}

We now consider the same system in the limit $N \to \infty$.
In this regime, it is convenient to normalize the coupling constants such that the action \eqref{eq:ON_Action} becomes
\begin{equation}
    S
    =
    \int_0^\beta d \tau \ \int d^{d-1}x \  \left[\frac{1}{2}(\partial_\mu \phi_i)^2+\frac{\widetilde \lambda}{N} (\phi_i \phi_i)^2+\sqrt{N} \, \widetilde h \, \delta^{d-1}(\vec x) \phi_1(\tau)\right]\,.
    \label{eq:actionLarge}
\end{equation}
With these conventions, we have $\widetilde\lambda_\star \sim \mathcal{O}(1)$ and $\widetilde{h}_\star \sim \mathcal{O}(1)$ \cite{Cuomo:2021kfm}.
The Hubbard-Stratonovich transformation of the action yields
\begin{equation}
    S
    =
    \int_0^\beta d \tau \ \int d^{d-1}x \  \left[\frac{1}{2}(\partial_\mu \phi_i)^2+\frac{1}{2}\sigma  \phi_i \phi_i+\sqrt{N}\,  \widetilde{h} \, \delta^{d-1}(\vec x) \phi_1(\tau)\right]\,.
    \label{eq:ActionLargeN}
\end{equation}
Integrating out the fields $\phi_i$, we obtain the effective action for $\sigma$:
\begin{equation}
    S_\text{eff} [\sigma]
    =
    \frac{N}{2}\operatorname{Tr}\log(-\Box +\sigma)
    -
    \frac{\widetilde{h}^2 N}{2}\int_{\mathcal M_\beta} d^{d}x \int_{\mathcal M_\beta} d^{d}y\, \delta^{d-1}(\vec x)\left(\frac{1}{-\Box +\sigma}\right)(x-y)\delta^{d-1}(\vec y)\,.
    \label{eq:effectivelargeN}
\end{equation}

In principle, the extremization of \eqref{eq:effectivelargeN} gives the one-point function $\vev{\sigma(\vec{x}) \Pm}_\beta$.
In the absence of the defect ($\widetilde{h}=0$), this procedure is straightforward and reproduces the known result for the thermal mass at large $N$ \cite{Sachdev:1992py,Iliesiu:2018fao}.
In the presence of the defect, the computation becomes more involved—even at zero temperature \cite{Allais:2014fqa}—but a combination of equations of motion and conformal symmetry constraints can be used to derive $\vev{\sigma(\vec{x}) \Pm}_\beta$ \cite{Cuomo:2021kfm}.

A similar approach applies in the finite-temperature case.
From the action \eqref{eq:effectivelargeN}, the equations of motion yield the differential equation
\begin{equation}
    \left(\frac{\partial^2 }{\partial |\vec x|^2}+\frac{d-2}{|\vec x|}\frac{\partial}{\partial |\vec x|}-\vev{\sigma(\vec{x}) \Pm}_\beta \right) \vev{\phi_1(\vec{x}) \Pm}_\beta
    =
    0 \,,
    \label{eq:differential}
\end{equation}
valid for $|\vec{x}| \neq 0$.
In the large $N$ limit, the scaling dimensions of $\phi_1$ and $\sigma$ are known to be \cite{Cuomo:2021kfm}
\begin{equation}
    \Delta_{\phi_1}
    = 
    \frac{d-2}{2} + \mathcal{O} \left(\frac{1}{N} \right)\,, \qquad
    \Delta_{\sigma}= 2 + \mathcal{O} \left(\frac{1}{N} \right)\,.
\end{equation}
The defect spectrum in the bulk-defect OPE of $\phi_{2}, \dots, \phi_{N}$ is known to all orders in $\varepsilon$ at large $N$, while for $\phi_1$ it is not fully known.
Nonetheless, using the known defect spectrum \cite{Cuomo:2021kfm}, one can write the first few terms in the OPE, corresponding to a low-temperature expansion of the correlator.
The thermal one-point function of $\phi_1$ in the presence of the defect reads
\begin{equation}
    \vev{\phi_1(\vec{x}) \Pm}_\beta
    =
    \frac{1}{|\vec x|^{\frac{d}{2}-1}}
    \Big[
    c_0
    +
    c_1\, z
    +
    c_2\, z^{1+\gamma}
    +
    c_3 \, z^{2} 
    +
    \mathcal O\left(z^{2+\epsilon}\right)
    \Big]\,,
    \label{eq:largeNphi1}
\end{equation}
at least in the range $3\le d\le 4$, where the leading terms are attributed to: the identity $\widehat{\mathds{1}}$, the tilt operator $\widehat{t}_{a}$, the scalar defect operator $\widehat{\phi}_1$, and the composite operator $\widehat{t}^{\,\,2}$, respectively.
Here $\widehat{\Delta} = 2 + \epsilon$ (with $\epsilon > 0$) denotes the dimension of the next lightest defect operator.
In the OPE \eqref{eq:largeNphi1}, $\gamma$ is the anomalous dimension of $\widehat \phi_1$, which in $d=3$ is $\gamma_{3d} = 0.541728\ldots$ \cite{Cuomo:2021kfm}, and $c_i$ are the OPE coefficients defined as in \eqref{eq:OnePoint_DefectOPE2}.

Note that other operators (e.g. $\partial^i \widehat \phi_1$) exist in the spectrum, but their one-point functions vanish due to $\mathrm{SO}(d-1)$ symmetry.
Also, $c_1 = 0$ since the tilt operator is an $\mathrm O(N-1)$ vector and its one-point function vanishes by symmetry.
Nonetheless, it is instructive to retain this term in the expansion.

The normalization of the identity contribution is fixed by the zero-temperature behavior of the correlator:
in $d=3$, one has $(c_{3d,0})^2 = N \times 0.558113\ldots$ \cite{Cuomo:2021kfm}, i.e. $c_0 = \lambda_{\phi_1 \widehat{\mathds{1}}}$.

Plugging \eqref{eq:largeNphi1} into the equation of motion \eqref{eq:differential} gives an expansion for the one-point function of the Hubbard-Stratonovich field:
\begin{equation}
    \vev{\sigma(\vec{x}) \Pm}_\beta
    =
    \frac{1}{|\vec x|^2}
    \left[
    \frac{(d-2)(4-d)}{4}
    +
    \frac{c_2}{c_0} \, \gamma (1+\gamma) z^{1+\gamma}
    +
    2 \frac{c_3}{c_0} z^2
    +
    o \left(z^{2}\right)
    \right]\,.
    \label{eq:1ptsigma}
\end{equation}
The first term corresponds to the zero-temperature result \cite{Cuomo:2021kfm}, while the second is the leading thermal correction due to the operator $\widehat \phi_1$.
The operator with $\widehat \Delta = 2$ also contributes, corresponding to the tree-level term in $\langle \phi_i \phi_i \Pm \rangle_\beta$, i.e., order $\mathcal{O}(\veps^0)$.

To fully determine the one-point functions \eqref{eq:largeNphi1} and \eqref{eq:1ptsigma} at order $\mathcal{O}(z^2)$, only $c_2$ and $c_3$ are needed.
These coefficients are generally hard to compute, even at large $N$.
In principle, they could be obtained by minimizing the effective action \eqref{eq:effectivelargeN}.
Here, we estimate them using the $\veps$-expansion results from the previous section.

Comparing \eqref{eq:largeNphi1} with \eqref{eq:finalphi1}, we find:
\begin{equation}
    c_2 = \lambda_{\phi_1\widehat{\phi}_1}\, \widehat{b}_{\widehat{\phi}_1} = \mathcal{O}(\veps^2) \,, \qquad
    c_3 = \lambda_{\phi_1 \widehat{t}^{\,\,2}}\, \widehat{b}_{\widehat{t}^{\,\,2}} = -\frac{\sqrt{N}}{2} \veps + \mathcal{O}(\veps^2)\,.
\end{equation}
Finally, we note that the equations of motion relate OPE coefficients in $\langle \sigma \Pm\rangle_\beta$ to those in $\langle \phi_1 \Pm\rangle_\beta$.
From the comparison above, one obtains
\begin{equation}
    \lambda_{\sigma \widehat{\phi}_1} = \frac{\gamma(1+\gamma)}{\lambda_{\phi_1 \widehat{\mathds{1}}}} \lambda_{\phi_1 \widehat{\phi}_1} \,, \qquad
    \lambda_{\sigma \widehat{t}^{\,\,2}} = \frac{2}{\lambda_{\phi_1 \widehat{\mathds{1}}}} \lambda_{\phi_1 \widehat{t}^{\,\,2}}\,.
\end{equation}
These relations are non-trivial to extract at zero temperature, where all one-point functions of local defect operators vanish (except for the identity).

\section{Summary of the chapter}

In this chapter, we applied the bootstrap problems and methodologies proposed in Chapters \ref{ref:dynamics} and \ref{sec:defectsBO} to the physically relevant and non-trivial case of the $\mathrm O(N)$ models in $2<d<4$. 

We began by reviewing the model from a Ginzburg–Landau perspective and motivated the use of the $\varepsilon$-expansion for computing correlation functions perturbatively near four dimensions. We also presented non-trivial agreements between different theoretical approaches, as well as with experimental data for the critical exponents of the 3d Ising model.

We first analyzed the large $N$ limit of these theories, which can be solved both exactly using the Hubbard–Stratonovich trick and numerically. This regime provides an ideal testing ground for comparing analytical and numerical methods in a non-trivial, yet analytically tractable, theory.

We then applied our numerical techniques to compute the two-point function $\langle \sigma \sigma\rangle_\beta$ in the 3d Ising model. Our results show excellent agreement with existing bootstrap and Monte Carlo simulations, up to a small deviation in the free energy, which remains within two standard deviations. We discussed this discrepancy and compared our method to other bootstrap approaches and Monte Carlo results.

Subsequently, we extended our analysis to predict the same correlator in the $\mathrm O(2)$ and $\mathrm O(3)$ models, providing new predictions for physical observables such as correlation functions, free energies, and one-point functions of various operators.

We also computed the free energy in the range $3 \le d < 4$ using zero-temperature input from the $\varepsilon$-expansion and showed how the free energy evolves across dimensions.

We then focused on the analytic bootstrap, computing the one-loop correlation function of two fundamental scalars in the $\mathrm O(N)$ model. The results show perfect agreement with both the Lorentzian inversion formula and the perturbative Feynman diagram calculation. We also discussed the connection between this approach and the OPE in momentum space, both in this context and more generally in perturbative settings.

Finally, we explored a hybrid analytical/numerical approach that incorporates thermal one-point functions obtained numerically in order to compute new correlation functions, in particular the correlator $\langle \epsilon \epsilon\rangle_\beta$ in the 3d Ising model. Although this method is less rigorous than the previous ones, we showed that the results are very promising and the approach highly efficient, suggesting it deserves further study in the future.

Furthermore, we perform an $\varepsilon$-expansion and a large-$N$ analysis for the magnetic line defect in the $\mathrm{O}(N)$ model, as an example of a conformal defect wrapping the thermal circle. This computation explicitly demonstrates that it is possible to extract both finite-temperature and zero-temperature data from the knowledge of a bulk one-point function. These computations provide an explicit example of such a correlator, represented as discussed in Chapter~\ref{sec:defectsBO}.

\chapter{Conclusions and outlook}
\label{sec:ConclusionsAndOutlook}

\section{Summary of the thesis and main results}
In this thesis, we have developed a novel, non-perturbative framework for the study of finite-temperature conformal field theories using both numerical and analytical bootstrap techniques. Building on the Matsubara formalism and exploiting the Kubo-Martin-Schwinger (KMS) condition together with the Operator Product Expansion (OPE), we have formulated a thermal bootstrap approach that allows one to extract non-perturbative information about thermal correlation functions from symmetry, analyticity, and consistency alone. These consistency conditions stem from the compatibility of the OPE with the KMS condition, which can be thought of as the thermal analog of the crossing equation in S-matrix theory or the associativity of the OPE in the conformal bootstrap.

The kinematic structure of thermal correlators was analyzed by studying the consequences of compactifying Euclidean time, identifying the broken and unbroken symmetries of the thermal manifold $S_\beta^1 \times \mathbb{R}^{d-1}$, and writing down the corresponding Ward identities. We also reviewed the analytic structure of thermal two-point functions, which provides a natural setting for the application of inversion formulas and dispersion relations.

We then presented a comprehensive analysis of thermal dynamics, including sum rules, consistency conditions, asymptotic behavior, and numerical implementations. Our results reproduce known features of free theories and two-dimensional models, and predict novel, nontrivial thermal one-point functions in strongly coupled models in $d > 2$. The framework was further extended to include conformal line defects wrapping the thermal circle, analogous to Polyakov loops in gauge theories. This setup is entirely new from the bootstrap perspective: we studied its kinematics and proposed a bootstrap formulation adapted to it.

In particular, we successfully applied our methods to the 3D Ising model, and extended them to study the XY and Heisenberg models in three dimensions. Results are provided in Figs.~\ref{Fig:IsingResults},~\ref{fig:OPE_Coefficients} and Tables~\ref{tab:3dIsingResults},~\ref{tab:3do2o3Results}. We also applied analytical bootstrap techniques to compute thermal two-point functions in the $\mathrm{O}(N)$ models in the $\varepsilon$-expansion, finding agreement with direct diagrammatic computations. Finally, we studied the magnetic line defect in the $\mathrm{O}(N)$ models using both the $\varepsilon$-expansion and large-$N$ methods.

Beyond the specific results, this thesis emphasizes a general strategy: that of adapting bootstrap techniques to the thermal setting. This opens up a new avenue for exploring quantum field theories at finite temperature using controlled, non-perturbative tools, and highlights the deep analogy between crossing symmetry and thermal periodicity. Furthermore, some of the techniques developed here—such as the hybrid numerical/analytical methods—may be of broader interest, with potential applications to the conformal bootstrap, (conformal) defects, deformed CFTs, moduli space dynamics, and beyond.

\section{Outlook}

The bootstrap approaches proposed in this thesis have already been applied to non-trivial and physically relevant theories, such as the $\mathrm{O}(N)$ models in three dimensions. Nevertheless, many directions remain open and offer exciting possibilities for future exploration using the tools developed here.

\paragraph{Spinning operators}
In this thesis, we focused on the bootstrap of two-point functions involving identical scalar operators. A natural generalization is to consider correlators involving spinning operators or non-identical scalars. In particular, one would like to study the two-point function of currents and the stress-energy tensor and explore integrals over the latter correlator. In the zero-temperature case, such integrals along null directions lead to energy correlators—quantities that would be extremely interesting to investigate at finite temperature.

\paragraph{Breaking conformality}
Another important direction is to study deformations away from the conformal point (as illustrated by the red dashed line in Fig.~\ref{fig:SchemeIntro}) in order to access thermal effects in more general quantum field theories. From an experimental perspective, it is often desirable to compute correlation functions and thermodynamic observables at arbitrary values of the coupling and temperature. Once the thermal CFT is under good theoretical control, it becomes natural to consider perturbations around it to explore the neighboring regions in theory space, particularly at finite temperature~\cite{future1}.

\paragraph{Accuracy and comparison with Monte Carlo and experiments}
This thesis presents new predictions for thermal one-point functions and the free energy in the $\mathrm{O}(N)$ models. A key objective for future work is to improve the precision of these estimates. As demonstrated in Chapter~\ref{chap:ONmodel}, a promising direction involves combining analytical and numerical techniques to enhance the accuracy of the results~\cite{NewAnalytic}. Moreover, the observables computed in this work are directly testable using Monte Carlo simulations and, in some cases, through experimental measurements—making this an especially compelling avenue for cross-validation and potential phenomenological impact.

\paragraph{Keldysh-Schwinger formalism and real-time evolution}
The bootstrap problem studied in this thesis is formulated within the Matsubara formalism, i.e.\ in terms of correlation functions defined on the Euclidean manifold with compact imaginary time, $S_\beta^1$. However, for many physical applications one is ultimately interested in the real-time evolution of finite-temperature systems. This can be achieved by extending the time coordinate to the complex plane: the imaginary part encodes thermal effects, as in the Matsubara approach, while the real part describes the physical time evolution of correlators. A natural target for future bootstrap developments is the computation of real-time observables, such as retarded, advanced, and Wightman correlators, at finite temperature. In the long term, this could allow for the bootstrap-based study of thermalization in isolated quantum systems.

\paragraph{Hydrodynamics and moments}
Closely related to the Keldysh-Schwinger formalism is the study of hydrodynamic behavior and transport in quantum systems. It would be extremely valuable to derive bootstrap bounds or even direct predictions for hydrodynamic observables, which are relevant across a broad range of contexts, from condensed matter and statistical physics to the physics of the quark-gluon plasma and holography. One possible route involves the analysis of integrated correlation functions—commonly referred to as \textit{moments} (for possible precise definitions see e.g. \cite{Dodelson:2023vrw,Dodelson:2024atp} —which, if suitably defined, may satisfy positivity constraints and could therefore be incorporated into bootstrap equations.

\paragraph{Eigenstate thermalization hypothesis}
Thermal correlation functions in equilibrium quantum systems may also be understood via the Eigenstate Thermalization Hypothesis (ETH)\cite{Deutsch:1991,Srednicki:1994mfb}. In conformal field theory, ETH suggests that thermal two-point functions can be approximated by heavy-heavy-light-light correlators of the form
\[
\langle \mathcal{O} \mathcal{O} \rangle_\beta \sim \lim_{\Delta \to \infty} \langle \mathcal{O}_\Delta| \, \mathcal{O} \, \mathcal{O} \,| \mathcal{O}_\Delta \rangle.
\]
If this holds, one could in principle incorporate ETH as an additional constraint in a bootstrap setup, thereby relating finite-temperature correlators to the heavy limit of four-point functions. This would provide a new perspective on the thermal bootstrap and connect it more directly to the standard conformal bootstrap framework.

\paragraph{AdS/CFT and black hole physics}
As emphasized in the introduction, the study of thermal conformal field theories is closely related to the physics of black holes in Anti-de Sitter space. In particular, strongly coupled quantum systems at finite temperature are holographically dual to black holes (or black branes) in AdS. Since the CFT is strongly coupled, the bootstrap approach developed in this thesis may offer a viable non-perturbative tool for studying holographic correlators. Arguably, the simplest and most well-studied example is $\mathcal{N}=4$ Super Yang-Mills theory, which is dual to gravity in AdS$_5$~\cite{NewAnalytic}. A natural long-term goal would be to compute the quasi-normal mode spectrum of the black hole from the thermal CFT data alone. Furthermore, more general black hole geometries—including those corresponding to finite-temperature CFTs on compact manifolds—can be probed by placing the boundary theory at both finite temperature and finite volume. It would be of great interest to generalize the bootstrap approach developed in this thesis to such setups.

\paragraph{Different manifolds and modularity}
As just mentioned, placing a CFT on $S^1_\beta \times S^{d-1}$ is especially relevant in holographic contexts. More broadly, however, it would be very interesting to extend the thermal bootstrap approach to CFTs defined on more general Euclidean manifolds. Among these, toroidal geometries stand out as natural settings for studying the implications of modularity in higher dimensions. It is expected that consistency conditions arising from modular invariance or background diffeomorphism invariance on such manifolds can lead to non-trivial constraints on the CFT data \cite{Cardy:1991kr,Allameh:2024qqp}, in analogy with the success of the modular bootstrap in two dimensions \cite{Cardy:1986ie}. Extending the current methods in this direction could thus reveal deep structural information about CFTs beyond flat space or thermal cylinders.

\bibliography{TidyBib.bib}
\bibliographystyle{utphys}

\end{document}